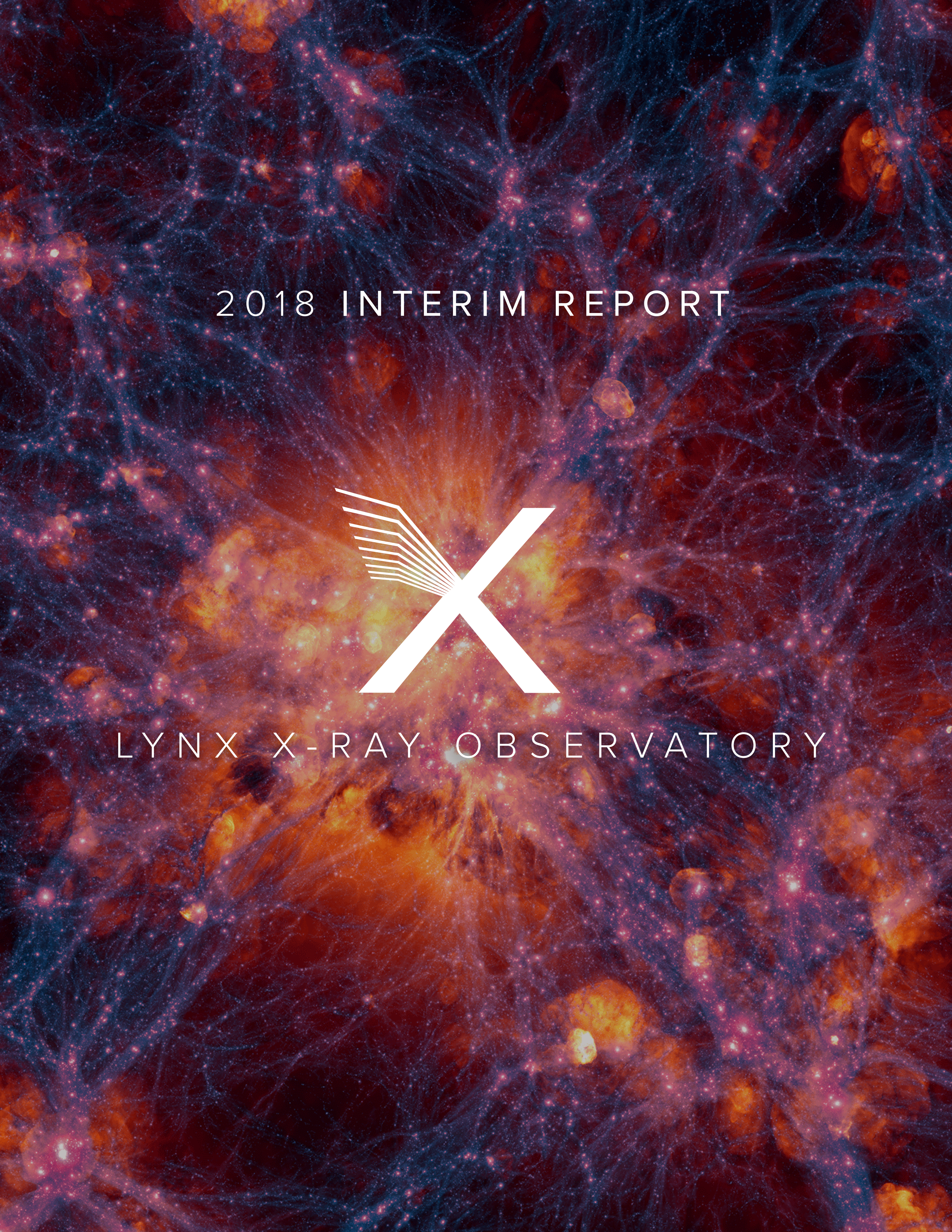

# 2018 INTERIM REPORT

# LYNX X-RAY OBSERVATORY

X-RAY OBSERVATORY

# LYNX

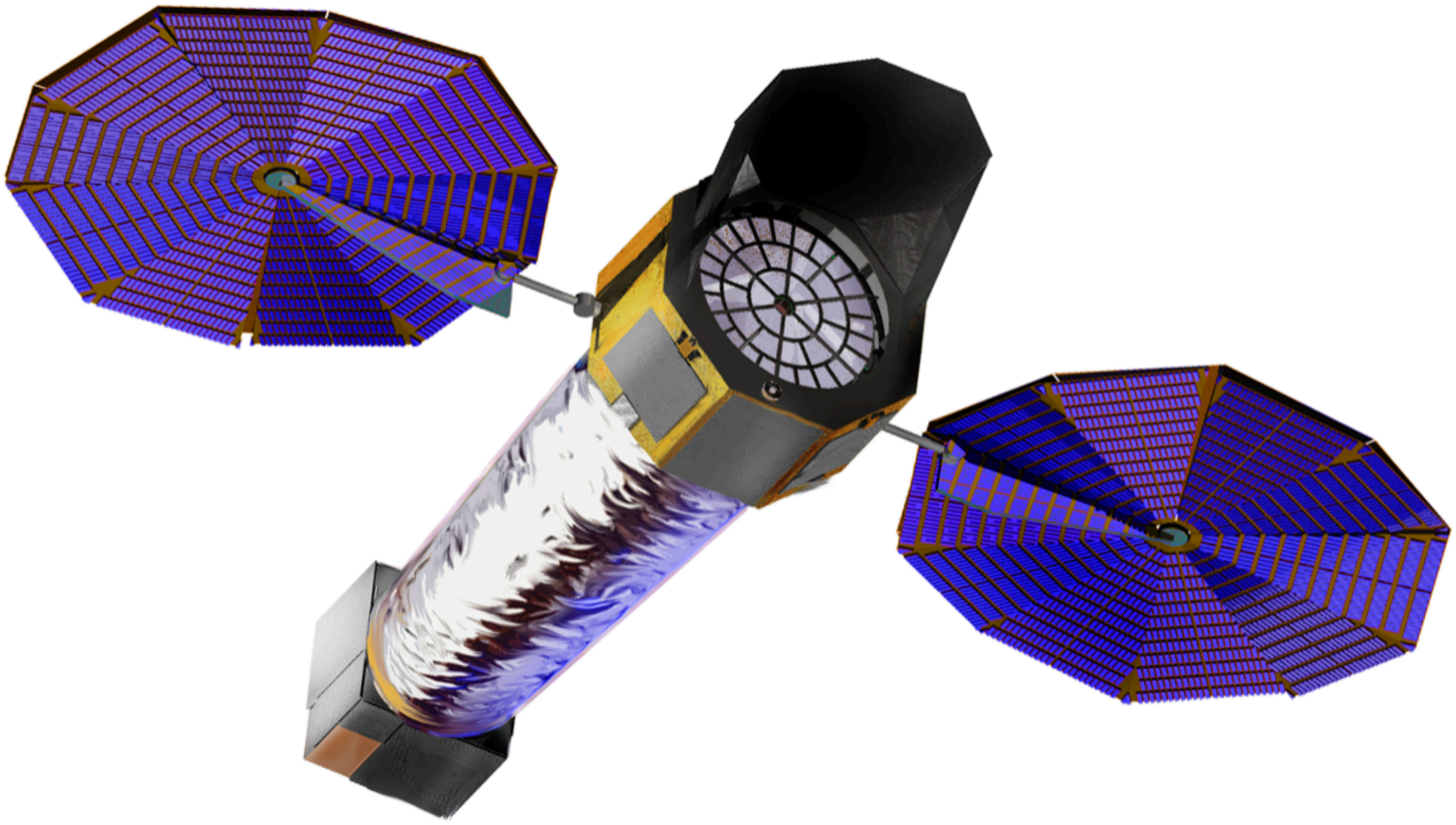

2020 Astrophysics
Decadal Mission Concept Studies

**_Lynx_ Interim Report**

August 2018

Science and Technology Definition Team (STDT) Chairs:

**Dr. Feryal Özel**
University of Arizona

**Dr. Alexey Vikhlinin**
Smithsonian Astrophysical Observatory

Submitted on behalf of the *Lynx* Team

# B. EXECUTIVE SUMMARY

*Lynx* is the X-ray observatory with radical leaps in capability that will **uncover the otherwise invisible Universe**

- to see **the dawn of black holes**
- reveal **what drives galaxy formation and evolution**
- unveil **the energetic side of stellar evolution and stellar ecosystems**

One of the exciting challenges of human discovery is understanding the complexities of the Universe in which we live. NASA Astrophysics is driven by three defining questions: How did we get here? How does the universe work? Are we alone? In the 2020's and beyond, a multi-wavelength and multi-messenger approach will be required to address these questions. X-ray observations are essential to this quest. The great power of X-ray measurements stems from the well-established fact that much of the baryonic matter, as well as the settings for some of the most active energy releases in the Universe, are primarily visible in the X-ray band. X-ray observations with a next-generation observatory are mandatory to gain a true understanding of the origins and underlying physics of the cosmos. The following on-going or near-future developments make such an observatory especially timely:

- Very massive, $\sim 10^9\ M_\odot$, black holes are being discovered at ever higher redshifts, currently reaching $z \approx 7.5$. The birth and early evolution of such supermassive black holes are a remarkable, yet poorly understood phenomenon, with strong impacts on the evolution of the first galaxies. The *James Webb Space Telescope (JWST)* will soon dramatically uncover the process of galaxy assembly and star formation in the early universe. After *JWST*, the next logical step in studies of the Cosmic Dawn will be to observe the nearly coeval Dawn of Black Holes.

- The stellar content in low-redshift galaxies is now extremely well characterized via large-scale projects such as the Sloan Digital Sky Survey *(SDSS)* and the crowd-sourced Galaxy Zoo. In a few years, the *Wide-Field Infrared Survey Telescope* (*WFIRST)* will provide sensitive near-infrared data for galaxy samples extending to higher redshifts. Cosmological numerical simulations are becoming ever more capable, making great strides toward reproducing realistic galaxies from the cosmological initial conditions, if the necessary ingredient of very powerful energy feedback is included. Direct detailed observation of the on-going feedback and its effect on gas in the galactic halos is the ingredient that is missing for completing the picture of galaxy formation and evolution.

- The emergence of the era of the multi-messenger astronomy following Laser Interferometer Gravitational-Wave Observatory (LIGO) detections of gravitational wave events from neutron stars and black hole mergers renews the emphasis on the endpoints of stellar evolution, particularly neutron stars and black holes in binary systems. Characterizing the binaries that lead to mergers, following up these events, and using all available tools to study compact object properties are of crucial importance in this era.

- Discovery and characterization of exoplanets is rapidly becoming a mature field. The emphasis will increasingly shift towards statistical characterization of the planet populations and the assessment of habitable conditions. The activity of the host star can significantly deplete planetary atmospheres, and at the same time may be required for primitive biochemistry. Studying the effects of stellar activity on habitability is especially important for planets around dwarf stars, the very population whose atmospheres will be accessible for studies in the 2020s with large, ground-based optical telescopes and *JWST*.

With these considerations in mind, we have developed *Lynx,* an extraordinarily powerful, next-generation X-ray observatory. *Lynx* is designed to provide **unprecedented X-ray vision into the otherwise "Invisible"**

**Universe** with unique power to directly observe the dawn of supermassive black holes, reveal the drivers of galaxy formation, trace stellar activity including effects on planet habitability, and transform our knowledge of endpoints of stellar evolution.

**The Dawn of Black Holes:** Massive black holes start to form as early as their host galaxies. *Lynx* will find the first supermassive black holes in the first galaxies detected by *JWST*, trace their growth from the seed phase, and shed light on how they subsequently co-evolve with the host galaxies. These young black holes are best observed in the X-ray band. Reaching into the seed regime in the early Universe requires X-ray sensitivities of $\sim 10^{-19}$ erg $s^{-1}$ $cm^{-2}$, which only *Lynx* can achieve.

**The Invisible Drivers of Galaxy Formation and Evolution**: The assembly, growth, and state of visible matter in cosmic structures are largely driven by violent processes that produce and disperse large amounts of energy and metals into the surrounding medium. In galaxies at least as massive as the Milky Way, the relevant baryonic component is heated and ionized to X-ray temperatures. Only *Lynx* will be capable of mapping this hot gas around galaxies and in the Cosmic Web, as well as characterizing in detail all significant modes of energy feedback. Essential observations will require high-resolution spectroscopy ($R \sim$5,000) of background active galactic nuclei (AGNs), the ability to image low surface brightness continuum emission, and $R \sim$2,000 spectroscopy of extended sources on arcsecond scales—all unique to *Lynx*.

**The Energetic Side of Stellar Evolution and Stellar Ecosystems**: *Lynx* will probe, with unprecedented depth, a wide range of high-energy processes that provide a unique perspective on stellar birth and death, internal stellar structure, star-planet interactions, the origin of elements, and violent cosmic events. *Lynx* will detect X-ray emission as markers of young stars in active star forming regions, study stellar coronae in detail, and provide essential insight into the impact of stellar X-ray and extreme ultraviolet flux and winds on the habitability of their planets. Images and spectra of supernova remnants in Local Group galaxies will extend studies of stellar explosions and their aftermath to different metallicity environments. *Lynx* will expand our knowledge of collapsed stars through sensitive studies of X-ray binaries in galaxies as distant as 10 Mpc and through detailed follow-ups of gravitational wave events. *Lynx* will greatly extend our X-ray grasp throughout the Milky Way and nearby galaxies by combining, for the first time, the required sensitivity, spectral resolution, and sharp vision to see in crowded fields.

We envision *Lynx* operating as a general observatory. Execution of the science in the three "pillars" outlined above will require many observing programs of a great variety. We expect that most, if not all, of those will be carried out as competed and peer-reviewed General Observer (GO) programs, with little or no time reserved for pre-selected key programs. Furthermore, our vision is that any data from directly allocated key projects, such as the deepest surveys, will be made available to all observers on a non-proprietary basis. This model has been extremely successful for great observatories such as *Hubble*, *Chandra*, and *Spitzer,* enabling the achievement of key objectives and tremendous science and discoveries beyond the initial motivations for the missions. While the pillars program we outline in this report is already very broad, the observatory is designed such that there will be substantial capabilities and observing time to execute a wide range of other GO programs, even those not anticipated today. Virtually all astronomers will be able to use *Lynx* for their own particular science."

Across the board, *Lynx* will be capable of making generational advances enabled by leaps in capability over NASA's existing flagship *Chandra* and the European Space Agency's planned *Athena* mission: 100-fold increase in sensitivity via coupling superb angular resolution with high throughput; 16 times larger field of view (FOV) for sub-arcsecond imaging; and 10–20 times higher spectral resolution for both point-like and extended sources. The keys for achieving these spectacular gains derive from the science payload:

**Mirror—**The primary science driver for the *Lynx* angular resolution is detecting early supermassive black holes at the seed stage or soon after. *Lynx* targets black holes with $M_{BH} \approx 10{,}000\ M_{\odot}$ accreting at the Eddington limit at redshift $z = 10$, which translates to an X-ray flux of $\approx 10^{-19}$ erg $s^{-1}$ $cm^{-2}$. To avoid source confusion at these fluxes, and to uniquely associate detected X-ray sources with *JWST* and *WFIRST* galaxies, requires on-axis angular resolution of 0.5 arcsecond (half-power diameter; HPD hereafter), and better than 1 arcsecond (HPD) across the FOV used for sensitive surveys.

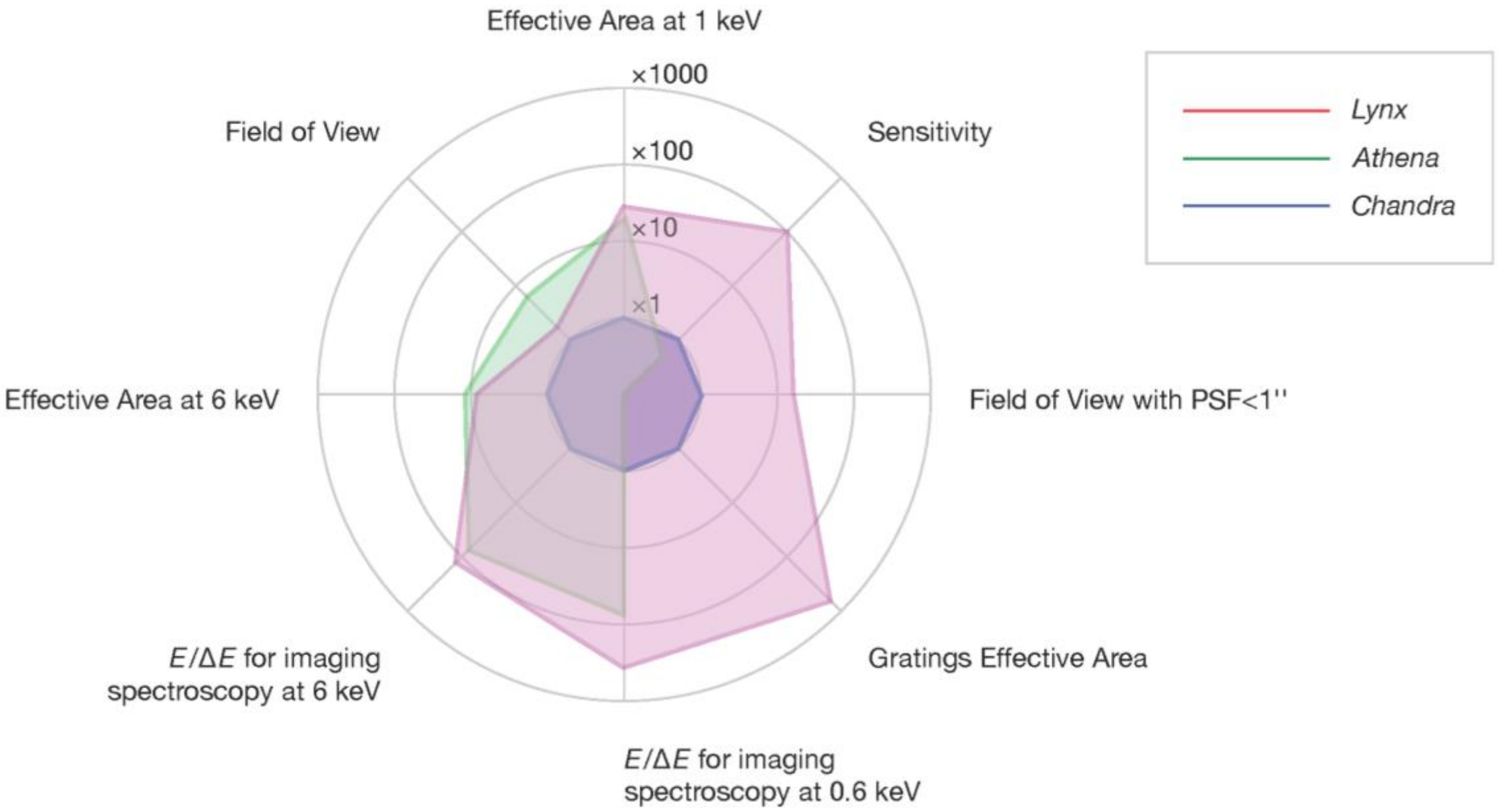

**Figure B- 1. The improvement of different capabilities from current (*Chandra*) and planned (*Athena*) X-ray missions to *Lynx*. The substantial leap in limiting sensitivity is achieved by the angular resolution of the mirrors in addition to their effective area, while the high-defintion imager, microcalorimeter and grating designs all contribute to the advances in high-resolution imaging and spectroscopic capabilities.**

*Lynx* deep surveys and many other programs need a large FOV with sub-arcsecond imaging. Here, the *Lynx* requirement is better than 1 arcsecond (HPD) point spread function (PSF) maintained to an off-axis radius of at least 10 arcminutes. This large gain over *Chandra*'s sub-arcsecond angular resolution, which is only out to 2.5 arcminutes, will be achieved via a combination of shorter mirrors, an optimized optical design, and a detector closely matching the optimal focal surface.

The mirror effective area, 2 $m^2$ at $E = 1$ keV, is sized such that the core program required for the three main pillars can be executed (mainly via GO proposals) in ≈50% of the observing time in a nominal five-year mission lifetime. Observing time beyond that envisioned for the core program would be allocated to peer-reviewed, competed programs across astrophysics, responsive to future discoveries and opportunities not even identified at present.

The minimum telescope capable of providing 2 $m^2$ of the soft X-ray band effective area has an outer diameter of 3 m and a focal length of 10 m. The *Lynx* focal length is identical to *Chandra*, while the mirror diameter is a factor of 2.5 larger. A 30-fold increase in the mirror effective area compared to *Chandra* will be achieved by using more, thinner, nested X-ray mirrors enabled by innovative technologies along with a larger input aperture.

In addition to a leap in the mirror effective area, *Lynx* is designed with a more capable suite of science instruments compared to *Chandra*.

**High-definition X-ray Imager—**Designed for high-resolution imaging and wide surveys, the High-definition X-ray Imager (HDXI) is comprised of an array of active pixel silicon sensors with 0.3-arcsecond pixels covering a FOV of 22 × 22 arcminutes, while closely following the optimal focal surface. The instrument provides moderate spectral resolution (~100 eV) over the 0.3–10 keV band and high frame rates to minimize pile-up and optical and ultraviolet loads, enabling thinner optical blocking filters for increased efficiency at lower energies, as well as fast timing measurements. The HDXI will be an instrument of choice for projects such as surveys, mapping diffuse baryons in the intergalactic and circumgalactic medium, studies of young star forming regions in the Milky Way, and responding to *Laser Interferometer Space Antenna (LISA)* triggers of supermassive black hole (SMBH) mergers, to name a few.

***Lynx* X-ray microcalorimeter—**The *Lynx* X-ray microcalorimeter (LXM) will provide non-dispersive spectroscopy with <3 eV energy resolution over the 0.2–7 keV band, and imaging with 1-arcsecond pixels over a 5 × 5 arcminute FOV. Major components of the *Lynx* pillar science, especially energy feedback studies, will require some additional capabilities that will be achieved by introducing two additional subarrays to the LXM. These arrays will share the focal plane and readout technology with the primary 5 × 5 arcminutes array. The required optimizations will be achieved via modifications of the pixel geometry and readout modes.

Located at the center of the main array, the enhanced imaging subarray will span the 0.2–7 keV band with 0.5-arcsecond pixels over a 1 × 1 arcminute FOV, providing an energy resolution of 2 eV. It will provide finer spectro-imaging required, e.g., to characterize the effects of AGN feedback on the interstellar medium of host galaxies and to measure the physical state of gas near the SMBH sphere of influence. The ultra-high spectral resolution subarray will cover the 0.2–0.75 keV band with 1-arcsecond pixels over a 1 × 1 arcminute FOV, providing 0.3 eV energy resolution required, e.g., for studies of supernovae-driven galaxy winds and density diagnostics in AGN outflows. This subarray will be placed to the side of the main array as an additional focal plane "stop."

A combination of the main array in the LXM and the two subarrays will provide a resolving power of $R \approx 2{,}000$, both in the soft band and near the Fe-K line complex at $E$ = 6.7 keV. Together, with excellent imaging capabilities, this will make the LXM an immensely capable instrument, substantially exceeding the capabilities planned for the *Athena* microcalorimeter. The *LXM* will serve as a workhorse for many core science programs and much of the general observatory science.

**X-ray Grating Spectrometer—**An even higher spectral resolution is required for absorption-line studies of diffuse baryons in galactic halos and the cosmic web, physics of stellar coronae, and assessing the impact of stellar activity on habitability of their planets. This capability will be provided by the X-ray Grating Spectrometer (XGS). The XGS grating arrays can be inserted into and removed from the optical path. These arrays will cover ~2/3 of the input aperture to provide ~4,000 $cm^2$ of effective area at the astrophysically important X-ray lines in the 0.2–2 keV band. The dispersed spectrum will be read out with a dedicated array of Si-based sensors. The resolving power will be fairly uniform across the band, at $R \approx$ 5,000 and possibly higher, at $R \approx$ 10,000.

**Mission approach—***Lynx* will operate as an imaging X-ray observatory—with a grazing incidence telescope and detectors counting individual X-ray photons. Post-facto aspect reconstruction leads to modest pointing precision and stability requirements, while enabling very accurate sky locations for X-ray events, telemetered to the ground along with their energy and timing information. This approach allows us to draw from decades of experience, particularly from efficient, flight-proven approaches, design choices, and mission operations software and procedures developed for *Chandra*. The *Lynx* Design Reference Mission has been designed to meet the science objectives of the future while capitalizing on this heritage where appropriate.

The *Lynx* spacecraft is built around the X-ray mirror assembly. The science instrument module is attached to the spacecraft by an optical bench. The LXM and HDXI detectors are placed on the movable platform, while the XGS readout array is at a fixed location on the instrument module. The *Lynx* subsystems for thermal control, guidance, navigation and control, and aspect share notable similarities with *Chandra*. Substantial updates will be needed for the power subsystem, and for communications due in part to a much larger, but still readily manageable, data volume.

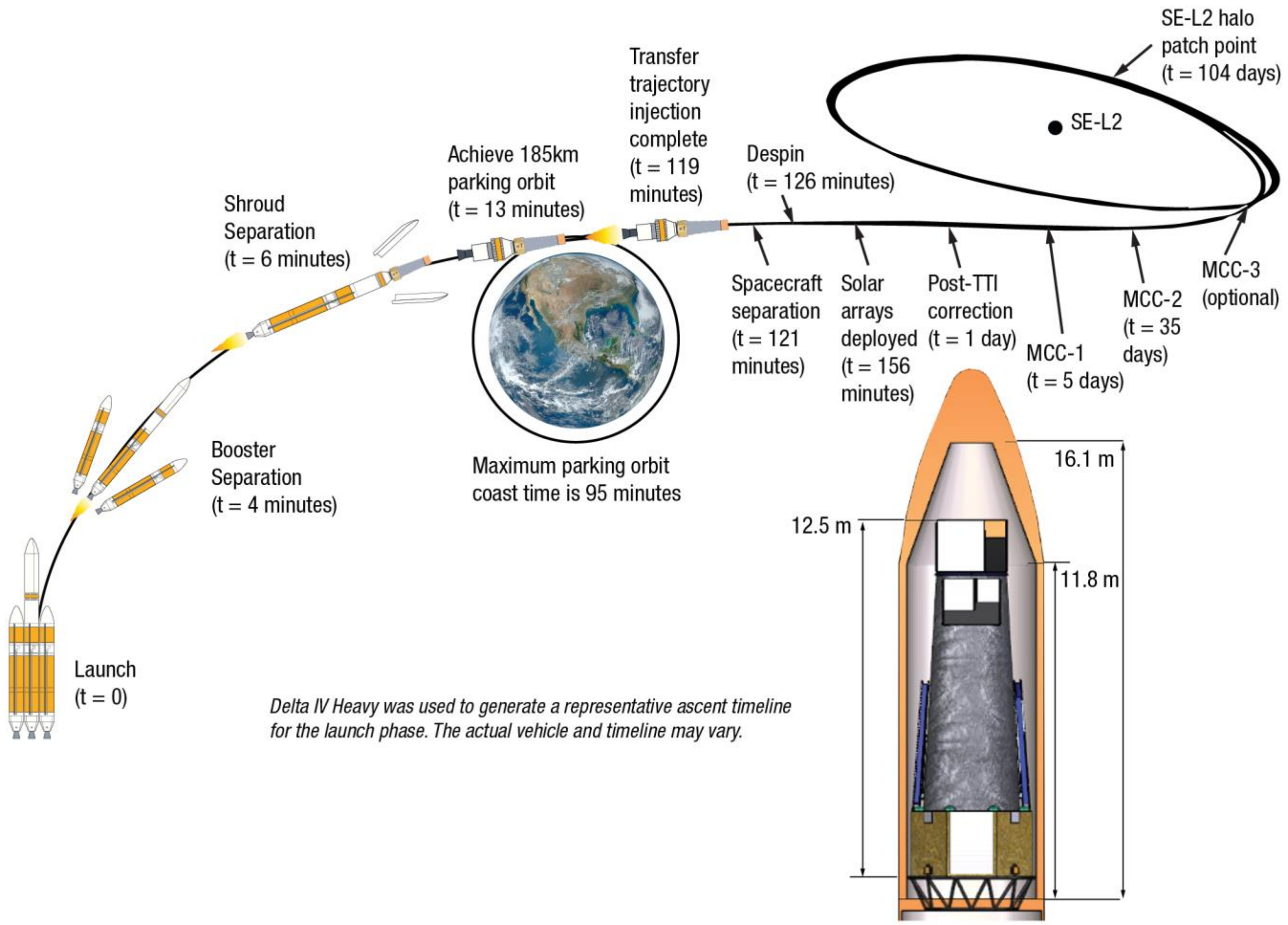

**Figure B-2. Launch to orbit timeline and *Lynx* X-ray Observatory inside of a heavy-class vehicle payload envelope. Guidance provided by NASA LSP.**

*Lynx* is designed for flexible, queued observing scheduling with the only restriction on the field of regard being the 45° Sun avoidance angle and ±5° anti-Sun angle. A review of *Chandra* mission operations shows that *Lynx* will be able to implement a similarly efficient (>85%) observing schedule. The Sun-Earth L2 orbit has been selected for *Lynx,* taking into account the efficiency of science operations, mission duration, radiation and thermal environments, communications, and launch. *Lynx* consumables (primarily propellant required for station-keeping at the L2 orbit) will have margins sufficient for a 20-year duration of extended mission without substantially impacting the overall mass budget.

This report discusses the *Lynx* science pillars and the capabilities required to accomplish that science, with the intention of executing most or all of these pillar programs through a competed, peer-reviewed General Observer process. The overall GO program will be very extensive, and our Final Report will illustrate the range of potential GO science with Lynx. The observatory is described along with the technology status of the optics and the science instruments. Draft technology roadmaps lay out the plans for advancing from current Technology Readiness Levels (TRL) to TRL 5 at the start of Phase A and TRL 6 at the start of Preliminary Design Review. These roadmaps, in addition to the extended heritage from the *Chandra* mission, provide great confidence that this "flagship" mission can be built and launched within the budget projected for the decade starting in the mid 2020s.

# C. TABLE OF CONTENTS

# D. *LYNX* SCIENCE PILLARS

## D.1 Overview and Motivation

In the 2020s and beyond, a multi-wavelength and multi-messenger approach will be required to address the three defining questions which drive NASA Astrophysics: How did we get here? How does the universe work? Are we alone? As we will argue in this report, X-ray observations are essential to this quest. The great power of X-ray measurements stems from the well-established fact that much of the baryonic matter, as well as the settings for some of the most active energy releases in the Universe, are primarily, and sometimes uniquely, visible in the X-ray band. X-ray observations with a next-generation observatory are crucial for gaining a true understanding of the origins and underlying physics of the cosmos.

As we survey the exciting developments of the past decade to formulate the open questions ahead of us, a few key developments jump out across all of astrophysics. We note a few here that ultimately motivate the development of a high sensitivity X-ray observatory.

- Very massive, $\sim 10^9\ M_{\odot}$, black holes are being discovered at ever higher redshifts, currently reaching $z \approx 7.5$. The birth and early evolution of such supermassive black holes (SMBH) are a remarkable, yet poorly understood phenomenon, with strong impacts on the evolution of the first galaxies. The *James Webb Space Telescope* (*JWST)* will soon dramatically uncover the process of galaxy assembly and star formation in the early universe. After *JWST*, the next logical step in studies of the Cosmic Dawn will be to observe the nearly coeval Dawn of Black Holes.
- The stellar content in the low-redshift galaxies is now extremely well characterized via large-scale projects such as the *Sloan Digital Sky Survey (SDSS)* and the crowd-sourced Galaxy Zoo. In a few years, the *Wide-Field Infrared Survey Telescope* (*WFIRST)* will provide sensitive near-infrared data for galaxy samples extending to higher redshifts. Cosmological numerical simulations are becoming ever more capable, making great strides toward reproducing realistic galaxies from the cosmological initial conditions, but have to make dramatic, unconstrained, and wildly varying assumptions about powerful energy feedback. Direct detailed observation of the on-going feedback and its effect on gas in the galactic halos is the ingredient that is missing for completing the picture of galaxy formation and evolution.
- The emergence of the era of the multi-messenger astronomy following *Laser Interferometer Gravitational-Wave Observatory (LIGO)* detections of gravitational wave events from neutron stars and black hole mergers renews the emphasis on the endpoints of stellar evolution, particularly neutron stars and black holes in binary systems. Characterizing the binaries that lead to mergers, following up these events, and using all available tools to study compact object properties are of crucial importance in this era.
- The discovery and characterization of exoplanets is rapidly becoming a mature field. The emphasis will increasingly shift towards statistical characterization of the planet populations and the assessment of habitable conditions. The activity of the host star can significantly deplete planetary atmospheres, and at the same time may be required for primitive biochemistry. Studying the effects of stellar activity on habitability are especially important for planets around dwarf stars, the very population whose atmospheres will be accessible for studies in the 2020's with large, ground-based optical telescopes and *JWST*.

These developments simultaneously expand the frontier of our knowledge of the Universe while highlighting the open questions that we still cannot answer. They serve as a guide, along with the 2010 astrophysics decadal survey and the NASA Astrophysics 30-year roadmap, for formulating the critical science questions across astrophysics to be addressed in the following decades. These key questions include (but are not limited to): How and when do the first black holes in the Universe light up? How do they grow and how do they affect their surroundings? How do feedback processes shape galaxies? What are the properties of the gas that reside outside of galaxies? How do black holes and neutron stars form? What are the evolutionary paths that lead to *LIGO* sources? What is the activity of young stars in active star forming regions and what is the impact of stellar X-ray and extreme ultraviolet (XUV) flux and winds on the habitability of their planets?

The best, and sometimes the only way, to address these questions is via observations with a sensitive, high angular resolution X-ray telescope. We will also show how an observatory like *Lynx* will maximize the scientific return from *JWST* and *WFIRST* to get a complete picture of galaxies across cosmic time, the formation of black holes at high redshifts, and the studies of stellar ecosystems in our present-day Universe.

This section lays out our core science goals and map them to key observations that are needed to address them. We will then trace them to specific observatory requirements and conclude the science section with the Science Traceability Matrix as well as the Notional Core Program that outlines the key observations.

Section E presents the Design Reference Mission that incorporates the elements that are needed to perform the range of key observations: a sub-arcsecond angular resolution, large field of view (FOV), large effective area X-ray mirror, a High-definition X-ray Imager (HDXI) that is designed for high-resolution imaging and wide surveys, a microcalorimeter that enables non-dispersive spectroscopy with <3 eV energy resolution over the 0.2–7 keV band, and an X-ray Grating Spectrometer (XGS) that will provide > R = 5,000 resolving power uniformly across the soft X-ray band. We will devote the remainder of the report to describing the observatory in detail along with the technology status of the optics and the science instruments.

## D.2 The Dawn of Black Holes

Supermassive Black Holes (SMBHs) in the centers of galaxies are a common phenomenon in the nearby universe. As observations of accreting black holes extended into higher redshifts (Mortlock 2011; Bañados et al. 2018) and tight relations between galaxies and their central SMBHs were discovered, a standard picture of co-evolution between the SMBHs and their hosts (e.g., Kormendy and Ho 2013), in which black-hole and galaxy growth profoundly influence one another, began to emerge. Furthermore, the discovery of quasars at z ~ 6–7 indicated that the nascent SMBHs must have come into existence extremely early on and have grown rapidly in order to acquire $10^9$ M$_\odot$ within a span of ~1 Gyr (Fan et al. 2001; Wang et al. 2015). Despite their prevalence and importance, however, the mechanisms for the appearance of the first black holes in the universe, their masses at birth, and their early growth processes remain largely unanswered by current astronomical data.

One can estimate a minimum "seed" mass by considering growth at the Eddington rate, which is a natural time-averaged upper limit to the growth rate of a black hole. The existence of $10^9$ M$_\odot$ black holes when the universe was less than 1 billion years old implies that seeds of 100–$10^5$ M$_\odot$ must have already formed by z=10. This constraint poses significant theoretical challenges to the formation models and, consequently, there exist several different channels with different predictions for the appearance of the first BHs in the universe. While Pop III, first-generation, metal-free stars were an early favorite, further studies showed that fragmentation during the star formation stage is efficient and is more likely to lead to stars with M<100 M$_\odot$, thus crimping the masses of black holes that could form from these stars. The collapse of rapidly inflowing gas has also been proposed and can lead to larger seeds, provided that the gas does not cool and fragment. The latter condition is thought to be achievable if these halos are embedded in regions of strong UV radiation which shuts down star formation and the cooling and collapse of gas until the host halo grows relatively massive ($10^8$ M$_\odot$). A third option invokes the collapse of a nuclear compact star cluster with a mass of ~$10^5$ M$_\odot$ to form a single massive black hole. However, in order for this channel to produce a $10^5$ M$_\odot$ BH, an ultra-dense star cluster, much denser than the present-day globular clusters or central stellar bulges, is required. This, in turn, requires that fragmentation be delayed until the gas has reached very high densities (>$10^{10}$ $cm^{-3}$; see, e.g., Omukai et al. 2008) in order to form massive black hole seeds.

Several other questions are closely related to the formation mechanism of seeds and the early growth of black holes. First, what is the occupation fraction of central SMBHs as a function of redshift? Second, what is the mechanism by which gas is supplied to central black holes to fuel their growth? Third, what is the relative contribution of accretion vs. mergers to the early growth of black holes? Fourth, how early do the feedback processes from these central accreting black holes begin to shape the properties of the galaxies they reside in? Finally, what is the contribution of accretion light to the reionization of the universe at high redshifts?

As the next section will show in more detail, extending observations of black holes to very high redshifts through deep X-ray surveys and reaching sensitivities required to detect nascent SMBHs provides the best path for obtaining observational answers to these questions.

### D.2.1 Observing Early Black Holes

Addressing the questions of "What is the nature of black hole seeds?" and "How does first accretion proceed?" is best achieved through observations that are capable of detecting the earliest black holes and subsequently tracing their growth from their births to the peak of their growth around z=2. The observations required for providing breakthrough progress in the formation and early activity of black holes specifically need to reach fluxes corresponding to the accretion luminosity of seed-mass black holes close to the redshifts of formation.

Pushing to this mass limit at very high redshift provides the cleanest test of the physics of SMBH seed formation, since further growth of the SMBHs obscures differences in their origin. Furthermore, the X-ray luminosity function at low redshifts is plagued by degeneracies in black hole mass, Eddington fraction, and halo occupation fraction. Therefore, indirect methods based on the X-ray luminosity function at redshifts z ≈ 2–6, such as analyzing log $N$ - log $S$ (the strategy adopted by, e.g., *Athena*), is inadequate for uncovering the dawn of black holes. We instead focus on the direct detection of the accreting newly formed black holes and establish in this section the requirements for observations of black holes at birth.

#### D.2.1.1 Observability of SMBH seeds

We consider SMBH seeds with masses of ~$10^4$–$10^5$ $M_\odot$ at z ~7–10. These are the most massive BHs that can form in the collapse of early gas clouds and stars, or via any of the several other alternative mechanisms considered in high-redshift galaxies (see Volonteri 2010; Tanaka and Haiman 2009). Pushing to this mass limit at very high redshift will provide the cleanest test of the physics of SMBH seed formation, since further growth of the SMBHs can potentially obscure differences in their origin. As a result, a detection of these seeds while they still have low mass are more powerful than, for example, statistical studies at a much higher mass range.

We do not consider even smaller masses because at the fluxes corresponding to ~$10^4$ $M_\odot$, observations begin to suffer from the X-ray luminosity "floor" set by early X-ray binaries. This is especially true of star-forming galaxies with typical expected high-mass X-ray binary populations. The high-energy spectral cutoffs of many X-ray binaries may mitigate but not completely eliminate the effects of such contamination.

For the X-ray emission from BH seeds, we adopt a power-law spectral shape with a photon index of Γ = 1.8. Assuming Eddington-limited accretion, we derive the corresponding rest-frame unabsorbed 2–10 keV X-ray luminosity limit at z=10 to be $1.2\text{x}10^{41}$ erg/s for a black-hole mass of $10^4$ $M_\odot$, corresponding to a flux of $1.1x10^{-19}$ erg/s/cm$^2$. Assuming a bolometric correction of 0.1, appropriate for the low-mass SMBHs, this translates to a bolometric luminosity of $L_{bol} = 1.2x10^{42}$ erg/s.

When we take into account the effect of absorption by making conservative assumptions about intrinsic absorption with a column density $N_H=10^{24}$ cm$^{-2}$ and solar abundances, as well as Galactic absorption applicable for the *Chandra* Deep Field-South, the flux limit allows *Lynx* to still reach black hole masses around ~$2x10^4$ $M_{\odot}$.

*This flux target sets the sensitivity requirement for the Lynx observatory. It also underscores why the Lynx capabilities, especially angular resolution and a large effective area in the soft X-rays, are necessary for meeting this science objective (Figure E-1 in §**E**).*

### D.2.2 Key Program for Observing Early Black Holes

In order to convert the detectability of high redshift black holes into telescope requirements and to observing time for key programs, we estimate a space density for such black holes. We first do this in the most model-independent way, based on dark matter halo abundances, so that the observational requirements do not depend on specific models of seed formation.

For the hosts of the ~$10^4$ $M_{\odot}$ black holes, we assume either an "optimistic" or "pessimistic" total mass scale. An "optimistic" halo mass is $10^8$ $M_{\odot}$, corresponding to $O(10^{-3})$ of the baryons locked up in the black hole. A "pessimistic" halo mass is $2\times10^{10}$ $M_{\odot}$, based on combining the $M_*$ vs $M_{halo}$ relation (Kormendy and Ho 2013, see their Figure 25; Behroozi et al. 2016) with $M_{BH} = 10^{-3}\times M_{bulge}$ (Kormendy and Ho 2013), and assuming $M_{bulge}$ ~$M_*$. The standard halo mass function model (Jenkins et al. 2001) computed for concordance cosmological parameters gives the number density of hosts in these mass ranges $n(10.2 < \log(M/M_{\odot}) < 10.5) = 10^{-3}$ Mpc$^{-3}$ and $n(8 < \log(M/M_{\odot}) < 8.3) = 2.0$ Mpc$^{-3}$ at z=8.5. Taking the product of the black hole occupation fraction and the duty cycle to be $f_{occ}\times f_{duty} = 0.1$ for the $M_{halo}$~$2\times10^{10}$ $M_{\odot}$ hosts and $10^{-4}$ for the $10^8$ $M_{\odot}$ ones, these translate into $10^{-4}$ Mpc$^{-3}$ for the X-ray luminous BHs. In other words, to detect ~1,000 BHs if they live in a fraction $10^{-4}$ of ~$10^8$ $M_{\odot}$ halos or in a fraction 0.1 of the $10^{10}$ $M_{\odot}$ halos, *Lynx* would need to survey close to 1 deg$^2$ to a flux limit of $1.1x10^{-19}$ erg/s/cm$^2$.

These model-independent numbers agree quite well with a wide range of cosmological simulations that incorporate a range of black hole seed formation scenarios into a variety of available hydrodynamic algorithms and implementations. As one example, a space density of $10^{-4}$ Mpc$^{-3}$ for the range of black hole luminosities of interest is indicated in recent high spatial resolution cosmological hydrodynamic simulations that focus on black hole formation physics at high redshifts using the code Ramses and account for BH formation in dense and low-metallicity environments (Habouzit et al. 2017). This space density translates into a sky density of ~ 700 deg$^{-2}$ when multiplied by the volume between z = 8–9, closely matching the estimate obtained from halo occupation calculations.

This result indicates that a survey of 1 deg$^2$ down to a flux of $1.1x10^{-19}$ erg/s/cm$^2$ would allow us to have four bins each with 70–700 objects at z = 8–9, meeting the requirements for usefully constraining the X-ray Luminosity Function (XLF). To minimize and assess the effects of cosmic variance, this square degree field can be spread across 3–5 distinct fields that are widely separated on the sky (e.g., Driver and Robotham 2010; Moster et al. 2011). Each field will then be 0.20–0.33 deg$^2$.

### D.2.2.1 Why *Lynx*?

In these deep surveys, angular resolution is of paramount importance and uniquely enables the high-redshift BH science. To illustrate this point, we show in **Figure D-1,** showing a 2 arcminutes × 2 arcminutes *JWST* deep field simulated using an Illustris-TNG light cone. As discussed above, X-ray sensitivity of $10^{-19}$ erg/s/cm$^2$ is required to find and resolve the first SMBHs in the first galaxies detected by *JWST*. To avoid source confusion at such low fluxes and associate X-ray sources with unique *JWST* and *WFIRST* counterparts requires better than 1 arcsecond angular resolution. *Lynx* will detect ~350 discrete sources in a deep exposure of the 2 arcminutes × 2 arcminutes region and will be able to identify essentially all with unique optical/infrared (IR) counterparts. With its 5 arcseconds point spread function (PSF), *Athena* (right panel) almost entirely misses a large population of normal galaxies which leads to very substantial source confusion. Many *Athena* detections cannot be associated with specific *JWST* galaxies. *Athena* confusion limit is ~200× above the sensitivities needed for studies of the black hole seeds.

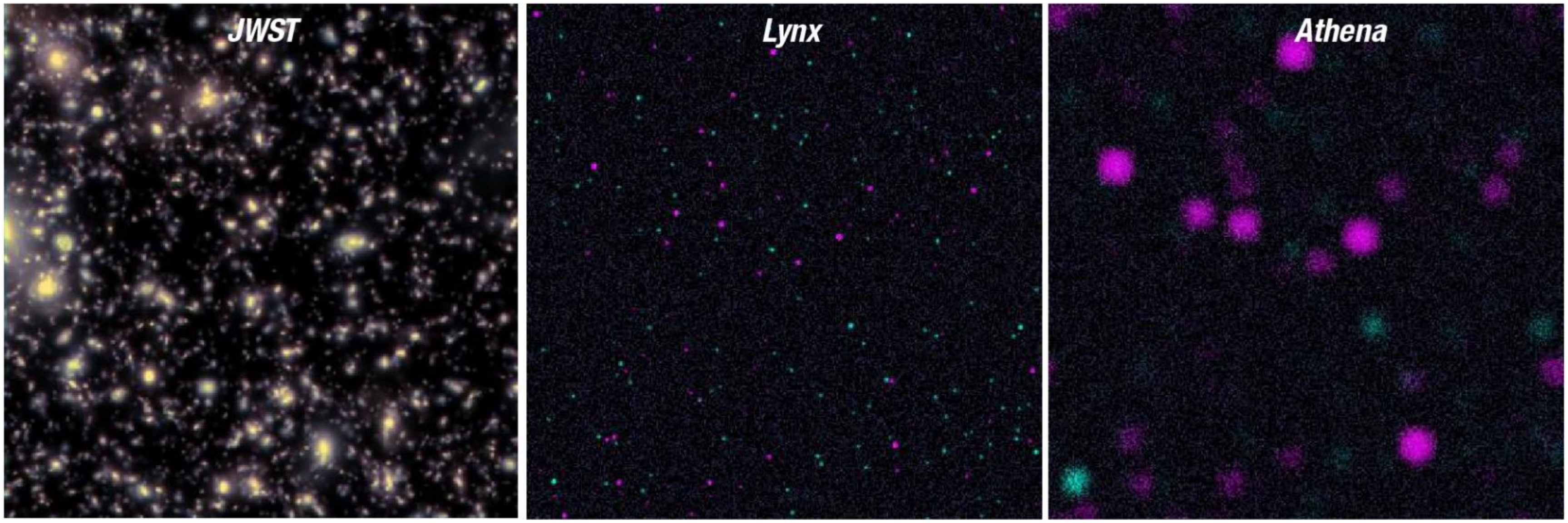

**Figure D-1. Simulated 2 arcminutes × 2 arcminutes regions from a deep *JWST* survey (using a light cone from the Illustris-TNG simulation output), 4 Msec *Lynx* survey, and 4 Msec *Athena* survey. In the X-ray images, colors code populations of AGN (purple) and moderate-redshift normal galaxies (green). *Lynx* will be able to reach ~$10^{-19}$ erg/s/cm$^2$ sensitivities needed for studies of the black hole seeds in the first galaxies detected by *JWST*. At 5 arcseconds PSF, *Athena's* confusion limit is ~200× above this threshold. This illustrates the importance of angular resolution in the *Lynx* design.**

### D.2.2.2 *Lynx* Observing Strategy and Required Sensitivity

The survey strategy that will be employed by *Lynx* makes use of the expected deep *JWST*/*WFIRST*/optical surveys in two key respects.

First, *JWST*/*WFIRST* surveys will provide the fields for the *Lynx* deep surveys for uncovering the dawn of black holes. As discussed earlier, the observations of the first black holes require the *Lynx* sensitivity to reach $1 \times 10^{-19}$ erg/s/cm$^2$ in the soft X-ray band, which can be achieved when searching for point sources at known positions. Thus, *JWST* and *WFIRST* deep surveys will provide the fields as well as reliable counterpart identifications for z=7–10 X-ray sources over the square degree survey area. These Mid-Infrared Instrument (MIRI) / NIR data must be sufficiently deep to sample galaxies with sufficiently low star formation rates such that their X-ray binary populations do not overwhelm the X-ray emission from seed black holes with masses of a few $10^8$ M⊙. If they are also accompanied by extremely deep high-resolution optical data (e.g., from the Large Synoptic Survey Telescope (LSST) Deep Drilling Fields), this will enable reliable selection of very high-redshift galaxies via photometric techniques. With these in mind, the deep *Lynx* surveys plan to make use of *JWST* fields such as those discussed in Mason et al. (2015) that have the following properties: a deep survey that reaches a magnitude of m = 32.0 over 40 arcminutes$^2$, a medium survey that reaches m = 30.6 over 400 arcminutes$^2$, and a shallow survey down to m = 29.3 covering 4000 arcminutes$^2$.

Second, the combination of *Lynx* data with the host properties obtained from the *JWST*/*WFIRST* imaging will provide the powerful determination of seed masses needed for the dawn of black holes science. In contrast, measuring the XLF alone yields degenerate constraints on the BH seed mass, Eddington-ratio distribution, and occupation fraction rather than uniquely discriminating between massive vs. low-mass seeds.

It is indeed possible to formulate a clear difference between "massive seeds" vs. "stellar remnant seeds" by using optical/IR imaging from *JWST*/*WFIRST* in combination with the X-ray data (e.g., Pacucci et al. 2015; Natarajan et al. 2017). In the massive-seed models, $10^{4-5}$ M⊙ black-hole seeds need to form in relatively low-mass hosts (~$10^{7-8}$ M⊙ halos) and efficiently consume a large fraction of the gas in the halo (e.g., Shang et al. 2010; Pacucci et al. 2015; Inayoshi et al. 2016). This process, however, is generally expected to occur only in a very small fraction of the $10^8$ M⊙ halos at z ~10. Realistic predictions yield a space density of $n_{BH} = 10^{-4}$ Mpc$^{-3}$ of such active BHs (but see, e.g., Agarwal et al. 2012; Habouzit et al. 2016 for a discussion of the uncertainty in that number).

As a result, a telltale feature is that the massive seeds would be strong outliers in the $M_{BH}$ /$M_*$ relation, with $M_{BH}$ very large compared to both $M_*$ and $M_{halo}$. This will remain the case until substantial further growth occurs, and the BH is incorporated into a more massive galaxy, acquiring a large (>$10^8$ M⊙) stellar component.

In contrast, in the "stellar-mass seed" scenarios, the same set of active $M_{BH} = 10^4$ M⊙ black holes are located in a larger fraction of more massive ($M_{halo}$ ~ $2 \times 10^{10}$ M⊙) halos, satisfying an $M_{BH}$ vs. $M_*$ relation similar to those observed for black holes in nearby galaxies. The discriminating feature of these models is that the black hole would be surrounded by a significant stellar cluster, with a mass of $M_*$ ~$2 \times 10^7$ M⊙. The optical/IR emission of this star cluster would be detectable by *JWST*, and likely dominate over the BH's emission in *JWST*'s NIR/MIRI bands. Thus, even though the space density of these drastically different BH formation scenarios would be similar, in the former case, we expect a much lower stellar-to-BH mass ratio, rendering the optical/IR starlight subdominant to the emission from the BH, and undetectable by *JWST*. Instead, the *JWST* NIR/MIRI bands would show the emission from the BH, with characteristic colors that are different from a stellar population (see, e.g., Figures 6 and 7 in Natarajan et al. 2017). These colors, together with the unusually large $L_X/L_{opt}$ ratio would be a clear diagnostic of the massive-seed scenario. A further diagnostic of the massive-seed scenario could be the presence of a bright nearby galaxy, a few kpc (or ~1 arcseconds at z ~10) away from the X-ray source. Such an ultra-close neighbor is a key ingredient in many of the proposed massive-seed models as a source of the ultraviolet (UV) radiation field (e.g., Dijkstra et al. 2008). Although this neighbor will likely merge with the BH host galaxy within a ~Myr, it may still be detectable as a distinct source on the sky accompanying a few percent of the X-ray sources (Visbal et al. 2014; Natarajan et al. 2017).

### D.2.3 Tracing the Growth of Black Holes

To understand the origin of SMBHs, it is essential to trace their evolution beyond the "seed" stage and up to the periods of maximal growth at z~1–2. It remains unclear whether *all* SMBHs emerge at high redshifts (z > 6), such that the high-mass (>$10^9$ M⊙) quasars we observe at these redshifts (e.g., Mortlock et al. 2011) are representative of only a subset of the full black hole mass function (BHMF) - or whether some low mass "seeds" begin significant growth phases only at lower redshifts so that there is strong evolution in the BHMF across the full redshift range from z=2 to 6. This question has implications not only in BH origins but also the role of BH feedback in galaxy evolution from "cosmic dawn" to "cosmic noon."

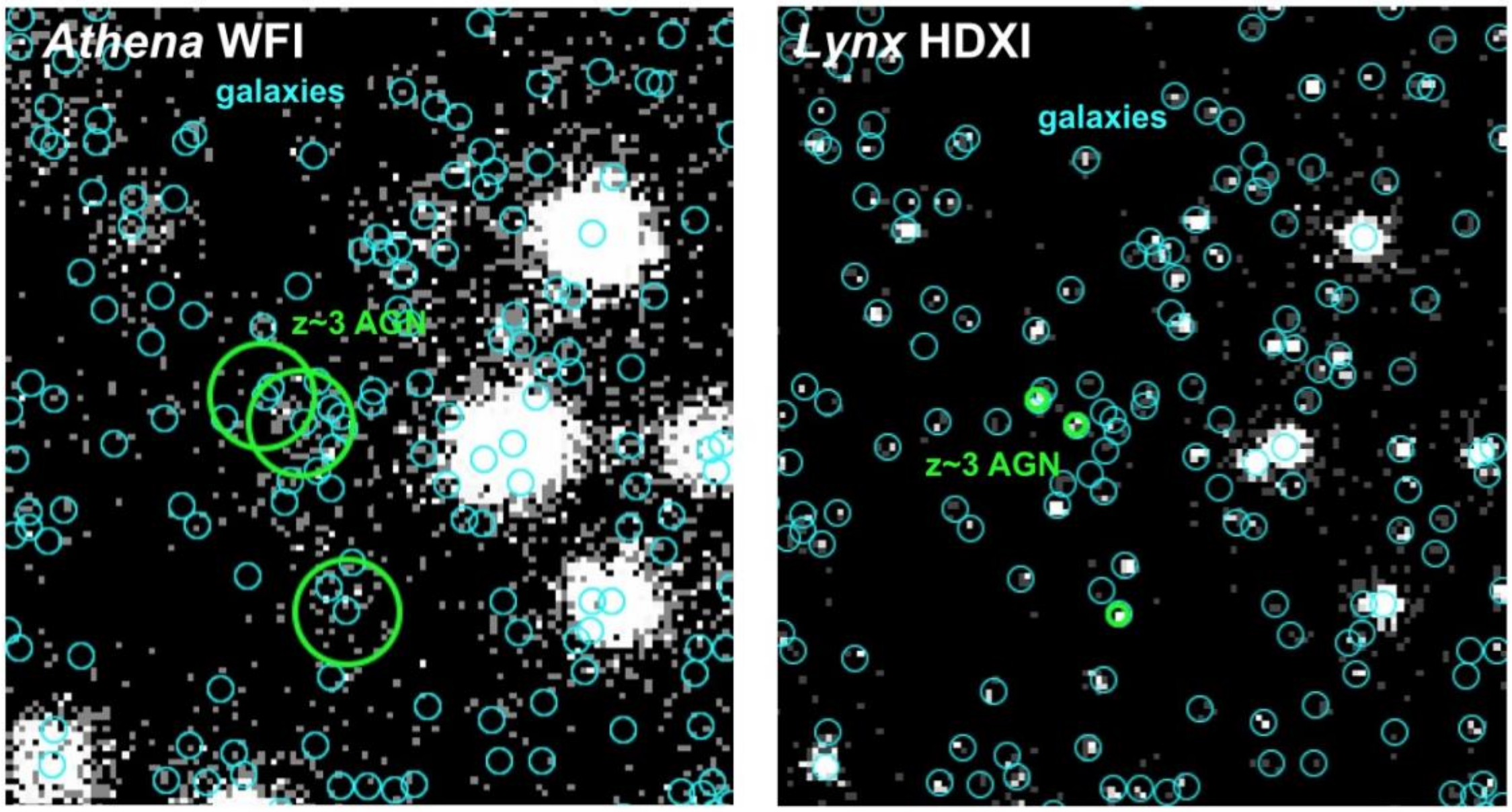

**Figure D-2. Mock observations of AGN at z=3 with at high spatial resolution with *Lynx*'s high-definition imager (HDXI) compared to the same field at a factor of ten poorer angular resolution. Individual sources are clearly detected in the former case while they suffer from severe source confusion in the latter.**

The aim of this observational program is to constrain the full XLF and Eddington ratio distributions down to low masses and luminosities across the full range in redshift from z=2–3 (redshifts below which will be well constrained by *Athena*) to z=6. Different models yield wildly divergent predictions for the BH mass function below M_BH = $10^8$ M⊙ at z > 3 (see, e.g., Kelly and Merloni 2012), so a measure of the XLF and/or comparison with galaxy properties requires us to probe a meaningful number of AGN below the "knee" in the BH mass function and the Eddington ratio distribution. As a fiducial number, we will take a mass limit of $10^7$ M⊙ and an Eddington ratio of 0.01. (Constraints below these values are currently completely unavailable with *Chandra*; see, e.g., Georgakakis et al. 2017).

Assuming an X-ray (2–10 keV) bolometric correction of ~30, which is appropriate for black holes of higher mass (Lusso et al. 2012), the luminosity threshold required is ~$3x10^{41}$ erg/s (conveniently, this is close to the limit expected for typical galaxy emission in host galaxies of this mass). For z=4, this yields a flux of $2x10^{-18}$ erg/s/$cm^2$ in the 0.4–2 keV band, requiring excellent soft response of the HDXI detector.

The X-ray luminosity function and Eddington ratio distribution are poorly constrained at these redshifts and luminosities, but a value of ~$10^{-5}$ sources/$Mpc^3$/dex is a reasonable estimate (see, e.g., Georgakakis et al. 2015). We, therefore, expect to obtain ~100 AGN per square degree in the luminosity range $10^{41-42}$ erg/s between z=3.5 and z=4.5. In order to obtain an Eddington ratio distribution, we need ~30–50 objects in this bin (i.e., 20% errors), thus requiring approximately 0.3–0.5 $deg^2$, which can be carried out within the deep survey region. As with the deep survey, the field is chosen to have deep optical/IR imaging.

A second, related issue is to obtain measurements of AGN clustering to determine how they populate dark matter halos. An effective technique for this measurement is the cross-correlation with galaxies. Simulations suggest that for galaxy densities that will be obtained with *JWST*, we require ~500 AGN to measure the cross-correlation with sufficient quality to, e.g., determine the satellite fraction to within a few percent. Given the observed luminosity function from Georgakakis et al. (2015), this will be possible over ~0.5–1.5 square degrees at z=2.5–4.5. The most effective program will be in the form of a "wedding cake," with very deep observations over 0.3–0.5 $deg^2$ (as described above) and shallower observations over a remaining ~1 $deg^2$. These fields can also be embedded in the deep survey dataset, as they do not need to be contiguous.

### D.2.4 Black Holes and Large-Scale Structure

#### D.2.4.1 Black Holes and Dark Matter Halos

Over the last few decades it has become clear that galaxy evolution is driven by both intrinsic processes, including the feedback of energy from the host galaxy's supermassive black hole, and processes relating to the host galaxy's local environment. However, there exist degeneracies between the properties of galaxies and their SMBHs such as their mass and properties of the local environment. For instance, massive galaxies are more likely to exist in dense environments. We are yet to understand in detail how environmental processes affect galaxy evolution over cosmic time, what the dominant mechanisms are for quenching galaxies and curtailing their growth and when were these processes important.

To disentangle environmental quenching and those of internal feedback mechanisms, we need to map the environments and host galaxy properties of SMBHs over cosmic time. The X-ray waveband offers a complete census of active BHs largely unbiased by host galaxy properties. *Athena* will provide a nearby view of active BHs but only the high spatial resolution of *Lynx* can access the earliest BH populations and the majority of accretion power in the Universe.

*Lynx*'s large field-of-view, grasp and superb spatial resolution allow us to statistically measure the clustering of active SMBHs from their birth through the peaks of AGN activity to the evolved halos of today. We can directly compare this AGN evolution with that of their host galaxies discovered by *JWST*, *Euclid*, *WFIRST* and *LSST*. Beyond statistical populations *Lynx* will identify the earliest forming groups and clusters, enabling us to probe the link between AGN and large-scale-structure directly, following the evolution of galaxies hosting active SMBHs as they fall into the gravitational potential of massive halos and are stripped of the gas which has fueled their growth.

#### D.2.4.2 Black Holes in Clusters

*Lynx* will enable us to directly measure both the distribution of AGN in dense environments, and, through observations of the hot gas (as described in **§D.2**), the properties of those environments simultaneously. *Lynx*, in combination with *Euclid*, *WFIRST* and *LSST*, will uncover how and when AGN and their host galaxies are quenched by environmental processes as a function of the environment properties such as total halo mass, radius and redshift.

The superb spatial resolution over the HDXI FOV is critical for the AGN detection, disentanglement from the diffuse cluster background, and source matching with *LSST*/*Euclid*/*WFIRST*/*JWST*. A large effective area and good soft response are also paramount to enable study of the highest-z targets. The proposed FOV of HDXI is ample for measurements of cluster AGN out to much beyond the virial radius in even the most massive clusters at $z > 2$.

This program requires targeted follow-up of the densest environments at $z > 2$, either as a survey or an archival program. In the absence of spectra or photo-z measurements, the power in the measurement is achievable by increasing the number of observed clusters. Naive simulations (such as drawing randomly from a distribution of clusters with $M > 5 \times 10^{13}\ M_\odot$ at $z < 3$, not optimized to test a model) assuming ~500 cluster fields with short (20 ks) exposures lead to more than an order of magnitude improvement in accuracy. For example, for the redshift dependence in the number of AGNs, parameterized as $(1+z)^\alpha$, the evolution index $\alpha$ will be tightly constrained to ±0.18, compared to the currently unconstrained range ~±2. Similarly, a currently loosely constrained mass cluster mass dependence, $M_{500}^\beta$ with $\beta \sim \pm 0.5$, will be established precisely, $\beta \sim \pm 0.04$. This will enable quantitative comparison of the evolution of AGN and star-forming host galaxies in dense environments. If fields with multi-wavelength data (*WFIRST* / *JWST* / *Square Kilometer Array (SKA)*, in addition to *LSST* and *Euclid* coverage) were instead targeted, even the naive program (i.e., not optimized in M and z) would be greatly more efficient.

These observations directly measure the halo occupation distribution (HOD) of AGN in massive clusters out to z~3 providing an anchor for AGN HOD models.

### D.2.5 Synergies

The crucial advances *Lynx* will make in the black hole key science program will be significantly enhanced by exploiting synergies with other space observatories and ground-based telescopes. Conversely, the X-ray observations will bring out the full potential of a large number of key observing programs planned for these future facilities. We explore here a few of these key synergies.

#### D.2.5.1 Synergies with *JWST* and *WFIRST*

*Lynx* is the natural observatory to complement and enhance the science breakthroughs that *JWST* and *WFIRST* are expected to lead to. This will proceed in both directions: *Lynx* is designed to benefit to the maximum extent from the deep fields that will be surveyed by *JWST* and *WFIRST* and, at the same time, enable insights into the black holes, galaxy formation, and Galactic science that is only possible by probing those fields in X-rays to reveal physical processes and components that is only possible with X-rays. The specific synergies were detailed in the discussion of *Lynx* observing strategy in **§D.1.2.2**, as well as for addressing the BH seed question and for tracing AGN growth in **§D.1.2** and **§D.1.4**.

### D.2.5.2 Synergies with Ground-Based Observations of Cosmic Reionization

Mapping out the epoch of reionization is one of the key frontiers of astrophysics in the coming decade. The history of the reionization and the nature of sources is the science objective of 21-cm surveys that aim to map out the neutral hydrogen present in the universe during the epoch of reionization and beyond.

Early X-ray heating from the first black holes in the universe plays a prominent role in the history of reionization, both heating and partially ionizing the intergalactic medium (IGM). X-rays escape more easily from their host galaxies into the IGM. Because of their reduced interaction cross sections with the surrounding gas compared to Lyα photons, they can also travel further, heating and ionizing the IGM as much as 10–1,000 Mpc from the source. Finally, since black holes and early X-ray sources can have a different spatial distribution than other ionizing sources, they contribute significantly to the temperature fluctuations in the IGM.

One of the most efficient tools to probe the thermal state of the IGM at high redshifts is the radio signal of neutral hydrogen with the rest-frame wavelength of 21 cm. This signal is sensitive to is a large number of astrophysical and cosmological processes; in particular, properties of the first X-ray sources such as their bolometric luminosity, spectra, spatial distribution, as well as the growth of the population in time strongly affect predictions for the neutral hydrogen emission from redshifts $z \sim 6-28$.

Cross correlating X-ray signals obtained by *Lynx* observations with the 21 cm tomography that will be obtained by SKA and Hydrogen Epoch of Reionization Array (HERA) can provide a direct probe of the fluctuations of the X-ray background that is made up of X-ray sources that fall below even *Lynx*'s sensitivity limit at very high redshifts. To enable such a cross correlation, the X-ray and radio observations need to be carried out on similar spatial scales. The preferred survey strategy will be to map out a significant fraction of a SKA field with 10–15 pointings of *Lynx* in order to cover spatial scales up to ~200 co-moving Mpc.

### D.2.5.3 Synergies with *LISA*

**Probing Both Channels of Black Hole Growth**

Mergers, in addition to accretion, are also thought to contribute to the growth of the first black holes in the universe. While at z=2–3, the Soltan-Paczynski argument shows that the typical SMBH acquired most of its mass via (radiatively efficient) accretion, at high-z, the relative contribution of accretion and mergers to the growth of $10^{4-6}$ Msun black holes is unknown.

While *Lynx* is designed to see the former channel, the detection of gravitational waves resulting from mergers by the planned space-based gravitational wave observatory *Laser Interferometer Space Antenna* (*LISA)* provides an additional exciting avenue for obtaining a census of black holes at high redshift and completing the picture of their physical properties. It is worth emphasizing that the mass-thresholds at $10^4$ $M_\odot$ at z=10 are the same for the *Lynx* and *LISA* observatories. Thus, *LISA* can help map out the relative role of growth by mergers as a function of redshift while *Lynx* can map out the (potentially dominant) mode of growth by accretion as well as provide information about the early black hole feedback on galaxies by looking for the direct X-ray signatures.

**Simultaneous X-ray and GW Chirps**

Numerous arguments suggest that a SMBH binary, merging in the *LISA* band, will produce an X-ray "chirp" that accompanies the gravitational wave (GW) signal. SMBHs with masses of $10^6$ Msun (to which *LISA* is the most sensitive) are predominantly located in the bulges of gas-rich disk galaxies (Kormendy and Ho 2013). Such galaxies undergo several mergers in a Hubble time, thereby producing the SMBH binaries that ultimately coalesce in the *LISA* band. During these galaxy mergers, both gas (e.g., Barnes and Hernquist 1996) and the SMBHs are driven to the new nucleus. The most natural scenario is therefore that merging *LISA* binaries are born in gaseous environments, similar to those believed to trigger bright quasar activity.

Accretion onto a binary SMBH, on scales much larger than the binary separation, should be similar to that for a solitary SMBH, i.e., via a geometrically thin, optically thick Shakura-Sunyaev disk. The gas-dynamics near the binary is modified: gas is expelled from the vicinity of the binary, out to roughly twice the binary separation (Artymowicz and Lubow 1994). However, numerous recent (magneto) hydrodynamical simulations have all concluded that the SMBHs continue to accrete at an undiminished rate, via narrow accretion streams (e.g., Artymowicz and Lubow 1996; MacFadyen and Milosavljevic 2008; Cuadra et al. 2009; Roedig et al. 2011; Nixon et al. 2011; Shi et al. 2012; D'Orazio et al. 2013; Gold et al. 2014; Farris et al. 2014). Importantly, the strong, quasar-like accretion continues even in the late, run-away stages of the merger, when the binary inspiral is driven by the GWs (Farris et al. 2015; Tang et al. 2018).

During the binary accretion process, the SMBHs are endowed via their own "minidisks." These minidisks are mutually truncated by the tidal forces of the companion SMBH, so that the sizes of the circumprimary and circumsecondary minidisks are roughly $r_1=0.27\, q^{-0.3} a$ and $r_2=0.27\, q^{0.3} a$, respectively, where $q$ is the mass ratio of the SMBHs and $a$ is the binary separation (Paczynski 1977; Roedig et al. 2014). A key point is that when the binary enters the *LISA* band, the tidal truncation radii are ~100 gravitational radii. While small, this region is approximately 10 times larger than the size of the region where X-rays are produced by quasars (Haiman 2017; see, e.g., Reynolds and Nowak 2003;

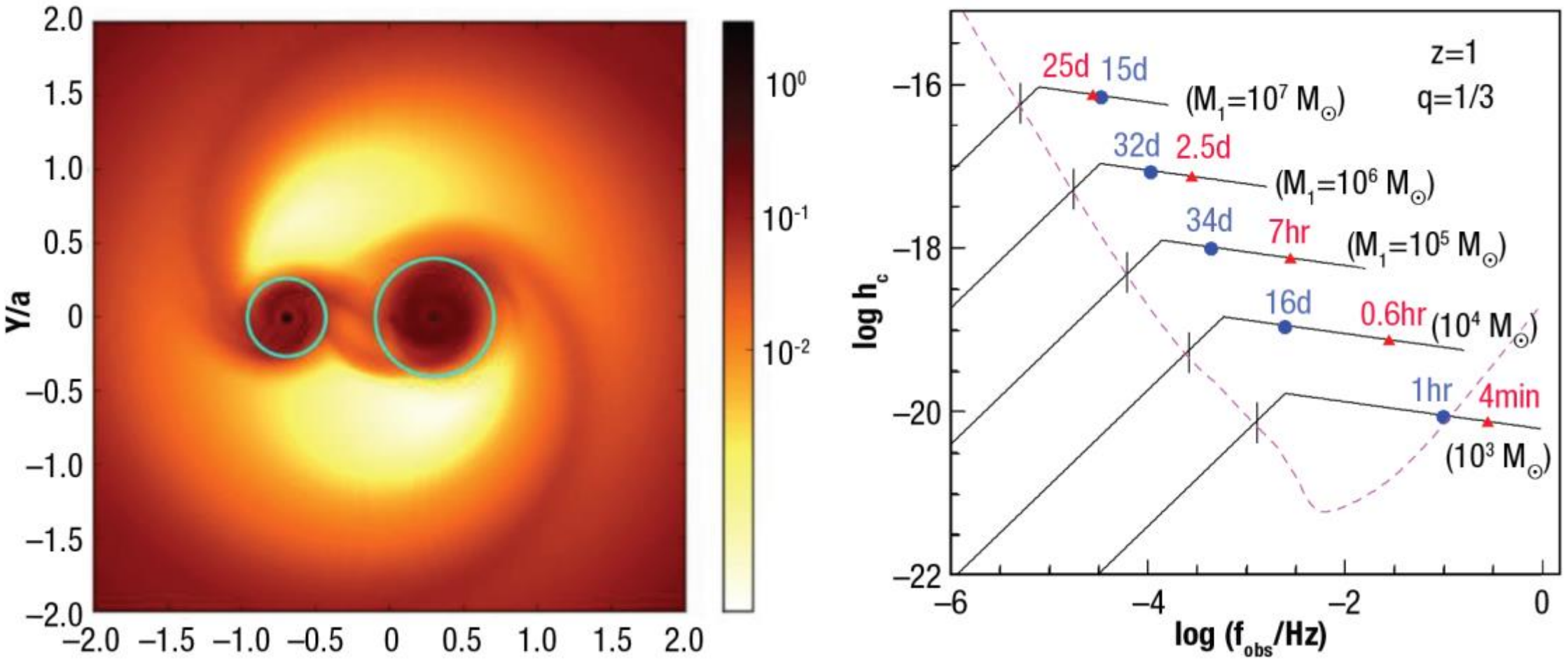

**Figure D-3. (Left) The surface density of an accretion disk near a binary SMBH, on a logarithmic scale. Cyan circles indicate the expected tidal truncation sizes of the circumprimary and circumsecondary disks. The binary has a mass ratio of $q=M_2/M_1=0.43$. (Adapted from Farris et al. 2014). (Right) Tracks across the *LISA* band of binaries at z=1 with mass ratios of q=1/3 and different primary masses $M_1$, as labeled. The break in the characteristic strain $h_c$ marks 5 years prior to merger. Along each track, the marks correspond to the times when: (i) the binary enters the *LISA* band (vertical line), (ii) the sky localization of a typical binary reaches an accuracy of 10 $deg^2$ (blue circle), and (iii) the tidal truncation radius of the circumprimary disk becomes smaller than 10 $R_g$ (red triangle). The dashed (magenta) curve shows *LISA*'s sensitivity, assuming a configuration with six links, 2 million km arm length, and a mission lifetime of five years (Klein et al. 2016). (Adapted from Haiman 2017.)**

Miniutti and Fabian 2004; Dai et al. 2010, Jimenez-Vicente et al. 2015; Guerras et al. 2017 for various constraints on the size of the X-ray emitting region).

The natural expectation, then, is that the merging SMBHs produce bright, quasar-like X-ray emission throughout its GW-driven inspiral in the *LISA* band, until their separation is so compact that the gas in their X-ray emitting regions are tidally stripped. This occurs only when the binary separation drops below ≈20 gravitational radii or, for a $10^6$ $M_\odot$ binary, 2–3 days prior to the merger.

Because the binary's orbital velocity in the *LISA* band is v/c=O(0.1), relativistic Doppler modulations and lensing effects will inevitably imprint periodic variability in the X-ray light-curve, at the tens of percent level, tracking the phase of the orbital motion, and of the GWs, unless the binary is face-on (Haiman 2017). For nearly equal-mass binaries, the accretion rates onto the individual SMBHs will also be strongly modulated at the binary's orbital period, yielding even stronger, order-unity variability, independent of inclination (Tang et al. 2018). The GWs themselves can localize a typical *LISA* binary (with mass of $10^6$ $M_\odot$ at redshift z=1) on the sky to within several square degrees, several weeks before the merger (Kocsis et al. 2008; Lang and Hughes 2008; McWilliams et al. 2011). Monitoring this area, *Lynx* will be able to uniquely identify a near-Eddington, quasar-like, X-ray source, with periodic flux variability. The period will be initially O(hr), but will decreasing, tracking the GW chirp. Between sky localization (a few weeks before merger) and tidal truncation (a few hours before merger), the GW and X-ray chirp signals could be observed in tandem, for several hundred cycles.

Detecting the X-ray chirp accompanying the GWs will help uniquely identify the electromagnetic (EM) counterpart of the *LISA* source, and enable a range of new science that requires EM counterparts (see Phinney 2009). Furthermore, a comparison of the phases of the GW and EM chirp signals will help break degeneracies between system parameters, and probe a fractional difference Δv in the propagation speed of photons and gravitons as low as $\Delta v \approx 10^{-17}$ c.

## D.3 Revealing the Invisible Drivers of Galaxy and Structure Formation

One of the central topics in astronomy is the emergence of galaxies and their concentrations, and a relationship between "normal" baryonic matter and the dark matter whose gravitational pull dominates the development of the large-scale structures in the Universe. In the generally accepted paradigm, the assembly, growth, and state of visible matter in the cosmic structures is driven, in addition to gravitational pull from dark matter, by violent processes which produce and disperse large amounts of hot gas and metals into the circumgalactic (CGM) and IGM. This paradigm consistently emerges from modern cosmological simulations of galaxy formation. **Figure D-4** shows an example derived from the Illustris simulation. Plots like this make it obvious that the distribution and evolution of the visible stellar material locked in galaxies reflects only the *consequences* of the overall process; the direct imprint of the drivers is in the CGM and IGM. These drivers leave little or no signatures in the optical and IR spectral bands, hence we term them as *Invisible Drivers of Galaxy and Structure Formation*.

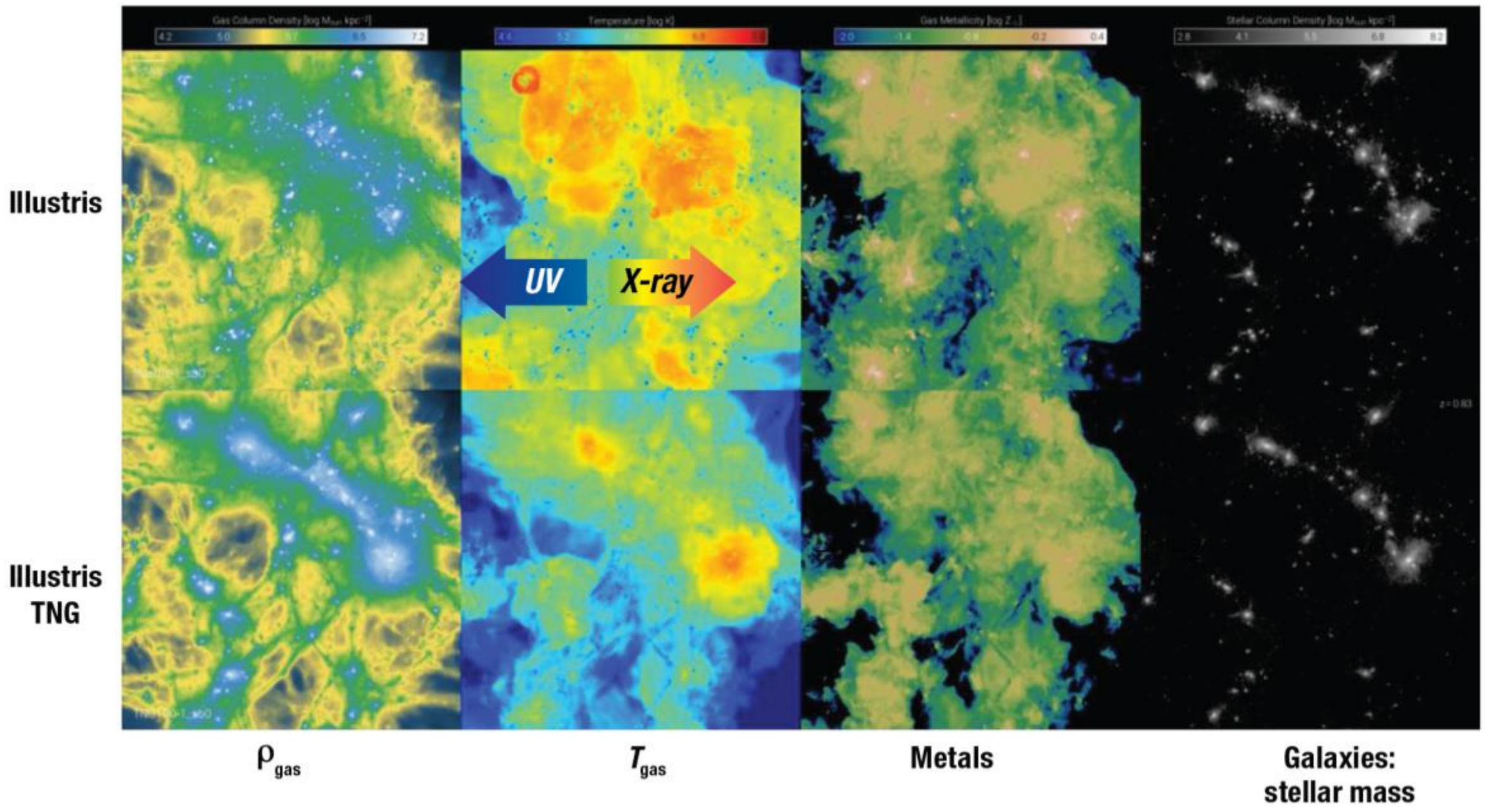

**Figure D-4. Snapshot from the Illustris simulation of galaxy formation, which shows the distribution of stellar material and a temperature map of the intergalactic gas at the same epoch in a 100×100 Mpc simulation box. Galaxy formation models, such as Illustris, successfully explain the basic properties of assembly and Star Formation Rate (SFR) history in galaxies only at the expense of significant feedback into the gas. The physics of feedback and its impact on the galaxy assembly is poorly understood because the observations of the relevant gas phases are scarce or non-existent. Two simulation runs shown here (Illustris and Illustris-TNG) use the same numerics, but very different prescriptions for feedback. The differences are striking in the distribution of gas, especially in the phase heated to X-ray temperatures; there is little or no consequence of the feedback details on the properties of stellar material locked in the galaxies. Hence, the OIR data reflect only the *consequences* of the overall process, while the direct imprint of the drivers is in the CGM and IGM. *Lynx* is designed to be able to take direct detailed measurements in all key locations in this large-scale gas distribution.**

The gas in the CGM and IGM has low densities and high temperature, and observations of this gas are scarce and incomplete (see **§D.3.2** below). Some of the major components of this gas or the processes predicted to exist and play a key role, have not been observed. The picture of galaxy formation will not be complete until we fully understand the energy feedback and its effects on the assembly of galaxies.

A generic prediction of galaxy formation models is that in galaxies with a mass similar to that of our own Milky Way and higher, the relevant baryonic component is heated and ionized to X-ray temperatures. Needed observations rely on high-resolution spectroscopy and the ability to detect low surface brightness continuum emission, and on a capability to map large areas in the sky in the OVII, OVIII, and other important spectral lines. Studies of feedback require a capability to take high spatial and spectral resolution data around and deeply within the galaxies.

### D.3.1 What Will *Lynx* Bring?

Exquisite spectral and angular resolution of *Lynx* will make it a unique instrument for mapping the gas around galaxies and in the Cosmic Web. It will be capable of measuring the energetics and statistics of most of the relevant feedback modes; new unique insights on the physics of feedback to inform theoretical models. These observations will provide conclusive insights into the major gaps in our understanding of the galaxy formation:

**Extended CGM halos**

The thermodynamic state and chemical composition of hot gas in galactic halos should bear an indelible record of effects of assembly history and feedback that shaped galaxy formation. *Lynx* will be able to probe hot CGM to a large fraction of the virial radius in $M_{tot} \gtrsim 3 \times 10^{12}\, M_\odot$ through direct imaging in emission, and for $M_{tot} \sim 1 \times 10^{12}\, M_\odot$ or below in absorption using bright background AGNs. This program is described in more details below in **§D.3.2**.

**Cosmic Web filaments**

The Cosmic Web filaments represent the typical environment for most of the "field" galaxies. Hot gas in the Cosmic Web filaments is expected to be a major reservoir of baryons in the low-redshift Universe, which remains largely unexplored. It contains a fossil record of strong feedback, the material ejected outside of the galaxy virial radii, and therefore its thermodynamic state and metallicity can serve as a crucial constraint for the galaxy formation models. *Lynx* will detect the Cosmic Web in absorption against virtually every X-ray bright background AGN in its spectroscopic program. More importantly, *Lynx* will have a sensitivity to *map* the hot gas in the Cosmic Web above an overdensity threshold of ~30. This program will require an HDXI survey with total exposure of a few Msec over an ~10 deg$^2$ region projected on now low-$z$ superclusters which can be selected, e.g., from *Sloan Digital Sky Surveys (SDSS)*. This does not have to be a stand-alone survey. It can be combined with projects aimed, e.g., at studies of the AGN evolution at moderately high redshifts.

**High redshift probes**

Detailed X-ray observations of the hot gas in galactic halos and Cosmic Web with *Lynx* are possible only at relatively low redshifts. These observations will already play a central role and provide constraints on the galaxy formation models. However, it obviously will be worthwhile obtaining additional constraints on the hot IGM near the epoch of the peak cosmic star formation, at $z = 2-3$. *Lynx* will be able to approach this from two angles, by observing high-redshift galaxy clusters and groups.

*Lynx* will be able to derive the basic parameters (e.g., density profiles, average temperature) in galaxy groups down to a mass scale of $M_{tot,500} = 2\times10^{13}\, M_\odot$ at $z>3$. This is not too far separated from galactic mass scale, and so strong feedback operating in galaxies will generally result in the observable effects in the intragroup medium.

As galaxies accrete onto galaxies clusters, the gas in their halos gets shock-heated, compressed, and mixed with the gas from halos of a large number of other galaxies. These processes can completely erase information on the thermodynamic state of gas prior to accretion on the cluster, but not on its chemical composition. The metallicity measurements in the outer cluster regions provide a strong constraint on the average metallicity of CGM at $z > z_{cluster}$, and hence on the feedback history. Recent *Chandra*, *XMM*, and *Suzaku* results point to a universal metallicity of ~0.3 $Z_\odot$ in the galaxy cluster outskirts, strongly suggesting that this represents a mean level of chemical enrichment of the IGM by $z = 0$ (e.g., Werner, N. et al. 2013).

*Athena* aims at making metallicity measurements in clusters to $z = 1$. Approximately 60% of stars observed at $z = 0$ are formed by $z = 1$ (Fig. 11 in Madau and Dickinson 2014), so some trends in the metallicity of clusters can be expected but should still be weak (as currently hinted by *Chandra* data; Ettori et al. 2015, McDonald et al. 2016, Mantz et al. 2017). *Lynx* will be able to make spatially resolved metallicity measurements in clusters to $z = 3$. The mean stellar mass density at $z \geq 2$ is less than 10% of the local value. Therefore, we expect dramatic redshift trends in the chemical composition of the cluster gas between $z = 1$ and 2 and beyond. Observations of this trend, uniquely possible with *Lynx*, will provide an excellent handle on the galaxy wind feedback right at the epoch of the peak cosmic star formation. The necessary cluster samples will be provided by the upcoming large-area X-ray and Sunyaev-Zeldovich (SZ) effect surveys.

**Feedback signatures in gas near and within galaxies**

The state of hot gas in the immediate vicinity and within galaxies is a key signature of the ongoing and recent energy feedback episodes. On scales from 100 pc to tens of kpc, *Lynx* will observe galaxy-scale winds driven and supernova (SN) and stellar feedback, including detailed measurements of their energetics and constraining their launching mechanisms (**Figure D-5**). These measurements will require ~1 arcseconds angular resolution and an ~0.3 eV spectral resolution in the X-ray microcalorimeter.

On 100 pc–1 kpc scales, *Lynx* will be able to resolve and characterize extended narrow emission line regions in nearby AGNs, providing a key diagnostic for shock excitation and thus a critical handle on where, and how much of, the AGN outburst energy is dissipated in the interstellar medium (ISM). Another diagnostic of the AGN energy feedback on galactic scales will be *Lynx* observations of AGN-inflated bubbles in the hot ISM of nearby elliptical galaxies. Both types of observations require sub-arcsecond imaging capabilities in the soft X-rays, unique to *Lynx*.

On the smallest scales, *Lynx* will track the hot interstellar medium in the active star forming regions in the Milky Way and nearby galaxies. Observations of the interaction of young, hot ISM with surrounding dense molecular clouds will be key for completing our understanding of how energy feedback from star formation *locally* shuts down new star formation **(§D.3.1**). Additional handles on the star formation feedback will be provided by *Lynx* observations of large samples of young supernovae remnants in the Local Group galaxies, establishing a comprehensive view of relationship between the recent (past several hundred years) SN activity and star formation **(§D.3.3**).

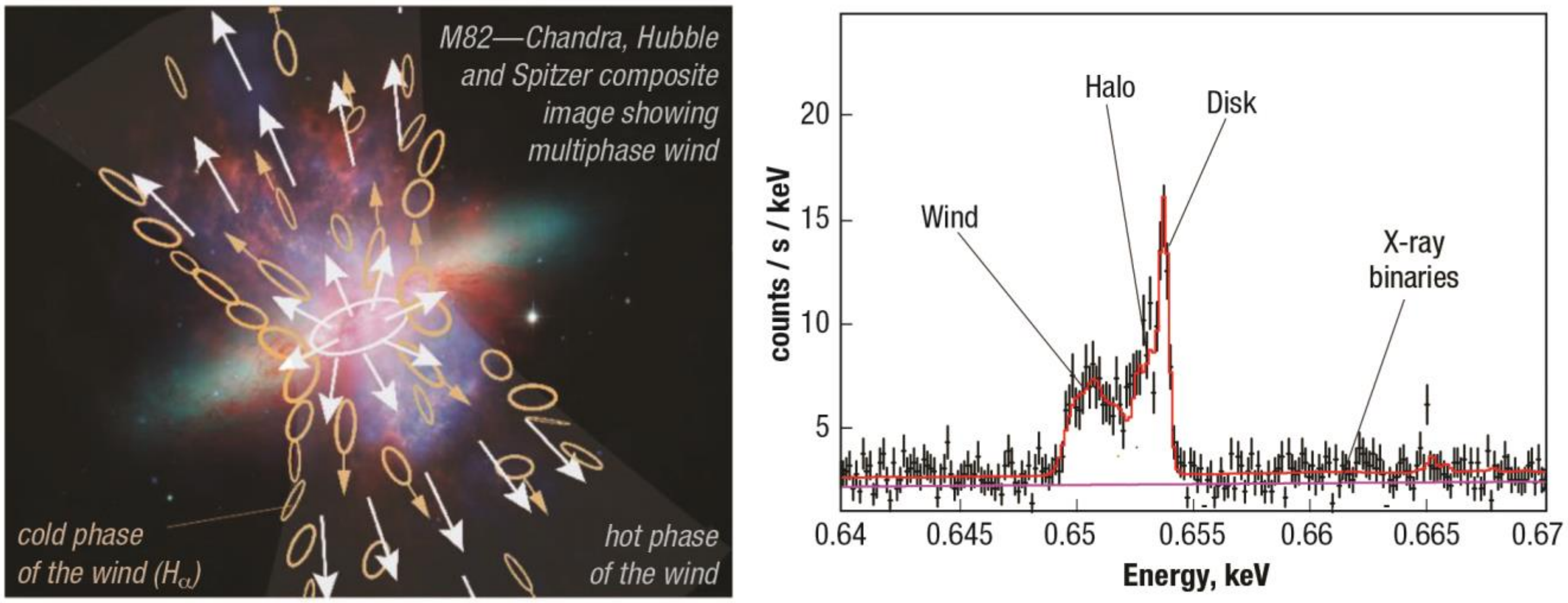

**Figure D-5. Schematic of the starburst-driven galaxy wind and the simulated *Lynx*. Cold gas entrained in the outflow is directly seen in UV absorption and molecular gas studies, but most energy is in the hot "fluid." The properties of the hot phase also reflect the details of the wind acceleration mechanism, such as the role of cosmic rays. The spectrum on the right shows a simulated observation of the OVII emission line using a 50 ksec *Lynx* ultra-high resolution microcalorimeter array ($\Delta E$ = 0.3 eV). Wind properties are derived (*credit*: E. Hodges-Kluck et al.) from a toy wind model based on Thompson et al. (2016) prescription with a mass-loading factor $\beta$ = 0.2. The model includes a 6×2 kpc diameter exponential hot disk ISM with $T$ = 0.4 keV, rotation velocity $v_{rot}$ = 250 km s$^{-1}$, and a cold exponential disk absorber. Hot halo follows a beta-model with $T$ = 0.2 keV, abundance of 0.3 Solar, and scale heights of 5×4 kpc. X-ray binaries are distributed with a power-law XLF along the disk. Velocity structure in lines is due to superposition of components and thermal broadening.**

**AGN mechanical feedback**

Central black holes are believed to be an important or dominant source of feedback in massive galaxies. The mechanical mode of feedback due to AGN-driven winds, warm absorbers, etc., is very easily coupled to the ISM, but the wind energetics remain relatively poorly understood compared to the feedback from electromagnetic radiation of relativistic jets. AGN winds are known to span a wide range of ionization states, but it is likely the high-ionization, ultra-fast outflows (UFOs) that are most relevant for galactic feedback. The main diagnostic of these flows is the blue shifted K-shell lines of FeXXV and FeXXVI at 6–9 keV rest frame. These observations will be a major topic for X-ray microcalorimeters on *Athena* as well as *Lynx*. However, the current quality of X-ray data generally does not allow one to independently establish the distance and density of gas in the wind[1], making the derived energy and momentum flux in the wind uncertain by an order of magnitude or more. High-resolution soft X-ray spectra with *Lynx* gratings will provide access to density diagnostics in the AGN winds, such as density-sensitive line doublets, changes in ionization state in response to the X-ray ionizing flux fluctuations, and longer-term correlations of the outflow ionization state with the black hole luminosity. These measurements will greatly reduce uncertainties in the measurements of the instantaneous mechanical output of the AGNs.

[1] Easily measured X-ray ionization parameter $\xi_X = L_X/(n_e R^2)$ constrains a degenerate combination of density and distance. The situation is further complicated by an unknown clumpiness of the outflow.

**Plasma physics effects in dissipation of the AGN feedback**

*Chandra* observations of cool cores in the nearby galaxy clusters show that the dissipation of energy from AGN outbursts in the intracluster medium is a remarkably complex process, where plasma microphysics (e.g., viscosity, dissipation of turbulence, heat conduction) plays a major role. *Hitomi* observations of the Perseus provide a glimpse of the power of high resolution spectroscopy, even with coarse angular resolution, for these studies. The X-ray Astronomy Recovery Mission (*XARM)* and then *Athena* will bring the X-ray spectroscopic capabilities to the next level and provide superb measurements of the total energy of gas motions in a large sample of galaxy clusters. However, as *Chandra* experience convincingly shows, the key to better understanding the microphysics of the AGN energy dissipation lies in the ability to resolve structures down to the Spitzer mean free path scale, $\lambda_{\rm Sp}$. *Chandra* is often limited by statistics, not angular resolution, in probing gas on the $\lambda_{\rm Sp}$ scales. *Lynx* will eliminate this problem. More importantly, it will provide gas velocities on a similar angular scale, providing a third dimension to the data. New handles on the plasma physics effects provided by *Lynx* observations of nearby galaxy clusters will be used to inform the "subgrid" treatment of the AGN feedback in numerical models of galaxy formation.

**Fueling and triggering of AGN feedback**

*Lynx* angular resolution will be sufficient to determine the gas state at or near the sphere of influence of SMBHs in nearby galaxies. While it is likely that the accretion flow within this radius significantly deviates from the Bondi solution, $\dot{M}_{\rm Bondi}$ can still serve as a useful proxy for the instantaneous accretion rate on the black hole. *Chandra* observations indeed reveal a good correlation between the derived Bondi accretion rate and AGN jet power for a small sample of nearby elliptical galaxies (Allen et al. 2006). The Bondi-Hoyle-Lyttleton accretion rate is typically used as a proxy for black hole feedback in numerical models of galaxy formation with a free efficiency factor, e.g. $\dot{E}_{\rm feed} = \varepsilon_f\,\varepsilon_r\dot{M}_{\rm Bondi}c^2$ in Springel et al. 2005. The efficiency factor $\varepsilon_f$ cannot be determined from first principles and can be uncertain by approximately an order of magnitude, with the correspondingly uncertain consequences for predictions of the AGN feedback effect on galaxies. Dramatically better sensitivity and new spectral gas diagnostics available with *Lynx* will make it possible to derive $\dot{M}_{\rm Bondi}$ much more reliably and in a larger sample of galaxies. These observations will be used to guide subgrid parameterizations of the AGN feedback in numerical models.

Of these topics, the most demanding and one of the most information-rich measurements are those of the hot halos around low-redshift Milky Way-type galaxies. We consider this program in more detail below.

### D.3.2 *Lynx* Observations of the State of Gas in Galactic Halos

As discussed above, a detailed picture of the state of diffuse gas in galactic halos is a critical missing piece in our understanding of *Invisible Drivers* of the galaxy and large-scale structure formation. It is critical that future observations probe the CGM over a large range of radii, and map its emission in addition to probing individual lines of sight in absorption:

- At small radii, the CGM properties are likely to be sensitive to recent stochastic events, such as starbursts and individual wind outflows, while at larger radii, $\gtrsim 0.2r_{200}$, the thermodynamic gas profiles should reflect the imprint of the overall integral effects of feedback during galaxy evolution and the steady state processes that govern the evolution of the gas halo. Near the virial radius, the halo properties probe the regions where structure formation "meets" galaxy formation, and accretion of fresh material onto galaxies and its virialization occurs.
- Theoretical predictions and currently available UV observations indicate that the gas halos around galaxies are strongly multiphase. The structure and physical origin of such multiphase gas is unknown, but it is likely that different phases correspond to different processes operating within the halo in the course of galaxy formation. The multiphase, nonazimuthally symmetric structure of the halos can be completely washed out in stacked data. Therefore, it is crucial to obtain imaging observations of individual objects, in both continuum and individual line emission.

- Determinations of specific entropy have been crucial to our understanding of the hot gas in galaxy clusters. Specific entropy will be a crucial diagnostic of the cooler CGM, as well. Determination of entropy as a function of radius requires measurements of both gas density and temperature. In the X-rays, one also needs to separate out the gas metallicity because density and temperature derived from X-ray spectra of low-temperature plasma are strongly degenerate with metallicity. Of course, the distribution of metals in the CGM by itself is an important diagnostic of *Invisible Drivers*.

**Target galaxy mass scale for CGM studies**

How far down in the mass scale should we extend the observations of hot gas in galactic halos? A large body of empirical data on galaxy stellar populations points towards a profound change in the trends of stellar population with mass near $M_{tot,200} \approx 10^{12}\ M_\odot$. Examples include the $M_{star} - M_{tot}$ diagram (Kravtsov et al. 2014) with a knee near $M_{tot} = 10^{12}\ M_\odot$; a roll-over in the galaxy stellar mass function at $M_{star} \approx 10^{10.5-11}\ M_\odot$, corresponding to $M_{tot} \approx 10^{12}\ M_\odot$ (Bernardi et al. 2017); strong rearrangement in the color vs. stellar mass diagram around $M_{tot} \approx 10^{12}\ h^{-1}\ M_\odot$ (Schawinski et al. 2014). Transition in the trends of galaxy properties around $M_{tot} \sim 10^{12}\ M_\odot$ is likely related to changes in the dominant modes of feedback (Croton 2006), cooling of hot gas, or changes in the gas accretion patterns (Keres et al. 2009), i.e. exactly the missing pieces in the galaxy formation picture. Therefore, $M_{tot,200} = 10^{12}\ M_\odot$ represents a natural low-mass target for CGM studies with *Lynx*. Our feasibility assessments reported below show that *Lynx* indeed should be able to characterize the CGM at, or close to, this mass threshold.

**Current and future information from non-X-ray observations**

The most common probe of the CGM so far has been through using UV absorption lines in the spectra of background AGNs. Recent results (e.g., Prochaska et al. 2017) indicate that a significant fraction of baryons within the inner ~100 kpc (0.4−0.5 $r_{200}$) of Milky Way-like halos are in a cool, $10^4$ K phase, but this phase declines significantly as one approaches $r_{200}$ (Borthakur et al. 2016). The structures in the cold phase—cold clouds, filaments, and extended disc—with a low volume filling factor must co-exist with a hot, tenuous phase, which current and future approved instrumentation cannot detect around the haloes of normal spiral and elliptical galaxies. Therefore, the key questions won't be answered by UV data alone: At what radius does the hot phase become dominant over the cold phase? How far from hydrostatic equilibrium is the hot phase, and what leads to the deviations from hydrostatic equilibrium—turbulence, bulk flows, rotation, and/or feedback? The formation, survival, and possible destruction of the cool phase out to $r \sim 0.5 r_{200}$ and beyond depends on its hydrodynamical interactions with the hot phase. Hence, the fate of the cold structures depends on the detailed thermodynamic state of the hot halo. The relevant questions here are:

- Do cold structures accrete onto the galaxy?
- Are they shredded by turbulence or rotation?
- Does feedback prevent this accretion?

The answers depend on knowing the detailed thermodynamic state of the hot halo. The cool phase (filaments, clouds, and extended discs) is not spherically symmetric, so individual objects must be observed to understand the asymmetries of the hot phase and how they relate to the properties cool structures.

Another non-X-ray diagnostic of the hot gas in CGM is sensitive SZ effect measurements. Detection of individual galactic halos in SZ is impossible even with future-generation experiments, but detections of the stacked thermal SZ effect have already been made with Planck (Collaboration et al. 2013), and kinetic SZ effect may have been detected with Atacama Cosmology Telescope (ACT; Schaan et al. 2016). These detections are forecasted to become routine in the upcoming CMB "Stage 3" and "Stage 4" experiments (Battaglia et al. 2017), although the focus is on objects with mass ~$10^{13}$ $M_{\odot}$, significantly higher than what should be targeted (see below). Angular resolution also will be an issue for this work. The optimal redshift range for stacking SZ signal is at $z$ ~0.5 or above. At this redshift, a 1 arcsecond beam of the "Stage 4" CMB experiments corresponds to 350 kpc, too coarse to constrain the structure of the CGM within the virial radius even in a stack.

To summarize, the ongoing and future CGM studies using the UV absorption lines and SZ effect are important, but they in no way remove the need to obtain detailed data for the hot phase of CGM in individual objects.

**Feasibility assessment for direct imaging of the CGM with *Lynx***

Our assessment of observing feasibility and *Lynx* performance requirements was based on analyzing mock observations generated from outputs of several modern numerical simulations of galaxy formation, including EAGLE, FIRE, MUFASA, Illustris-TNG, and Agertz and Kravtsov zoomin simulations. These simulations all aim at reproducing galaxy stellar populations, but use different numerical models, subgrid physics, and prescriptions for feedback. We use these outputs as a representative sample of what a sensitive X-ray observatory such as *Lynx* can see in the galactic halos; different simulations lead to a consistent picture.

Galaxy halos are the brightest in the soft X-ray band, $E < 0.7$ keV because of their low temperatures. However, the emission is dominated by a small number of bright spectral lines (notably, OVIII and OVIII transitions). The contrast of the CGM continuum emission relative to the unavoidable foreground from the Milky Way halo is low, making CGM density measurements in the soft band impossible, except for the very inner radii. At higher energies, $E > 0.7$, the CGM spectrum has a stronger continuum component, and the foreground Milky Way halo emission is much weaker. Therefore, gas density can be derived using the CGM emission in this energy band. Overall, we find that simultaneous solid detections of the CGM flux in three spectral bands, 0.4–0.7, 0.7–1.05, and 1.05–1.5 keV, are sufficient for ≈10% determination of gas density, and ~20% determination of temperature and metallicity. Detections in only two bands constrain a degenerate combination of density, metallicity, and temperature. Detection only in the soft band provides a measure of the O lines flux, which is hard or impossible to convert to thermodynamic quantities. Therefore, X-ray observations aiming at a detailed characterization of the CGM must have the sensitivity for solid detections in all three of the 0.4–0.7, 0.7–1.05, and 1.05–1.5 keV bands.

Numerical models we analyzed all predict that the expected X-ray emission from the CGM of the Milky Way-type galaxies at large radii is very faint. It is certainly well below the *Chandra* and *XMM-Newton* limits. Its detection is challenging even for the next-generation X-ray missions. The limiting factor is the low expected contrast of the CGM emission relative to the astrophysical and instrumental backgrounds, which leads to an unavoidable level of systematics. In addition, imaging of the CGM at $E \gtrsim 0.7$ keV is severely affected unless most of the Cosmic X-ray background is resolved into discrete sources, for which arcsecond resolution is required. Mock simulations of long exposures show that the residual background fluctuations from sources below the *Athena* confusion limit are at least a factor of 10 above the Poisson noise-dominated residual fluctuations for *Lynx*. In addition, *Athena* mirrors will not be protected from stray light, which will introduce additional large-scale nonuniformities in the background. These two factors will introduce severe fundamental limitations to *Athena's* ability to map diffuse gas in galactic halos and Cosmic Web filaments. In contrast, *Lynx* will be limited almost exclusively by statistical noise. Overall, we project that mapping of the CGM can be accomplished with *Lynx* to at least half the virial radius in galaxies with mass as low as $\approx 3\times10^{12}$ $M_{\odot}$. A sample of what *Lynx* will be able to observe in galactic halos is shown in **Figure D-6**.

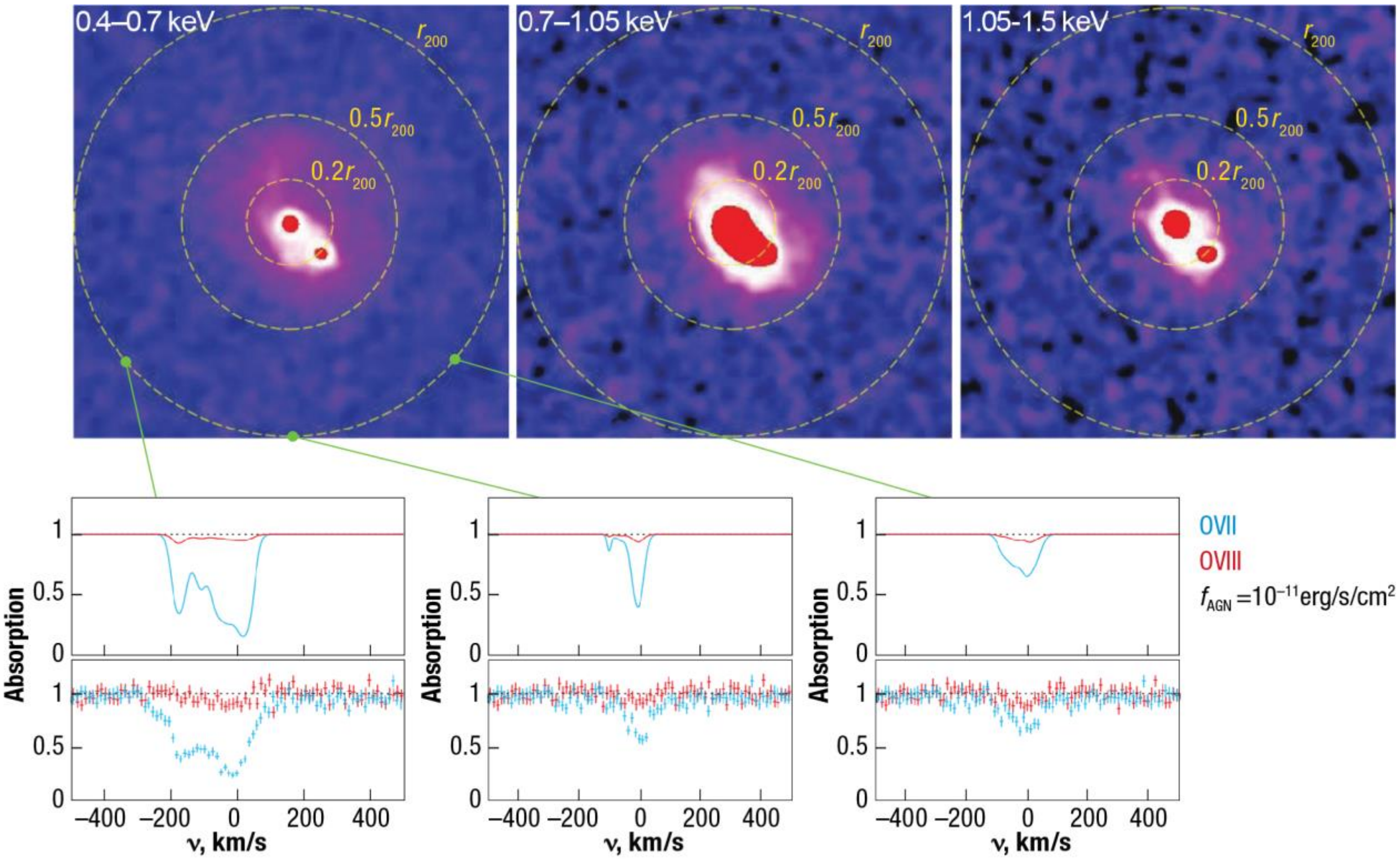

**Figure D-6. Mock *Lynx* observations of a $M_{tot}$ = $3\times10^{12}$ $M_{\odot}$ galaxy at $z$ = 0.03 generated using the output from EAGLE simulation. The *top panels* show 500 ksec HDXI images in three energy bands (the virial radius, $r_{200}$, corresponds to 9 arcminutes). In addition to the CGM signal, they show the realistic noise from astrophysical and instrumental backgrounds, detection and removal of background point sources, etc. The CGM emission can be mapped out to ~$0.5r_{200}$ in all three bands, revealing the halo density, metallicity, and temperature structure. The *bottom panels* show simulated OVII and OVIII absorption spectra (300 ksec XGS observations, $f_{AGN}=10^{-11}$ erg/s/cm$^2$) at three arbitrary lines of sight located near the virial radius. Solid lines show the ideal absorption spectra derived from simulations output, including thermal broadening and random velocities which data points indicate what *Lynx* will observe in those cases with XGS at spectral resolution $R$=5,000. Note strong variations of absorption spectra for different lines of sights, emphasizing the complex kinematic and multi-phase structure of the CGM. *Lynx* will provide very high signal-to-noise measurements of the OVII line, and solid detections of OVIII. A joint analysis of the OVII and OVIII lines will constrain the temperature distribution of random velocities and temperatures in each of the lines of sight.**

**X-ray absorption line studies of CGM**

Numerical simulations outputs also enable us to assess how far out and down in the mass scale the hot CGM can be probed via absorption line studies. EAGLE outputs, in particular, provide spectral line profiles including full thermal broadening and kinematic information. Examples of predicted absorption spectra for sight lines at the virial radius of $M_{tot}$ = $10^{12}$ $M_{\odot}$ galaxies are shown in **Figure D-6**. Overall, the mock analysis shows that OVII absorption should be routinely detectable with *Lynx* gratings in this regime; OVIII is also detectable in many cases. The OVIII/OVII ratio is a sensitive temperature diagnostic, so detection of both lines should constrain the CGM temperature rather well. This is an important characteristic of the CGM thermodynamic state, even though the gas density cannot be derived from the oxygen absorption lines.

The kinematic structure of OVII and OVIII is complex, and it can be characterized rather well in the stronger detection cases, opening a door to a new diagnostic of the hot CGM in $L^*$ galaxies. Because of the complex, non-Gaussian structure of the line-of-sight velocity distribution, the proper characterization of the lines requires the spectral resolving power at least matching the expected thermal width of the oxygen lines, $R \approx 5{,}000$.

To assess how many absorption line measurements of the CGM halos are possible with *Lynx*, we used the RASS-6dFGS catalogue of bright AGNs (Mahony et al. 2010) as a representative sample of X-ray bright AGNs, and then integrated the halo mass function to determine how many sight lines can be explored for a given total observing time. We find that a 5 Msec survey of 80 X-ray bright AGNs should yield ~30 detections, such as those shown in **Figure D-6,** for galaxies within $\Delta \ln M = 1$ of $M_{tot} = 10^{12}\ M_{\odot}$—together with many more detections of oxygen absorption in higher-mass systems and in the Cosmic Web filaments.

## D.4 The Energetic Side of Stellar Evolution and Stellar Ecosystems

A wide range of high-energy processes provide a unique perspective on stellar birth and death, internal stellar structure, star-planet interactions, the origin of elements, and violent cosmic events. The role of *Lynx* is to greatly extend our X-ray grasp throughout the Milky Way and nearby galaxies by combining, for the first time, the required sensitivity, spectral resolution, and sharp vision to see in crowded fields. *Lynx* will be able to detect X-ray emission as markers of young stars in active star forming regions and study stellar coronae in detail. Essential insights will be provided into the impact of stellar X-ray and extreme ultraviolet flux and winds on the habitability of their planets. Images and spectra of supernova remnants in Local Group galaxies will extend studies of stellar explosions and their aftermath to different metallicity environments. *Lynx* will expand our knowledge of collapsed stars through sensitive studies of X-ray binaries in galaxies as distant as 10 Mpc and through detailed follow-ups of gravitational wave events.

### D.4.1 Young Star-Forming Regions

Young stars display enhanced X-ray emission due to their youth and the concomitant rapid rotation, and the X-rays are an unambiguous marker for young stars with ages t<~600 Myr. Preibisch et al. (2005) find that 97% of optically visible late-type F-M cluster stars in the Orion Nebula are detected in the X-rays with *Chandra*. X-ray data is a great complement to the well-known infrared excess selection criteria. The IR excess reveals young stellar objects still containing their natal material or with circumstellar disks, but this method becomes ineffective after disks have dissipated (Torres et al. 2006; Evans et al. 2009), which is expected to happen at t=10–70 Myr. Combined X-ray and IR data thus is a great tool for studies of the protoplanetary disks dissipation, while sensitive X-ray data alone can provide a uniquely complete sample of young stars for follow-up studies. This is especially true for the young star clusters because they are obscured (up to 100–200 magnitudes of extinction), and *Lynx* can probe populations with large column densities ($N_H > 10^{23}$ cm$^{-2}$) that are not observable at optical and UV wavelengths.

X-ray observations of young star clusters provide a view on how the star formation-driven feedback operates at its basic scale. Young stars alter their environments through direct and dust-processed radiation fields, ionization fronts, stellar winds, and supernovae (SNe). This stellar feedback produces the multiphase structure of the ISM (McKee and Ostriker 1977), regulates star formation in giant molecular clouds (GMCs; e.g.,

Zuckerman and Evans 1974, Krumholz and Tan 2007), possibly drives turbulence in GMCs (Mac Low and Klessen, 2004), and disrupts and destroys GMCs on tens of Myr timescales (Matzner 2002, Krumholz et al. 2006). Fast stellar winds and SNe carve out large cavities, called superbubbles, that sweep up material from the surrounding medium and are filled with tenuous, hot (~$10^7$ K), shock-heated gas (e.g., Castor et al. 1975; Weaver et al. 1977; Chu and Mac Low 1990; Rogers and Pittard 2014). Diffuse X-ray emission associated with the collective effects of stellar winds and SNe has been observed in the star-forming regions of the Milky Way and external galaxies (e.g., Townsley et al. 2006, Güdel et al. 2008), and a significant fraction of the SN/wind energy and mass may end up in Galactic outflows/winds (e.g., Li et al. 2017).

The environments in which stars form are crucial pathways to understanding our astrophysical origins. Star formation sows the seeds for planet formation and future planetary habitability. Solar systems are integrated ecosystems, and the next generation advances in grappling with this topic require advances in understanding how the processes at work before a star is born, and during the course of its pre-main sequence and main-sequence lifetime, impact future planet formation and planetary habitability. Accretion processes dominate early in the star's life, while magnetic activity continues from stellar birth to old age. Magnetic activity manifestations are controlled by the spatial distribution and properties of magnetic fields. This is not a steady state and conditions change with time due to small-timescale heating events. Precision probes of the near stellar environment through X-ray spectroscopy provide one of the few means to determine magnetic field structure.

### D.4.1.1 Breakthrough progress with *Lynx*

**Census of young stars.**

*Chandra's* sensitivity is sufficient to probe the entire stellar mass scale, from brown dwarfs up to massive O and WR stars, only out to ~Orion Nebula's distance, 410 pc. This severely limits the range of environments available for study. Many young star clusters with a much higher SFR than Orion Nebula are located in the Carina-Sagittarius spiral arm at $d$ ~5 kpc. *Chandra* lacks the sensitivity at these distances, and *Athena* will be source-confused in cluster cores already beyond $d$ ~250 pc. Reaching into Carina-Sagittarius spiral arm's star clusters will also be problematic with *Gaia* for two reasons: the 3D space motions will only be available for relatively bright stars with $m$ <16, and because *Gaia* will be confusion-limited in regions with stellar densities >750,000 per square degree. *Lynx* sensitivity and angular resolution, by contrast, will be fully sufficient to study young clusters in the Carina-Sagittarius arm at $d$ ~5 kpc.

**Diffuse X-rays from young, hot ISM.**

Large uncertainties remain on the energy and momentum injection from star clusters and how these properties vary as a function of e.g., star cluster mass, age, environment, and metallicity. One major reason for these uncertainties is a dearth of observational constraints on the mass in the diffuse, hot phase of the ISM, and where and how stellar wind and SN energy are deposited to the surroundings. Measurements of these parameters from a diverse sample of star clusters in the Milky Way and in the Local Group are crucial to develop physically-motivated star and galaxy formation models (e.g., Dobbs et al. 2014; Hopkins et al. 2014) and to improve the predictive power of simulations (e.g., Scannapieco et al. 2012). *Lynx*'s sensitivity is required to detect diffuse X-ray emission in Milky Way and Local Group. In particular, typical temperatures and luminosities of the hot plasma shock-heated by stellar winds and SNe in HII regions are $kT_X \approx$ 0.1–0.8 keV and $L_X \sim 10^{31}$–$10^{35}$ erg s$^{-1}$ (e.g., Chu et al. 1995, Oey 1996, Jaskot et al. 2011). The large dispersion in X-ray luminosities arises because wind/SN energy can be lost through several mechanisms, e.g., radiative cooling, thermal conduction, dust heating, physical leakage of hot gas from the HII shells (e.g., Rosen et al. 2014). The sensitivity of *Lynx* is required to detect the X-ray dim ($L_X$ ~$10^{31}$ erg/s$^{-1}$) superbubbles out to the distances of M31 and M33 with modest exposures and several thousands of moderate X-ray luminosity superbubbles beyond the Local Group (to $D$ <30 Mpc). The sub-arcsecond spatial resolution of *Lynx* is crucial to disentangle point sources from the diffuse emission and thus to assess accurately the energetics of the diffuse component.

**Physics of accretion and coronal emission in young stars.**

Under the standard magnetospheric accretion scenarios, the accreted material should come crashing onto the stellar surface with velocities of up to several hundred km $s^{-1}$, heating the plasma to X-ray temperatures. Much of that emission is reprocessed and thus, e.g., infrared observations can measure the total accretion rate. However, to develop a physical model of the accretion flow, the X-ray diagnostics are by far the best. In this area, the improvements in observational capabilities offered by *Lynx* are drastic. The current capabilities of the *Chandra* gratings have only about 2.8 $cm^2$ of collecting area at OVIII, a key ion. The effective area of *Lynx*/XGS (~4,000 $cm^2$) means that, statistically, **a 1 ks *Lynx* observation will surpass a 500 ks *Chandra* observation.** *Chandra* has obtained spectra for only a handful of T Tauri stars in its 20 years of operation. *Lynx* will extend our reach to the Taurus-Auriga star forming region and even Orion, greatly extending our scientific grasp into new age and mass ranges.

The high-resolution X-ray spectra obtained with *Lynx* gratings will provide multiple temperature and density diagnostics required to study the accretion flow and separate out the coronal emission expected to originate at lower densities and multi-temperature (including very hot, tens of MK) plasma. An example of such a study was performed by Brickhouse et al. (2010) using a deep *Chandra* spectrum of the young accreting star TW Hya X-ray spectra (**Figure D-7**). They showed that the coronal emission is fed by the accretion process. This mechanism is very unlike the origin of stellar coronae on dwarf stars or stars without disks, and suggests a fundamentally different stellar dynamo in accreting systems compared to non-accreting systems.

For such studies, *Lynx* will be able to go well beyond traditional density diagnostics, such as the helium-like triplets of OVII, Ne ix, and Mg xi. The density sensitivity of Fe L-shell transitions is a good independent density diagnostic, but detection of these lines requires high sensitivity and high spectral resolution, only available with *Lynx*. Their use enables density constraints at hotter temperatures than probed by the helium-like triplets, enabling a better separation between accretion and corona. The ability to velocity-resolve the accretion contribution from the coronal contribution in lines with temperature sensitivity in both components is a key advance for a next-generation X-ray spectrograph. Line broadening allows determination of turbulent velocities, an important constraint on accretion theories. Brickhouse et al. (2010) derived a turbulent velocity of 165±18 km $s^{-1}$ in TW Hya. *Lynx* spectral resolution is expected to be at least 5× *Chandra*'s, leading to the corresponding improvement in our ability to probe the velocity structure.

Observations of TW Hya reveal a different absorption column for Ne ix compared to OVIII. Since the accretion streams are absorbing the emission in our line of sight, it is important to measure the series lines—He α, He β, and He γ—to measure the column density and rule out a resonance scattering interpretation. Since the higher order lines are factors of ten or more weaker than the strongest resonance lines, this work requires high spectral resolution.

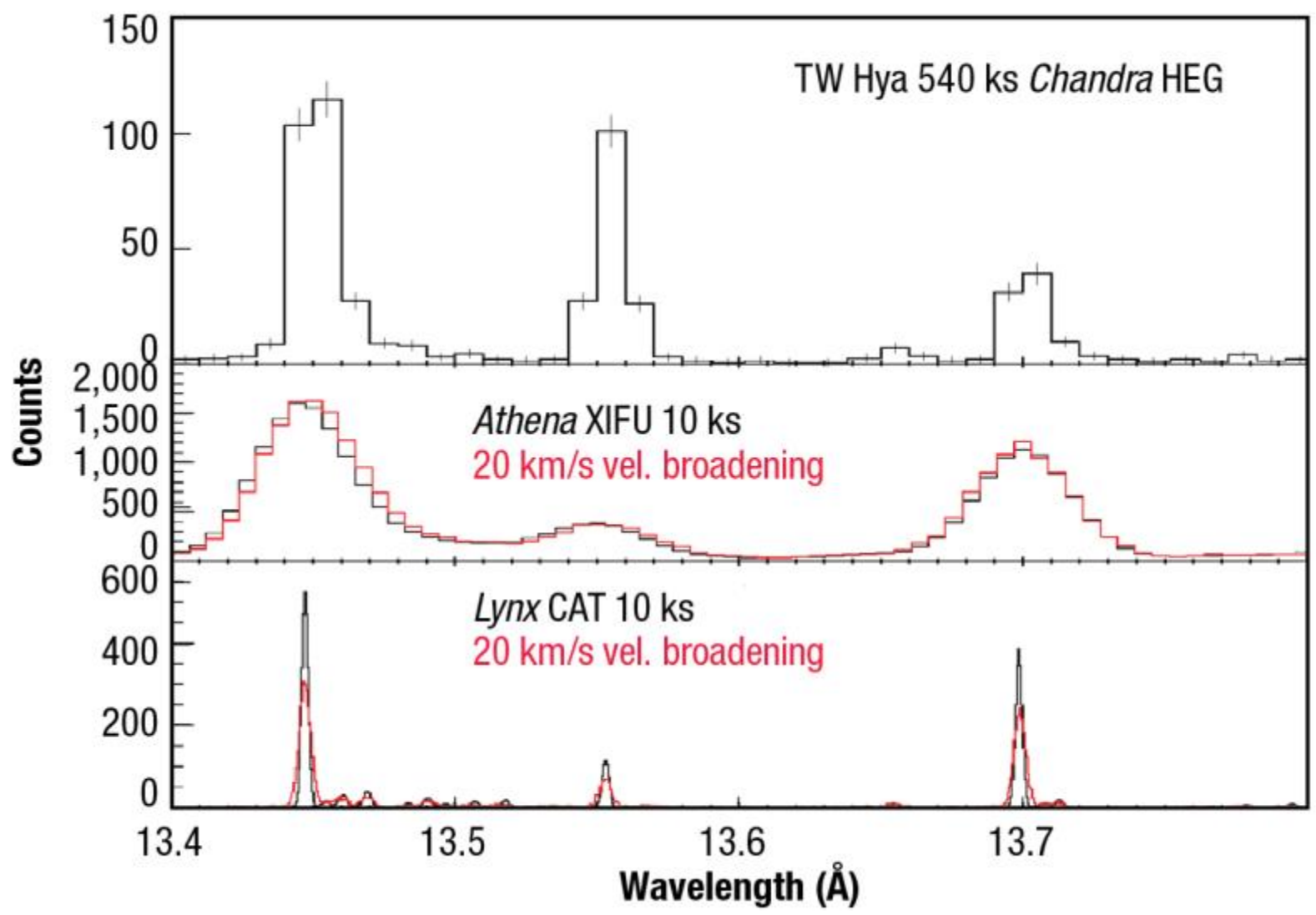

**Figure D-7. Top: *Chandra* High Energy Grating (HEG) spectrum of the accreting T Tauri star TW Hya, from a combined 540 ks of publicly available data (TGCAT; Huenemoerder et al. 2011) in the region of the Ne IX triplet. Brickhouse et al. (2010) analyzed 489 ks of data from this object, finding evidence for the expected accretion shock- and stellar coronal-related emissions, in addition to a third region of the stellar atmosphere: a very large extended volume of warm post-shock plasma. Bottom panels: Simulated *Athena* X-ray Integral Field Unit (XIFU) and a *Lynx* XGS spectrum, respectively, obtained in 10 ks each of exposure time. The red curves use the same input model but include an additional 20 km $s^{-1}$ velocity broadening of the emission lines, as would be expected from turbulent broadening associated with the shock. This difference is clearly visible with the R = 5,000 resolution grating of the XGS. In contrast, the *Athena* spectrum has numerous blends in the neighborhood of the strong lines, and the resolution is insufficient to establish a local continuum needed for accurate line flux determinations).**

For studies on corona emission, the multiple temperature and density line diagnostics provided by *Lynx* will enable measurements of the coronal emission measure distribution in the wide temperature range for stars of different mass and age. The same type of measurements is typically used for remote sensing of the Solar corona. Being able to relate the processes occurring on young stars to those of our closest star, the Sun, is vitally important for understanding our own astrophysical origins and how we trace back the Sun in time. Our current knowledge of thousands of exoplanet hosts, coupled with future, even grander expectations for exoplanet discovery, needs to be connected with understanding of our astrophysical origins.

### D.4.2 Stellar Activity and its Impact on Planet Habitability

The existence and structure of the outer atmospheres of stars in the low-mass half of the Hertzsprung-Russell (HR) diagram are among the most pressing issues still unsolved in stellar evolution. For stars in the cool half of the HR diagram, stellar coronal activity and flaring are ubiquitous. Understanding the origin of stellar coronal emission is important for our knowledge of the stars, as well as for probing the environments the stars create around them. X-ray observations have not been tools for exoplanet studies until now, but will be in the future as a more refined understanding of the impact the star has on its near stellar environment is needed. For instance, recent work by Garraffo et al. (2016, 2017) using magnetospheric modeling of M dwarfs has demonstrated the impact on habitability of close-in exoplanets. This may require a revision to the standard definition of a habitable zone based on severe space weather induced by Alfvén surface crossing for close-in exoplanets. While it is currently an open case whether the best candidates for habitable planets will be around G, K, or M stars, it is clear that understanding the role of the star's ionizing radiation, stellar winds, and energetic particles will be of central importance in assessing the case for habitability.

While it is nearly certain that the activity of a host star can affect the planet's habitability, our current understanding of this issue is very incomplete. The role of *Lynx* will be to: 1) dramatically improve our knowledge of the stellar XUV flux and its evolution, and 2) constrain the power of stellar winds and coronal mass ejection (CME) events.

The fundamental problem with assessing the effects of stellar XUV emission is that the strongest impact on the planet atmospheres is due to the extreme UV (EUV) emission, which is often difficult to observe directly due to absorption. High resolution observations with *Lynx* gratings will resolve triplet, satellite, and dielectric recombination lines from N, O, Ne, Mg, Fe (K, L, and M-shells). These X-ray measurements can be used to project a large portion of EUV flux (Drake et al. 2017).

Additional handles on the relation between the EUV flux and easily observed X-rays will be provided via an improved, physically-motivated models of stellar coronae. X-rays trace coronal structure directly: current state-of-the-art models use reconstructions of photospheric magnetic field to extrapolate to what the coronal magnetic field distribution will look like (e.g., Donati and Landstreet 2009 and references therein), or use dynamo simulations to predict what the large-scale structure of the corona should look like (Cohen et al. 2017). The ability to provide detailed coronal constraints, beyond an X-ray flux, to temperatures, densities, abundances, and velocities, are crucial observational inputs to constraining models predicting large-scale coronal structure, and elucidating smaller scale, active region-size coronal emission (Hussain 2012).

The impact of stellar XUV emission cannot be understood without taking into account its variability and evolution. The XUV flux generally scales with stellar rotation, except in the saturated regime at early ages, and since rotation decreases with time, this is a function of time as well. A star's X-ray emission evolves much more steeply at early ages than other wavelengths, indicating that it is a much larger factor in planetary irradiation at early times than later. Whether a planet retains its atmosphere after a range of evolutionary time depends fundamentally on the host star's rotational evolution and thus, nonlinear change in XUV flux with time.

Magnetic reconnection flares are an ubiquitous occurrence among stars near the main sequence with an outer convection zone. The dramatic impulsive increases in X-ray luminosity (factors of up to thousands; Osten et al. 2010, 2016) can significantly increase the irradiation planets experience, particularly at early times when the star's overall magnetic activity is enhanced. Young solar analogs experience stellar superflares $10^4$ times more energetic than the largest solar flare at a rate of at least once a week (Wolk et al. 2005). The rate of stellar flaring varies with the star's age, and it affects the energetic particle flux the planet experiences. A detailed study of how the characteristics of stellar flares change with time, in particular, maximum X-ray temperature and X-ray luminosity, are necessary ingredients for realistic upgrades to the current models of the evolution of planetary irradiation with time. Clarifying the role of energetic particles in stellar flares not only confirms that stellar flare processes really are analogs of solar flares, it also provides powerful constraints on the impact of flares on exoplanetary conditions, due to the severe impact implied by scaling from solar results (Segura et al. 2010). *Lynx* will obtain the necessary data via monitoring thousands of stars in its surveys of young star forming regions, and via detailed spectroscopy of M-dwarfs.

Measurements of magnetic structure will also contribute to development of the coronal models. Stars in the saturated regime may have different magnetic dynamos at work than in the unsaturated regime (c.f. work of Wright et al. 2011). Different dynamos have different magnetic field configurations, different XUV flux levels and configurations of coronal temperature and density, and different stellar wind properties. Recent results have shown that stellar twins are not magnetic twins (Kervella et al. 2017, Barnes et al. 2016, Kochukhov and Lavail 2017, Lynch et al. 2017). The detailed magnetic-field distributions cannot be predicted a priori from determination of a star's spectral type, mass, radius, rotation period, nor even integrated magnetic field strengths. Coronal properties flow from magnetic properties, as the high energy coronal emission is produced by plasma trapped in closed coronal loops.

Even if the XUV irradiation is precisely known, its effect on the planet's atmosphere is hard to predict from first principles. Models show the atmospheric loss rate probably has a highly non-linear dependence on the incident stellar XUV flux (Johnstone et al. 2015). *Lynx* will provide a unique empirical calibration of the atmospheric mass loss via transit X-ray spectroscopy. The transit of the hot Jupiter HD189733b was detected through X-ray absorption by oxygen in *Chandra* observations by Poppenhaeger et al. (2013), who found that the scale height of X-ray absorbing gas was higher than suggested by optical and UV transits. Hot Jupiters and similar giant close-in planets are important for improving theory and models describing atmospheric loss. X-ray absorption measures gas *bulk chemical composition* along the line-of-sight—in this case in the transitioning exoplanet atmosphere backlit by the host star's corona. Such measurements are unique to the X-ray range, but only the very closest hot Jupiters are accessible with *Chandra* and *XMM-Newton*. *Lynx* will be able to observe HD 189733b-like transits out to 140 pc, a factor of more than 300 improvement in survey volume over current missions. Combination with optical/IR data will provide a powerful probe for clouds and hazes that can confuse IR spectroscopic analyses (Sing et al., 2016). By co-adding observations of many transits, the *Lynx* calorimeter could also open such studies to larger habitable planets, such as super Earths around nearby M dwarfs (see **Figure D-8**).

Stellar winds originate from open magnetic field lines threading the corona. Stellar wind mass loss is also critical to planetary atmospheric escape. Results in our own solar system have demonstrated the impact of the solar wind on Mars' early evolution (Lammer et al. 2003), with photo dissociation of atmospheric water, followed by charge exchange between solar wind ions and atmospheric constituents, and sputtering carrying off a significant amount of water in the early history of Mars. So far, we have only indirect measures of mass loss in the cool half of the HR diagram. The results of Wood et al. (2005, 2010) appear to indicate a rise of stellar wind mass loss with increasing surface X-ray flux ($dM/dt \sim F_X^{1.35}$), up to a point beyond which much lower values of wind mass loss are inferred. The so-called wind dividing line implies very different conditions experienced by the early Sun, which is not able to be extrapolated from the star's surface X-ray flux. This mass loss is critical to the atmospheric escape process when the stellar winds are high. CMEs are the unsteady component of the Sun's mass loss, and while for the Sun this contributes <10% of the total stellar mass loss, the rate of CMEs has an important role in the potential habitability of exoplanets (Lammer et al. 2007). CMEs are poorly constrained in stellar environments, with few observational tools to perform systematic diagnoses of their occurrence and properties (Osten and Wolk, 2017). One exception may be the unique flare of Proxima reported in (Haisch et al. 1983), wherein increases in absorbing column N$H$ during the flare were inferred from a decrease in X-ray flux at low energy. If this interpretation is correct, the deduced N$H$ $\sim 10^{20}$ $cm^{-2}$ would be consistent with a solar-type CME with scale about a stellar radius expanding at 500 km $s^{-1}$.

Direct detections of steady stellar winds, or constraints provided by upper limits thereto, are a crucial missing ingredient to the recipe for how stars influence their environment. Charge exchange emission from the interaction of the stellar wind with the local interstellar medium provides one possible observational route. Wargelin and Drake (2001) provided an upper limit to the mass loss rate of Proxima using the expected signature of a cloud of charge exchange emission from interaction of the stellar wind with the local interstellar medium. Current observational efforts to provide a routine detection method for stellar CMEs at other wavelength regions are coming up short (e.g., Crosley et al. 2016, Crosley, and Osten 2018) and require more sophisticated observing tools to make progress.

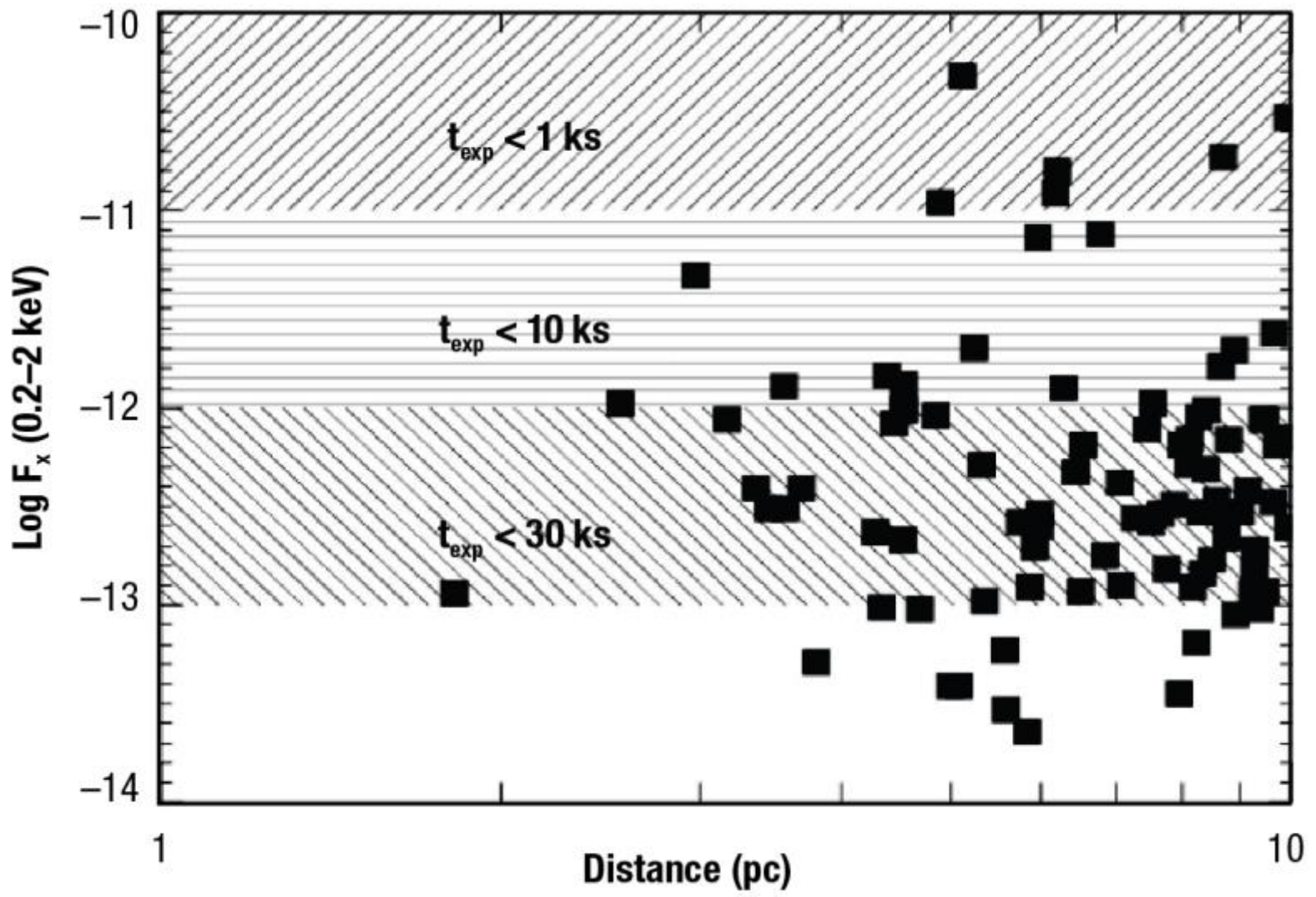

**Figure D-8. Grasp of *Lynx* grating spectroscopy for M dwarfs in the solar neighborhood. X-ray fluxes for M dwarfs taken from Stelzer et al. (2013); for 96 of 163 M dwarfs within 10 pc, there is an X-ray counterpart. Simulations of expected counts in the bright OVIII line at 19Å, both with and without a velocity broadening of 20 km/s, for typical coronal parameters, reveal the exposure time required to be able to deduce coronal structures via line broadening analysis. Exposure times for flux levels of $10^{-14}$ erg cm$^{-2}$ s$^{-1}$ are 80–100 ks. The number of M dwarfs with X-ray fluxes below $10^{-13}$ erg cm$^{-2}$ s$^{-1}$ is likely incomplete due to sensitivity limitations of the Roentgen Satellite (ROSAT) All-Sky Survey. A total of 77 M dwarfs have X-ray fluxes greater than $10^{-13}$ erg cm$^{-2}$ s$^{-1}$, and their spectra could be surveyed in a systematic fashion for line broadening with the *Lynx* observatory in an exposure time less than about 1.85 Megaseconds.**

### D.4.3 Endpoints of Stellar Evolution: Supernova Remnants

Supernovae (SNe) play an essential role in the Universe. Metals synthesized during the explosion chemically enrich galaxies, supplying fodder for dust and the next generation of stars (e.g., Woosley et al. 2002, Fukugita and Peebles 2004). Their shock waves plow through the ISM for thousands of years, accelerating particles to extreme energies (~$10^{15}$ eV) and amplifying magnetic fields up to a hundred times that of the ISM (see e.g., Reynolds 2008). The shocks also heat surrounding gas and impart momentum, altering the phase structure of the ISM (McKee and Ostriker 1977), shaping galaxies (e.g., Governato et al. 2010), and driving kpc-scale galactic winds (e.g. Veilleux et al. 2005).

Although hundreds of SNe are found each year at optical wavelengths by dedicated surveys (e.g., Law et al. 2009, Leaman et al. 2011, Holoien et al. 2017), they are often too distant to resolve the SN ejecta and the immediate surroundings of the exploded stars (e.g., 1 arcsec corresponds to ∼50 pc for distances of 10 Mpc). Studies of the closest SNe, such as SN 1987A (McCray and Fransson 2016), have advanced the field tremendously, but our understanding of SN progenitors and details of the explosion dynamics are hampered by the infrequency of nearby events.

Young supernova remnants (SNRs) offer the means to study SN explosions, dynamics, and shocks at sub-pc scales, and they are an important tool to explore the relationship between compact objects and their explosive origins. Consequently, observations of SNR morphologies, kinematics, and chemical abundances are crucial to test and constrain three-dimensional SN simulations (e.g., Lentz et al. 2015, Janka et al. 2016, Roberts et al. 2016, Müller et al. 2017). SNRs are observable across the electromagnetic spectrum for up to $10^5$ years after the explosions, and over 500 SNRs have been identified in the Milky Way and Local Group galaxies (e.g., Badenes et al. 2010, Sasaki et al. 2012, Green 2014, Maggi et al. 2016, Garofali et al. 2017). Metals synthesized during explosions are shock-heated to $\sim 10^7$ K temperatures, and thus X-ray telescopes are the only means to probe the bulk of SNR ejecta material.

### D.4.3.1 Breakthrough Progress with *Lynx*

SNRs are optimal targets to explore long-standing open questions in stellar evolution, stellar environments, and supernova explosions. In the context of core-collapse SNe (explosions marking the ends of the lives of stars >8 Msun), perhaps the biggest issue being debated is the mechanism of the explosion. Several scenarios have been proposed, such as neutrino-driven convection (Herant et al. 1994, Burrows et al. 1995, Kifonidis et al. 2000, Blondin and Mezzacappa 2006, Foglizzo et al. 2006, Scheck et al. 2008) or magnetically-driven explosions (Burrows et al. 2007). Type Ia SN progenitor systems (see review by Maoz et al. 2014) have also not been identified definitively, i.e. whether they arise from white dwarfs with non-degenerate companion stars (the single-degenerate [SD] scenario) or from two white dwarfs (the double-degenerate [DD] scenario). Finally, the connection between compact objects (neutron stars [NSs] and black holes) and explosions is also uncertain. For example, what determines whether a Core Collapse (CC) SN produces a NS or a black hole (Sukhbold et al. 2016)? Are black holes only produced in failed explosions, or do some SNe yield black holes (Heger et al. 2003, Adams et al. 2017)?

As high-fidelity SN simulations incorporate more physics, SNRs will be the primary means to test and improve models (e.g., Wongwathanarat et al. 2013, 2017). X-ray observations are crucial as they probe the heavy nucleosynthetic products of the explosions, and they can detect the thermal and non-thermal emission from young neutron stars. Elucidation of the three-dimensional structure of SNR ejecta, including the kinematics and localization of intermediate-mass (O, Ne, Mg) and heavy metals (Si, S, Ar, Ca, Cr, Mn, Fe), will constrain how elements produced by different burning processes are ejected in explosions. Abundance ratios and plasma properties (ionization state, temperature, density) will be important tools to assess the dynamical state of the SNRs, shock-heating, and the progenitor stars' natures. For breakthrough progress, these measurements must be done for a large sample (dozens to hundreds) of SNRs to explore the diverse classes of explosions and how they depend on e.g., galactic environments.

Although *Chandra*, *XMM-Newton*, and *Suzaku* have advanced our understanding of SNe and SNRs tremendously, certain limitations of these facilities have precluded breakthrough progress in several areas. Most notably, a X-ray microcalorimeter with arcsecond spatial resolution and eV energy resolution is necessary to map ejecta metals in three dimensions in the Milky Way and Local Group SNRs. CCD energy resolution is insufficient to resolve He-like and H-like line complexes of ions at X-ray wavelengths, and grating spectrometers are only useful when objects have minimal angular extension (e.g., SN 1987A: Dewey et al. 2008; SNR 1E 0102.2−7219: Flanagan et al. 2004) or the observation focuses on individual bright, isolated knots (e.g., Cassiopeia A: Lazendic et al. 2006; G292.0+1.8: Bhalerao et al. 2015).

Before its untimely demise, *Hitomi* observed the Large Magellanic Cloud (LMC) SNR N132D for 3.7 ks (Hitomi Collaboration et al. 2017), and it readily detected line complexes of S, Ar, and Fe. The Fe emission was highly redshifted at arcminutes 800 km $s^{-1}$, but no blueshifted component was detected, suggesting that the Fe-rich ejecta was ejected asymmetrically. N132D is not especially asymmetric in morphology (e.g., Lopez et al. 2011), yet the addition of the third dimension of data reveals a bulk asymmetry that was not evident in two dimensions. Thus, this brief *Hitomi* observation was a tremendous proof of the power of a microcalorimeter in SNR studies.

The capabilities of the *Lynx* microcalorimeter are required to produce three-dimensional maps of ejecta in dozens of young SNRs in the Milky Way (including Cas A, G292.0+1.8, G11.2−0.3, G15.9+0.2, RCW 103, Kes 79, W49B, RCW 86, Tycho, Kepler, SN 1006, 3C 397, G344.7−0.1, G272.2−3.2). These maps can be compared directly to the predictions of three-dimensional SN models. In addition, SNR expansion and NS velocities can be measured easily with the sub-arcsecond resolution of *Lynx* via proper motions and previous *Chandra* imaging: e.g., a velocity of v = 1,000 km $s^{-1}$ corresponds to 0.4 arcseconds in a time baseline of 10 years for a distance of D = 5 kpc.

*Lynx*'s sensitivity is necessary to study in detail large extragalactic populations for the first time. Particularly, in the LMC and Small Magellanic Cloud (SMC), imaging and microcalorimeter observations will reveal the three-dimensional structure, ejecta mass and composition, shock-heating properties, particle acceleration characteristics, and possibly associated NSs. In the LMC and SMC, *Lynx*' spatial resolution is crucial to spatially-resolve ejecta knots and CSM clumps or the angular extent of sources generally (e.g., SN 1987A will be ~3 arcseconds across in the 2030s; Orlando et al. 2015). In M31 and M33, SNRs have sizes of ~3–30 arcseconds (~10–100 pc); *Lynx*'s capabilities are required to obtain >1,000 net counts in a 100 ks HDXI exposure for several dozen sources, sufficient to do detailed spectral modeling and characterize the hot plasma properties (temperature, ionization timescale, and metal abundances). Generally, *Lynx* will be able to resolve and obtain spectra from most SNRs within 1 Mpc, and this sample will be large enough to do statistical comparison between populations, e.g. compare SNR properties between galaxies, in different galactic environments and metallicities.

Identifying NSs at the centers of the Milky Way and some extragalactic SNRs requires the capabilities of *Lynx*. In particular, the microcalorimeter will have sufficient energy resolution and sensitivity to detect lines from NS chromospheres in Galactic SNRs. By comparison, with *Chandra* observations, it is difficult to distinguish whether a coincident point source is a NS or merely a background object. Additionally, *Lynx* will find or set limits on Cassiopeia A-like NSs (called central compact objects (CCOs)) in Magellanic Cloud SNRs using modest ('50 ks) exposures. To date, only one possible detection of thermal emission of a CCO has been reported in a Magellanic Cloud SNR, 1E 0102.2−7219, using a ~390 ks *Chandra* exposure (Vogt et al. 2018).

### D.4.4 Endpoints of Stellar Evolution: Compact Objects

X-ray binary (XRB) populations are shaped by stellar and binary evolution, so observations can test key model ingredients such as common envelope phase and SN kicks. Evolutionary paths for some classes of *LIGO* sources, such as NS-NS mergers, almost certainly pass through a high-mass X-ray binary stage. *Chandra* observations of nearby galaxies mostly probe the upper end of the XRB luminosity function, and therefore lack the range to discriminate among various models. *Lynx*'s capabilities are required to reach $L_X$ ~100× fainter than *Chandra*, providing the needed dynamic range. For a given $L_X$, *Lynx* will cover ~1,000× greater volume, increasing the number of galaxies to build up samples of rare objects. In addition to testing the evolutionary paths to *LIGO* sources, the increased statistics of the X-ray binary populations in nearby galaxies will allow us to address the following outstanding questions:

- Currently, there is an observational "gap" between the masses of neutron stars and black holes in X-ray binaries. This is not predicted by current stellar evolution theory, so it may be due to selection effects. Probing down to low luminosities ($L_X \sim 10^{34}-10^{35}$ erg $s^{-1}$) is critical for detecting flares from quiescent black holes and neutron stars in nearby galaxies to determine the true mass distribution of the compact objects. This will then refine theories on how black holes form from SN explosions.

- While the X-ray luminosity of X-ray binaries in galaxies known to be proportional to SFR (for high-mass X-ray binaries with lifetimes of ~$10^7$ years) and to stellar mass for low-mass X-ray binaries (with lifetimes of ~$10^9$ years), star formation history and metallicity are also crucial factors. Namely, at low metallicities stellar winds are not as strong, allowing for higher-mass stars and black holes to be formed. Thus, $L_X$ is inversely proportional to metallicity. Also, the expected X-ray luminosity of XRBs is a function of the age of the system as the amount of mass transfer from the companion star varies. Both of these effects result in young, low-metallicity systems being much more X-ray bright than older, higher-metallicity systems. X-ray binary populations in nearby galaxies studied today by *Chandra* tend to be older, higher-metallicity systems with the exception of low-metallicity dwarf galaxies that do not have sufficiently high SFR and stellar mass to have a detectable number of XRBs. *Lynx*'s capabilities are required to detect the X-ray binary populations in (relatively) nearby galaxies with conditions similar to early universe galaxies, e.g., Lyman-break Analogs, where currently *Chandra* and *XMM-Newton* are only probing the total X-ray luminosity and Ultraluminous X-ray (ULX) populations.

Long-term monitoring of the kilonova GW170817 with *Chandra* has proven instrumental for constraining the possible outcome of this remarkable event. *Lynx* will provide a similar capability matching the increased sensitivity of the future ground-based gravitational wave detectors.

## D.5 General Observatory Program

The science in the three pillars described above has both great depth and breadth. While aiming at three over-arching topics, it covers broad swaths of astronomy on all distance scales. We expect that most of this science will be accomplished via competed, peer-reviewed General Observer (GO) programs. Moreover, the capabilities of *Lynx* enable a dynamic and exciting GO program that will have a significant impact and enable discoveries in all areas of astronomy and astrophysics. Virtually all astronomers will be able to use *Lynx* for their own particular science. In the Final Report, we will provide much more information illustrating potential GO programs beyond these three science pillars.

## D.6 Science Traceability Matrix

The *Lynx* science traceability matrix (**Foldout FO1**) has been derived using the performance requirements for observing programs required to execute the science in the three main pillars. Listed in the matrix are the main drivers for each of the major performance requirements. For reference, we also provide a list of other core science programs relying on each of the capabilities, but typically with less demanding requirements. Additional notes on the mirror and science instruments requirements are given below.

**Mirror—**The prime science driver for *Lynx* angular resolution is detecting early supermassive black holes at the seed stage or soon after. The required X-ray sensitivity is $\approx 10^{-19}$ erg s$^{-1}$ cm$^{-2}$. To avoid source confusion at these fluxes, and to uniquely associate detected X-ray sources with *JWST* and *WFIRST* galaxies requires an angular resolution of 0.5 arcseconds (HPD), and better than 1 arcsecond (HPD) across the FOV used for sensitive surveys.

Many *Lynx* programs need a large FOV with sub-arcsecond imaging. Here, the *Lynx* requirement is better than 1 arcsecond (HPD) PSF maintained to an off-axis radius of at least 10 arcminutes.

The mirror effective area, 2 $m^2$ at $E$ = 1 keV, is sized such that the core program required for the three main pillars can be executed (mainly via GO proposals) in ≈50% of the observing time in a nominal five-year mission (see section **§D-6** and **Figure D-9**).

**HDXI—**The High-definition X-ray imager needs pixels < 0.33 arcseconds to adequately sample the mirror PSF and a FOV ≥ 20 arcminutes × 20 arcminutes to meet the requirement for surveys and imaging objects with large angular extent.

**LXM—**The *Lynx* X-ray microcalorimeter will provide non-dispersive spectroscopy, required by many core science programs and much of the general observatory science. The main array needs < 3 eV energy resolution over the 0.2–7 keV band, and imaging with 1 arcsecond pixels over a 5 arcminutes × 5 arcminutes FOV, driven equally by requirements of the SNR, galaxy cluster, and ISM observations.

Major components of the core program, especially energy feedback studies, will require even more demanding specialized capabilities. These will be achieved by introducing two additional subarrays to the LXM.

The enhanced imaging subarray with 0.5 arcseconds pixels, 1 arcminutes × 1 arcminutes size, and an energy resolution of 2 eV over the 0.2–7 keV band is required to characterize the effects of AGN feedback on the ISM of host galaxies and to measure the physical state of gas near the SMBH sphere of influence. The ultra-high spectral resolution subarray with 1 arcsecond pixels and 1 arcminute × 1 arcminute size will provide 0.3 eV energy resolution in the 0.2–0.75 keV band required, e.g., for studies of supernovae-driven galaxy winds and density diagnostics in AGN outflows.

**XGS—**A spectral resolution higher than what is achievable with X-ray microcalorimeter technology in the foreseeable future is required for absorption-line studies of diffuse baryons in galactic halos and in the cosmic web, physics of stellar coronae, and assessing the impact of stellar activity on habitability of their planets. This capability will be provided by the X-ray Grating Spectrometer. The XGS applications are broad but the performance requirements are driven by absorption line studies of diffuse baryons in galactic halos. Approximately 4,000 $cm^2$ of mirror plus gratings effective area is required at the astrophysically important X-ray lines in the 0.2–2 keV band, especially OVII and OVIII absorption lines. A resolving power of $R$ ≈ 5,000 is needed to match the expected thermal line widths, and $R$ ≈ 10,000 is desired.

We present in D7 a notional observing plan, following the Science Traceability Matrix, that shows exposure times for the core programs required to execute the three major science pillars for a 1 keV effective area of 2 $m^2$. We also discuss how the exposure times motivate the choice of the effective area.

X-RAY OBSERVATORY LYNX

# Science Traceability Matrix

# FO1

Table F01-1

| Science Goal | Science Objective | Key observations and physical parameters | Measurement requirements | Mirror and Instrument Requirements: Instrument | Mirror and Instrument Requirements: Property | Mirror and Instrument Requirements: Value | Mission Functional Requirements |
|---|---|---|---|---|---|---|---|
| **See the Dawn of Black Holes** | **Observe progenitors of supermassive black holes at their seed stage at z=10** | Detection of black holes in z=6–10 galaxies down to a mass limit of Mlim=10,000 Msun over a volume with $10^3$–$10^7$ potential host galaxies | Surveys with flux limits [0.5–2 keV]:<br>• $1.6\times10^{-19}$ erg/s/$cm^2$ over 1 $deg^2$<br>• $7\times10^{-20}$ erg/s/$cm^2$ over 400 $arcminutes^2$ | Mirror+ HDXI | Angular Resolution (HPD) | <1 arcseconds across the field | Operate and survive in the science orbit, with a minimum observing efficiency of 85%, for the duration of the 5-year mission |
| | | | | | Grasp @ 1 keV | ~600 $m^2$ $arcminutes^2$ | |
| | | | | | Imager pixel size | 0.33 arcseconds | |
| **Reveal Invisible Drivers of Galaxy and Structure formation** | **Determine the state of diffuse baryons on galactic halos to guide the galaxy formation models** | Direct imaging observations of 15 low-z galaxies with $M{\sim}3\times10^{12}\ M_{sun}$ | Reach 10% accuracy for derived thermodynamic parameters of hot gas at 0.5 $r_{200}$ | Mirror+ HDXI | Effective Area @1 keV | 2 $m^2$ | Accommodate payload in launch vehicle |
| | | | | | Field of view | 10 arcminute radius | |
| | | | | | Spectral resolution (FWHM) @ 1 keV | 60 eV | Provide data collection that is sufficient for uninterrupted observations by all science instruments |
| | | | | | Particle background @ 0.5–2 keV | < 0.0005 cnt/s/$arcmin^2$/keV | |
| | | Characterization of hot halos beyond the virial radius in 60 galaxies with mass $10^{12}$–$10^{13}\ M_{sun}$ at z=0-1 | Observe 80 bright AGN sight lines to reach the sensitivity of 1 mA for OVII and OVIII absorption lines | XGS | Spectral Resolving Power | 5,000 | |
| | | | | | Effective area @ 0.3–0.7 keV | 4,000 $cm^2$ | Provide pointing attitude control and knowledge consistent with sub-arcsecond imaging, and stability consistent with 1 arcminute FOV |
| | **Establish the Energetics, Physics, and the Impact of Energy Feedback on Galactic Scales** | Spatially and spectrally resolve the structure of starburst-driven winds in low-redshift galaxies | Measure the outflow velocity profile in 20 galaxies with 100 km/s accuracy, and derive the momentum and energy flux | LXM / ultra high-resolution subarray | Spectrometer pixel size | 1 arcsecond | |
| | | | | | Energy resolution (FWHM) @ 0.2–0.75 keV | 0.3 eV | |
| | | | | | Spectrometer subarray size | 1 arcminute × 1 arcminute | |
| | | Determine the effects of AGN energy feedback on ISIM, and determine the physical state of gas near the SMBH sphere of influence in nearby galaxies | In 30 nearby galaxies, resolve extended emission line regions, AGN inflated bubbles, and characterize the thermodynamic state of gas with 10% precision at or close to the Bondi radius from the central black hole | Mirror + LXM / enhanced spatial resolution subarray | On-axis angular resolution | 0.5 arcseconds (HPD) | |
| | | | | | Spectrometer pixel size | 0.5 arcseconds | |
| | | | | | Energy resolution (FWHM) @ 0.6–7 keV | 3 eV | |
| | | | | | Spectrometer subarray size | 1 arcminute ×1 arcminute | |
| **Unveil the Energetic Side of Stellar Evolution and Stellar Ecosystems** | **Constrain SN explosion physics, origin of elements, and a relation between SN activity and local environment** | Survey of young SNR in the Local Group galaxies | Measure spatial structure of SNRs in spectral lines of individual elements, and in non-thermal emission | LXM / main array | Spectrometer pixel size | 1 arcsecond | |
| | | | | | Spectrometer field of view | 5 arcminutes × 5 arcminutes | |
| | | | | | Energy resolution (FWHM) @ 0.6–7 keV | 3 eV | |
| | | | | | Effective area @ 6 keV | 1,000 $cm^2$ | |

**See the Dawn of Black Holes**

Massive black holes start to form as early as their host galaxies. *Lynx* will find the first supermassive black holes in the first galaxies detected by *JWST*, trace their growth from the seed phase, and shed light on how they subsequently co-evolve with the host galaxies. These young black holes are best observed in the X-ray band. Reaching into the seed regime in the early Universe requires sensitivities ~$10^{-19}$ erg/s/$cm^2$, which only *Lynx* can achieve.

**Reveal Invisible Drivers of Galaxy and Structure formation**

The assembly, growth, and state of visible matter in cosmic structures are largely driven by violent processes that produce and disperse large amounts of energy and metals into the surrounding medium. In galaxies at least as massive as the Milky Way, the relevant baryonic component is heated and ionized to X-ray temperatures. Only *Lynx* will be capable of mapping this hot gas around galaxies and in the Cosmic Web, as well as characterizing in detail all significant modes of energy feedback. Essential observations will require high-resolution spectroscopy (R~5,000) of background AGNs, the ability to detect low surface brightness continuum emission, and R~2,000 spectroscopy of extended sources on arcsecond scales—all unique to *Lynx*.

**Unveil the Energetic Side of Stellar Evolution and Stellar Ecosystems**

*Lynx* will probe with unprecedented depth a wide range of high-energy processes that provide a unique perspective on stellar birth and death, internal stellar structure, star-planet interactions, the origin of elements, and violent cosmic events. Lynx will detect X-ray emission as markers of young stars in active star forming regions, study stellar coronae in detail, and provide essential insight into the impact of stellar XUV flux and winds on habitability of their planets. Images and spectra of supernova remnants in Local Group galaxies will extend studies of stellar explosions and their aftermath to different metallicity environments. Lynx will expand our knowledge of collapsed stars through sensitive studies of X-ray binaries in galaxies as distant as 10 Mpc and detailed follow-ups of gravitational wave events. Lynx will greatly extend our X-ray grasp throughout the Milky Way and nearby galaxies by combining for the first time the required sensitivity, spectral resolution, and sharp vision to see in crowded fields.

## D.7 Notional Observing Plan for Core Science Program

**Table D-1** below shows exposure times for the core programs required to execute the science of the three major pillars, estimated for a mirror configuration with a soft-band effective area of $2m^2$. **Figure D-9** shows how the total exposure time scales with effective area.

**Table D-1. Notional core science programs.**

* Highlighted in bold are the programs determined to be performance drivers for the mission (see the Science Traceability Matrix in **FO1**)

| Science Theme | Program | Typical observations | Instrument | Total exposure |
|---|---|---|---|---|
| The Dawn of Black Holes | **1.1 Origin of SMBH seeds*** | Surveys over 1 $deg^2$ to depth fx = 1.6 x $10^{-19}$ [0.5–2 keV] plus a deeper survey over 400 $arcmin^2$ to fx =7 x $10^{-20}$ erg/s/cm2 | **HDXI** | 23 Msec |
| | 1.2 Growth of supermassive black holes from Cosmic Dawn through Cosmic Noon to the Present. Relation between AGN and environments. Triggering and quenching AGNs. Relationship to star formation activity | Survey down to fx = 2 x $10^{-18}$ over up to 2 $deg^2$ | **HDXI** | 2 Msec |
| The Invisible Drivers of Galaxy and Structure Formation | **2.1 State of diffuse baryons in galactic halos — direct imaging*** | Survey of ~ 15 low-redshift isolated (spiral) galaxies, pushing 10% thermodynamic (gas density) measurements to 0.5 r500 for M~3 x $10^{12}$ and to r200 for M~1 x $10^{13}$ | **HDXI** | 7.5 Msec |
| | **2.2 State of diffuse baryons in galactic halos — absorption line spectroscopy*** | Observe ~80 AGN sightlines (fagn ~ 1 x $10^{-11}$) to detect ~60 absorption line systems in the foreground galaxy halos. Detection limits for absorption lines are EW ~3mA and down to 1mA for r>r200. Same sightlines are used to characterize the gas in the Milky Way Halo. | **XGS** | 5 Msec |
| | 2.3 State of gas and feedback measurements in high-redshift galaxy clusters and groups | Gas temperature, density and metallicity profiles in ~30 clusters and groups at z>2 | **LXM** / main array | 6 Msec |
| | 2.4 Characterization of the first galaxy groups at z=3–4 | **HDXI observations of ~10 high-z galaxy groups** | **HDXI** | 2 Msec |
| | 2.5 Spectroscopic survey of AGN to determine energetics of the AGN feedback | Soft-band spectroscopy with R >1,000 down to 0.2 keV to measure density-sensitive spectral features. 3 Msec XGS, LSM / ultra-high-resolution array | **XGS, LXM** / ultra-high-resolution subarray | 3 Msec |
| | 2.6 Characterize the supply side of AGN energy feedback | Measure thermodynamic state of diffuse gas near the Bondi radius of SMBHs in nearby elliptical galaxies | **LXM** / enhanced spatial resolution subarray | 1 Msec |
| | **2.7 Measure the energetics and effects of AGN feedback on galactic scales*** | Observe AGN-inflated bubbles in the ISM of low-redshift elliptical galaxies | **LXM** / enhanced spatial resolution subarray | 2 Msec |
| | | Spectro-imaging of extended narrow emission line in nearby spiral galaxies | **LXM** / enhanced spatial resolution subarray | 2 Msec |
| | **2.8 Understand the energetics and mechanics of the supernovae-driven galactic winds*** | Observe galaxy winds in ~20 objects, with the ability to characterize velocities <100 km/s on arcsecond scales. 2.5 Msec LSM / ultra-high spectral resolution array | **LXM** / ultra-high-resolution subarray | 2.5 Msec |
| | 2.9 Galaxy cluster-scale feedback | LXM observations of nearby galaxy clusters to constrain plasma physics effects in the cluster cores | **LXM** / main array | 2 Msec |
| The Energetic Side of Stellar Evolution and Stellar Ecosystems | 3.1 Stellar coronal physics, impact of stellar activity on planet habitability, accretion on young stars | Spectroscopic survey of 80 stars within 10 pc | XGS | 2 Msec |
| | | Transit spectroscopy of planets around dwarf stars down to super-earth regime | XGS, LXM / ultra-high-resolution subarray | 1 Msec |
| | 3.2 Young forming regions | Surveys to detect entire mass distribution of stars in active star forming regions to d = 5 kpc | HDXI | 2 Msec |
| | **3.3 Endpoints of stellar evolution: SNRs*** | Targeted observations of the youngest SNRs in the Milky Way, up to ~50 objects | LXM / main array | 2 Msec |
| | | Statistics and typing of SNRs in different environments in nearby galaxies | LXM / main array | 1 Msec |
| | 3.4 Endpoints of stellar evolution: X-ray binary populations | Survey of X-ray binary populations and ISM in nearby galaxies. 2 Msec LXM | HDXI, LXM | 2 Msec |

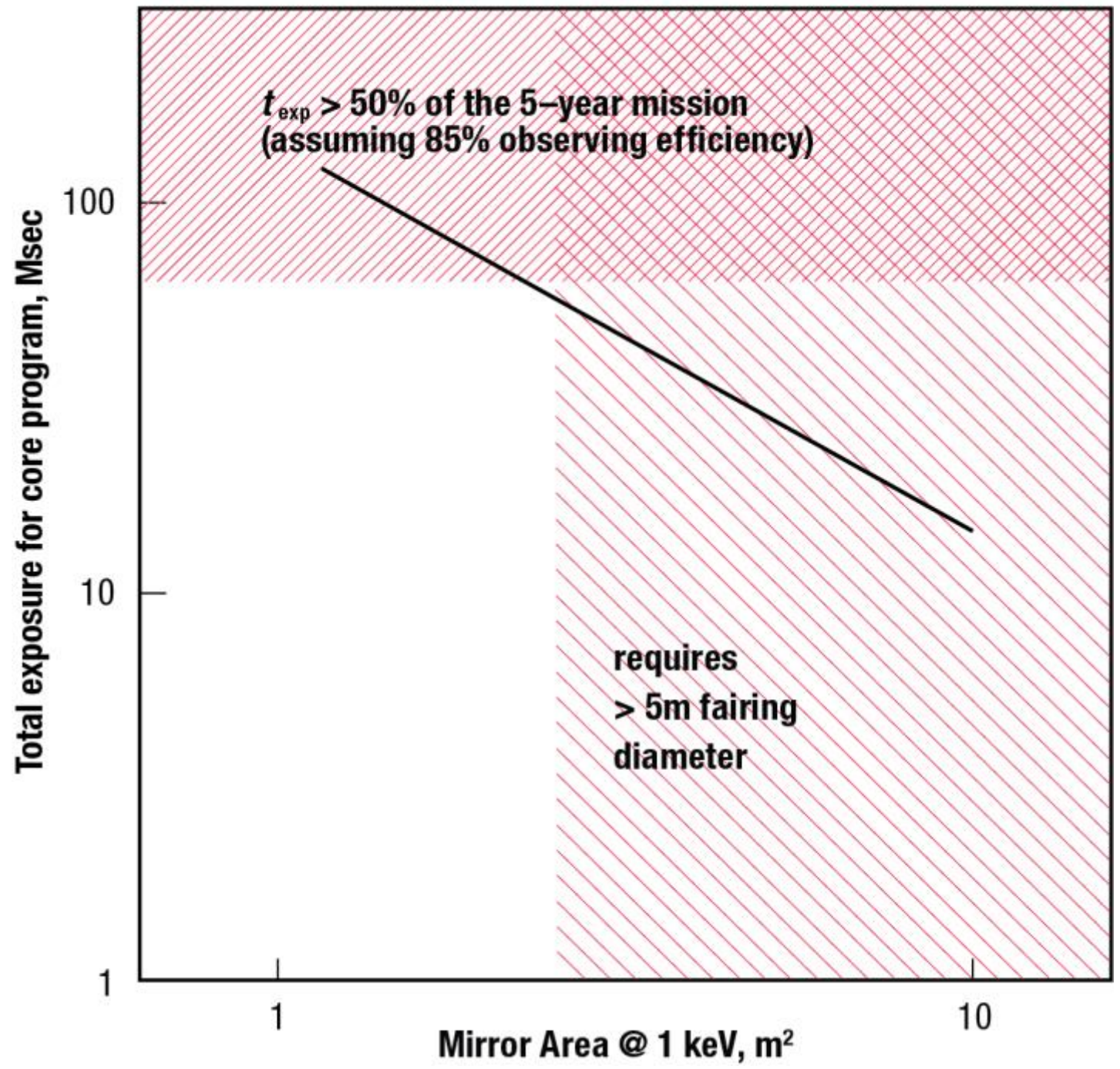

**Figure D-9. Total exposure time of the notional core science program, as a function of the mirror effective area at $E = 1$ keV. An effective area of 2 $m^2$ represents an appropriate design choice because such a telescope can fit inside the dynamic envelope of the standard 5-m fairings, while still leaving open room for a dynamic and exciting GO program that will have a significant impact and enable discoveries in all areas of astronomy and astrophysics.**

# E. THE *LYNX* DESIGN REFERENCE MISSION

The *Lynx* observatory will have unparalleled X-ray vision, necessary for revealing the evolution and physics of the cosmos. With leaps in capability over NASA's existing Flagship *Chandra* and ESA's planned *Athena* mission, *Lynx* will have a 50-fold increase in sensitivity via coupling superb angular resolution with high throughput, 16 times larger FOV with arcsecond or better imaging, and 10–20 times higher spectral resolution for both point-like and extended sources.

## E.1 Introduction to Design Reference Mission

Transformational science, such as that described in **§D,** can be carried out with a realizable mission that capitalizes on the use of heritage (systems, processes, and operation) and redundancy, and minimizes the number of technologies requiring development. The *Lynx* Design Reference Mission (DRM) details a notional payload and mission definition designed to flow down from the primary *Lynx* science objectives (**FO1**); namely to directly observe the dawn of supermassive black holes, reveal the drivers of galaxy formation, trace stellar activity including effects on planet habitability, and transform our knowledge of endpoints of stellar evolution. *Lynx* mission parameters are summarized in the Mission Traceability Matrix (**FO2**). The concept provides a path for optimizing observing time, achieving a stable observing environment, and meeting all spacecraft, ground system, and operational objectives.

*Lynx* is a Flagship-class X-ray observatory that will launch in the early- to mid-2030s on an expendable or a recoverable Heavy lift vehicle and will operate in a halo orbit around the SE-L2 libration point. The mission is baselined for a 5-year life, with consumables sized to extend the life to 20 years. *Lynx* will operate as a general observatory, allowing for a variety of observing programs that includes, but is not limited to the science detailed in the broad "pillars" outlined in the Science Traceability Matrix (**FO1**). Any data from directly allocated key projects, such as the deepest surveys, will be made available to the community on a non-proprietary basis.

## E.2 Observatory Overview

The observatory is comprised of the telescope and spacecraft elements. The telescope system includes a large area, sub-arcsecond grazing incidence X-ray mirror assembly, and three science instruments that together provide wide field imaging with high angular resolution and high spectral resolution for both extended and point-like sources. *Lynx* spacecraft systems are designed using high-Technology Readiness Level (TRL), low risk, and heritage components. The spacecraft provides station-keeping and attitude control for stable orbital operations, on-board autonomous safing activities, power for all phases of observatory operations, thermal stability during scientific observations, on-orbit science and observatory housekeeping data collection and storage, data uplink and downlink communications via the Deep Space Network (DSN), and contamination control for the mirrors and X-ray detectors to limit degradation of the scientific observations throughout *Lynx's* operational life.

### E.2.1 Telescope

The *Lynx* science program requires an observatory that is significantly more capable than NASA's existing Flagship *Chandra* and ESA's planned *Athena* missions, driven foremost by the needs of the Dawn of the Black Holes science pillar. The mirror requirements for this program are discussed in **§D.6.** On-axis angular resolution of ~0.5 arcseconds (half-power diameter (HPD) is required to avoid source confusion at the faintest fluxes and to uniquely associate X-ray sources with high-redshift optical and near-IR galaxies. A mirror effective area of 2 m² at E=1keV and a FOV with better than ~1 arcsecond imaging extending to at least 10 arcminutes off-axis is required to adequately sample the population of black hole seeds at high redshift. A telescope with these parameters will be able to survey at much lower flux thresholds and with a much greater speed than any other existing or planned X-ray observatory (**Figure E-1**).

The optimal way to implement an observing program that requires such long exposures of specific sky locations while individually registering every detected event is to reconstruct *post-facto* an X-ray image from an accurate history of the spacecraft pointing direction. On-board absolute pointing accuracy is constrained by the detector characteristics and is 10 arcseconds (3-sigma) (**§E.2.1.3)**, while a pointing stability of 0.17 arcsecond/s per axis (3-sigma) is constrained by the spacecraft gyros and reaction wheels, and allows for readout intervals as long as 1 second with no impact on image reconstruction. This permits 0.2-arcsecond HPD diameter image reconstruction and 1 arcsecond root mean square (RMS) absolute astrometric accuracy.

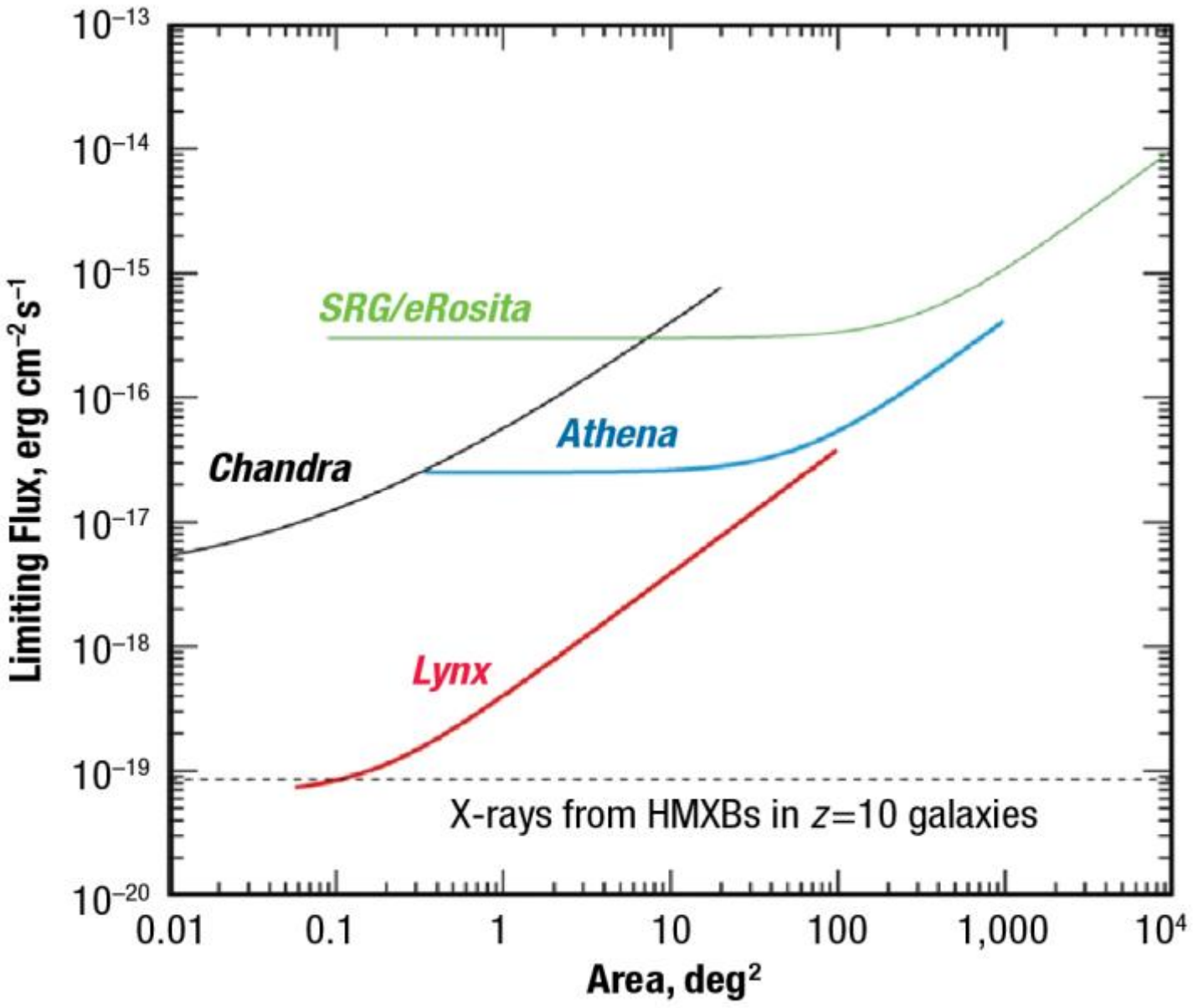

**Figure E-1. Limiting point source flux, for 15 Ms of observations, as a function of the area scanned for different partitioning of those 15 Ms. Athena and SRG/eRosita/SRG become limited by source confusion at a much higher limiting flux than *Lynx*.**

*Lynx* requires a detector capable of direct imaging of black hole seeds in deep extragalactic fields, taking full advantage of the *Lynx* mirror assembly angular resolution and FOV. In addition, this detector must have excellent soft X-ray spectral response to these (redshifted) sources and moderate spectral resolution across the 0.3–10 keV band to deduce thermodynamic properties of the hot gas in galactic halos (**Figure D-6, core program 2.1; hereafter, the programs are referenced by numbers given in the table in §D.7**) and other extended objects. A silicon active pixel sensor array of fine pixel scale (0.3 arcseconds) and large format (23 arcminutes x 23 arcminutes FOV) is capable of meeting these requirements.

*Lynx* also requires a detector of ~3 eV (non-dispersive) spectral resolution that can also spatially resolve Active Galactic Nucleus (AGN) feedback signatures from surrounding hot gas and jets in galaxies, groups, and clusters on 1 arcsecond or finer scales (**programs 2.3, 2.6, 2.7, 2.8**). This detector must resolve starburst-driven winds in low-redshift galaxies at high spectral resolution (0.3 eV) over ~1-arcminute FOVs at 1 arcsecond imaging resolution (**Figure D-5, program 2.8**); map metallicity gradients (5 eV resolution over 5 arcminute FOV) in circumgalactic, group, and galaxy cluster fields (**program 2.3**); and survey young Supernova Remnants (SNRs) in Local Group galaxies (**program 3.3**). This can be accomplished with a science-driven microcalorimeter array design customized to meet all these combinations of spectral, spatial, and FOV requirements.

*Lynx* must also have an instrument capable of characterizing warm gas in galactic halos out to their virial radius (**Figure D-6, program 2.2**). This can be accomplished by absorption line studies of background AGNs requiring a high-spectral resolution (R=5,000) dispersive grating spectrometer capable of 1 mÅ sensitivity in key absorption lines of OVII and OVIII.

Thus, *Lynx* includes a high angular-resolution, large effective area mirror assembly tightly coupled to a science instrument suite that is capable of fine imaging, high-resolution dispersive grating spectroscopy at low energies, and high-resolution imaging spectroscopy across the *Lynx* waveband, in order to ensure that all *Lynx* science goals are met. **Figure E-2** provides a top-level view of the *Lynx* telescope configuration.

Two of the science instruments, the High Definition X-ray Imager (HDXI) and the *Lynx* X-ray Microcalorimeter (LXM), along with their electronics and radiators, are mounted on a moveable platform that is part of the Integrated Science Instrument Module (ISIM). This translation table assembly permits either instrument to be placed on-axis depending on the scientific objectives of a particular observation. A focus mechanism on the translation table allows for fine focus adjustment. The high-spectral-resolution, dispersive X-ray Grating Spectrometer (XGS) focal plane detector assembly is mounted in a fixed location on the ISIM offset from the optical axis to intercept the dispersed spectrum regardless of whether the HDXI or LXM is the primary focus. The X-ray Mirror Assembly (XMA), grating arrays, and aft contamination cover are mounted onto a common structure that makes up the *Lynx* Mirror Assembly (LMA). The grating arrays covers ~2/3 of the LMA aperture and can be retracted when not in use.

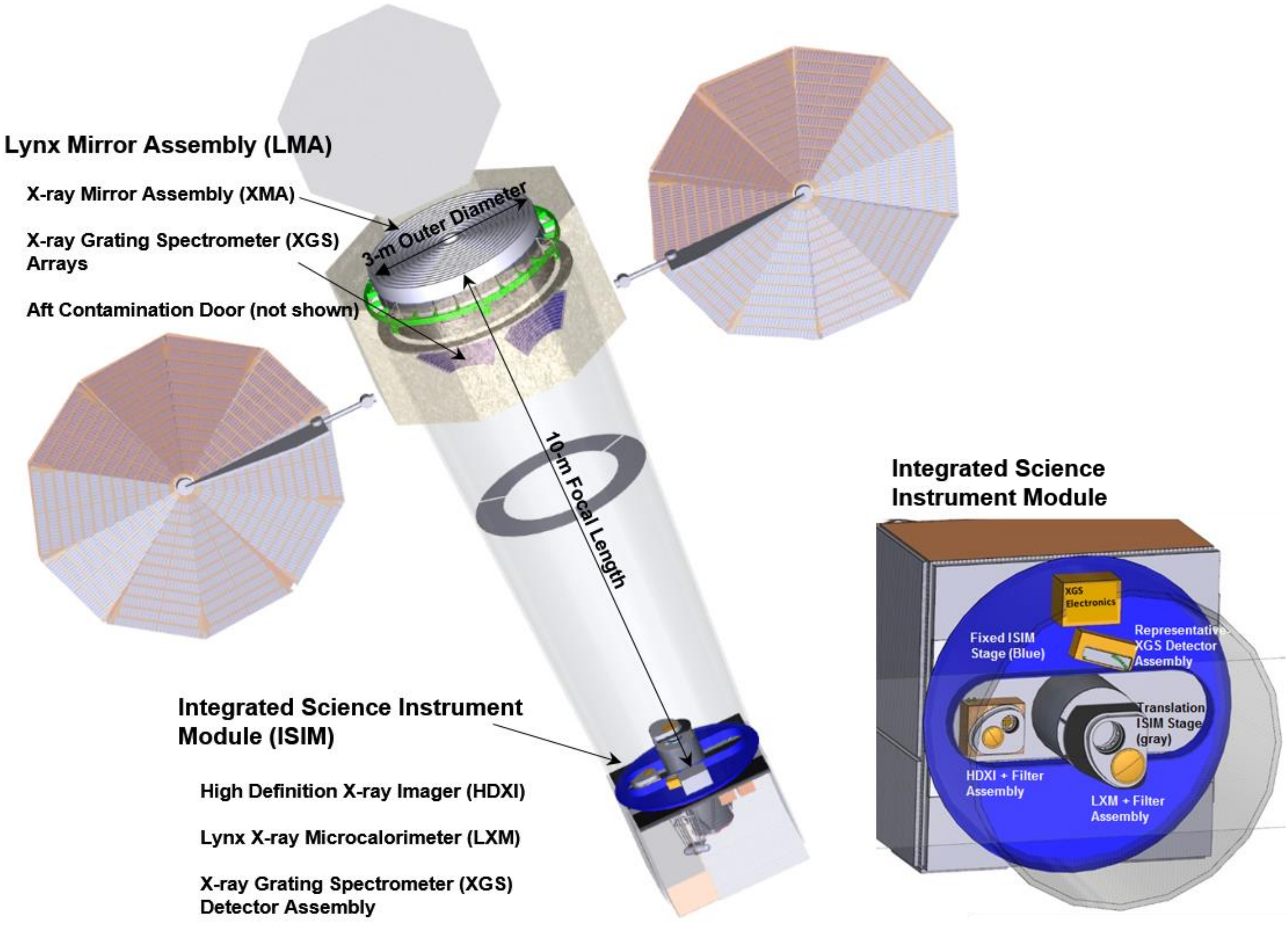

**Figure E-2.** ***Lynx*** **X-ray Telescope configuration.** ***Lynx*** **has a 10-m focal length and consists of a high-resolution, 3-meter diameter, large area X-ray mirror assembly with pre- and post-collimators (XMA) surrounded by the spacecraft bus and complemented by an instrument suite that includes a low-noise megapixel imaging array (HDXI); a large-format, high-spectral-resolution, small-pixel array (LXM),**

**and high-efficiency, high-spectral-resolution, dispersive spectrometer (XGS). Detectors for all three instruments are mounted onto the ISIM.**

A non-reflecting pre-collimator, or baffle, at the entrance to the XMA reduces the heat which the mirror radiates to deep space. In addition to the diffuse X-ray background, there is also a particle background at SE-L2. The low-energy particles can be focused by the optics onto the detectors inducing false signals and causing damage over time. To mitigate this, a magnetic broom, similar to that used on *Chandra*, is used to divert these particles out of the optical path. Detailed background estimates for SE-L2 are being carried out by the *Athena* team (Lotti, et al. 2017) and will be directly applicable to *Lynx*.

#### E.2.1.1 X-ray Mirror Assembly

The large effective collecting area required for *Lynx* is achieved by nesting a few hundred rings of co-aligned, co-axial mirror pairs to optimize the available entrance aperture. Thinner mirror substrates allow for greater nesting of mirror pairs and larger effective area while simultaneously reducing mass per collecting area. In addition, careful selection of a high atomic number mirror coating material(s) and thicknesses can further enhance the collecting area. The outer diameter of the XMA design is 3-m; sufficient to achieve the *Lynx* science requirements while allowing for maximum flexibility in the choice of launch vehicle and fairing size. Fundamental XMA requirements are listed in **Table E-1**.

**Table E-1. XMA key requirements matched to *Lynx* science goals.**

| XMA Parameter | Requirement | Science Drivers | Reference(s) |
|---|---|---|---|
| Energy Range | 0.3–10 keV | Sensitivity to high-z sources drives the best achievable efficiency in the soft band. The High-energy coverage enables observations that constrain the intracluster medium and cluster core metallicities (e.g., Fe Kα emission line at 6.7 keV). | DBH, GFE, SESE |
| Angular Resolution | 0.5 arcsecond HPD on-axis<br><br>< 1 arcsecond HPD across the FOV | Detecting early supermassive black holes at the seed stage or soon after requires an X-ray sensitivity of ≈ $10^{-19}$ erg $s^{-1}$ $cm^{-2}$. To avoid source confusion at these fluxes, and to uniquely associate detected X-ray sources with *JWST* and *WFIRST* galaxies. Other examples include mapping diffuse baryons in galactic halos and in star forming regions, energy feedback studies, and post-merger evolution of GW sources. | P 1.1, P 1.2, P 2.2, P 2.7, P 2.8, P 3.2 |
| Grasp (Effective Area * FOV for <1 arcsecond HPD)<br><br>Field of View<br><br>Effective Area @ 1 keV | ~600 $m^2$ arcminutes$^2$<br><br>10 arcminutes radius<br><br>2 $m^2$ | Supermassive black hole deep surveys; tracing the growth of BHs through time (from z≈10); mapping diffuse baryons in emission in galactic halos and cosmic web; identifying young stars in active star forming regions; survey *LISA* triggers for electromagnetic counterparts. | P 1.1, P 1.2, P 2.2, P 3.1 |

Note: DBH – Dawn of Black Holes, GFE – Invisible Drivers of Galaxy Formation and Evolution, SESE – The Energetic Side of Stellar Evolution and Stellar Ecosystems; P #.# denotes Programs listed in Table D.1 Core Programs

The large FOV of *Lynx*, expressed in terms of the light-grasp for sub-arcsecond imaging, is achieved with the Wolter-Schwarzchild (W-S) optical prescription, which eliminates coma aberration (Chase and Van Speybroeck 1973; Aschenbach 2009; Saha et al. 2014; 2017), along with mirror elements with substantially smaller ratio of length to focal distance than *Chandra*.

There are multiple on-going NASA-funded X-ray mirror technologies that are capable of meeting *Lynx* requirements. The *Lynx* team has undertaken a trade study that assessed those technologies that are the most mature. These are full shell, adjustable segmented, and silicon meta-shell optics technologies. This trade study considered the ability of each technology to meet the *Lynx* science requirements and their capacity to overcome technical challenges and meet programmatic mission constraints. Output from the *Lynx* optics trade study includes a detailed optical design and technology maturation plan for each of the three technologies that will be included in the *Lynx* final report submitted to NASA in 2019[2].

Members of the *Lynx* STDT, industry partners, independent consultants, and members from the full shell, adjustable segmented, and silicon meta-shell optics technology teams carried out this assessment. The result is a recommendation from this team to the STDT to use the silicon meta-shell optics assembly architecture to focus the design for the final report and to include full shell and adjustable segmented optics as feasible alternates. Following formal acceptance of this recommendation, a higher-fidelity LMA that includes the XMA, grating arrays, and inner contamination door, will be designed using this technology and integrated into the observatory notional configuration. The project life cycle schedule and costing will be refined and provided to NASA in the *Lynx* final report for submission to the 2020 Astronomy and Astrophysics Decadal Survey. The assessment related to alternate feasible technologies will be documented in the final report as appropriate.

The detailed design for the integrated LMA will include consideration for contamination control throughout the project life cycle. Molecular and particulate contaminants on X-ray optical surfaces can degrade their performance, while changes in contamination levels can compromise calibration stability. Subsequent to calibration, contamination covers forward and aft of the mirror assembly will allow a dry nitrogen purge on the ground, with the covers remaining closed until the completion of a post-launch outgassing phase. In operation, the *Lynx* thermal control subsystem maintains the XMA at approximately room temperature, higher than the surrounding subsystems to minimize particulate and molecular adhesion to mirror surfaces.

#### E.2.1.2 High-Definition X-ray Imager

Silicon-based X-ray imaging spectrometers have flown on nearly every X-ray observatory since 1999, including *Chandra* (Advanced CCD Imaging Spectrometer), *XMM-Newton* (Struder 2001), *Suzaku* (X-ray Imaging Spectrometer), and *Swift* (X-ray Telescope) (Garmire et al. 2003, Koyama et al. 2007). For X-ray observations in the energy range probed by *Lynx*, active pixelated silicon-based sensors offer high readout rates, low noise, high broad-band quantum efficiency, and minimal cross talk compared to CCDs. The *Lynx* HDXI uses an array of pixelated Complementary Metal Oxide Semiconductor (CMOS) based active sensors providing a wide FOV (0.15 deg$^2$) with high angular resolution (~0.3 arcsecond pixels) optimally matched to the telescope point spread function (PSF). HDXI requirements and their traceability to *Lynx* science are summarized in **Table E-2**.

[2] Subsequent to submitting this Interim Report for review, the optics trade study was completed and a primary architecture was chosen to focus the design for the final report. However, except where necessary (see, e.g., Section E.3.1 Luanch Vehicle), all three feasible designs are carried through this public release of the Interim Report as originally written.

**Table E-2. HDXI key requirements matched to *Lynx* science goals. Derived requirements represent instrument design parameters optimized for a Flagship General Observer program.**

| HDXI Parameter | Requirement | Science Drivers | Reference(s) |
|---|---|---|---|
| Energy Range | 0.3–10 keV | Detection of black holes in z=6–10 galaxies down to a mass limit of $M_{lim}$=10,000 $M_{Sun}$ over a volume with $10^3$–$10^7$ potential host galaxies; Studies of hot ISM and stellar feedback in active star forming regions in the milky Way; Understanding SN ejecta kinematics and localization | DBH, GFE, SESE |
| Field of view | 23 arcminutes x 23 arcminutes | SMBH deep survey efficiency; Direct imaging of hot gas in halos to $R_{200}$ for nearby galaxies; Survey of young SNR in the local Group galaxies | P 1.1, P 1.2, P 2.1, P 3.3 |
| Pixel size | 16 µm x 16 µm (0.33 arcseconds) | Adequately samples the mirror Point Spread Function (PSF); Point source sensitivity; Resolve AGN from group emission; Constraining mass loss on the closest stars; Spatially resolving SNR ejecta-knots and clumps in the Circumstellar Medium | P 1.1, P 1.2, P 2.4, P 3.2 |
| Energy Resolution | 60 eV (FWHM) @ 1 keV | Direct imaging of hot gas in galactic halos in continuum and line emission; Characterization of the first galaxy groups at z≈3–4 | P 2.1, P 2.4 |
| Read Noise | ≤ 4 $e^-$ | Improves low energy detection efficiency and energy resolution | derived |
| Full-field count rate capability | 8,000 ct $s^{-1}$ | Reduces dead time for bright diffuse sources (e.g. Perseus and Cas A) | derived |
| Frame Rate<br>Full-field<br>Window (20x20 pixels) | <br>> 100 frames $s^{-1}$<br>>10,000 windows $s^{-1}$ | <br>Maximizes the low-energy throughput<br>Minimizes the background and enables fast timing studies | derived |

Note: DBH – Dawn of Black Holes, GFE – Invisible Drivers of Galaxy Formation and Evolution, SESE – The Energetic Side of Stellar Evolution and Stellar Ecosystems; P #.# denotes Programs listed in Table D.1 Core Programs

The HDXI focal plane consists of 21 abutted 1024 x 1024-pixel advanced silicon sensors tiled to approximate the curved focal surface of the telescope. The 21 sensors are housed in a vacuum enclosure **(shown in Figure E-)** at a nominal, passively cooled, operating temperature of -60°C. The vacuum housing also encloses a filter mechanism as well as $Fe^{55}$ calibration sources. A thin Al filter is directly deposited on each sensor to provide a minimum level optical blocking at all times. These filters affect the low-energy throughput and must be designed to allow the *Lynx* science requirements to be met.

Digitized sensor data passes through event recognition processers (ERPs) that filter non-X-ray events and empty pixels, thereby reducing the telemetered data volume by 3 to 4 orders-of-magnitude. Processing is under the control of detector electronics units (DEUs) commanded from the spacecraft. The HDXI will also have a high-speed windowing mode in which a small region (20 x 20 pixels) of a sensor can be read out at 10-kHz. This high-speed windowing mode will be used to reduce pileup (which occurs when the timing resolution of the system is incapable of distinguishing between two or more incident photons that arrive close in time) from bright sources and allow high-resolution timing measurements.

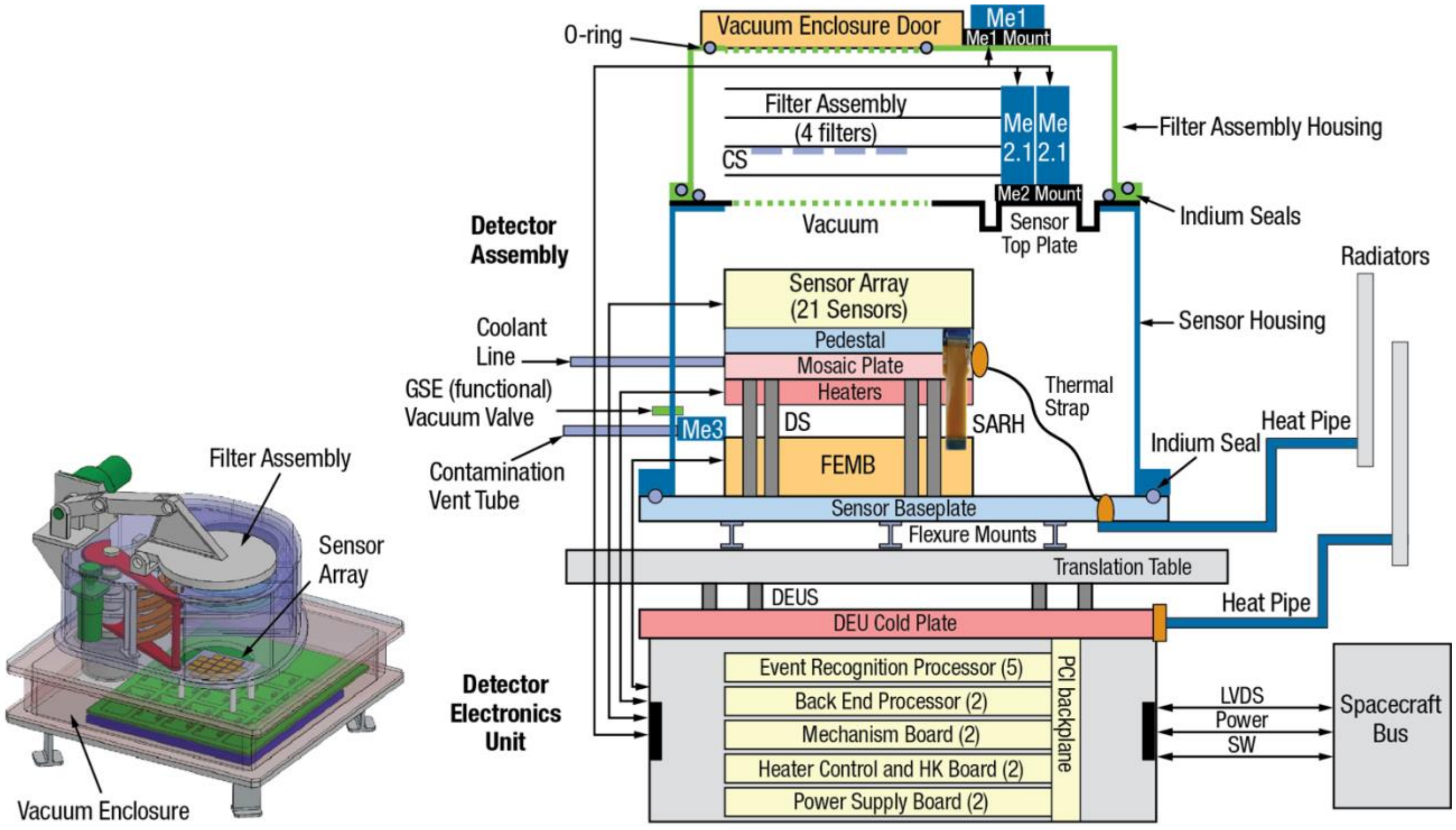

**Figure E-3. (Left) Schematic and (Right) block diagram of the *Lynx* HDXI system. The HDXI DEU and all radiators are attached to the ISIM translation table. Flex cables are used between the translation table and spacecraft bus to provide power, control, and data. The filter mechanism allows for individual filters or a calibration source to be inserted in front of the sensor array.**

Contamination on the sensor array throughout mission lifetime will be minimized by careful temperature control and venting of the HDXI filter and detector assemblies and surrounding structure. Contamination control is necessary for detection longevity, as accumulation over time may negatively impact the low energy detector response. The filters will be held at room temperature to decrease contamination from the rest of the observatory. The filters and mechanisms will be kept clean and the filter-housing interior will be coated with a Molecular Absorber Coating (MAC). During detector bakeout, when the instrument temperature is raised to 'bake out' any contaminants, the instrument will be translated to one side of the translation table assembly, at which position the detectors will view a plate coated with MAC affixed to the stationary part of the ISIM. This plate will be in the direct line-of-sight of the detector, minimizing contamination accumulation. Keeping the detector warmer than the rest of the spacecraft during cruise will prevent outgassing materials from contaminating the detector. An early bakeout in this configuration will drive off any water or other contaminants that accumulate during launch processing. A comprehensive contamination control plan that includes these bakeouts is being formulated as part of this study.

### E.2.1.3 *Lynx* X-ray Microcalorimeter

The *Lynx* LXM provides high spectral resolution, high spatial resolution non-dispersive imaging spectroscopy over a broad energy range (0.2–10 keV). LXM is composed of a large array of X-ray sensors that determine the energy of individual incident photons by precisely measuring the temperature rise caused by each. It will meet the many different science-driven performance requirements of *Lynx* with different sub-regions of the detector array (**Table E-3**). These arrays are identified as the "Main Array," the "Enhanced Main Array," and the "Ultra-High-Resolution Array." These arrays offer a range of solutions, trading spatial and spectral resolution, along with FOV and count rate capability.

**Table E-3. LXM requirements and traceability to the *Lynx* science goals.**

| **Main Array** | **Requirement** | **Science Drivers** | **Reference(s)** |
|---|---|---|---|
| Energy Range | 0.2–7 keV for 3 eV | Low temperature (<1 keV) thermal emission or low energy non-thermal sources | GFE, SESE |
| Field of view | 5 arcminutes x 5 arcminutes | Encompasses full SNRs, galaxies, and clusters of galaxies for high-res imaging and spectroscopy | P 2.3, P 2.9, P 3.3 |
| Pixel size | 1 arcsecond x 1 arcsecond | Arcsecond scale features in shocks and filaments, and point sources in crowded regions (XRBs and stars) | P 2.3, P 2.9, P 3.3, P 3.4 |
| Energy Resolution | 3 eV (FWHM) | Map the structure of stellar driven winds in galaxies (typical velocities 100–1,000 km/s) | P 2.3, P 3.3 |
| **Enhanced Main Array** | **Requirement** | **Science Drivers** | **Reference(s)** |
| Energy Range | 0.2–7 keV for 3 eV | Low temperature thermal emission or low energy non-thermal sources | GFE, SESE |
| Field of View | 1 arcminute x 1 arcminute | Encompasses AGN Jets, centers of galaxies, and cores of clusters of galaxies | P 2.6 |
| Pixel Size | 0.5 arcseconds x 0.5 arcseconds | Sub-arcsecond features in shocks and filaments, and point sources in crowded regions (XRBs and stars); Study of distribution of AGN within and around groups/clusters; removing AGN; study of thermodynamic properties of cluster gas; feedback in groups and clusters | P 2.6, P 3.3, P 3.4 |
| Energy Resolution | 3 eV | Energetics and dynamics of plasmas, Supermassive Black Hole growth (resolving Fe K line profiles), AGN feedback at peak (typical velocities ~ 100 km/s), and temperature and velocity profiles within the Bondi radius for a sample of Supermassive Black Holes | P 2.6, P 2.7 |
| **Ultra High-Res. Array** | **Requirement** | **Science Drivers** | **Reference(s)** |
| Energy Range | 0.2–0.75 keV | Faint diffuse baryons in emission, such as in galactic halos | GFE, SESE |
| Field of View | 1 arcminute x 1 arcminute | Encompasses the hot gas around galaxy halos | P 2.5, P 2.8 |
| Pixel size | 1 arcsecond x 1 arcsecond | Arcsecond scale features in shocks and filaments, and point sources in crowded regions (XRBs and stars) | P 2.5, P 2.8, P 3.1 |
| Energy Resolution | 0.3 eV (FWHM) | Velocities/turbulent broadening down to ~50 km/s (outflows and thermal velocities) | P 2.5, P 2.8 |

Note: DBH – Dawn of Black Holes, GFE – Invisible Drivers of Galaxy Formation and Evolution, SESE – The Energetic Side of Stellar Evolution and Stellar Ecosystems; P #.# denotes Programs listed in Table D.1 Core Programs

The count rate capability is much higher for the Ultra High-Resolution Array than for the Main and Enhanced Main Arrays, and is a derived value based on the intrinsic design of the arrays. The Ultra High-Resolution Array count rate capability ranges from 80 cps to 1,000 cps per 1 arcsecond pixel, depending on the required energy resolution (lower resolution allows for higher count-rate capability). A trade will be made during this study regarding the count rate versus readout capability considering science capability, cost, and schedule.

A study is also being carried out to determine the *Lynx* payload trades associated with extending the energy range to ~15–20 keV. The science case for this is broad, and includes improving black hole spin measurements (e.g., Risaliti et al. 2013), expanding our understanding of the X-ray emitting corona associated with accreting black holes (e.g., Fabian et al. 2017), improved measurements of feedback (e.g., Nardini et al. 2015), studies of obscured AGN (e.g., Lansbury et al. 2014, Bauer et al. 2015), X-ray reverberation mapping to uncover the geometry of the central engine (e.g., Kara et al. 2015), and studies of ultraluminous X-ray sources (e.g., Walton et al. 2018). LXM is ideally suited for this, in that a higher energy response can be achieved with no modification of the current instrument design, though modification to the XMA, via additional mirrors or multilayer coating, would be required. Additional considerations such as the mirror manufacturing (cost and schedule) and calibration also need to be examined.

*Lynx* has baselined Transition-Edge Sensors (Irwin and Hilton 2005), single pixel and multi-pixel, for this study because of its relatively high maturity level compared to other thermometer technologies. Multi-pixel TESs, or hydras, allow for some degree of thermal multiplexing to reduce the number of TESs that need to be read out. This multiplexing allows for wider focal plane coverage, or finer sampling of the FOV for the same coverage, without a commensurate increase in the number of wires or readout components. Large arrays of these sensors are read out using microwave Superconducting Quantum Interference Device (SQUID) resonators on High-Electron Mobility Transistor (HEMT) channels. Minimal to no energy resolution degradation from the readout is expected (Mates et al. 2017). The layout for the three arrays is shown in **Figure E-**, and indicates the number of pixels and HEMTs required for each array. The maximum assumed number of HEMTs powered on at a single time is 16.

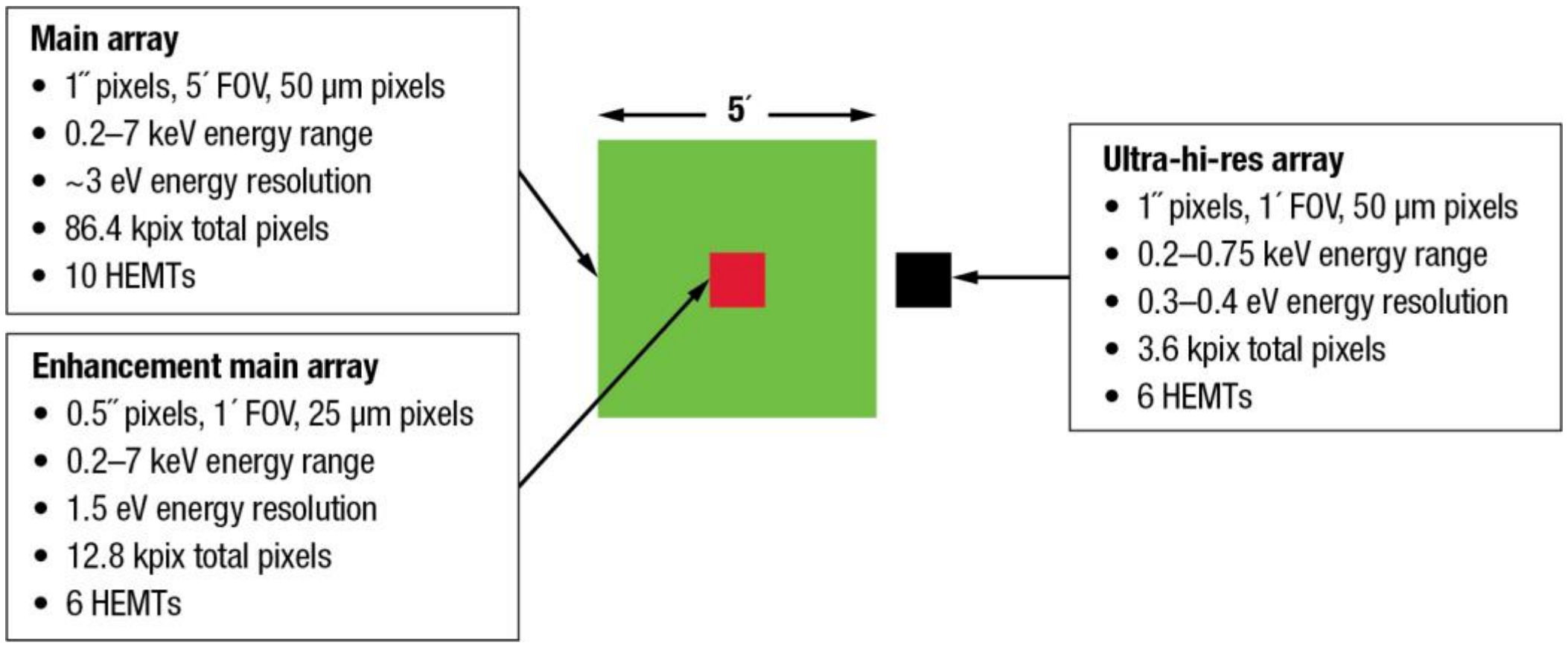

**Figure E-4. Baseline layout of the LXM array. There are three different regions of the array corresponding to three different pixel types, each with different properties as labeled.**

The LXM will be cooled to 50 mK using a cryostat that takes advantage of heritage (*Hitomi*) and mature cryocooler technologies. Cooling to 4K will be achieved via a thrust-tube type design mounted in a fashion similar to that used for *Spitzer*. There exists a wide variety of cryocooler systems that are at various maturity levels. To determine the optimum solution for *Lynx*, a trade study of cryocooler options is being performed to trade maturity, mass, power, complexity, and cost, and will be completed well in advance of the submission of the *Lynx* Final Report. Vibration and thermal isolation between LXM and the ISIM are currently being designed to minimize impacts on the telescope performance. The cryocooler compressor, rotating valve, and all moving parts are supported on a separate stand from the cryostat to minimize vibrational coupling into it. Multistage adiabatic demagnetization refrigerators further cool the array to 50 mK (Tuttle et al. 2017). Details of the LXM cryostat and read-out electronics are shown in **Figure E-** and the focal plane assembly design and block diagram are shown in **Figure E-.**

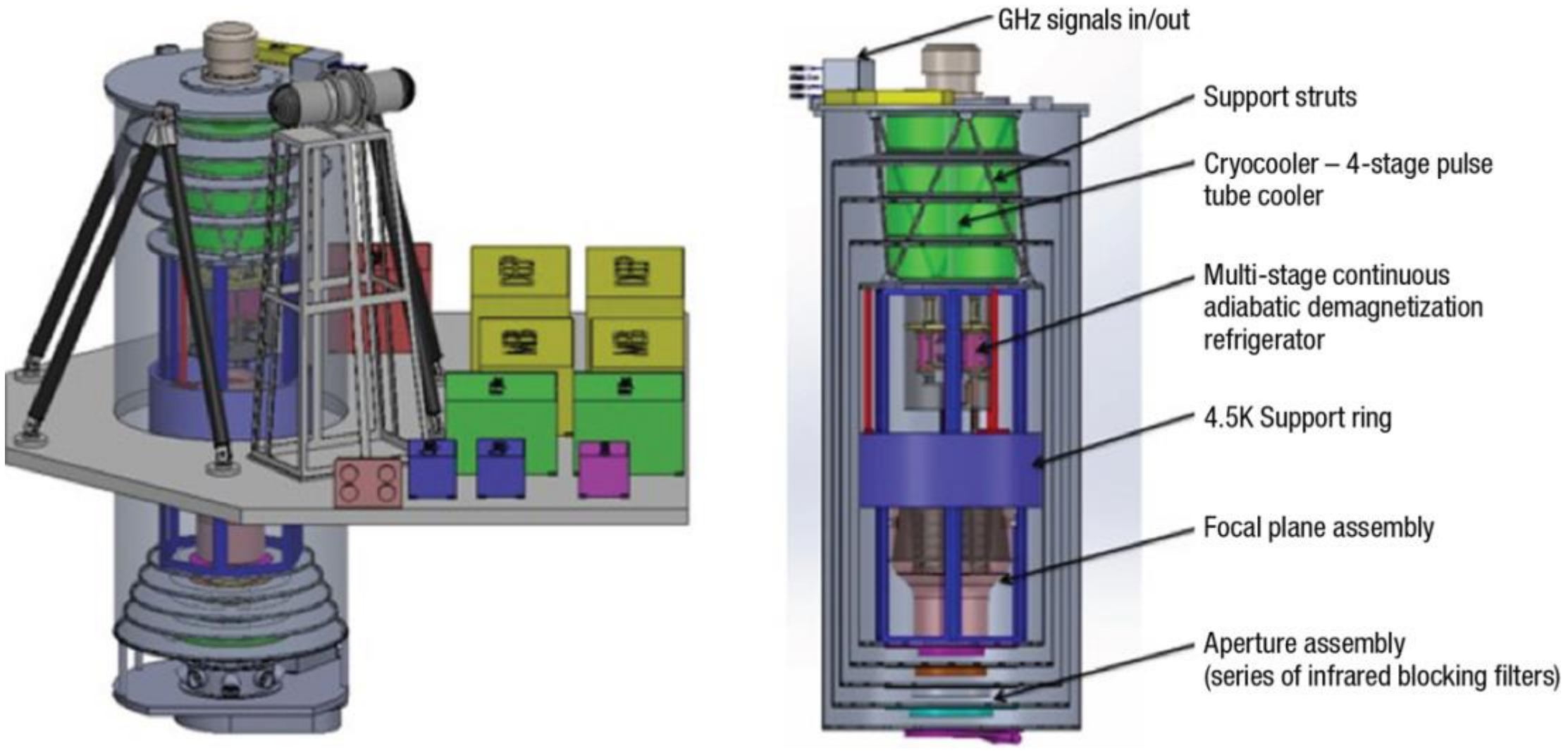

**Figure E-5. (Left) Design of the LXM cryostat and read-out electronics. (Right) Details of the design shown through a cross-sectional view of the cryostat.**

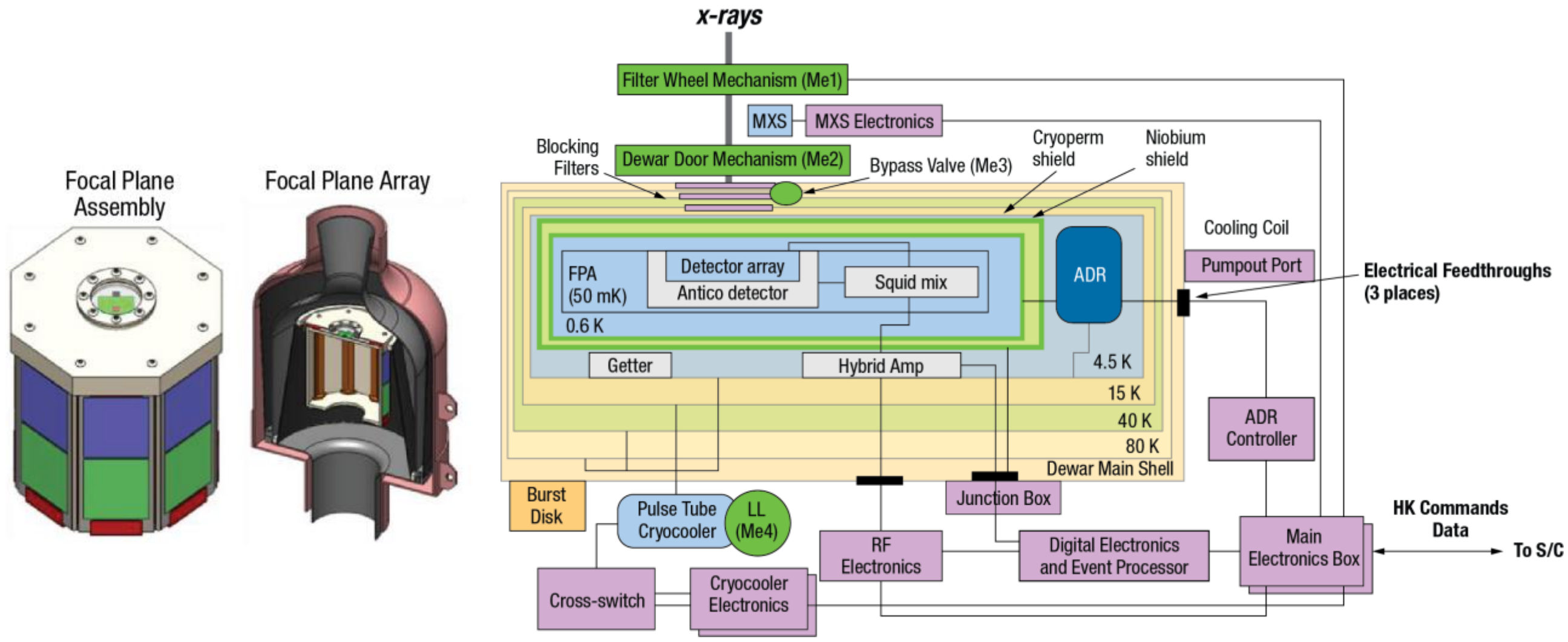

**Figure E-6. (Left) Preliminary LXM focal plane assembly design. The Main, Enhanced Main, and Ultra High-Resolution Arrays are visible. (Right) Block diagram of the LXM.**

The LXM focal plane assembly also includes an assembly with a modulated X-ray source (MXS) that is capable of providing pulsed X-ray lines at multiple energies and is similar to that used on *Athena's* X-IFU (Barret et al. 2016) and *Hitomi's* SXS (de Vries et al. 2017) for in-flight calibration. Infrared (IR)/optical blocking filters necessary to block long-wavelength photons from being incident on the microcalorimeter array and creating noise are also included. At the same time, these filters need to be thin enough to maximize X-ray throughput. An engineering study is planned to determine the optimum IR blocking filter design to provide a high system (filter + detector) Quantum Efficiency (QE) across the *Lynx* bandpass, sufficient for meeting *Lynx* science observation goals (§**D**).

Compared to the planned *Athena* X-ray Integral Field Unit (X-IFU; Smith et al. 2016), the LXM Main Array has the equivalent diameter FOV, but with a smaller pixel size, necessary to exploit the XMA HPD. This fine resolution will permit *Lynx* to observe sub-arcsecond-scale features in clusters and jets, and minimize source confusion in crowded fields. The use of hydras for thermal multiplexing reduces the number of TESs that need to be read out to be comparable to the X-IFU, allowing LXM to leverage the X-IFU readout layout (similar wire density and flex cable technologies). Because the focal plane for LXM is similar in size to that of X-IFU, the mechanical, thermal, magnetic shielding, anti-coincidence detector, IR filter, and MXS designs will also be leveraged. Many of these elements also have heritage from the microcalorimeters on *Astro-E* and *Astro-H* (Kelley et al. 2000, Kelley et al. 2016).

### E.2.1.4 X-ray Grating Spectrometer

The XGS will provide high-throughput, high-resolution spectra at soft energies (0.2–2 keV). Grating arrays are mounted just aft of the XMA along the optical path with actuators that allow them to be inserted into the optical path when in use (**Figure E-**). The *Lynx* XGS has a spectral resolving power requirement of R >5,000 ($\lambda/\Delta\lambda$) over the 0.2–2.0 keV energy band, with an effective area > 4,000 $cm^2$ @ 0.6 keV. For comparison, the Reflection Grating Spectrometer on *XMM* has a resolving power of R = 150–800 over the 0.33–2.5 keV band and an effective area of ~150 $cm^2$ @ 0.83 keV (see *XMM-Newton* Users' Handbook) and the High-Energy Transmission Grating on *Chandra* provides spectral resolving power up to R = ~65–1,070 (over the 0.4–10 keV range) and an effective area of 59 $cm^2$ @ 1keV (The *Chandra* Proposers' Observatory Guide). The XGS will contribute to the *Lynx* science goals as detailed in **Table E-4**.

Two different gratings technologies are currently under study for *Lynx*: Off-Plane Reflection Gratings (OP-XGS—Miles et al. 2017, DeRoo et al. 2017) and Critical Angle Transmission Gratings (CAT-XGS—Heilmann et al. 2016, Heilmann et al. 2017). Both technologies meet *Lynx* requirements, and both are at similar maturity levels and have *XMM-Newton* Reflection Grating Spectrometer and *Chandra* High- and Low-Energy Transmission Grating heritage, respectively. Advantages and disadvantages will be detailed for each, regarding required spacecraft resources, and technology maturation cost, and schedule. A *Lynx*-specific technology maturation plan will be generated for both grating types to be included in the *Lynx* Final Report.

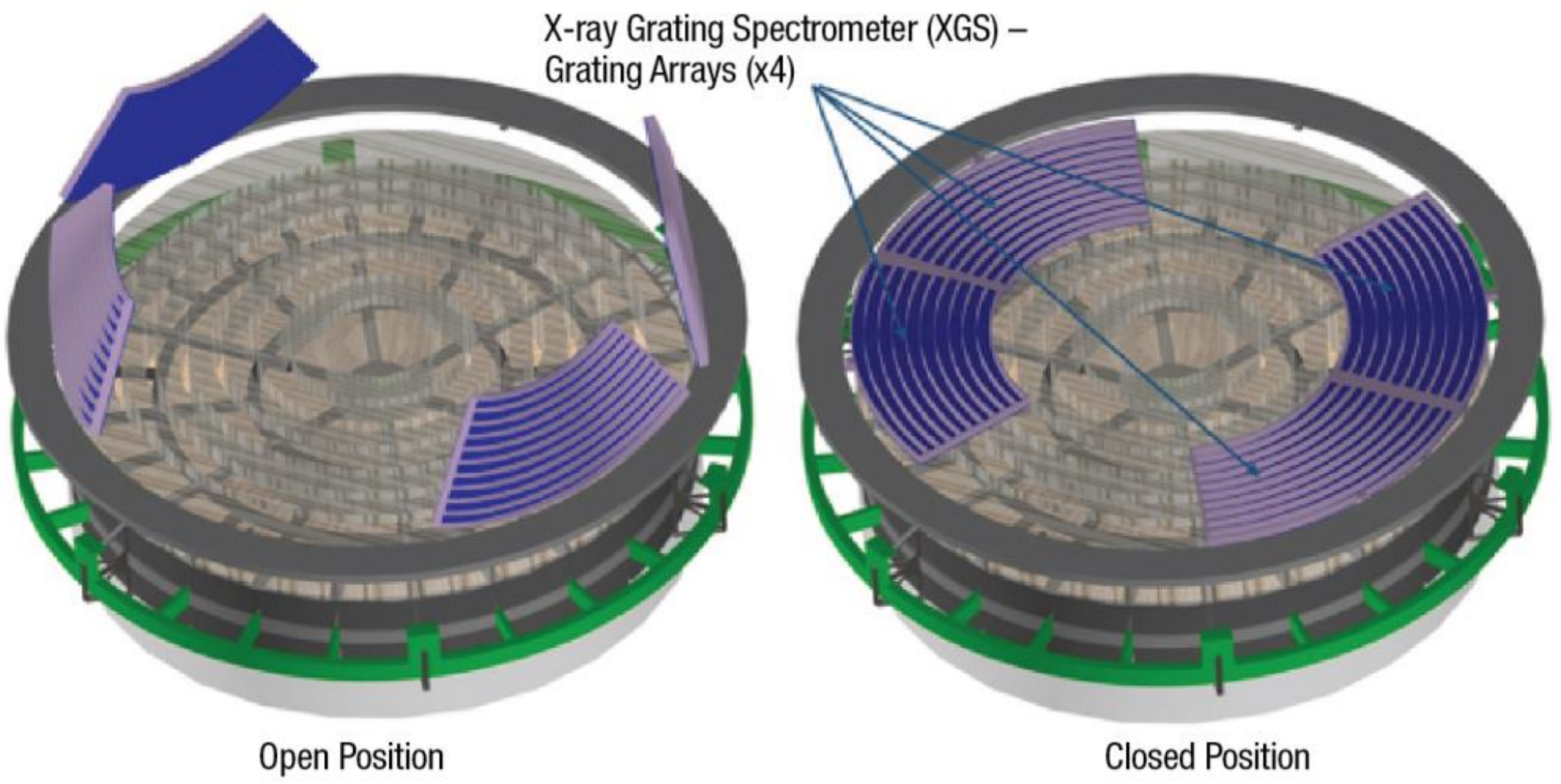

**Figure E-7. Open (Left) and Closed (Right) low-fidelity design for XGS grating arrays shown in blue in relation to the XMA shown in gray. The LMA (that includes the XMA and grating arrays) mounting structure is shown in green. The grating arrays are placed out of the optical path when not being used. The number of arrays and array layout are still being assessed.**

**Table E-4. XGS key requirements and traceability to *Lynx* science goals.**

| XGS System | Requirement | Science Drivers | Reference(s) |
|---|---|---|---|
| Energy Range | 0.2–2.0 keV | The strongest lines lie in this energy band: the K-shell transitions of key elements from C to Si, plus Fe L-shell, and weaker density-sensitive lines | GFE, SESE |
| Effective area | 4,000 cm$^2$ @ 0.6 keV | Absorption line spectroscopy (OVII/OVIII lines) of galactic halos near the virial radius and of the Cosmic Web; Obtain systematic assessment of the processes occurring in young stars, extending the observable age and mass range; In-depth exploration of coronal structures on cool stars with a range of ages | P 2.2, P 3.1 |
| Spectral Resolving Power, R | 5,000 @ 0.6 keV (R=10,000 Goal) | Resolve line shapes to their thermal width; measure dynamics, turbulence, and redshifts; Spectrally resolve coronal structures on cool stars, accessing bulk motions and velocity broadening | P 2.5, P 3.1 |
| Line-spread function width | 1 arcsecond | See spectral resolving power | derived |
| **XGS Readout** | **Requirement** | **Science Drivers** | **Reference(s)** |
| Readout Pixel size | 16 µm x 16 µm | Required to achieve the spectral resolving power | derived |
| Readout noise (rms) | ≤ 4 $e^-$ | Improves low energy quantum efficiency | derived |
| Readout Energy Resolution | ~80 eV @ 277 eV | Allows for grating order separation | derived |

Note: DBH – Dawn of Black Holes, GFE – Invisible Drivers of Galaxy Formation and Evolution, SESE – The Energetic Side of Stellar Evolution and Stellar Ecosystems; P #.# denotes Programs listed in Table D.1 Core Programs

The XGS detector assembly consists of multiple advanced silicon sensors enclosed in a vacuum housing and employs on-sensor optical/UV blocking, but is mounted directly to the fixed ISIM rather than to the translation table. These detectors will use the same technology as that of HDXI, and follow the same technology development path. Observatory resources required for these detector arrays are similar for each grating type.

The mechanical layout of the detectors within an assembly is different for each grating technology. The OP-XGS sensors are arranged in an arc consistent with the blaze angle and diffraction for reflection gratings and includes eight sensors for measuring the spectral performance and one sensor for measuring the zero order (**Figure E-**, Left). The CAT-XGS detector assembly includes 18 sensors arranged on the focal plane in a linear configuration, consistent with the diffraction for transmission gratings (**Figure E-**, Right). A focusing mechanism will be directly integrated into the housing of the detector assembly. Even though the number of sensors and layout is different for each grating type, the mechanical, thermal, and electrical designs are similar (**Figure E-**).

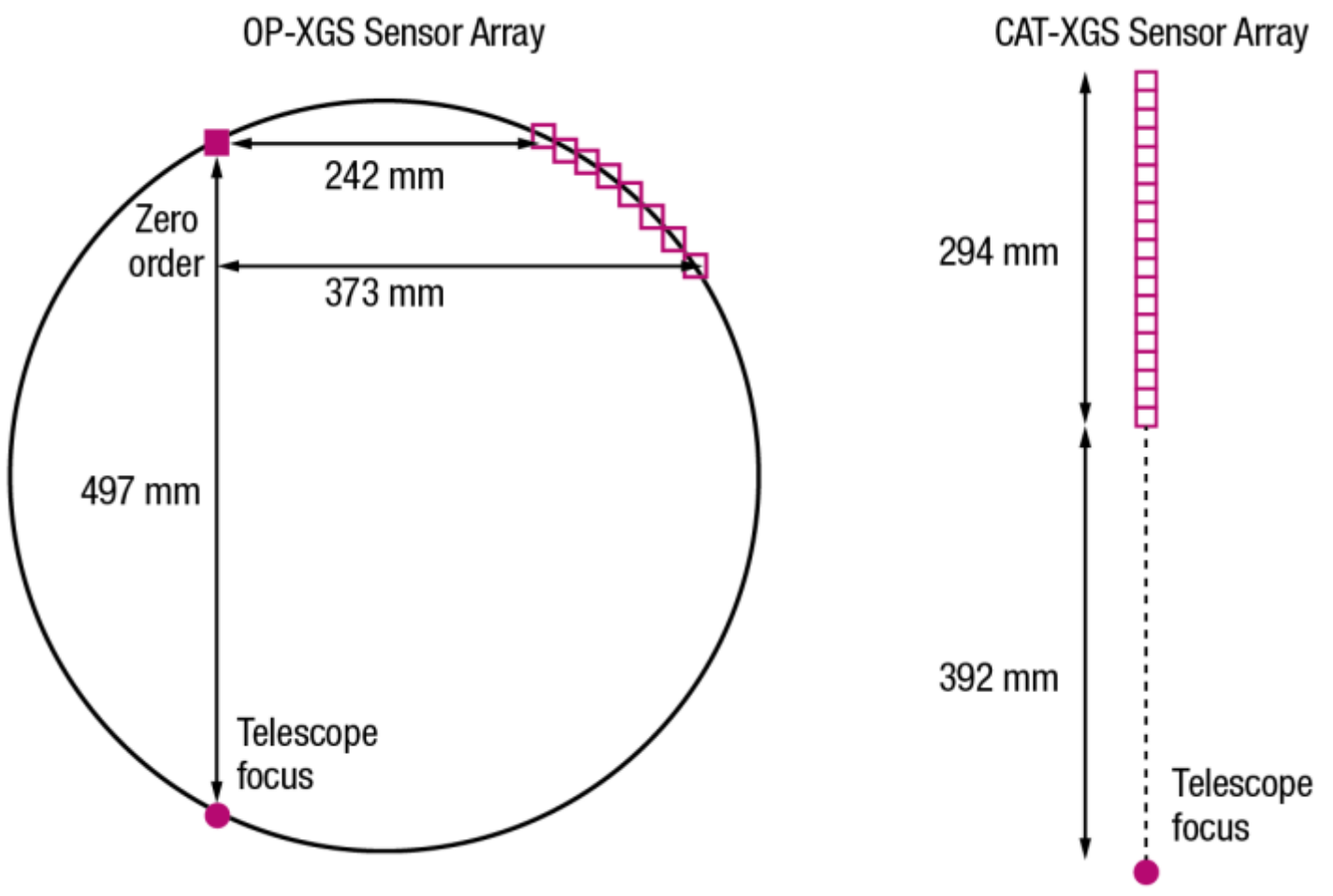

**Figure E-8. (Left) OP-XGS sensor layout and (Right) CAT-XGS sensor layout indicated by the small boxes and relative to the telescope focus.**

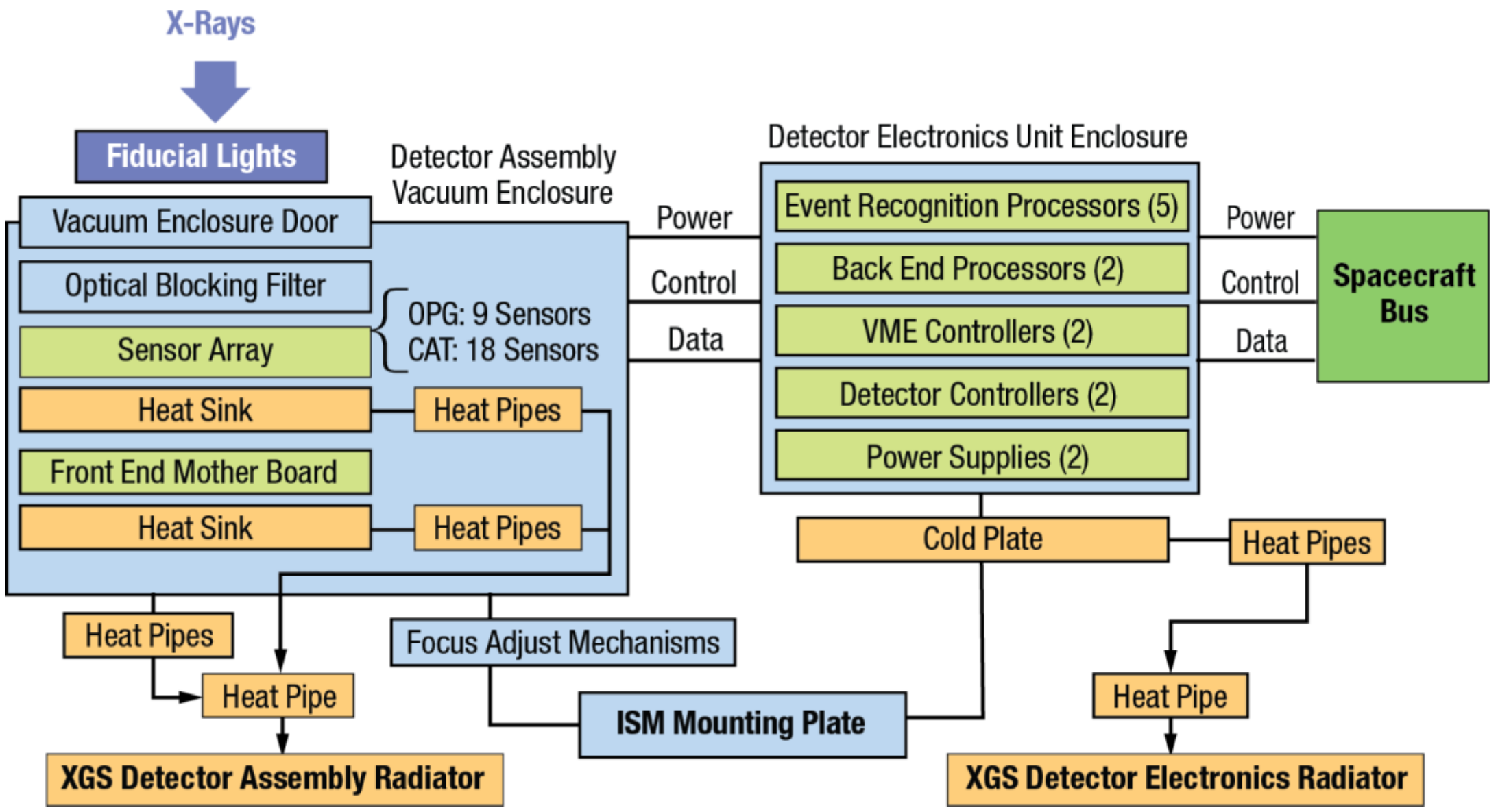

**Figure E-9. XGS detector assembly block diagram. The same architecture is applicable to both grating types, though the number of sensors is different for each.**

As mentioned in **§E.2.1.1**, fore- and aft-contamination covers over the XMA and the grating arrays to allow for a dry nitrogen purge on the ground. These doors will remain closed until the completion of a post-launch outgassing phase. On orbit, the XMA and grating arrays will be kept at higher temperatures than the surrounding subsystems to minimize particulate and molecular adhesion to mirror and grating surfaces. A detailed thermal control system is being generated as part of this study.

### E.2.1.5 Optical Bench Assembly

The optical bench assembly (OBA) is a non-deployable, tapered-cylinder structure that connects the spacecraft and LMA forward section to the aft ISIM and focal plane instruments. It is made of a lightweight composite (M46J) structure, supported by ring stiffeners. In addition, the OBA integrates the LMA to the spacecraft and provides mounting for a magnetic broom assembly used to divert charged particles away from the focal plane detectors to reduce particle background and damage. The OBA and the magnetic broom have *Chandra* heritage. Moderate temperature control of the OBA is important to maintain the focal length, and to minimize misalignments between the LMA and the focal plane science instruments. Analyses are underway to assess the temperature and gradient requirements, and subsequent effects on the telescope performance.

A trade study is also underway to assess the possibility of using an extendable metering structure that would allow *Lynx* added flexibility in selecting a launch vehicle, which could result in cost savings for the mission. The current *Lynx* mass requires a 'Heavy-class' launch vehicle; however, an 'Intermediate-class' launch vehicle and co-manifested Space Launch Systems (SLS) are being considered. This study will trade cost and launch vehicle availability with risk.

### E.2.2 Spacecraft

The spacecraft bus includes all necessary subsystems to enable the scientific and operational functionality of the observatory (**Figure E-10**). The design of the bus and individual subsystems is considered robust at this stage, with extensive use of low risk, high TRL, and heritage components, and the application of industry-standard margins throughout. A brief summary of the spacecraft subsystems is provided below.

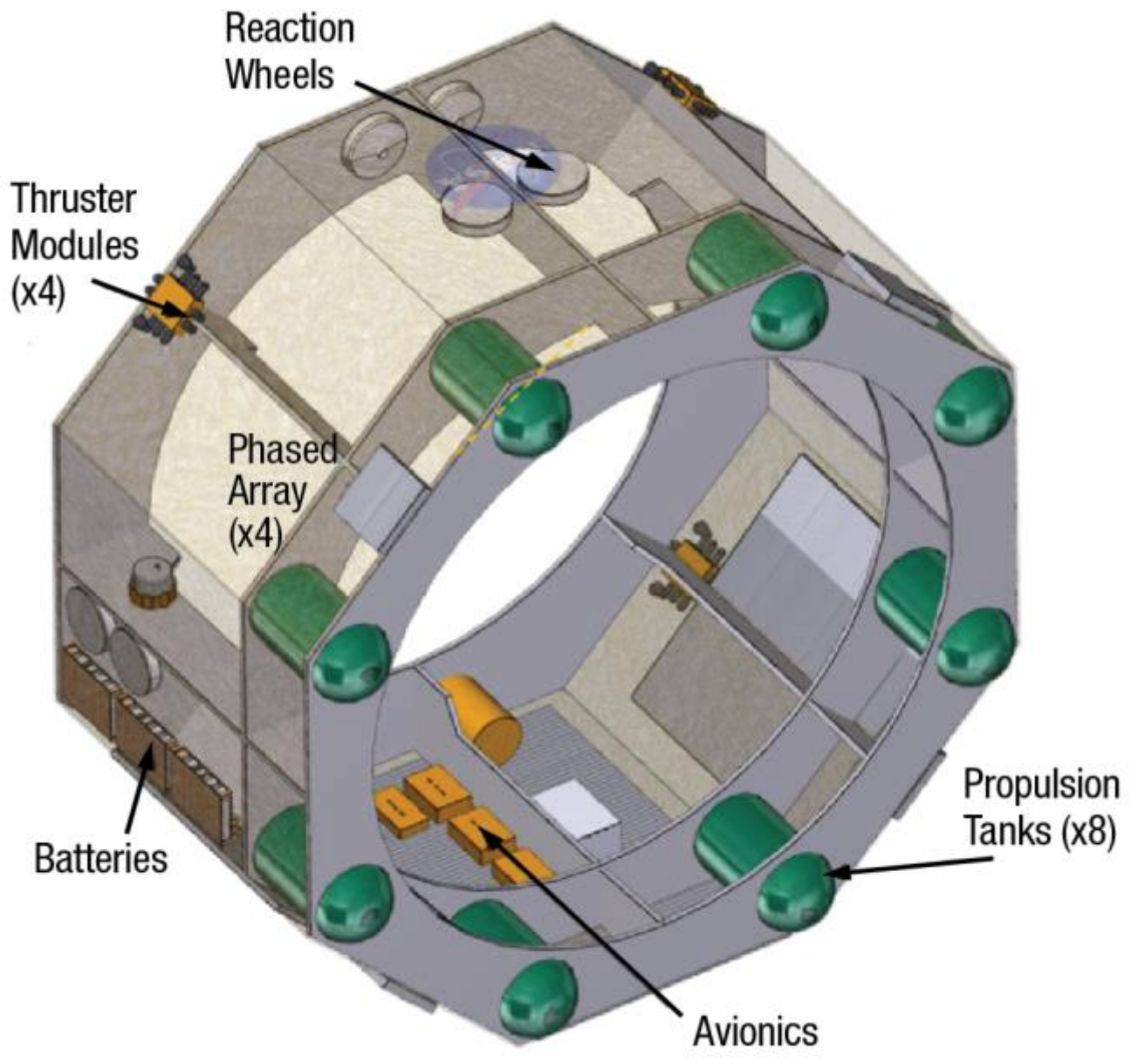

**Figure E-30. Schematic of the *Lynx* spacecraft bus. All spacecraft elements are high maturity level and are readily available.**

#### E.2.2.1 Propulsion

Mission analyses to determine delta-v are based on the *James Webb Space Telescope (JWST)* and the *International X-Ray observatory (IXO)*, while estimates of despin and momentum unloading propellant mass are provided by GN&C. The propulsion subsystem is a monopropellant blowdown system utilizing hydrazine as fuel and gaseous nitrogen as the pressurant and can be realized with existing flight-proven components. The current design utilizes eight ATK 80397-1 propellant tanks loaded with a total of 458 kg of propellant, four Northrop Grumman MRE-15 main engines and eight (plus eight redundant) MRE-1.0 RCS/ACS thruster modules. The propulsion system is sized to provide the required delta-v to reach the SE-L2 orbit and perform an initial despin with sufficient margin to perform station keeping and periodic momentum unloading maneuvers for a maximum of 20 years on orbit.

#### E.2.2.2 Guidance, Navigation, and Control

The GN&C system provides the mission-critical spacecraft attitude determination and pointing control. More specifically, this system achieves the required 10 arcseconds absolute pointing accuracy and stability of +/- 0.17 arcsecond per sec, per axis, necessary to implement the *Lynx* observing program. This system is designed similarly to that of *Chandra* and uses six Rockwell Collins TELDIX RDR 68-3 reaction wheels sized to counteract environmental disturbance torques. This architecture allows the observatory to carry out a 90° slew maneuver in approximately 50 minutes, excluding settling time of ~5 minutes. A maximum continuous observation time of $10^5$ seconds is possible, with pauses for momentum unloading as necessary. Longer continuous observation times can be scheduled with appropriate momentum management. Unloading of reaction wheel momentum due to environmental disturbance torques (primarily due to solar pressure) is assumed to occur once the reaction wheels reach a total momentum capacity of 85%.

The SE-L2 orbit is free from Earth and Moon shadowing, allowing for uninterrupted observing of any target. However, no *Lynx* science instrument can tolerate direct solar radiation, so viewing is restricted to angles larger than 45° from the direction toward the Sun. This restriction makes about 15% of the sky inaccessible at any given moment but no part of the sky remains inaccessible for more than three months of each year. In addition, the spacecraft and instrument designs take advantage of the hot and cold sides of the observatory to locate radiators, fuel lines, etc., which imposes a constraint on the observatory roll angle (rotation about boresight) of approximately +/- 15° to prevent impingement of direct sunlight on these surfaces.

The Pointing Control and Aspect Determination (PCAD) is part of the GN&C system that maintains *knowledge* of the spacecraft orientation in order to ensure health and safety, controls the maneuvers required to orient the telescope FOV to successive desired celestial targets, and holds each target attitude for the commanded duration. In addition, the PCAD provides position and orientation data in the science data stream, so that the ground aspect solution can reconstruct the relative X-ray image of the sky to within 0.2 arcsecond diameter rms, and provides absolute location on the sky to within 1 arcsecond rms radius. The PCAD system is based on the *Chandra* design heritage and the current design includes a Ball Aerospace star-tracker, three Honeywell Miniature Inertial Measurement Units, two Adcole Coarse Sun Sensors, and two Adcole Fine Sun Sensors, and is capable of tracking on the order of 8 to 10 images with 1s–4s readout.

To hold the target attitude, the PCAD subsystem uses a star camera with a 2 $deg^2$ FOV to acquire and track specific (cataloged), ~$6^{th}$–$10^{th}$ magnitude stars near the field of the target, inertial reference units (gyroscopes) to monitor drift rates, and reaction wheels commanded to spin, as needed, to compensate for disturbance torques (due primarily to solar wind and radiation pressure) and to provide the maneuvering to science targets. In addition, sets of fiducial lights are mounted near the HDXI and LXM, and a system of reflectors directs fiducial light to the star camera through registration points on the LMA. This hardware is used to monitor the detector to LMA alignment and stability. The star camera images the fiducial light pattern along with bright guide stars in the field. The post-facto aspect solution makes use of the guide star positions, the fiducial light positions, and the integrated gyroscope rate data to compute the solution for the pointing direction, roll, and gyroscope biases, and interpolates the solution to the precise arrival time of each registered X-ray photon event, allowing each photon to be registered to its point of origin on the sky.

Ongoing work in the GN&C area includes a detailed stability analysis, refinement of disturbance torque analysis, and a precise determination of the allowable roll angle error for thermal management.

#### E.2.2.3 Power

The electrical power system is designed to generate, store, manage, and distribute required power to the observatory throughout all phases of operation via a combination of deployable solar panels and on-board energy storage. Energy storage is provided via five 28-V batteries with one additional battery for 1-fault tolerance. The batteries are sized to provide launch power (725 W) for 156 minutes from launch to the completion of initial checkout and solar array deployment. Two Ultra-Flex deployable solar arrays with a total area of 42 $m^2$ are utilized to provide sufficient total power and are articulated to allow for full sun illumination for any boresight orientation with respect to the sun. The arrays are sized to meet the 7.8 kW peak power requirement when the observatory is in Normal Mode, Configuration 3 (**§E.3.3.1)** with ~40% margin. The power system design takes into account degradation over a 20-year mission lifetime. Refinement of the electrical power system is ongoing.

#### E.2.2.4 Thermal

Thermal control and regulation of the LMA and OBA are critical to meeting observatory performance requirements and the *Lynx* science objectives. The spacecraft thermal subsystem is designed to maintain the spacecraft bus at an average temperature of 283 K, and the assumption is that the XMA steady state temperature will be maintained at a temperature at which it is assembled and aligned, and that is warmer than the spacecraft (293 K +/- 1 K) to minimize contamination. The current design assumes the use of a high-TRL thermal control approach with the use of Optical Solar Reflectors (OSRs), Aluminized Kapton, heaters, and radiators. OSRs are selected in lieu of Multi-Layer Insulation (MLI) to avoid the effects of Silverized Teflon degradation as observed on *Chandra*. The thermal control design for the LMA will depend on the specific technologies selected for the mirrors and the gratings. Development of a detailed integrated thermal model of the observatory, including the thermal control of the LMA and OBA, will be completed by the Fall of 2018 and is being supported by the *Lynx* Study Office-industry CAN partnership.

#### E.2.2.5 Avionics and Flight Software

The avionics equipment located in the *Lynx* spacecraft bus is designed to perform the functions of GN&C, thermal control, power switching for the scientific instruments, data storage, command management, and uplink and downlink operations. These systems will draw heavily from heritage (e.g. Jet Propulsion Laboratory (JPL) *Mars Orbiter*, *Chandra* (Schwartz et al. 2014), *Spitzer* (Gehrz et al. 2007), and *JWST* design. *Lynx* does not have any requirements that would prohibit the use of technologies that are readily available.

For the flight computer, a flight heritage system will be baselined, similar to that used for the JPL *Mars Orbiter,* that is designed for long life in the SE-L2 environment. Based on analysis of the science instrument designs, the calculated science data collection rate is 240 Gbits per day. The spacecraft bus design assumes up to 48 hours of science data storage, with a total data storage capacity of 1 Tbit at 1.4 Gbits per second.

The *Lynx* flight software includes software for the spacecraft and science instruments. The spacecraft software will reside on the redundant spacecraft flight computers. Flight software will control communications and data handling, PCAD, electrical power, thermal control, and is responsible for recognizing fault conditions and managing safe modes. Safe mode control will include a separate set of control processing electronics that operate with different software. The science instruments will include software that will reside on the electronics units developed by each science instrument provider. All *Lynx* flight software development will comply with NASA Software Engineering Requirements, per NASA Procedural Requirement (NPR) 7150.2B, as Class A, Safety Critical software.

#### E.2.2.6 Command and Data Handling

For communications with the ground, *Lynx* will utilize NASA's DSN to provide telemetry, tracking, and command (TT&C), ensuring high reliability and high data rate communications for downloading its science and spacecraft health data and uplinking commands. **Table E-5** summarizes the relevant data volumes.

**Table E-5. *Lynx* data volumes are modest and easily handled in the 2030s timeframe.**

| Source | Expected Volume | Comments |
|---|---|---|
| Science Data Collection Rate (maximum average) | 240 Gbits/day (2.78 Mbps) | Based on science objectives and known X-ray fluxes |
| Downlink Frequency | 1–3 times/day; 1 hour each | *Chandra*-like operations |
| Uplink Frequency | 1–3 times/day; 1 hour each | *Chandra*-like operations |
| Downlink Rate | 22.2 Mbps | 240 Gbits/day collection rate, downlinked 3 times per day for 1 hour per downlink |
| Uplink Rate | <1 Mbps | |

For the purpose of this study, it is assumed that a flight heritage communication system will be baselined, similar to that used on the *Mars Reconnaissance Orbiter*, which supports data volumes up to 270 Gbits/day. It is also assumed that the *Lynx* communication system will utilize Ka-band for science data downlink (per Space Communications and Navigation (SCAN) guidance) and X-band for low-rate telemetry and backup. In the current *Lynx* concept, high-flight heritage phased array antennae are assumed. Avoidance of pointing vibrations arising from use of gimbaled antennae provides more stability to the telescope during science operations.

The *Lynx* final report will include a breakdown, by instrument, for required data rates. Instrument data rates facilitate efficient ground-based calibration, which typically bounds these requirements over those set by the science observations (such as those listed in **Table D-1**).

### E.2.3 Observatory Assembly, Integration and Test and Calibration

Assembly, Integration and Test (AI&T) and calibration activities are performed to verify and validate system and project-level requirements during the *Lynx* build-up. The tests, along with other standard requirement verification methods, will be defined in detail in the *Lynx* Verification and Validation Plan and specific test planning documents, which will be developed during Phase A of the project. Early consideration of system-level AI&T activities provides insight and validation of what the critical path will be during Phase C of the project. Analysis of AI&T flows for key systems also drives understanding of efficiencies to reduce over-testing of flight hardware, ground support equipment (GSE) needs, identification of test facility needs, and potential facility upgrade costs. Development of the *Lynx* AI&T was based on that used for *Chandra* and *JWST* flight systems, and proposed *IXO* systems. The overall AI&T flow is provided in **FO-3,** and the timeline for observatory-level AI&T is provided in the project schedule in **§G**. AI&T planning details are provided in **§H**.

Calibration of the XMA and the science instruments will be done prior to observatory AI&T. During AI&T, a combination of sources (optical and X-ray) will be used to characterize the end-to-end performance. Calibration methodology and required facilities, or modification to facilities, for the XMA is dependent on the particular technology selected for the notional reference mission and its assembly. These plans, as well as plans for calibrating the science instruments, will be included in the *Lynx* Final Report.

### E.2.4 Integrated Analysis of *Lynx*

An integrated analysis of *Lynx* architecture has been initiated via an industry Cooperative Agreement Notice (CAN) partnership. This study will enable refinements to the current design, by considering the integrated system, and producing an error budget for the on-orbit payload performance. This error budget will be used to update the observatory design until all payload performance requirements are met. During this process, trades will be identified that aim at optimizing performance, cost, and schedule for the fully integrated system.

## E.3 Mission Overview

The *Lynx* mission is designed to maximize science-observing efficiency and provide an environment that will extend the mission lifetime well beyond the nominal 5 years, while employing heritage, rigorously designed systems, and standard operating practices.

### E.3.1 Launch Vehicle

The outer diameter of the LMA design is 3-m; sufficient to satisfy the *Lynx* science requirements (**§D**) while allowing for flexibility in the choice of launch vehicle and fairing size. Due to the uncertainty regarding the specific launch vehicle with the payload envelope and lift capability to launch *Lynx* to SE-L2 in the 2030s, NASA's Launch Services Program (LSP) has provided payload envelope, lift capability, and environments for generic vehicle class types (Intermediate class and Heavy class), as well as for the SLS vehicles. Per current LSP guidance, the maximum payload mass to SE-L2 for the intermediate-class launch vehicles in the 2030s is 6,500 kg, and for the heavy class launch vehicles is 10,000 kg.

The *Lynx* observatory mass for the configuration considered in this report is 7,875 kg. That configuration incorporated the heaviest among three X-ray mirror technologies under consideration (**§E.2.1.1**). After the mission configuration for this interim report was frozen, the *Lynx* team completed a comprehensive technical and programmatic assessment of all three technologies. The option selected for the Design Reference Mission in the final *Lynx* study report leads to a significantly lower mass (reduction of more than 1,000 kg) for the *Lynx* Mirror Assembly. Based on this information, the heavy class vehicle meets the requirement to launch *Lynx* to SE-L2 with a margin sufficient for this stage of mission design.

The *Lynx* concept team is also currently working with the SLS Program at MSFC to determine configuration options for launching *Lynx* on the SLS vehicle as a co-manifested payload. This option provides considerable launch vehicle cost savings for the *Lynx* mission as launch vehicle costs are limited to integration services only. The current *Lynx* configuration easily meets the 10,000-kg and 7.2-m-diameter co-manifested payload envelope, but not the 8.4-m length limitation. To enable this option, the *Lynx* team will perform a trade study to determine the cost and risk associated with utilizing an extendable optical bench such that the observatory will fit within the SLS co-manifested payload envelope or shorter launch vehicle shrouds.

### E.3.2 Launch to Orbit

*Lynx* is planning to launch in the early to mid-2030s and the current assumption is that *Lynx* will be integrated onto a Heavy-class (expendable or recoverable) vehicle that will launch from NASA KSC. Following a transfer trajectory insertion (TTI) maneuver, *Lynx* will be inserted into the 800,000-km semi-major axis halo orbit around the SE-L2 libration point. Several orbits were analyzed for *Lynx* (study details can be provided upon request), including SE-L2, Drift-away, Lunar Distant Retrograde Orbit (LDRO), *Chandra*-type Orbit (CTO), and Transiting Exoplanet Survey Satellite (TESS)-like. After careful consideration, SE-L2 was selected because it provides: (1) essentially no eclipsing, (2) a stable thermal environment, (3) avoidance of trapped radiation belts, (4) high observing efficiency ≥ 85%, and (5) moderate fuel and propulsion requirements relative to some of the other orbits considered. The observing efficiency is the percentage of real-time *Lynx* will spend on science observations and takes into account estimated times for slewing, thermal and vibrational stabilization, and calibration.

The launch to orbit timeline and delta-v budget is shown in **Figure E-1.** This timeline assumes launch on a Delta IV Heavy vehicle. Even though the Delta IV Heavy is not expected to be available in the 2030s, it is assumed to be representative of expected capability (and not necessarily cost) of the generic Heavy-class vehicles in the 2030s.

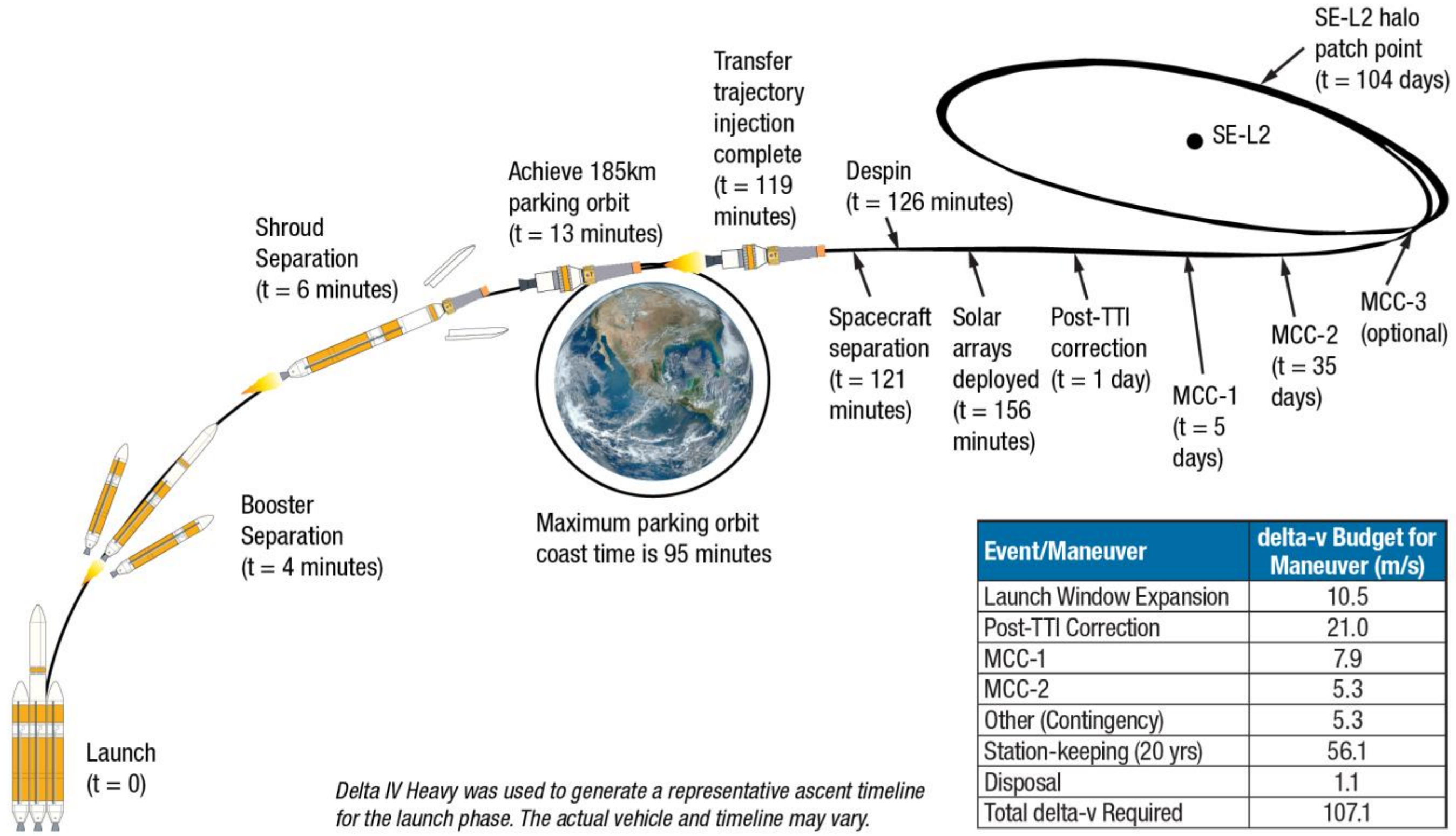

| Event/Maneuver | delta-v Budget for Maneuver (m/s) |
|---|---|
| Launch Window Expansion | 10.5 |
| Post-TTI Correction | 21.0 |
| MCC-1 | 7.9 |
| MCC-2 | 5.3 |
| Other (Contingency) | 5.3 |
| Station-keeping (20 yrs) | 56.1 |
| Disposal | 1.1 |
| Total delta-v Required | 107.1 |

**Figure E-11. Launch to orbit timeline and delta-v budget.**

For the first 156 minutes following launch, *Lynx* will rely on batteries to power minimal spacecraft systems, survival heaters for critical telescope elements, and to deploy the solar panels. Orbit progression for five years is shown in **Figure E-11.** The estimated time to reach SE-L2 is 104 days. During this time, the spacecraft and telescope systems are powered up, allowed to outgas, and undergo system checks and initial calibration. Early orbit operations schedules are being developed for each of the telescope systems, with an integrated element that takes into account contamination mitigation during the outgassing and checkout phases. The delta-v budget from launch through 20 years of operations is 107.1m/s, which includes station keeping to provide a stable orbit.

Once on-orbit at SE-L2, power provided to the instruments is consistent with their Normal Mode operation. If there are anomalies which threaten loss of control of the spacecraft, a hierarchy of safing operations will ensure a safe orientation and power-positive observatory. Non-essential systems may be powered down, redundant systems switched on, until the spacecraft is stabilized in a power-positive orientation. During normal operations, the on-board avionics system will collect data from the three instruments and spacecraft at a maximum average collection rate up to 240 Gbits/day. Data will be downlinked via the DSN up to three times per day, at a rate of 22.2 Mbps.

### E.3.3 Operations

All *Lynx* science instruments are photon-counting detectors that accumulate event-based time, position, and energy data. This data is accumulated and temporarily stored on-board, along with engineering data, before being periodically telemetered to the ground where it is archived, processed, and distributed to the scientific community. This section describes the fundamental on-orbit and ground support operations necessary to carry out the *Lynx* science program.

#### E.3.3.1 On-Orbit Operations

**Normal Operating Mode**

Following on-orbit activation and checkout, *Lynx* will be primarily in Normal (also called Science) Mode conducting an autonomous pre-planned program of celestial observations. A typical scientific observing timeline includes a series of maneuvers between targets, target acquisition, and data collection. Typical data collection times for *Lynx* are expected to range from ~10 ks up to ~100 ks per target pointing with slews to new targets requiring 1–3 ks. In this mode, the focal plane science instruments are either in a data collection or standby configuration, and the observatory attitude is maintained by the PCAD system under control of the onboard computer **(§E.2.2.2).** Normal spacecraft operations such as switching focal plane instruments, instrument calibrations, momentum unloading, ground contacts, and recorder data playback all take place in Normal Mode.

The translation table assembly portion of the ISIM is used to place either the HDXI or the LXM on the mirror optical axis. Independently, the XGS grating array can either be inserted into the optical path to intercept a portion of the focusing X-rays or retracted. When inserted, the XGS-dispersed spectrum is directed onto the XGS detector array while the non-dispersed portion remains focused onto the focal plane instrument. Therefore, there are four observing configurations available for science observations:

(1) XGS gratings inserted and LXM at the primary focus (XGS+LXM)
(2) XGS gratings inserted and HDXI at the primary focus (XGS+HDXI)
(3) HDXI as the primary with gratings retracted (HDXI-only)
(4) LXM as the primary focal plane instrument with gratings retracted (LXM-only)

A focusing mechanism located on the ISIM allows for adjustment for either the LXM or HDXI. The XGS detector assembly incorporates its own focus adjust mechanism. The instrument configuration is implemented during the slew to the target.

All Normal Mode operations are preplanned using a scheduling process that seeks to maximize the time on-target while accommodating all necessary spacecraft operations. The mission schedule plan will be used to generate spacecraft and instrument commands, which are then uplinked to the spacecraft and stored. A sufficient number of commands will be loaded to assure autonomous Normal Mode operation for 72 hours. Stored command loads can be interrupted and updated as needed to accommodate Target of Opportunity (ToO) requests (and emergency situations). It is anticipated that ToO requests may require up to 24 hours to initiate and review new command sequences, depending on spacecraft (thermal, power, momentum, and pointing) constraints, minimization of maneuver error, and the frequency of ground contact.

On-orbit calibration observations are performed as part of the Normal Mode science operations. A set of standard celestial targets will be determined for calibration use. These targets will be periodically observed to monitor LMA, science instrument, and aspect system performance. In addition, HDXI and LXM contain in situ calibration sources that are mounted on their filter assemblies. These will allow calibration data acquisition when those instruments are not carrying out celestial observations. As with other X-ray observatories such as *Chandra*, time allocations for calibration is expected to be about 5% of available observing time early in the mission and decrease to 2–3% subsequently.

**Safe Mode**

A hierarchy of safing actions will be defined in which the computers are switched off and Control Processing Electronics holds the vehicle in a safe, power-positive orientation. Safing actions are initiated autonomously by on-board detection of one or more preset limit sensor violations and may result in the spacecraft assuming a predetermined safe attitude while awaiting ground instructions **(§E.2.2.6)**. The spacecraft is designed to be able to survive in Safe Mode indefinitely. Anomalies will be recognized automatically by alert software and relayed to mission operations personnel (**§E.3.3.2**). The anomaly response will ensure that the spacecraft and instruments are safe, and then develop and implement an appropriate recovery plan to return to Normal Mode observations.

### E.3.3.2 Ground Operations

All science and observatory data will be received and all commands to *Lynx* will be generated by an integrated team comprised of Flight, Science, and Ground Operations personnel. The team is responsible for the spacecraft health and safety, carrying out all observational programs, monitoring and performing necessary maintenance, and retrieving and transmitting all data for processing, archiving, and distribution.

The Flight Operations Team schedules, plans, and generates spacecraft command sequences, uplinks and verifies spacecraft commands, and monitors real time data during communications with *Lynx*. The Science Operations Team is responsible for planning the mission schedule sequence, including specifying the science instrument configuration for each observation and on-orbit scientific instrument monitoring and calibration. The Ground Operations Team is responsible for supporting and maintaining all ground support hardware and software facilities used for scheduling, commanding, data flow, archiving, and communications. This includes facility infrastructure upkeep, network integrity, and facility security. A trade will be performed to assess the staffing of the facility with one shift per weekday plus dependence on automated alerts for anomalies during off-hours versus the current *Chandra* approach of staffing 2 shifts per day, 7 days per week, without depending on automated alerts during communication times with the observatory. Performance, risk, and cost will be the primary factors applied during this trade study. Because the mission operations facility is the sole interface to *Lynx*, redundant critical systems will be provided for at a physically separate site.

Communication to *Lynx* from the mission operations facility will be through the DSN (**§E.2.2.6**). One hour of telemetering during one to three daily contacts are envisioned during Normal Mode operations. Following data receipt and quality check, engineering data will be forwarded to the flight operations team for monitoring and health checks while all science and ancillary data will be transmitted to science personnel for further analysis. A schematic of the *Lynx* data flow is shown in **Figure E-2.** Science personnel services include standard data processing (with scientific validation and verification), archiving of data products and distribution to the *Lynx* community, maintenance and distribution of calibration products and analysis tools, and archival search and retrieval services. Science personnel also perform long-term mission planning by optimally scheduling targets provided by the *Lynx* user community. Because *Lynx* is a facility-class mission, proposed observations will be subject to peer review to maximize the scientific return of the *Lynx* observing program. This General Observer (GO) program approach has been successfully executed by past large missions such as *Hubble, Chandra*, and *Spitzer* to list a few. An open GO program can achieve the key science envisioned at conception while also enabling new directions of inquiry and unimagined new discoveries in the future.

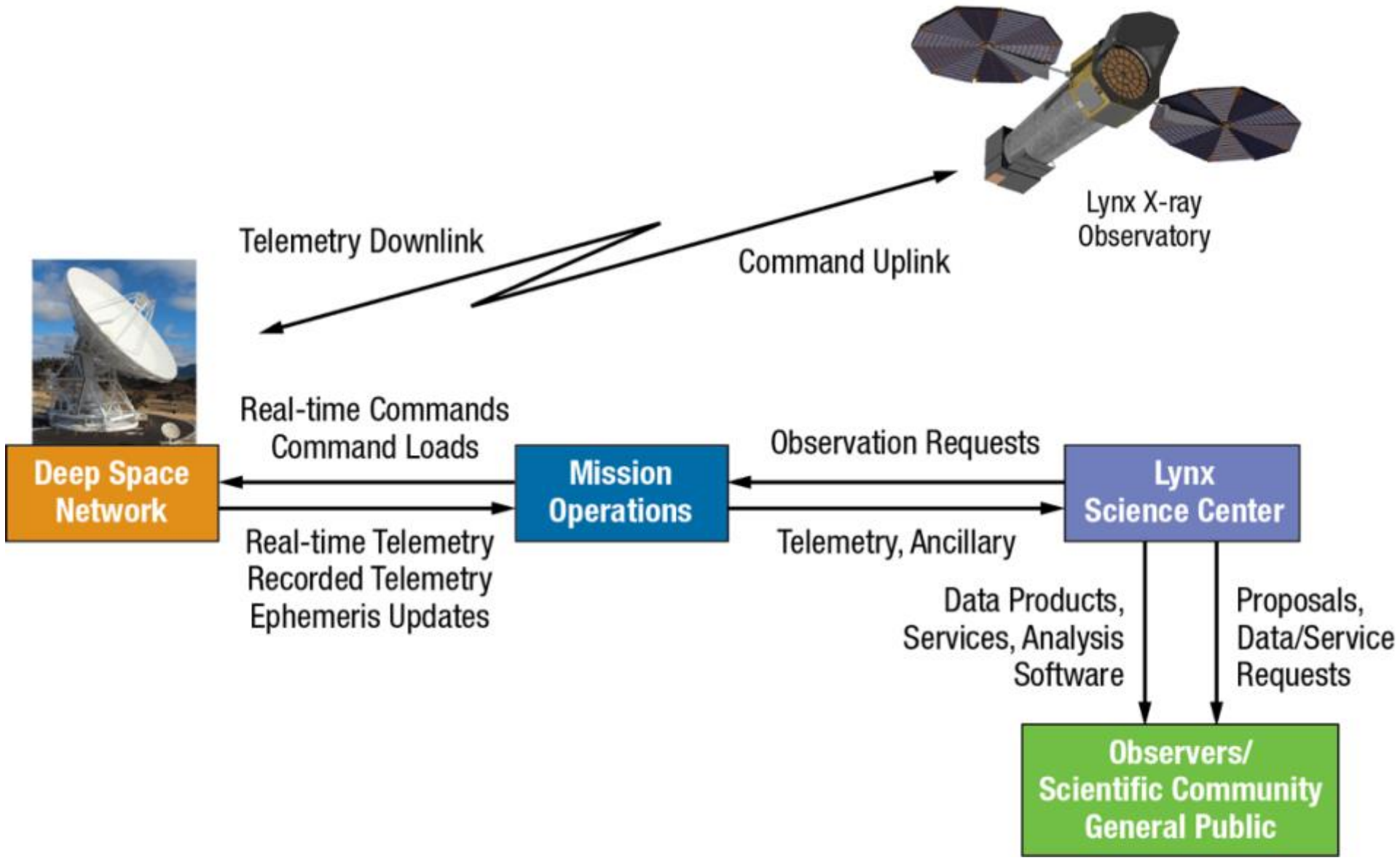

**Figure E-12. Architecture showing the physical flow of data from the *Lynx* Observatory, through the DSN to the *Lynx* Science Center, and ultimately to the observers, the general scientific community, and the public.**

# Mission Implementation

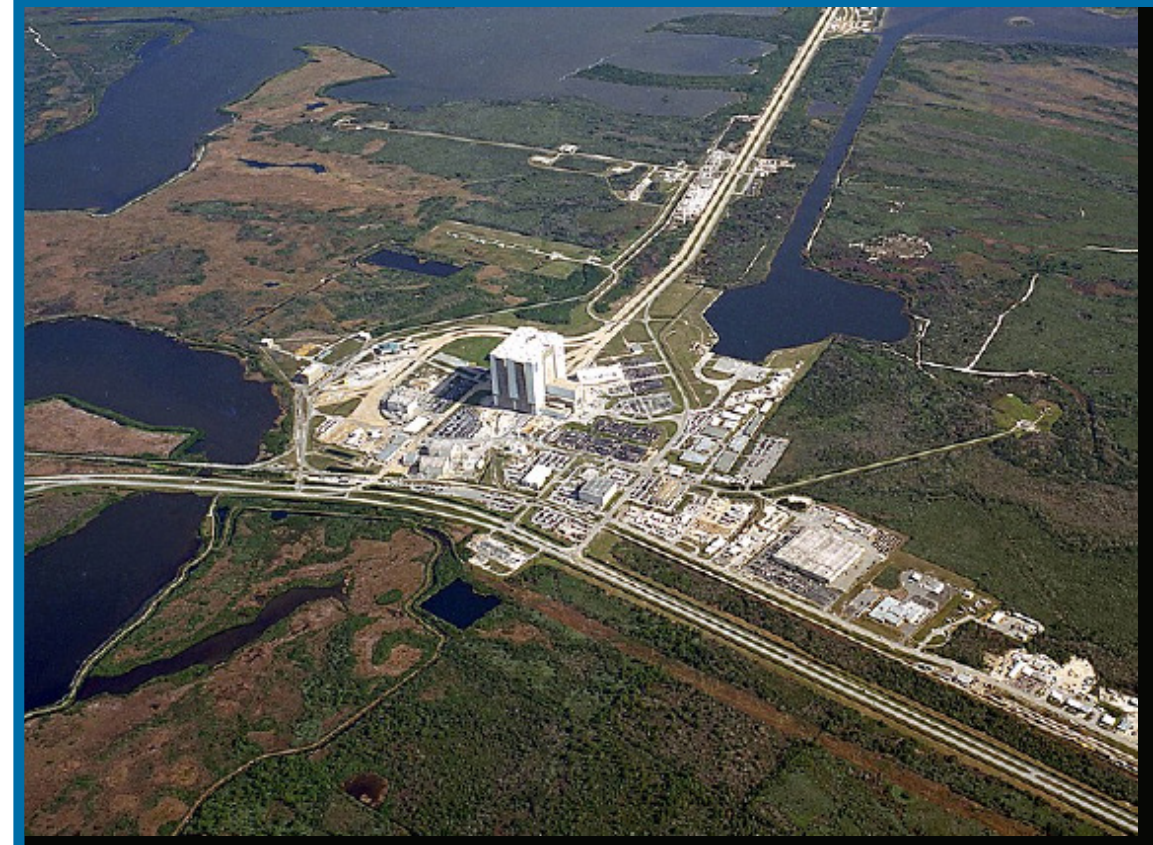

**Figure F02-1.** *Lynx* will launch from KSC in the mid 2030s for a five-year mission, extendable to 20 years.

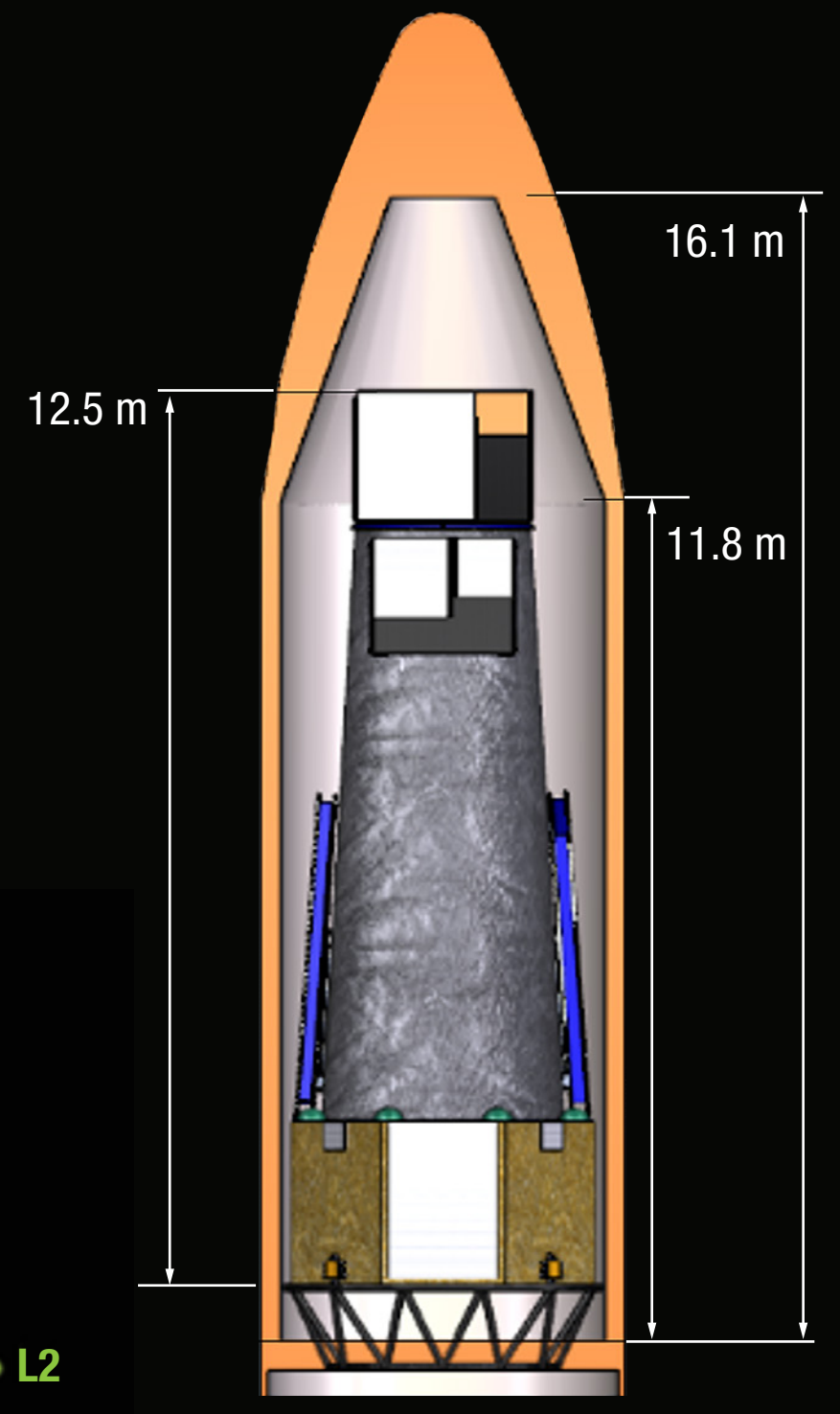

**Figure F02-2.** *Lynx* X-ray Observatory inside of a heavy-class vehicle payload envelope. Guidance provided by NASA LSP.

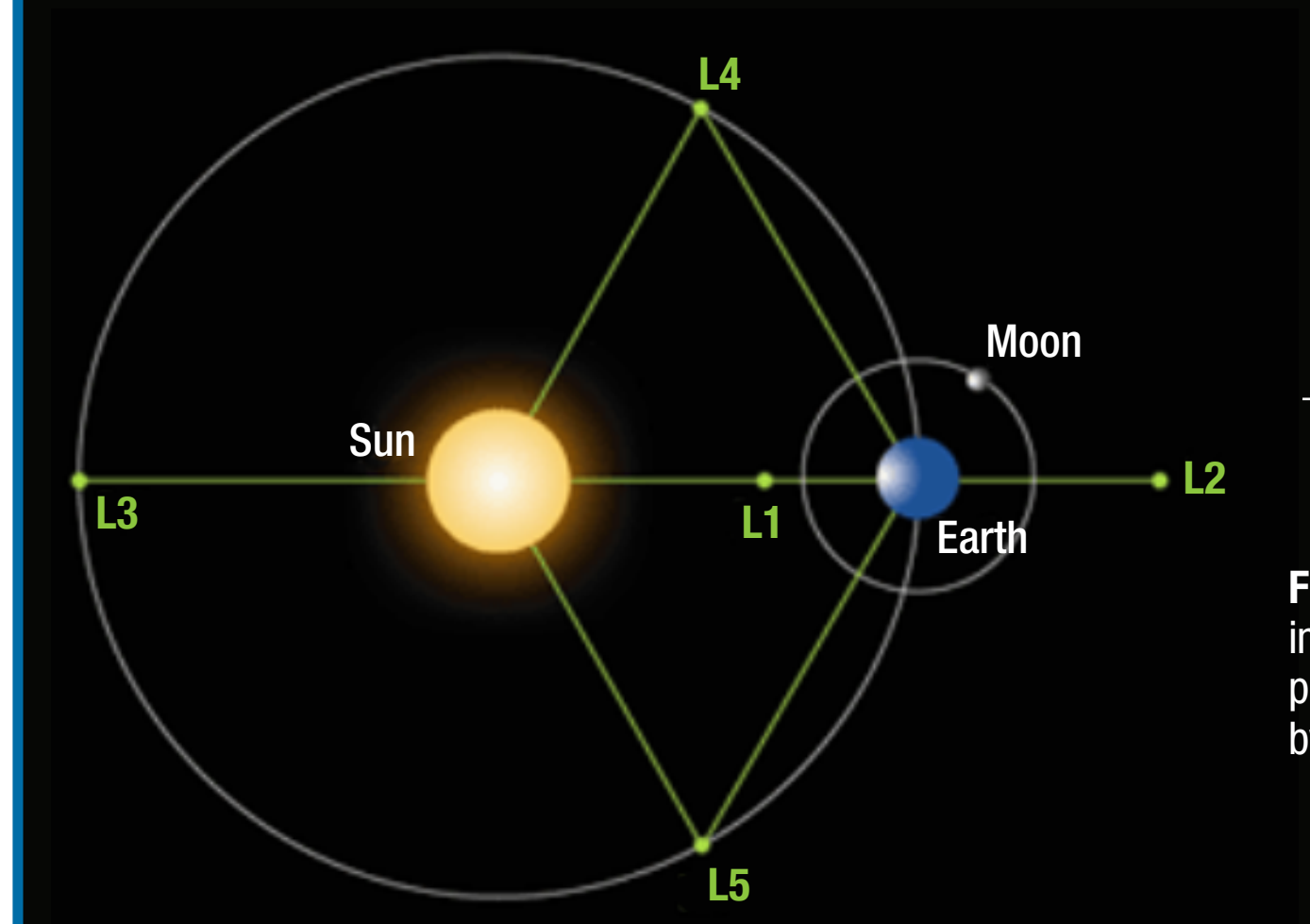

**Figure F02-3.** *Lynx* will operate in the SE-L2 orbit, selected for maximum suitability from science- observing, thermal, environment, and communications considerations over DAO, LDRO, CTO and TESS-like orbits.

**Table F02-1.** Mission Traceability Matrix.

| Mission Functional Requirements | Mission Design Requirements | Spacecraft Requirements | Ground System Requirements | Operations Requirements |
|---|---|---|---|---|
| Minimum observing efficiency 85% | • DRM requires 2.5 years to complete<br>• Place in SE-L2 halo orbit<br>• Propulsion to reach SE-L2 | • Control attitude<br>• Maneuver between celestial targets | • Efficient mission planning and target sequencing<br>• Derive aspect solution | • 1–20 targets/day<br>• Continuous data collection for 1–100 ks/target |
| Operate and survive in science orbit | • Solar power with battery storage<br>• Propulsion for momentum-unloading maneuvers, station keeping, and EOL disposal<br>• S/C bus surrounding XMA to ease thermal management of XMA<br>• Leverage hot/cold sides of observatory for thermal management<br>• Assume Risk Class A<br>• Minimal on-orbit servicing<br>• SE-L2 radiation and particle environment | • Distribute needed power; peak <8 kW<br>• Provide for communication with existing ground stations (DSN)<br>• Maintain operating temperatures within required limits<br>• Place instruments at optimum focus | • Monitor health and safety of observatory<br>• Provide for on-orbit calibration | • Maintain 45 degree Sun avoidance<br>• Command instruments for data collection and standby<br>• Restrict roll angle to manage thermal environment<br>• Store commands for up to 72 hr autonomous operation |
| Accommodate payload in launch vehicle | • Design for NASA-provided LV per LSP recommendations<br>• Design to survive launch: Fit w/in static and dynamic envelope defined by LV; exceed minimum modal frequency requirements for LV | • Battery power until solar array deployment | • Plan initial on-orbit activation and checkout | • Maintain optics and instruments in low-power mode prior to solar array deployment<br>• Instrument initial V&V |
| Provide data collection that is sufficient for uninterrupted observations by all science instruments | • Use existing DSN ground station for communications<br>• Ka-band for science data downlink (per SCAN guidance)<br>• X-band for command and engineering up/downlink | • 6 Mbps maximum data collection rate<br>• 1 Tbit onboard data storage<br>• Provide for 240 Gb downlink/day (3hrs @ 22 Mbps)<br>• Translate science instruments to/from primary aim point<br>• Insert/retract XGS<br>• Defocus as needed to manage high count rates | • Plan Normal (Science) Mode program<br>• Provide L0 data with <72 hrs latency<br>• Provide for data archival, retrieval, and distribution to public<br>• Plan instrument configuration to avoid excessive data collection | • Follow Normal (Science) Mode pre-planned observing program |
| 5-year mission | • Consumables for up to 20 years contingency | | | |
| Provide pointing attitude control and knowledge consistent with sub-arcsecond imaging, and stability consistent with 1 arcminute FOV | • Photon counting science instruments accumulate event-based time, position, and energy data | • Pointing attitude to 10 arcseconds absolute<br>• On-board knowledge 4 arcseconds<br>• Stability 0.17 arcseconds<br>• Chandra-like PCAD system | • Post facto image reconstruction consistent with 0.2 arcseconds RMS system accuracy<br>• Absolute celestial location to 1 arcsecond | • Monitor exact pointing attitude history and spacecraft alignment |

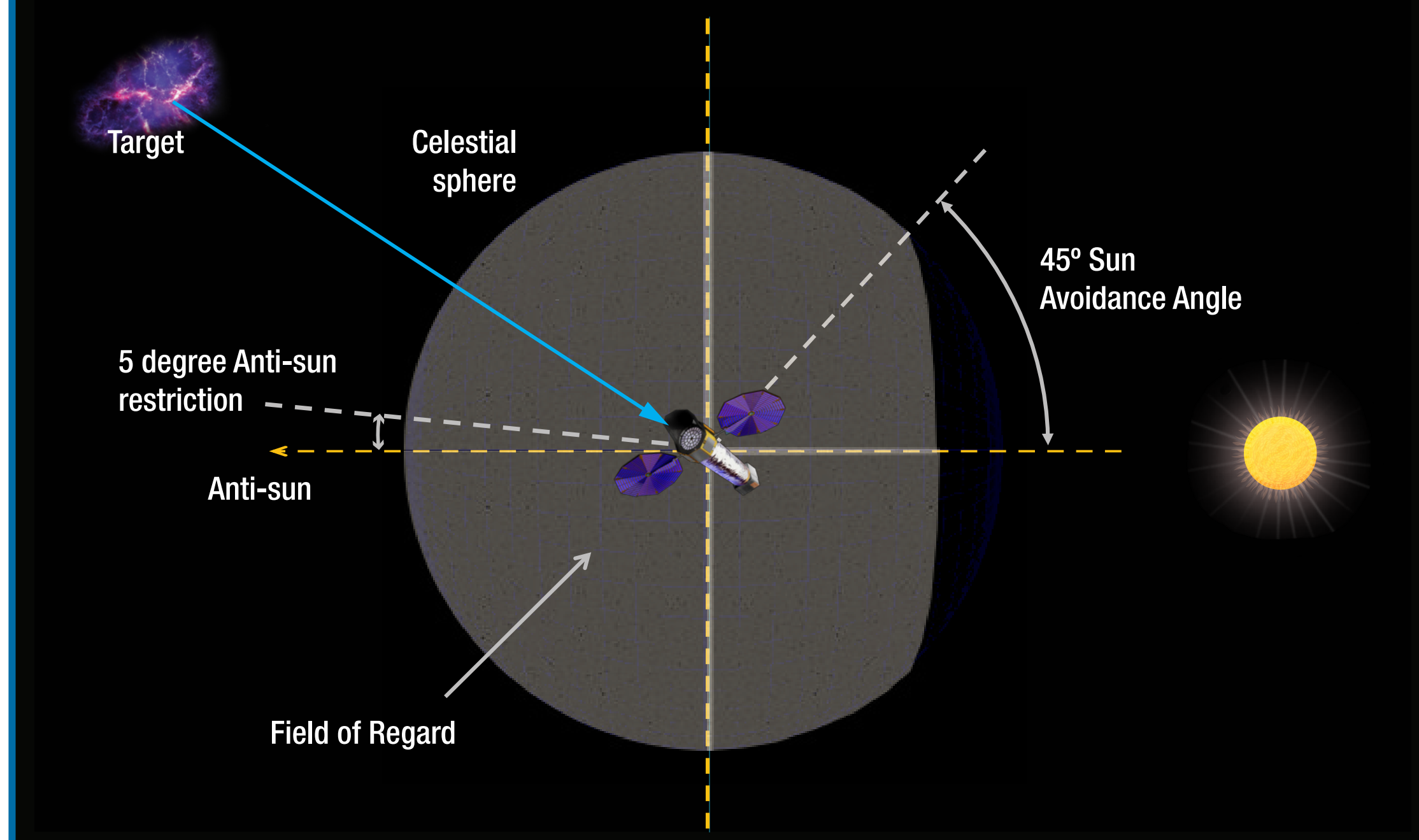

**Figure F02-4.** *Lynx* Field of Regard. *Lynx* can view the entirety of the celestial sphere less the 45 degree sun avoidance cone imposed by the sun shield that is in place to protect the sensors from solar impingement.

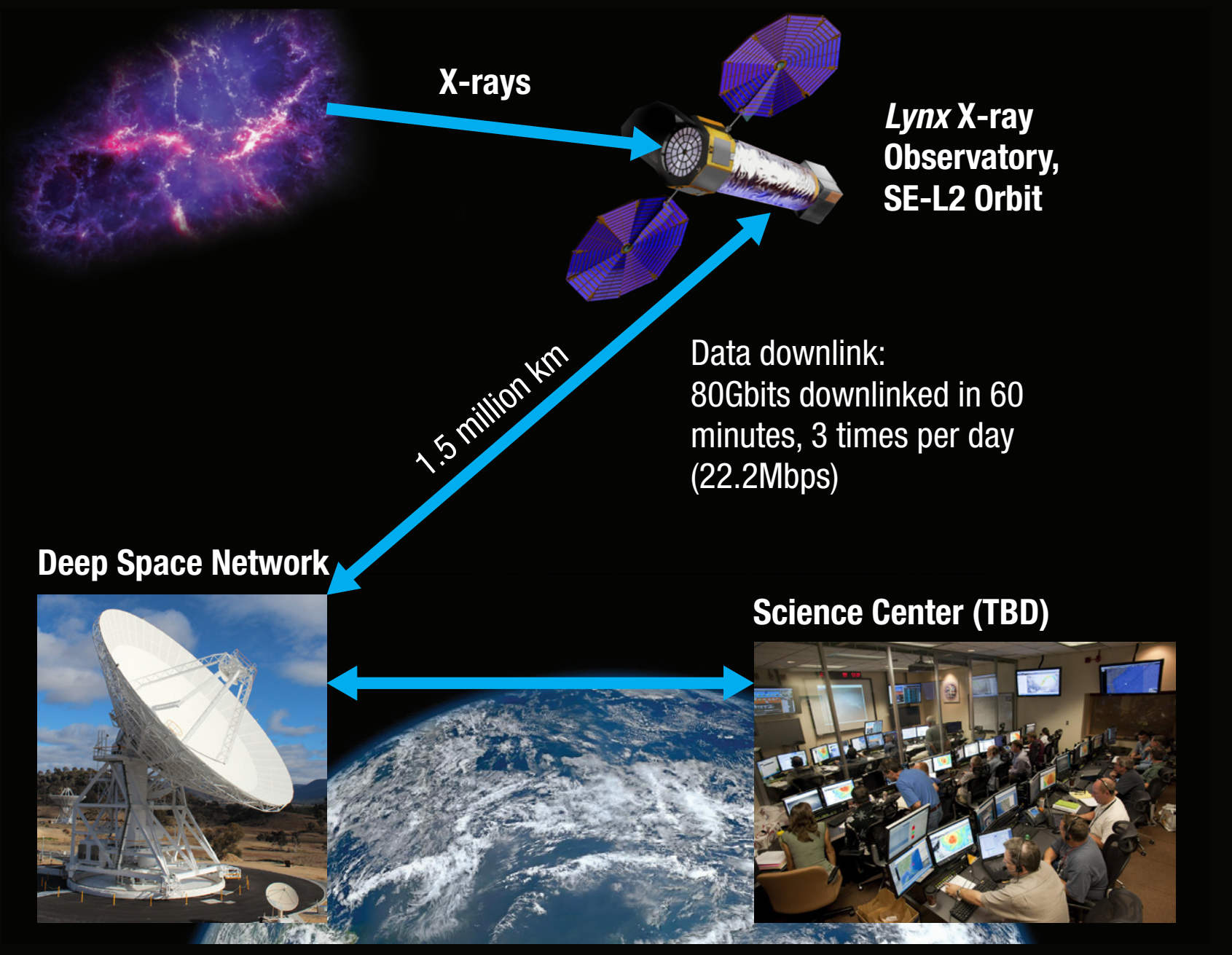

**Figure F02-5.** *Lynx* will use Deep Space Network (DSN) to provide communications between the spacecraft and ground operations. The DSN provides three sites, located in California, Spain, and Australia. DSN provides nearly continuous contact coverage during commissioning and meets the requirements to accurately determine the *Lynx* orbit and provide sufficient uplink and downlink telemetry rates.

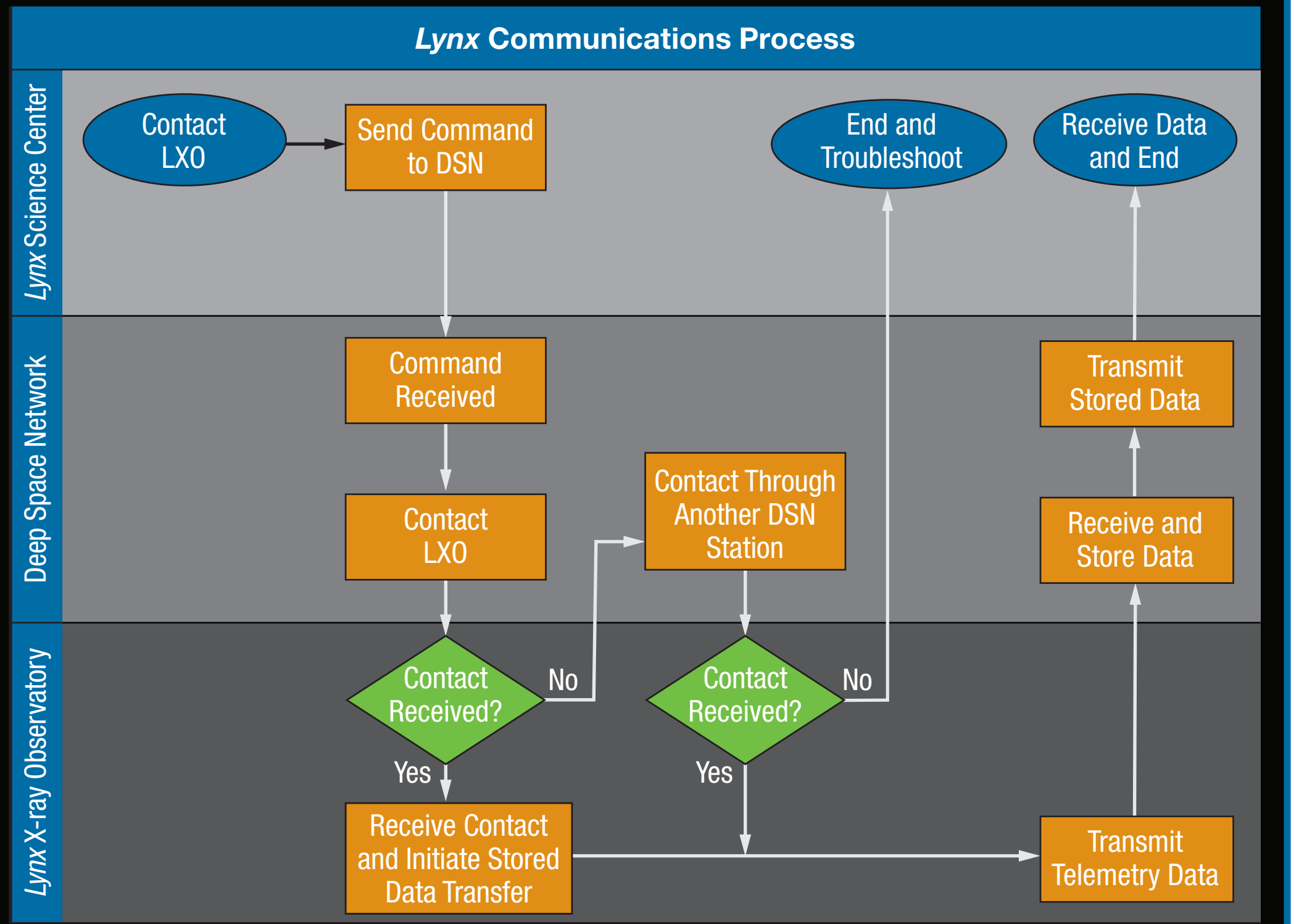

**Figure F02-6.** *Lynx* Communications Process.

# F. *LYNX* TECHNOLOGY DEVELOPMENT

**The *Lynx* optics are currently at a TRL 2–3 and the science instruments are at a TRL 3–4. It is expected that all enabling technologies will be at a TRL 4 prior to the Decadal decision in 2020. All enabling *Lynx* technologies will achieve a TRL 5 by Phase A and TRL 6 by Preliminary Design Review (PDR), and have well-understood and acceptable Technical and Programmatic risk and cost.**

## F.1 Summary Overview

*Lynx* will be by far the most capable of a series of dedicated missions operating in the 0.2–10 keV energy range using grazing-incidence optics to focus X-rays onto high-resolution imaging detectors and spectrometers (both imaging non-dispersive and dispersive). *Lynx* has a well-bounded architecture that makes use of a few key technologies to accomplish Flagship-quality science over a long mission lifetime.

Over the past decade, there have been significant advances to each of these key technologies; however, development beyond the current state of the art (SOTA) is required for each to meet the *Lynx* science requirements. **Table F-1** summarizes the key *Lynx* technologies, their current SOTA as defined by the Physics of the Cosmos Program Office, the specific *Lynx* requirements that drive their development, and the main challenges to advancing their TRL specific to the *Lynx* mission. **Table F-1** also shows the current TRL and the milestone(s) to mature to the next level as assessed by the most recent Physics of the Cosmos (PCOS) Annual Technology Report. **All *Lynx* key technologies have actively funded efforts and are expected to reach (or approach) a minimum TRL 4 prior to the Decadal decision in 2020.**

**Table F-1. *Lynx* enabling technologies requiring technology maturation. SOTA descriptions are from *Lynx*-independent PCOS Program Annual Technology Report.**

| Technology | State of the Art (PCOS) | *Lynx* Requirements | Challenges | TRL |
|---|---|---|---|---|
| High-resolution, large-area, lightweight X-ray grazing incidence mirror assembly 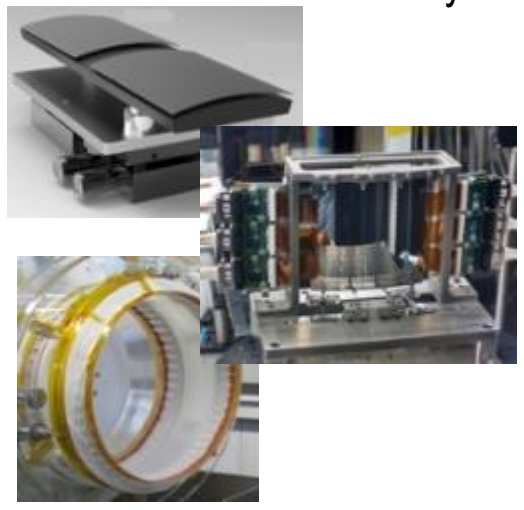 | •Less than 1 mm-thick uncoated and lab-mounted, achieved 3.7 arcseconds in full-illumination X-ray test[1*] (Recent results indicate 2.2 arcseconds).<br>•Simulations predict corrections of 7–10 arcseconds quality slumped glass segments down to 1–2 arcseconds single reflection HPD[2]<br>•Direct-polished fused silica monolithic full shell, 0.48-m diameter 2-mm thick; no recent tests[3] | •2 $m^2$ EA at 1 keV<br>•0.5 arcsecond on-axis, on-orbit system angular resolution<br>•Minimum grasp 600 $m^2$ arcminutes$^2$ at <1 arcsecond resolution at 1 keV<br>•3-m max outer diameter | •Low areal cost and mass<br>•Maintain thermal and structural integrity through coating, alignment, mounting and/or post-fabrication figure correction to preserve high-precision figure | Level: 2–3 → 4<br>•Technology must demonstrate 1 arcsecond performance (of single pair or full shell) through X-ray testing. |
| Fast, low-noise, megapixel X-ray imaging arrays 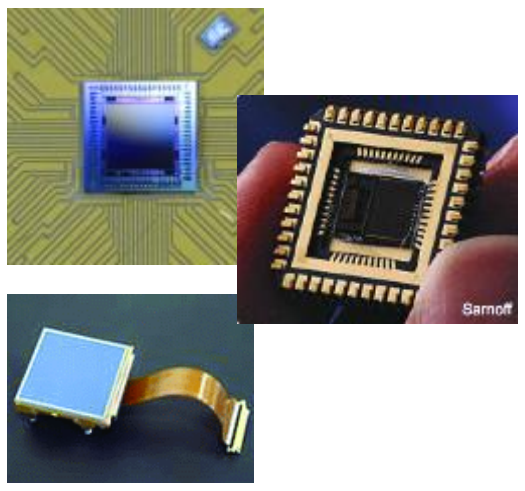 | •hybrid CMOS APS: at TRL > 6 but noise and low-E sensitivity need further development; currently have read noise 5.6 e-, energy resolution 156 eV @ 5.9 keV, small pixel (12.5 µm), large array[6]<br>•Monolithic CMOS APS: small pixel large format (1k x 1k), response below 1 keV, needs development on response above 1 keV and on high RTS noise[7] | •64-mm x 64-mm array with 16-µm pitch<br>•0.2–10-keV energy range at high QE<br>•[TBD count rate capability] | •Imaging arrays covering wide fields of view with excellent spatial resolution, fast readout, low noise, and moderate spectral resolution | Level: 3 → 4<br>•Breadboard demonstration of a single sensor with Event Processor Electronics and low-power ASIC. |
| | •Digital CCD with CMOS readout in development stage[8] | | | |
| High-efficiency high-energy-resolution X-ray grating arrays 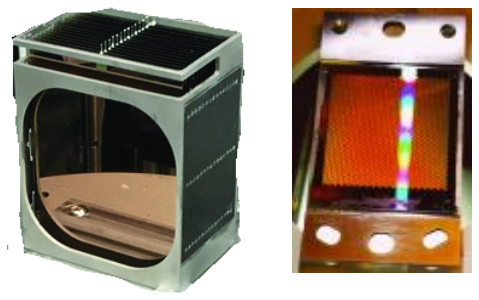 | • 32 mm x 32 mm, 4 µm depth achieve 30% absolute efficiency, R>5,000[9]<br>• X-ray tested 55–65% efficiency 0.2–1.3 keV[10] | •R>5,000 over 0.2–2 keV<br>•0.4 $m^2$ effective area at 0.6 keV | •Demonstrate fabrication scalable to large format grating arrays<br>•Improved diffraction efficiency | Level: 4 → 5<br>•Fabrication, mount, align, and X-ray test large-format array module<br>•Demonstration in operational environment |
| Large-format, high-spectral resolution, small pixel X-ray focal plane arrays 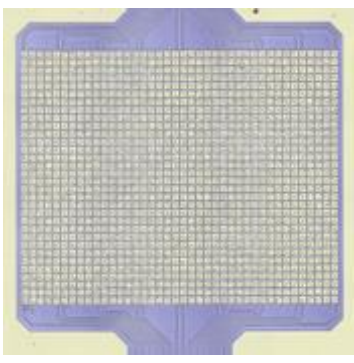 | • *Athena* prototype 6 x 32 and 32 x 32 arrays < 3 eV at 6 keV TRL 4 (ROSES/SAT) multiplexing w/SQUIDs demonstrated 130 pixels (TRL 4)[11]<br>• large arrays[12]<br>• high energy resolution[13]<br>• read out using microwave SQUID resonators | •~100k pixel total<br>•5 arcminutes x 5 arcminutes field main array optimized for 0.2-10 keV and 1 arcminute x 1 arcminute side array up to 1 keV<br>•0.5 arcsecond to 1 arcsecond (25–50 µm) pixels<br>•0.3 eV (side array) to 3 eV (main array) energy resolution | •Fabrication, wiring, and readout of large-format small pixel detector array | Level: 3 → 4<br>•Demonstrate the required performance (pitch, energy resolution, energy range) with required wiring density for all pixel types<br>•Demonstrate microwave SQUID circuitry and HEMT architecture w/ desired multiplexing properties |

* References noted in State of the Art column above are shown below. Full citation shown in Appendix H.4, References: 1 – Zhang et al. (2017), Chan et al. (2017); 2 – DeRoo et al. (2017); 3 – Civitani et al. (2017); 4 – Chan et al. (2014); 5 – Broadway et al. (2015); 6 – Hull et al. (2017); 7 – Kenter et al. (2017); 8 – Ryu et al. (2017); 9 – Heilmann et al. (2017); 10 – Miles et al. (2017); 11 – Kilbourne (2015-2017); 12 – Bandler et al. (2016); 13 – Lee et al. (2015)

A high-level summary of technology development milestones for all the key (optics and science instrument) technologies to achieve TRL 5 by KDP-A and TRL 6 by PDR is provided in the following. In most cases, there is more than one technological approach with the potential to reach the required science objectives. In these cases, each approach is addressed separately. The technology development plans assume KDP-A, the start of mission Phase A, occurs October 1, 2024 and PDR occurs 45 months following that gate (**Figure F-1**). This is consistent with the project schedule provided in **§G**.

| NASA Life-Cycle Phases | Approval for Formulation | Formulation | Approval for Implementation | Implementation | | |
|---|---|---|---|---|---|---|
| Life-Cycle Phases | Pre-Phase A: Concept Studies | Phase A: Concept and Tech Developmen | Phase B: Preliminary Design and Tech Completion | Phase C: Final Design and Fabrication | Phase D: System Assembly, Integration & Test | Phase E: Operations and Sustainment |
| Life-Cycle Gates | KDP-A 10/1/24 | KDP-B 10/1/26 | KDP-C 10/1/28 | KDP-D 6/1/35 | KDP-E 10/1/36 | |
| Project Life-Cycle Reviews | MCR 8/1/24 | SRR 8/1/26 | PDR 7/1/28 | CDR 9/1/30; SIR 3/1/35 | ORR 7/1/36 | |
| | | Subsystem Reviews | | | | |

**Figure F-1.** ***Lynx*** **Mission timeline consistent with NPR 7120.5. KDP – Key Decision Point, MCR – Mission Concept Review, SRR – Systems Requirement Review, PDR – Preliminary Design Review, CDR – Critical Design Review, SIR – Systems Integration Review, ORR – Operational Readiness Review.**

## F.2 Optics

The *Lynx* X-ray mirror assembly (XMA) must provide a large effective area and exquisite angular resolution over a large FOV. Anticipating future launch vehicle capabilities, the XMA and supporting structures must be designed to achieve low mass per unit collecting area, have the structural integrity to withstand launch conditions and the environment of space, and maintain its optical precision throughout the life of the mission.

Large effective collecting areas are achieved by nesting very large numbers of thin, lightweight mirror pairs that fully utilize the available aperture, resulting in a mirror assembly comprised of hundreds of full-circumference or thousands of segmented mirror elements. The high angular resolution will require precision polishing, alignment, and bonding of the mirror elements, as well as careful structural and thermal design to preserve the optical performance throughout calibration; observatory assembly, integration and test (AI&T); launch; and science operation. Alternatively, less precisely figured optics could be corrected to final precision after assembly using an adjustable segmented optics approach.

Three X-ray optic technology approaches have been identified that conceptually meet the *Lynx* effective area, angular resolution, and grasp scientific requirements while remaining within system-level mass and geometry contingency reserves. These are: (1) Silicon Meta-Shell Optics that use precision-polished monocrystalline silicon segments interlocked and bonded onto a central structural shell, (2) Adjustable Segmented Optics that utilize low-voltage piezo actuators to induce in-plane stress for figure error correction of thin slumped-glass segmented mirror elements and a modular construction, and (3) Full-Shell Optics that use directly fabricated, polished, and passively figure-corrected full-circumference mirror shells. All three optics developments are actively funded by NASA; Full-Shell Optics is also funded in Italy by the Italian Space Agency (ASI) at the Brera Astronomical Observatory (OAB), part of the National Institute for Astrophysics (INAF).

| Phases | | Pre-Phase A: Concept Studies | | Phase A: | Phase B: |
|---|---|---|---|---|---|
| Gates | | | KDP-A<br>10/1/24 | KDP-B<br>10/1/26 | KDP-C<br>10/1/28 |
| Reviews | SOTA<br>TRL: 3-4<br>3/18 | TBD | MCR<br>8/1/24 | SRR<br>8/1/26 | PDR<br>7/1/28 |

| TRL2 | TRL3 | TRL 4 | TRL 5 | TRL 6 |
|---|---|---|---|---|
| • Technology concept and/ or application formulated.<br>• Invention begins, practical application is identified by speculative, no experimental proof or detailed analysis is available to support the conjecture.<br>• Documented descriptions of the application/ concept that addresses feasibility and benefit.<br>• SOTA esceeds TRL 2 for all optics technologies considered | • Lab validation (metrology or X-ray) of mirror figure and, if applicable, the ability to correct mirror figure for single segment or for full shell azimuthal segment, must be demonstrated to within a factor of 6 of error budget allotment.<br>• Lab validation (early prototype-proof of concept) of mounting and essential hardware elements must be demonstrated to within a factor of 6 of error budget element.<br>• Models, analogies, or lab demonstrations validating other elements related to as-corrected mirror error contributions (e.g. coatings, thermal, g-release) must be made. | • Lab validation (X-ray test) of breadboard lab mount of at least one aligned mirror pair or full shell pair, coated, full illumination, must be demonstrated to within a factor of 3 of error budget roll-up for these elements (figure correction, coating, mounting, alignment—Mirrors have thicken and size consistent with point design).<br>• Models, analogies, or lab demonstrations validating other elements related to as-corrected mirror error contributions (e.g. thermal, g-release) must be made.<br>• Demonstration must be traceable, via error budget, to the on-orbit operational environment. | • Lab validation (X-ray test) of three aligned, coated, realistically mounted mirror pairs or full shell pairs (innermost, middle, and outermost) must be demonstrated to within a factor of 1.5 of error budget roll-up for these elements (mirrors have thickness and size consistent with point design).<br>• Assemblies must be tested in operational environment that includes vibrational and thermal.<br>• Models, analogies, or lab demonstrations validating other elements related to as-corrected mirror error contributions (e.g. module-to-module alignment if applicable, g-release) must be made.<br>• Demonstration must be traceable, via error budget, to the on-orbit operational environment. | • Critical environmental and X-ray test of a scalable flight-like protoype assembly that demonstrates that *Lynx* requirements are met. This demonstration must include all thermal control subsystems for testing. |

| Technology Drivers | • Mirror Fabrication | • Mounting, alignment and bonding<br>• Manufacturing |
|---|---|---|

**Figure F-2. Preliminary *Lynx* optics technology SOTA and TRL maturation path to TRL 6. The red box indicates that the current SOTA is between a TRL 2 and TRL 3 for all optics technology approaches considered for *Lynx*.**

### F.2.1 Silicon Meta-shell Optics

This design is based on a hierarchical meta-shell concept (McClelland et al. 2016, 2017) that uses monocrystalline silicon as the substrate for 100 mm x 100 mm x 0.5 mm mirror segments needed to meet *Lynx* performance. This material can be fabricated and polished to high quality because it is nearly free of internal stress (Zhang et al. 2012, 2013, and 2017). The technology uses traditional direct polishing methods, as were used to achieve *Chandra*'s 0.5-arcsecond angular resolution, and modern direct polishing techniques including stress-polishing and ion-beam figuring.

Recent full-illumination X-ray tests of an uncoated single mirror segment pair have demonstrated a 2.2-arcsecond HPD, thus establishing the silicon meta-shell optics technology at TRL 3. Other tests have demonstrated iridium coating, alignment, and bonding processes that do not degrade mirror pair figure beyond performance expectations.

**Technology Maturation Summary:** To advance to TRL 4, a single, coated, stiff-fixtured mirror pair of final size and thickness must demonstrate roughly 1 arcsecond (3x error budget for this element) performance under full X-ray illumination. This will require building and testing single-pair mirror modules, incorporating refinements in the mirror fabrication process of grinding, lapping, etching, polishing, trimming, and ion-beam figuring to achieve a uniform consistency in figure quality and micro-roughness. In parallel, another technological gate will be to align and bond mirror substrates using precision-machined spacers identical to those that will be used in the XMA configuration. This will require fabricating and machining or lapping monocrystalline silicon spacers to a height precision of 20-nm rms. To date, the process has been verified to be able to align mirrors with 0.1-arcsecond error in image centroid. The bonding process, using epoxy to permanently attach each mirror to its four spacers, will also be refined. Bonding has been verified with both optical metrology and X-ray testing to introduce no more than 0.5-arcsecond HPD to the currently achievable image of a pair of mirrors; vibration tests verify the process to be of sufficient structural integrity. In addition, advancement to TRL 4 will require demonstration of the basic functionality of an integrated multi-pair subassembly prototype. Advancement to TRL 4 is expected by early 2020.

To advance to TRL 5, an assembly of three or more co-aligned mirror pairs must be built, X-ray tested, and shown to achieve the required angular resolution performance (within a factor of 1.5 of final requirement). In addition, the mirror module must sustain spaceflight environmental tests, including vibration, thermal-vacuum, acoustic, and shock, yet still maintain its optical performance. The fabrication process listed above must be repeated but to the more exacting sub-arcsecond *Lynx* requirement. Advances in metrology will be needed to calibrate reference flats on interferometers and null lens optics to ensure optical metrology provides the necessary accuracy to reach this milestone; the alignment and bonding process must be extended to multiple mirror pairs. The completion of TRL 5 work will result in a concrete plan, including schedule and cost estimates, to establish TRL 6.

To achieve TRL 6, a meta-shell that is similar to or identical to a *Lynx* flight meta-shell will be constructed and fully tested to meet all requirements: point spread function (PSF), effective area, FOV, mass, vibration, thermal-vacuum, acoustic, and shock. In particular, the construction and testing process of this meta-shell would help establish the production and qualification processes of flight meta-shells. The completion of this TRL 6 demonstration will retire all risks, both technical and programmatic, including schedule, cost, and those associated with the logistics of building and testing all *Lynx* meta-shells.

**Development Considerations:** The silicon meta-shell optics approach relies on (with the exception of some metrology equipment) commercial off-the-shelf materials, components, and equipment for the manufacturing, assembly, and testing of a large number of mirror segments. The production of the XMA represents a significant manufacturing, quality control, and logistical challenge, but the silicon meta-shell approach is highly amenable to parallel production with several lines envisioned to run simultaneously to meet schedule and cost Program requirements. During technology development (as early as Pre-Phase A), consideration will be given to developing automated techniques for fabricating mirror substrates, coating, alignment, and bonding procedures, which can lead to additional cost and schedule savings. A detailed manufacturing and assembly plan will be developed for the *Lynx* final report.

### F.2.2 Adjustable Segmented Optics

The adjustable segmented optics approach uses a lithographically printed pattern of individually addressable electrodes over a thin continuous film of piezoelectric material to produce controlled, localized stress that acts as a bimorph when energized to induce mirror figure correction. The concept is to apply post-assembly corrections that compensate for fabrication, mounting, and bonding errors. The potential exists with this design for on-orbit figure correction of thermally-induced distortions by adding semiconductor strain gauges to monitor temperature changes. The design utilizes thin (0.4 mm thick), deformable slumped glass mirror substrates. A figure control test (DeRoo et al. 2017) of a 0.22 m radius-of-curvature cylindrical segment test mirror (100 mm x 100 mm) with 112 individually-connected piezo-actuators demonstrated the ability to deterministically correct mirror figure. Recent optical metrology tests indicated that corrected and measured differences between imparted figure correction and desired figure correction is about 0.5 arcsecond. The technology is currently at TRL 2–3.

**Technology Maturation Summary:** The adjustable segmented optics technology requires improvements in mirror fabrication, alignment, and mounting. The first mirror development milestone is an optical test of a single sub-scale[3] mirror segment that demonstrates successful electrical connectivity among the components, the stability of the piezoelectric bimorph effect, and a reduction in noise sensitivity. This will be followed by an X-ray test to demonstrate successful release layer coating of a high precision mandrel, stress compensation of piezoelectric processing to +/-10%, and reduction of mirror mounting deformations to <0.5-arcsecond post-correction. The milestone needed to reach TRL 4 will be to demonstrate X-ray performance of a full-scale mirror pair within a of 3 of error budget allowance and demonstrate row-column addressing circuitry. Steps to demonstrate the circuitry includes developing a lithography mask alignment and printing process for application to curved mirror segments, demonstrating the conductive paths through the insulating layer, full functionality and yield, and lifetime testing. To advance to TRL 5 will require X-ray testing a mirror module using full size mirror segments and achieving the *Lynx*-required angular resolution with piezo-actuator adjustment. The module must also survive appropriate operational thermal and vibrational environments.

The mirror alignment development is currently at TRL 2. An alignment metrology system must be constructed and tested for stability and repeatability and demonstrate independently verified accuracy. TRL 4 will be reached when alignment of a mirror segment pair is demonstrated using a full-scale alignment metrology system to 0.25 arcsecond accuracy and verified by an X-ray test. To attain TRL 5, the alignment scheme must be demonstrated with full size multiple mirror segments and performance verification via X-ray measurements.

The mirror mounting technology development stages will begin with structural analysis incorporating load conditions, gravity release, temperature sensitivity and the capacity to correct the mounted mirror segments. Demonstrations will begin with single sub-scale mirror segments and progress toward full-scale mirror segment pairs. TRL 4 will be met when the mounting design passes thermal, vibration, shock, and acoustic testing and is correctible to 0.1 arcsecond. TRL 5 will require demonstrating (including environmental testing) a representative mirror module consisting of several mirror segment pairs and mass surrogates in a flight-like module structure and integrating the mirror fabrication and alignment technologies developed previously.

**Development Considerations:** Manufacturing and testing a large number of mirror segments requires a robust plan that, like the silicon meta-shell optics, will include parallel production and the development of automated techniques for fabricating mirror substrates, coating, alignment, and bonding. The adjustable segments are larger than those for the silicon meta-shell design, and so will potentially require fewer parallel tracks. However, there are additional manufacturing steps associated with the piezo coatings. A detailed manufacturing and assembly plan will be developed for the *Lynx* final report.

### F.2.3 Full-Shell Optics

The directly fabricated full-shell optics path is based on the grinding, polishing, super-polishing, and final ion beam figuring or differential deposition correction of thin (1.7 to 4 mm) fused-silica full-circumference mirror shells. The technology relies on the intrinsic stiffness of full shells and adopting an ad hoc integration concept based on a Shell Supporting Structure (SSS) jig used for handling and support throughout manufacturing. The technology is currently at TRL 2–3.

[3] Current experimentation uses 100-mm x 100-mm segments whereas the final optical design for *Lynx* requires 200-mm long and up to ~400-mm wide segments.

**Technology Maturation Summary:** Realizing TRL 3 will require demonstrating, via metrology or X-ray testing, that an azimuthal segment of a full shell (primary or secondary surface) can achieve an HPD within a factor of 6 of the error budget allowed for fabrication figure errors, or full-shell illumination to about 2x this value. In order to accomplish this, improvements need to be made regarding the polishing operations required to reach the desired figure in the precision, accuracy, and repeatability of the metrology, and in improving the ion beam figuring performance.

Further improvements in the SSS, and understanding the implications of various error contributions caused by transferring the mount from polishing to assembly (particularly, out of roundness), are critical to achieving TRL 4. During this stage, metrology issues for a full shell to the level of 3x the final error allowance must be resolved and any coating errors must be mitigated. A design for a spider mount and demonstration of basic functionality will be required for shells of various representative diameters. X-ray testing of the shell before and after removal from its SSS will quantify any deformations caused by internal stresses induced in the fabrication process and verify intermediate metrology and theoretical predictions.

TRL 5 requires that alignment and integration of a mirror shell into a flight-like support structure is demonstrated and that full-shell performance at the level of 1.5x the error allowance has been achieved through X-ray testing in this medium-fidelity mount. Once the methods for alignment and integration into the support structure are satisfactorily determined, several shells of different diameters will be integrated and X-ray tested as a whole and subjected to appropriate operational thermal and vibrational environments, in order to bring the system to a TRL 5.

TRL 6 will require that *Lynx* angular resolution can be met using a flight-like mount and that scalability (multiple-shells) has been demonstrated. The entire assembly must then be tested in a flight-like environment.

**Development Considerations:** Manufacturing ~200 full shells requires a robust plan that includes consideration of automated polishing, coating, alignment, and bonding, as well as transport between each stage. The aspect ratio of the largest mirror shells makes them the most unwieldy and therefore the most difficult to produce, test, align, and mount. Thus, fabrication of large mirrors will be undertaken at an early stage. To do so, industrial resources capable of supplying mirror shells up to 3-m diameter will need to be identified, and polishing machines, diamond turning lathes, and the metrology systems capable of handling large optics will need to be procured. As with the other two optics technologies, a manufacturing plan will be provided for the *Lynx* final report, optimized for the project schedule and cost.

### F.2.4 Low-Stress Reflective Coatings

All three optics designs require successful X-ray testing with reflective coatings to achieve TRL 4. These coatings must exhibit high reflectance over the *Lynx* energy band-pass, impart little or no high-frequency surface roughness (to minimize X-ray scatter), and minimize coating-stress-driven substrate deformations (to preserve angular resolution). Control of coating-stress-driven deformations is a crucial technological challenge for thin substrates that are easily deformed by coating stress. Iridium single-layer, and boron-carbide/iridium bilayer films have been demonstrated to have high X-ray reflectance but they also induce high stress. Preliminary experimental work has demonstrated multilayer coatings that can provide higher reflectance and may provide lower stress than iridium single-layer or boron-carbide/iridium bilayer films. However, more research is needed to optimize the X-ray performance of such coatings and to demonstrate acceptable stress and stability. A variety of techniques have demonstrated iridium-based coatings, suitable for *Lynx,* having near-zero stress on flat substrates. However, figure preservation after coating on figured thin-shell substrates at the 0.5-arcsecond-level has not yet been achieved. Another technique is coating both sides of substrates to balance the induced stress. This has been shown to introduce no more than 0.5-arcsecond figure degradation in monocrystalline silicone segmented mirror pairs. For adjustable segmented optics, it is conceivable that the reflective coating stress can be balanced instead by the piezoelectric film, while full shell optics are intrinsically not as sensitive to coating stresses. An integrated (optics + coatings) technology maturation plan will be developed for one or more of the *Lynx* optics designs for inclusion in the *Lynx* Final Report.

## F.3 Instruments

*Lynx* science instruments are described in **§E.2**, and are the High-Definition X-ray Imager (HDXI), *Lynx* X-ray Microcalorimeter (LXM), and the X-ray Grating Spectrometer (XGS). The SOTA for each of these is at a TRL 3 or higher. HDXI and XGS have multiple feasible technology efforts that are currently funded. LXM is also currently funded and has elements that leverage developments from past and future relatively near-term flight missions, including *Hitomi*, *XARM*, and *Athena*.

### F.3.1 High-Definition X-ray Imager

The *Lynx* fast, low-noise megapixel X-ray imaging arrays will provide 0.3 arcsecond (<15 μm pitch) angular resolution over a ~23 x 23 arcminute$^2$ FOV at moderate energy resolution. The *Lynx* HDXI will utilize complementary metal-oxide semiconductor active pixel sensor technology to provide significant enhancements over the previous generation of CCD instruments with *Chandra*, *Swift*, and *Suzaku* flight heritage such as higher readout rates and lower power consumption.

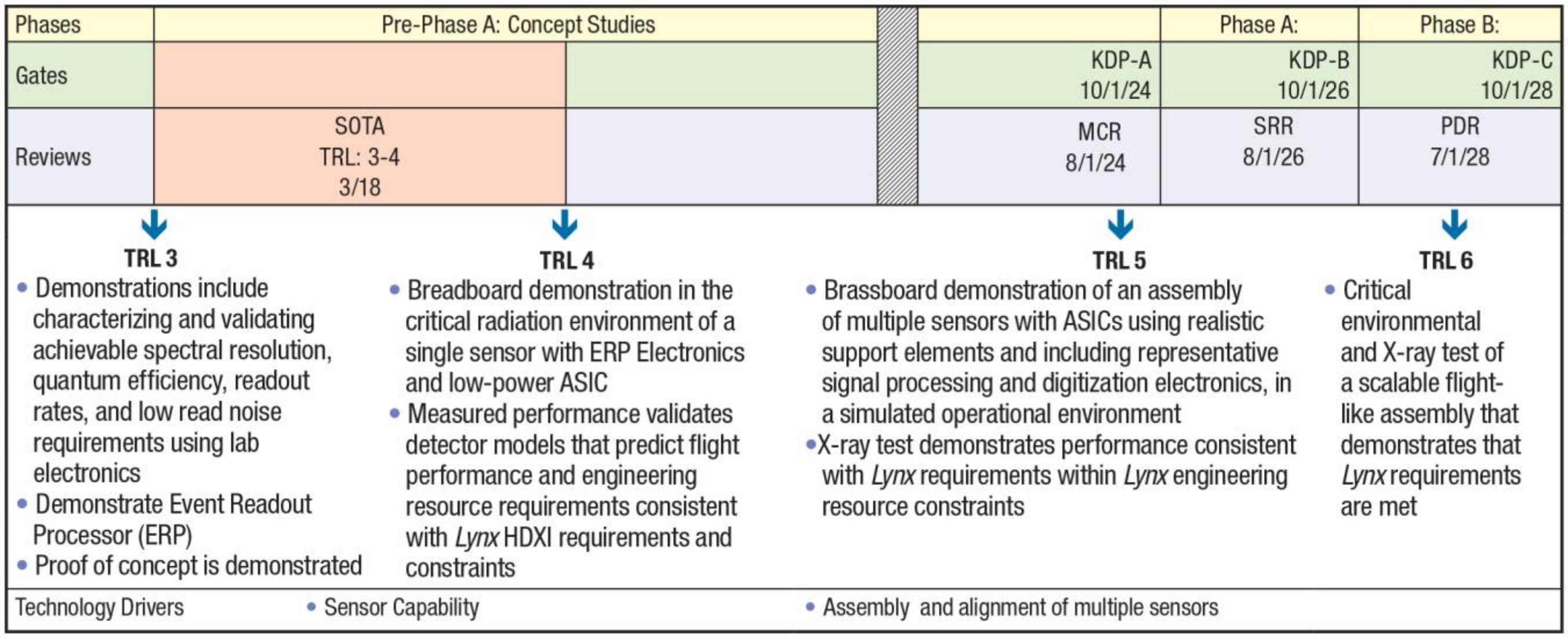

**Figure F-3. Preliminary HDXI SOTA and TRL maturation path to TRL 6. The red-shaded region indicates that the current SOTA is between a TRL 3 and TRL 4 for all HDXI technology approaches considered for *Lynx*.**

**Technology Maturation Summary:** There are several sensor technologies each capable of meeting many of the performance requirements for *Lynx* (pixel size, energy resolution, read noise, quantum efficiency at low and at high energy, and readout rates). These technologies differ primarily in their architecture but not in their functionality, and each has demonstrated proof of concept. Overall, these technologies are assessed at TRL 3 by the most recent PCOS Program Annual Technology Report with some aspects at higher TRL such as rapid readout electronics (TRL 4-5) and directly deposited optical blocking filters (TRL 5). Each of these technologies requires similar resources from the spacecraft and all three have similar development paths. A trade study will be carried out by the *Lynx* HDXI team to document the strengths and weaknesses of three of these technologies; a detailed technology maturation plan is being developed for each. Development of all three technologies is currently being funded in part through NASA competitive funding programs.

To advance to TRL 4, breadboard sensors must be characterized and modeled in sufficient detail that the design modifications needed to achieve required performance, within *Lynx* engineering resource constraints, can be clearly delineated. Device susceptibility to and tolerance of the critical on-orbit radiation environment must be fully understood.

To advance to TRL 5, brassboard sensors exhibiting X-ray performance consistent with *Lynx* requirements for spectral resolution, detection efficiency, frame rate, and visible/IR light rejection must be demonstrated with realistic analog signal processing elements.  Power consumption, scaled to a flight sensor configuration and operating temperature requirements, must be shown to be consistent with *Lynx* resource constraints.

To advance to TRL 6, flight-format sensors and support elements, in flight-like packages and installed in a scaled engineering model focal plane, must successfully pass representative vibration and thermal vacuum tests. Radiation tolerance of the flight-scale sensors and supporting signal processing elements must also be demonstrated.

**Development Considerations:** In addition to the technologies that drive the maturation gates, there are also low-risk subsystem technologies that require development. These technologies are necessary for HDXI to mature, and require resources to develop; they include the low-power Application Specific Integrated Circuits (ASICs), event processing electronics, and filter fabrication. These will be included as a necessary part of the detailed technology maturation plan included in the *Lynx* Final Report.

Low-power ASICs that can also meet the *Lynx* requirement for high-speed frame rate readout must be developed along with Field-Programmable Gate Array (FPGA)-based event processing boards (EPBs) that can also manage the high frame rates. Aluminized plastic filters are at TRL 9, as they have flight heritage. HXDI filter development aims to improve filter low energy response and study the utility of filters directly deposited onto HDXI sensors.

### F.3.2 *Lynx* X-ray Microcalorimeter

The *Lynx* broadband non-dispersive high-spectral resolution imaging spectrometer consists of a large array of X-ray sensors, cooled to 50 mK, that thermometrically determine the precise energy of individual incident photons. The LXM will require three different pixel types (pitch and thickness, energy resolution and range, pixel counts and multi-pixel sensors or 'hydras') to achieve its various science goals. Small X-ray microcalorimeter arrays have flown on several Astro-E and Astro-H missions. Developing large-format microcalorimeters and associated readout technology is funded by NASA. X-ray microcalorimeter technologies for *Lynx* have been assessed at TRL 3 in the 2017 Physics of the Cosmos Program Annual Technology Report.

| Phases | Pre-Phase A: Concept Studies | | | Phase A: | Phase B: |
|---|---|---|---|---|---|
| Gates | | | KDP-A 10/1/24 | KDP-B 10/1/26 | KDP-C 10/1/28 |
| Reviews | SOTA TRL: 3-4 3/18 | | MCR 8/1/24 | SRR 8/1/26 | PDR 7/1/28 |

| TRL 3 | TRL 4 | TRL 5 | TRL 6 |
|---|---|---|---|
| • Analytical studies and lab demonstrations that indicate proof of concept include characterizing and validating multi-pixel TES-detectors relevant to *Lynx* requirements and pixel arrays<br>• Demonstrate feasibility of readout architecture | • Demonstrate the required performance (pitch, energy resolution, energy range) with appropriate wiring density for all pixel types<br>• Demonstrate required microwave SQUID circuitry with appropriate noise, and resonator widths and spacings<br>• Demonstrate basic functionality of the HEMT amplifier architecture | • X-ray testing of focal plane assembly that includes a hybrid array of all required pixel types with required wiring scale, heat sinking, and performance; readout of required pixel types with required density of resonators; and appropriate HEMT architecture with cabling<br>• Assembly includes cryogenic system and realistic support structure<br>• X-ray test demonstrates feasibility of achieving *Lynx* performance requirements in simulated operational environment | • Critical environmental and X-ray test of a scalable flight-like assembly that demonstrates that *Lynx* requirements are met |

| Technology Drivers | • Sensor Arrays and Readout | • Focal plane assembly |
|---|---|---|

**Figure F-4. Preliminary LXM (SOTA) and TRL maturation path to TRL 6. The red box indicates that the current SOTA is between a TRL 3 and TRL 4.**

**Technology Maturation Summary:** To advance to TRL 4, the required performance of all pixel types, with the requisite wiring density, will need to be demonstrated. This includes demonstrating that small (25 μm to 50 μm) pitch 25-pixel hydras can achieve 3 eV energy resolution and maintain position sensitivity, and demonstrating single pixels on a 50 μm pitch with 0.3 eV resolution. In addition, the microwave SQUID readout circuitry and code division multiplexing must demonstrate the appropriate noise and resonator widths and spacings over a many GHz range. To achieve this also requires demonstrating undegraded readout at reduced multiplexing factors and readout at relevant multiplexing factors. Demonstrating code division multiplexing that: (a) provides appropriate noise, switching speeds, input channel count, and circuit yield in isolation, and (b) provides undegraded readout of extended array sensors when integrated with a microwave SQUID resonator readout is also required. Finally, to advance to TRL 4, a high-electron mobility transistor amplifier architecture will be needed that demonstrates basic functionality (meeting noise, gain, and linearity specifications) without excessive power dissipation.

Advancement to TRL 5 will require X-ray testing, in a simulated operational environment, of focal plane assemblies. These assemblies should be fabricated on single substrates that include arrays of all the required pixel types, with wiring to scale, and appropriate heat dissipation performance. The readout circuitry will need to be demonstrated with the required density of resonators and relevant multiplexing factors while maintaining low noise levels. The signal cabling will need to combine high bandwidth, low loss, and low thermal conductivity without imposing crosstalk between output channels with separate amplifiers. Such an assembly will require a cryogenic system and realistic support structure.

To advance to TRL 6 will require the development of a flight-like assembly that is tested in a critical environment. This testing will include radiation, thermal, vibrational, and acoustical testing of the assembly. All scaling issues regarding thermal and electrical multiplexing must be solved in order to reach TRL 6.

**Development Considerations:** LXM low-risk subsystem technologies that require development resources include a cryogenic system capable of cooling the sensor array to 50 mK, and adequate magnetic shielding of the microcalorimeter array and the readout circuits. Many cryostats have been developed for space-based applications, and cryocoolers that integrate into them are under development. As *Lynx* develops, the cryocooler definition will improve and the vibration requirements will be refined. *Lynx* will design a cryostat that meets performance requirements and is optimized for mass, power, cost, and risk. Mainly engineering tasks, Mu-metal, and superconducting magnetic shield designs need to be developed and verified, similar to those being developed for the *Athena* X-IFU, to provide adequate magnetic shielding of the micro-calorimeter array and the readout circuits.

### F.3.3 X-ray Grating Spectrometer

There are two viable paths under study for achieving the scientific requirements for wavelength-dispersive diffraction grating spectrometers for *Lynx*: the Critical-Angle Transmission X-ray Grating Spectrometer (CAT-XGS; Heilmann et al. 2016, 2017) and the Off-Plane X-ray Grating Spectrometer (OP-XGS; Donovan et al. 2018, Applied Optics 57, 454). Both consist of an objective grating array positioned just downstream of the focusing optics (to maximize spectral dispersion at the focal plane), and a readout detector array in the focal plane that is offset from the telescope's imaging focus. The most recent PCOS Program Annual Technology Report has assessed both XGS technologies at TRL 4. Both are actively being developed through NASA's Strategic Astrophysics Technology program. Dispersive reflection grating spectrometers have flown on *XMM-Newton*, transmission grating spectrometers have flown on *Chandra* and CAT gratings are under development for the *Arcus* Mid-Ex currently in Phase A study. The technology maturation plan for the readout detector array will follow that of the HDXI and is not discussed further here.

A trade study will be carried out by the *Lynx* team to document the successes of and challenges to each XGS technology, and a detailed technology maturation plan is being developed for both. Depending on the outcome of the trade, which will include science requirement considerations, spacecraft resource needs, and programmatics (schedule and cost), one or both of these technologies will be considered for the *Lynx* payload and DRM for more detailed integration design, schedule, and costing. A detailed technology maturation plan for both technologies will be included in the *Lynx* Final Report.

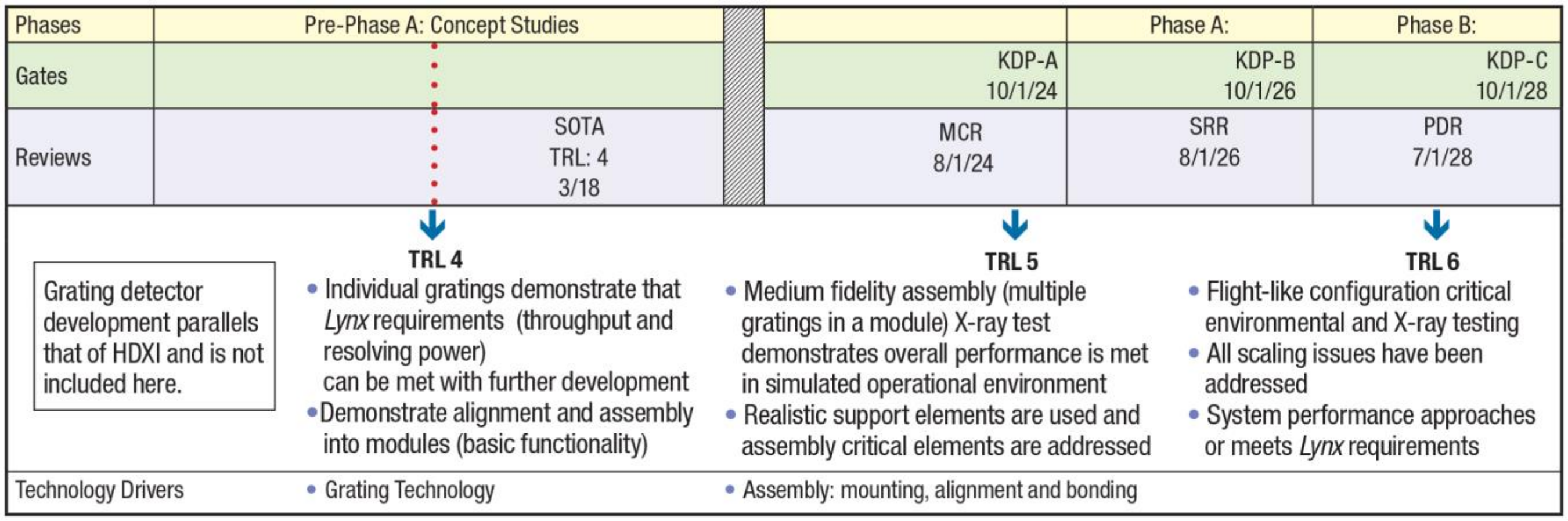

**Figure F-5. Preliminary XGS SOTA and TRL maturation path to TRL 6. The red dashed line indicates that the current SOTA is at TRL 4 for both grating technology approaches being considered for *Lynx*. The XGS detector development parallels that of HDXI and is not included here.**

**Technology Maturation Summary:** To advance to TRL 5, an assembly of multiple gratings in a module, using realistic support elements, will be X-ray tested (in vacuum) to demonstrate that overall performance (alignment, throughput, and efficiency) is met in a simulated flight environment. For the grating surfaces, this will require developing the lithography and etching process to minimize period variation, extending the fabrication to the larger flight-like grating surface area, improving the imprinting technology to ensure that the replicated grating profiles are within tolerances, and shaping and coating the grating surfaces. For the assembly, this will require an alignment strategy (and associated metrology) and module design that preserves the performance, and utilizes materials for the module structures that are thermally and mechanically compatible with the grating elements.

To advance to TRL 6, a flight-like configuration, with grating array geometry corresponding to the *Lynx* mirror design, will be environmentally and X-ray tested and all scaling issues will have been addressed. For the environmental testing, flight-like mechanical, actuator, and thermal subsystem interfaces will be designed and built. X-ray testing will require an optical assembly to provide the correct X-ray beam incidence geometry.

**Development Considerations**: Design considerations for XGS, and hence the development path, depends critically on the optical design and state of the readout detectors. These considerations do not drive the gates for maturing the XGS grating technology, but do need to be defined and developed prior to meeting TRL 5.

The optical design drives the XGS design, as it quantifies the correlation between resolving power and increasing azimuthal telescope aperture coverage, grating size, grating misalignments, and other factors. The XGS detector array will use the same CMOS-based APS technology that is used for HDXI. Sensor array geometry and readout characteristics must be defined and integrated onto the *Lynx* ISIM fixed table, without interference from LXM and HDXI.

# G. PROGRAMMATICS

*Lynx* is a Flagship NASA mission, currently in concept phase, that will be managed and executed following the NASA Procedural Requirements (NPR) for a Category 1, Risk-Class A, Single Project Program. The Project will deliver the telescope and spacecraft elements for launch of the observatory in the 2030s, maintain a low risk posture throughout development and operation, and adhere to the project budget to ensure mission success.

## G.1 Overview

*Lynx* is designed to carry out an ambitious science program, as expected for a Flagship class mission, while establishing and maintaining a low to moderate risk posture. This approach is intended to ensure a launch in the mid-2030s at a relatively moderate cost for a Flagship mission, thereby permitting a diverse NASA astrophysics program in the same decade. *Lynx* is a Category 1, Single Project Program as defined in NPR 7120.5, and classified as Risk Class A per NPR 8705.4, and will be under the authority of the NASA Associate Administrator (AA) and the Science Mission Directorate (SMD) Associate Administrator. The project will reside within the NASA SMD Astrophysics Division and overall project management responsibilities will be assigned to the selected lead NASA Center.

The project will perform life-cycle reviews for a Single-Project Program in accordance with the project management processes defined in NPR 7120.5E, with the systems engineering requirements in NPR 7123.1B. The reviews will be conducted by an independent Standing Review Board (SRB) that will provide recommendations on the ability of the project to proceed through the prescribed Key Decision Points (KDPs) and life-cycle phases. Decision authority for *Lynx* rests with the NASA AA and the SMD AA. The Technical Authority decision will rest with the lead NASA Center Director, and concurrence will be provided by the NASA Chief Engineer.

## G.2 Risks and Risk Mitigation

The *Lynx* team has identified and ranked the five top project risks and has defined the likelihood (L) and consequence (C) of risk occurrence on a scale of 1–5, with 1 being the lowest probability and 5 being the highest. Project level risks are defined as those that have the potential to change the project level technical or programmatic baseline. In addition to project risks, each subsystem will also carry risks at that level. For the purpose of the Interim Report, only the project level risks are summarized below. The final report will include subsystem risks, primarily as they relate to technology development. The project risk list is shown in **Table** G-1**.** The 5 x 5 risk chart for these risks is provided in **Figure G-1.** The risk standard scale for consequence and likelihood is consistent with Goddard Procedural Requirements (GPR) 7120.4D, Risk Management Reporting. The identified risks fall under the general categories of technology maturation, manufacturability, and mass exceedance. Detailed information on specific technology development plans are found in **§F**.

**Table G-1. Summary of top *Lynx* project risks.**

| Risk | Title | L | C | T | S | $ |
|---|---|---|---|---|---|---|
| 1 | Optics Technical Maturation | 4 | 3 | X | X | X |
| 2 | LXM Technical Maturation | 3 | 3 | X | X | X |
| 3 | Optics Manufacturing | 3 | 3 |  | X | X |
| 4 | LXM Assembly | 2 | 3 |  | X | X |
| 5 | Launch Vehicle Availability | 1 | 4 | X | X | X |

L = likelihood of occurrence; C = consequence; T = technical risk, S = schedule risk, $ = cost risk

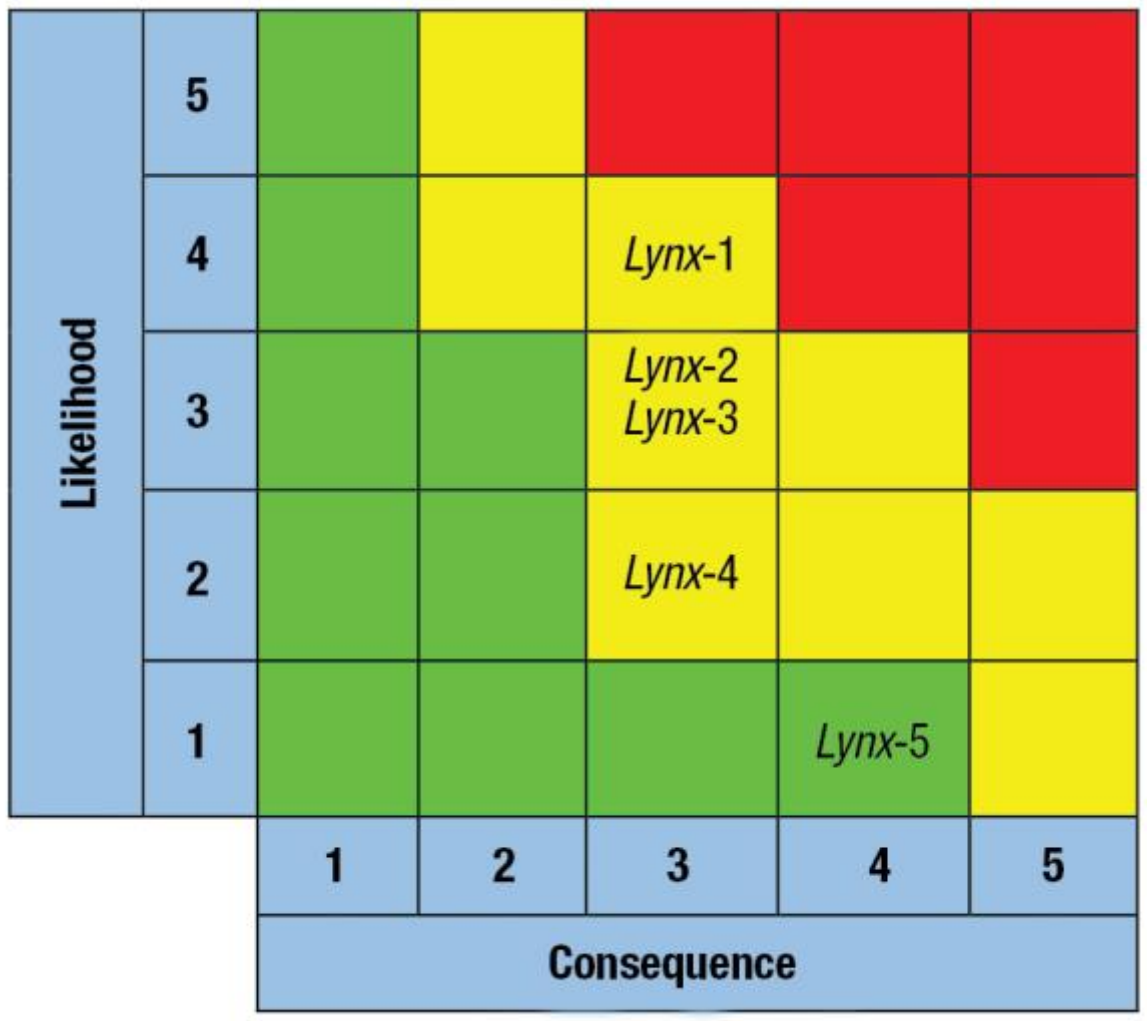

**Figure G-1. *Lynx* risk ranking.**

**Risk 1—**If the X-ray mirrors are unable to achieve requisite technology maturation and performance, there is a risk that mission science and/or technology development cost and schedule will be compromised.

**Mitigation:** A technology maturation plan **(§F)** is under development for 3 different feasible, actively-funded X-ray mirror technologies currently under consideration by the *Lynx* team. Each of these has a set of unique risks and requires a tailored mitigation plan. These plans will be updated and laid out in detail prior to submission of the final concept study report in 2019. Technology maturation impacts to cost and schedule are being determined and incorporated into the plan. A detailed mitigation strategy will be developed once these are complete. In addition to application of conservative cost and schedule reserves, periodic reviews will be carried out as needed to ensure that developmental goals are being met.

**Impact:** Reduced science capability or increased cost and schedule for technology development.

**L x C:** 4x3

**Risk 2—**If the *Lynx* X-ray Microcalorimeter (LXM) is unable to achieve requisite technology maturation and performance, there is a risk that mission science and/or technology development cost and schedule will be compromised.

**Mitigation:** A technology maturation plan **(§F)** is being detailed for the LXM and the rest of the science instruments. Of the three instruments, the LXM is the least mature, but includes several subsystems and technologies that are similar to those on the *Athena* X-ray Integral Field Unit (X-IFU) instrument. This includes Electromagnetic Interference (EMI) / Electromagnetic Compatibility (EMC) shielding, focal plane mechanical design, IR blocking filters, and sensor wiring arrangements. Technology developments from the X-IFU will be leveraged as applicable for the LXM. Additionally, several industry studies have been initiated to investigate the LXM cryogenic design to identify the solution space (mass, volume, and complexity vs cost) for this already mature subsystem. In addition to application of conservative cost and schedule reserves, periodic reviews will be carried out as needed to ensure requisite development milestones are met.

**Impact:** Reduced science capability or increased cost and schedule for technology development.

**L x C:** 3x3

**Risk 3—**If the X-ray mirrors and assembly cannot be fabricated within the projected timescale, project schedule margin will be eroded at the risk of increased project life cycle cost.

**Mitigation:** For each feasible optics system design, an early study of manufacturability and production of the mirror segments, mounting systems, and module fabrication has been initiated through industry partnerships. These studies will identify areas to reduce the overall development schedule for this portion of the project critical path and provide validation of project assumptions to be revisited upon completion. Pending technology development funding availability, prototype development will take place to more clearly identify the approach for manufacturing, mounting, and module fabrication. Schedule margin of 10 months has been added to the project schedule critical path through calibration.

**Impact:** Critical path schedule duration and increased project cost.

**L x C:** 3x3

**Risk 4—**If the LXM and its subsystems and components cannot be fabricated, assembled, tested, and integrated within the projected timescale, project schedule margin will be eroded at the risk of increased project life cycle cost.

**Mitigation:** Detailed manufacturing plans for the LXM will be developed over the course of the next year. Manufacturing plans from the *Athena* X-IFU will be leveraged, as applicable, for the LXM. Schedule margin of 10 months has been added to the project schedule critical path through calibration.

**Impact:** Critical path schedule duration and increased project cost.

**L x C:** 2x3

**Risk 5—**If the overall *Lynx* system mass budget exceeds the payload mass launch limit for a heavy class vehicle to reach SE-L2, either a reduction in scope to reduce mass or a different class launch vehicle will need to be considered.

**Mitigation:** The *Lynx* team has developed and will maintain a mass budget for the concept that will be refined over the next year. The mass budget utilizes conservative industry standard mass growth allowances for this stage of development. As the design matures, trades on mass growth versus science capability will be made, if needed, to ensure the observatory remains within the launch vehicle payload envelope (with margin). Based on the current concept design, *Lynx* will require a heavy-class launch vehicle, as the total mass of the observatory exceeds the 6,500 kg mass limit for an intermediate-class vehicle and is below 10,000 kg with sufficient margin for this phase of the study (as defined by the Launch Services Program and described in **§E**). If the observatory mass with margin exceeds the heavy-class vehicle limit, use of NASA's Space Launch System as a primary payload will be considered. If during the course of design maturation, the payload mass encroaches on the heavy lift capability to SE-L2, mass reductions will be considered including, but not limited to, reducing the effective area.

**Impact:** Reduced science capability or increased mission cost.

**L x C:** 1x4

## G.3 Schedule

For the purpose of the Interim Report, the *Lynx* team developed a preliminary project life cycle schedule from the start of technology development funding through launch. The schedule, shown in **Figure G-2**, identifies the project-level milestones and KDPs consistent with NPR 7120.5 for a Single Project Program. The schedule identifies dates for these milestones and includes the assumed critical path. The project-level milestone dates were determined with an understanding of the development complexity of the *Lynx* subsystems. The schedule was compared against project schedules with similar developments such as *Chandra*, *Athena*, and *WFIRST*. The schedule includes preliminary development inputs provided by the science instrument and optics development teams, while assuming technology development funding availability and a profile similar to *WFIRST*, and the preliminary project life cycle cost estimate (not presented for the Interim Report). While detailed life-cycle schedules and costing for the optics and science instruments are still in development, enough information is available to define approximate milestone dates for key elements of the *Lynx* project and the critical path. A total of 13 months of schedule margin has been added to the critical path activities, consistent with guidance from NASA/SP-2010-3403, Schedule Management. The duration of key phases for the project are presented in **Table G-2,** and key event dates are presented in **Table G-3.** Dates presented in this schedule are considered preliminary, and further refinement of this schedule will occur throughout the remainder of the concept study.

In the project schedule, the development timeline for the *Lynx* X-ray mirror assembly (XMA) specifically adopts the preliminary development schedule for the silicon meta-shell optics technology. This approach provides the most conservative project life cycle given the large number of replicated elements required in the assembly. Following the *Lynx* mirror assembly trade study, as discussed in **§E**, the development schedule for the selected optics design will be included in the project schedule. Given the silicon meta-shell optics development schedule, the XMA is on the critical path through manufacturing and delivery for testing and calibration. With the Design, Development, Test, and Evaluation (DDT&E) complexity of the LXM, this instrument drives the schedule of the Integrated Science Instrument Module (ISIM) integration, testing, and delivery for calibration of the instrument suite with the mirror assembly. Note that the grating array will be calibrated with the XGS readout prior to delivery of the readout for integration with the ISIM. Following calibration of the XMA to the science instruments and ISIM, it will be integrated with thermal pre-collimator and the grating array to become the *Lynx* Mirror Assembly (LMA). After mirror and instrument calibration, the remaining critical path goes through the AI&T of the X-ray telescope, and integration with the spacecraft and launch vehicle.

This schedule assumes a Pre-phase A start in October 2021 for technology development of the optics and science instruments. The start of pre-Phase A activities is consistent with the assumed timing of the 2020 Decadal Survey decision in ~June 2020 (TBR), and NASA Astrophysics Projects Division (APD) funding for technology development. The selection of all technologies and the final *Lynx* architecture decision is planned to take place no later than four months prior to Mission Concept Review (MCR) in order to allow for a detailed mission, cost, and schedule analysis. The MCR is planned for August 2024, with KDP-A planned for October 2024 and the start of Phase A. At KDP-A, all selected *Lynx* technologies will be at TRL 5, per the Technology Roadmap defined in **§F**. Phase A is planned to take 25 months, during which time the acquisition planning for the major contracting efforts takes place. It is assumed the *Lynx* project will include a prime contract for spacecraft development and overall observatory integration, and will likely include subcontracts for the XMA and ISIM DDT&E efforts, similar to the *Chandra* and *JWST* projects. The Acquisition Strategy Meeting is assumed to take place in January 2026, prior to the *Lynx* project System Requirements Review (SRR), which is planned to take place in July 2026. KDP-B follows four months later in November 2026.

Phase B is expected to take 23 months, culminating in the mission PDR, at which time all *Lynx* technologies will be TRL 6. The Prime, XMA, and ISIM contracts will be finalized during this phase, allowing for the purchase and delivery of long lead items. The XMA, given its long fabrication and assembly schedule, is planned to begin tooling and conduct its Critical Design Review (CDR) during Phase B to allow for early start of system fabrication, assembly, and test. During this Phase, the spacecraft SRR and PDR are expected to take place, with the project-level PDR following in July 2028. KDP-C follows three months later in October 2028.

Phase C is expected to take 80 months during which time all of the major sub-assemblies will complete their critical design and integration and test phase. The project CDR is planned to take place in September 2030. During Phase C, the AI&T of the XMA and ISIM will be completed, and the subsystems will be delivered for test and calibration. Calibration is expected to take about eight months to complete. Following calibration, the XMA will be integrated with the thermal pre-collimator and grating array to create the LMA. This assembly will be integrated with the Optical Bench Assembly and the ISIM to create the X-ray telescope. This sequence of AI&T activities is anticipated to take 15 months. Prior to the end of X-ray telescope AI&T, the project-level System Integration Review (SIR) will take place to ensure all necessary steps to initiate integration of the X-ray telescope and spacecraft have been completed. SIR is anticipated for March 2035.

Phase D is expected to start in June 2035 and take 17 months to complete. During Phase D, the X-ray telescope and spacecraft will be integrated and tested to become the *Lynx* X-ray Observatory. This activity is expected to take 12 months to complete, after which time *Lynx* will be delivered to the NASA Kennedy Space Center launch integration facility for integration with the launch vehicle, which is expected to take about six months. The mission Operational Readiness Review will take place about four months prior to launch in July 2036. The launch readiness date is planned for December 2036.

Following launch, the mission will enter Phase E and initiate on-orbit operations following test and checkout during the 104-day cruise to SE-L2. Phase E is planned to take 60 months, although *Lynx* is designed with expendables for 240 months of life to accommodate the potential for substantial extension of the operating mission.

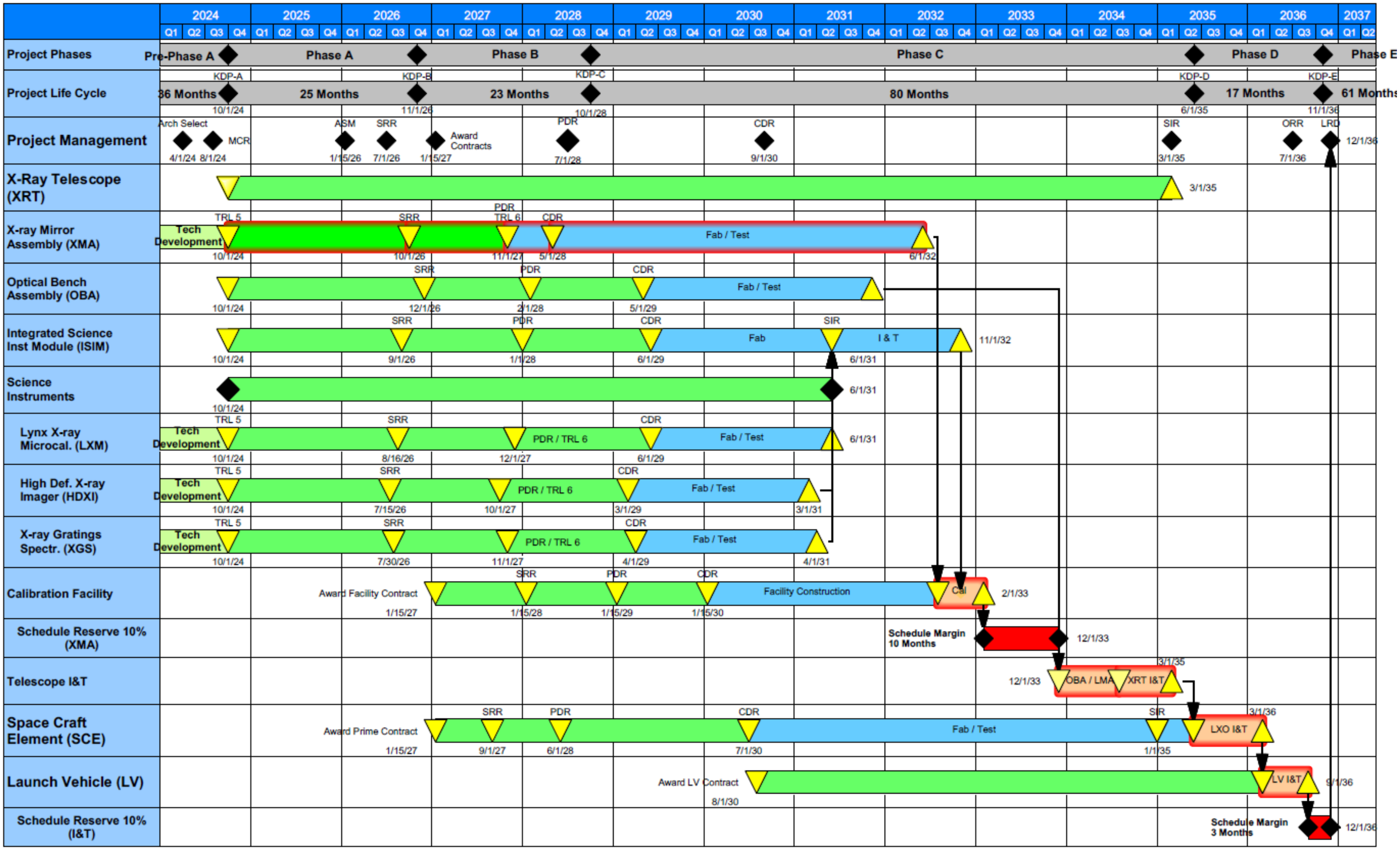

Figure G-2. *Lynx* Project Schedule

**Table G-2. Key Phase Duration Table.**

| Project Phase | Duration (months) |
|---|---|
| Pre-Phase A – Technology Development | 36 |
| Phase A – Conceptual Design | 25 |
| Phase B – Preliminary Design | 23 |
| Phase C – Detailed Design | 80 |
| Phase D – Integration and Test | 17 |
| Phase E – Primary Mission Operations | 60 |
| Start of Phase A to SRR | 21 |
| Start of Phase B to PDR | 20 |
| Start of Phase C to CDR | 23 |
| Start of Phase C to SIR | 77 |
| Phase A to XMA delivery to calibration | 92 |
| Phase A to ISIM delivery to calibration | 97 |
| Phase A to LXM delivery to ISIM | 80 |
| Phase A to HDXI delivery to ISIM | 77 |
| Phase A to XGS delivery to ISIM | 78 |
| Telescope Integration and Test | 15 |
| Schedule Margin | 13 |
| Total Project Development Duration (Phase A – LRD) | 146 |
| Primary Operations (Phase E) | 61 |

**Table G-3. Key Event Dates.**

| Project Milestone | Milestone Date |
|---|---|
| Technology Development / Start of Pre-Phase A | 10/1/2021 |
| Architecture Decision | 4/1/2024 |
| MCR | 8/1/2024 |
| KDP-A / Start of Phase A | 10/1/2024 |
| SRR | 7/1/2026 |
| KDP-B /Start of Phase B | 11/1/2026 |
| PDR | 7/1/2028 |
| KDP-C / Start of Phase C | 10/1/2028 |
| CDR | 9/1/2030 |
| Delivery of XMA to calibration | 6/1/2032 |
| Delivery of ISIM to calibration | 11/1/2032 |
| Delivery of LXM to ISIM integration | 6/1/2031 |
| Delivery of HDXI to ISIM integration | 3/1/2031 |
| Delivery of XGS to ISIM integration | 4/1/2031 |
| SIR | 3/1/2035 |
| KDP-D / Start of Phase D | 6/1/2035 |
| ORR | 7/1/2036 |
| KDP-E / Start of Phase E | 11/1/2026 |
| LRD | 12/1/2036 |
| End of Primary Mission | 12/1/2041 |

## G.4 Cost

During this phase of the study, detailed spacecraft, telescope, and mission design studies were performed that included preliminary cost estimates. These estimates will be refined during the next phase of the study and incorporated into the mission-level estimate to be included in the final report submitted in 2019.

### G.4.1 Cost Estimation Methodology

Cost estimation and analysis for the *Lynx* spacecraft and instruments are being conducted using the industry-standards Project Cost Estimating Capability (PCEC) Cost Model (replaced NAFCOM), the SEER® hardware model, the PRICE® TruePlannng Space Missions model, and the PRICE® hardware model. PCEC contains over 43 planetary and earth orbiting spacecraft missions and provides estimates at the subsystem level. All PCEC missions are documented within the Cost Analysis and Data Requirements (CADRe) database. PCEC is a parametric model developed and maintained by MSFC Engineering Cost Office to estimate the cost of spacecraft, launch vehicles, and human space flight systems, and provides full visibility into all calculations and statistics. PCEC is capable of being calibrated to an analogous system and, in this case, *Chandra* will be used to provide calibration factors for the PCEC estimate; comparisons will be made against other flagship telescope mission costs as well. SEER® and PRICE® are commercial cost models that also employ large spacecraft databases as tools for cost estimation.

Separate estimating runs will be utilized to estimate at the subsystem and component levels with the SEER® model. SEER® provides a risk analysis capability and will be used with the component estimating run to apply reserve based on risk. For all other models, a simple reserve of 30% will be applied (reserve is not applied to fee). All estimates will cover the *Lynx* project work breakdown structure (WBS) shown in **§H.2.3**; it is based on the NASA standard WBS with the only exclusion being the cost for the launch vehicle. The launch vehicle costs will be a pass-through based on study team input and a NASA Launch Services Program estimate. Estimates will be validated using historical technical and cost data from the Resource Data Analysis and Retrieval (REDSTAR) database to determine whether the particular cost produced by the estimates is reasonable and within family for the environment and type of mission.

General ground rules and assumptions (GRAs) for the *Lynx* cost estimate process are listed below (specific GRAs will be added as needed).

- All costs are in fiscal year (FY) 2020 dollars, based on NASA inflation tables.
- Subsystem costs are limited to prime contractor incurred costs, exclusive of prime contractor fee. This includes the prime contractor subcontract management costs.
- Subcontractor costs, including subcontractor fee, is considered prime contractor incurred cost.
- Program management, system engineering, and safety and mission assurance are estimated at the system level (spacecraft or instrument) and at the project level in WBS 1.0, 2.0, and 3.0.
- Cost estimates are based on the project schedule, as presented in **§G.3,** and predicted mass and power in the Master Equipment List (MEL) / Power Equipment List (PEL). The schedule includes standard critical path margins, per the NASA Schedule Management Handbook. The MEL/PEL utilizes industry standard mass and power growth allowances.
- Parametric models will be utilized to estimate the cost for Phases B to D, assuming all technologies are at a minimum of TRL 6 at PDR.
- Phase A will be estimated using historical database missions (approximately 5% of the Phase B–D cost).
- Phase E estimates assuming a 5-year mission life (with consumables for 20 years) will be developed using the Mission Operations Cost Estimation Tool (MOCET) and compared to the actuals from the first five years of the *Chandra* mission.
- All instruments and spacecraft will be estimated as near Earth.

- Non-recurring costs associated with the DDT&E effort will encompass the period from the beginning of full scale development (start of Phase B) through the beginning of production of the first flight article. Recurring costs associated with the flight article are excluded from DDT&E. Costs associated with test articles and testing are included in DDT&E.
- Recurring costs associated with the flight unit production will include the period beginning with the start of production, (including the cost of any long lead procurements) and ending with the delivery of the first unit. Flight unit costs represent only the costs of the first unit to fly.
- Facilities costs will be included, as necessary, for major facility builds or upgrades.
- Individual subsystem totals will contain all hardware, engineering labor, and manufacturing costs charged to that subsystem, including all management, engineering, testing, and assembly functions. This includes the labor, materials, specific test equipment, and ground support equipment associated with the integration of the components or assemblies in the subsystem.
- A standard 10% fee is assumed for the spacecraft.
- A 9% fee is assumed on the instruments based on the average fee on historical instrument systems. The fee on instruments may change based on university or laboratory developments.
- For SEER®, a 15% factor (based on flight system costs) for program support is assumed at the project level for WBS 1.0 (program management), WBS 2.0 (systems engineering) and WBS 3.0 (safety and mission assurance).
- For PCEC and PRICE®, a cost estimating relationship is used to calculate the program management, systems engineering, and safety and mission assurance cost. This cost is split between flight system (40%) and program level (60%).
- A 5% vehicle integration factor (based on flight system costs) is assumed for all models.
- The spacecraft estimates will be developed based on a standard protoflight philosophy—1 flight unit, 55% factor (based on subsystem costs) for non-flight qualified prototype units.
- The instrument estimates will be developed based on a modified protoflight philosophy—a full-up engineering development unit— with a 55% factor (based on subsystem cost and historical data on non-flight qualified prototypes), 10% factor (based on subsystem costs) for engineering test units (flight qualified), and 10% factor (based on subsystem costs) for spares. This results in a 65% prototype factor versus the standard 55% factor for a protoflight design approach.
- The grassroots estimate developed by Goddard Space Flight Center for the LXM will be utilized.
- PRICE® TruePlanning Space Missions and SEER®-H estimates developed by MSFC for the High Definition X-ray Imager and X-ray Grating Spectrometer instruments will be utilized.
- PRICE® TruePlanning Space Missions and SEER®-H estimates developed by MSFC for the X-ray Optics and grating arrays will be utilized.

### G.4.2 Cost Validation (TBD for Interim)

### G.4.3 Cost Risks (TBD for Interim)

# H. APPENDICES

## H.1 Meet The Team

**Community Chairs of the Science and Technology Definition Team (STDT)**

- Feryal Özel, Professor of Astronomy, University of Arizona
- Alexey Vikhlinin, Deputy Associate Director, High Energy Astrophysics Division, Harvard-Smithsonian Center for Astrophysics

**Members of the STDT**

- Steven Allen, Professor of Physics, Stanford University
- Mark Bautz, Associate Director, Massachusetts Institute of Technology Kavli Institute
- W. Nielsen Brandt, Verne M. Willaman Professor, Pennsylvania State University
- Joel Bregman, Professor of Astronomy, University of Michigan
- Megan Donahue, Professor of Physics & Astronomy, Michigan State University
- Zoltan Haiman, Professor of Astronomy, Columbia University
- Ryan Hickox, Associate Professor of Physics & Astronomy, Dartmouth College
- Tesla Jeltema, Assistant Professor of Physics & Biological Sciences, UC Santa Cruz
- Juna Kollmeier, Astronomer, Carnegie Institution for Science
- Andrey Kravtsov, Professor of Astronomy & Astrophysics, University of Chicago
- Laura Lopez, Assistant Professor of Astronomy, The Ohio State University
- Piero Madau, Professor of Physics & Biological Sciences, UC Santa Cruz
- Rachel Osten, Assistant Astronomer, Space Telescope Science Institute
- Frits Paerels, Professor of Astronomy, Columbia University
- David Pooley, Assistant Professor of Physics & Astronomy, Trinity University
- Andrew Ptak, Astrophysicist, NASA/Goddard Space Flight Center
- Eliot Quataert, Thomas and Alison Schneider Professor of Physics, University of California Berkeley
- Christopher Reynolds, Professor of Astronomy, University of Maryland
- Daniel Stern, NuSTAR Project scientist, Jet Propulsion Laboratory

**Lynx Study Scientist, Study Manager, and Deputy Study Scientist**

- Jessica Gaskin, Astrophysicist, NASA Marshall Space Flight Center
- Karen Gelmis, *Lynx* Concept Study Manager, NASA Marshall Space Flight Center
- Doug Swartz, Deputy Study Scientist, Universities Space Research Association/Marshall Space Flight Center

***Lynx* Study Office**

- Alex Dominguez, NASA Marshall Space Flight Center
- Kevin McCarley, NASA Marshall Space Flight Center
- Robbie Holcomb, NASA Marshall Space Flight Center
- Teresa Brown, NASA Marshall Space Flight Center
- J. Brent Knight, NASA Marshall Space Flight Center
- Mark Freeman, Harvard-Smithsonian Center for Astrophysics
- Dan Schwartz, Harvard-Smithsonian Center for Astrophysics
- Eric D. Schwartz, Harvard-Smithsonian Center for Astrophysics
- Harvey Tannanbaum, Harvard-Smithsonian Center for Astrophysics

**Engineering Support Team**

- Quincy Bean, NASA Marshall Space Flight Center
- Tyrone Boswell, NASA Marshall Space Flight Center
- Spencer Hill, NASA Marshall Space Flight Center
- Randy Hopkins, NASA Marshall Space Flight Center
- Jack Mulqueen, NASA Marshall Space Flight Center
- Andrew Schnell, NASA Marshall Space Flight Center
- Rob Suggs, NASA Marshall Space Flight Center
- Stever Sutherlin, NASA Marshall Space Flight Center
- Mike Baysinger, Jacobs
- Pete Capizzo, Jacobs
- Leo Fabinski, Jacobs
- Jay Garcia, Jacobs
- J. Brent Knight, Jacobs
- Justin Rowe, Jacobs
- Mark Zagarola, Creare
- Deepnarayan Gupta, Hypres
- Greg Feller, Lockheed Martin
- Jeff Olson, Lockheed Martin
- Dave Frank, Lockheed Martin
- Ben Zeiger, Luxel
- Jon Arenberg, Northrop Grumman
- Charlie Atkinson, Northrop Grumman
- Bill Purcell, Ball Aerospace
- Steve Jordan, Ball Aerospace

- Ted Mooney, Harris Corporation
- Lynn Allen, Harris Corporation
- Mike McEachen, Orbital ATK

**Science Support Team**

- John ZuHone, Harvard-Smithsonian Center for Astrophysics
- Grant Tremblay, Harvard-Smithsonian Center for Astrophysics
- Reinout van Weeren, University of Leiden
- Tom Aldcroft, Harvard-Smithsonian Center for Astrophysics
- Arnold Rots, Harvard-Smithsonian Center for Astrophysics
- Pat Slane, Harvard-Smithsonian Center for Astrophysics
- Larry David, Harvard-Smithsonian Center for Astrophysics
- Paul Plucinsky, Harvard-Smithsonian Center for Astrophysics
- Scott Wolk, Harvard-Smithsonian Center for Astrophysics
- Jonathan McDowell, Harvard-Smithsonian Center for Astrophysics

**Ex-Officio observers of the STDT**

- Terri Brandt, NASA/Goddard Space Flight Center (acting Program Office Chief Scientist (PCOS))
- Daniel Evans, NASA Headquarters (*Lynx* Program Scientist)
- Robert Petre, NASA/Goddard Space Flight Center (Chief, GSFC X-ray Astrophysics Laboratory)
- Randall Smith, Smithsonian Astrophysical Observatory (*Athena* Liaison)

**International ex-officio observers of the STDT**

- Peter Jonker, SRON & Radboud University Nijmegen (SRON appointee)
- Brian McNamara, University of Waterloo (CSA appointee)
- Kirpal Nandra, Max Planck Institute for Extraterrestrial Physics (DLR appointee)
- Giovanni Pareschi, INAF Brera Observatory (ASI appointee)
- Arvind Parmar, ESTEC, Noordwijk (ESA appointee)
- Gabriel Pratt, CEA Saclay (CNES appointee)
- Makoto Tashiro, Saitama University (JAXA appointee)

***Lynx* Science Working Group (SWG)**

The SWGs will be responsible for identifying outstanding science questions, developing a compelling science case, and aiding the STDT with producing a mission concept that best addresses these questions. We welcome members both internal and external to the X-ray community, at all career stages, and from domestic and foreign institutions.

- Cycles of Baryons in and out of Galaxies
- First Accretion Light in the Universe

- Evolution of Structure and AGN populations
- Physics of Plasmas
- Physics of High Density Matter, Compact Objects, and Accretion
- Physics of Feedback
- Stellar Life cycles
- X-rays in the Multi-wavelength, Multi-Messenger Era

**_Lynx_ Optics Working Group (OWG)**

The goal of the OWG is to assist the STDT in demonstrating that a credible and feasible path exists to fabricate an X-ray telescope to support the *Lynx* mission. In order for the mission to be capable of realizing the science envisioned by the STDT, the STDT seeks assistance from the X-ray optics community—including experts from academia, industry, and research institutions, in identifying potential approaches for creating the X-ray mirrors and all related technologies (e.g., alignment and mounting techniques, thermal controls, and metrology) required to assemble individual optical components into a large-area, satellite-borne X-ray telescope.

**OWG Co-Chairs**

- Mark Schattenburg, MIT Kavli Institute for Astrophysics and Space Research
- Lester Cohen, Harvard-Smithsonian Center for Astrophysics

**Optics Design Team Leads**

- Kiranmayee Kilaru, USRA/NASA MSFC
- Giovanni Pareschi, INAF-Brera Astronomical Observatory
- Paul Reid, Harvard-Smithsonian CfA
- William Zhang, NASA Goddard Space Flight Center

**_Lynx_ Instrument Working Group (IWG)**

The goal of the IWG is to support the STDT in defining the science instruments required for a compelling and executable mission. The IWG will help the STDT translate science goals into technical instrument requirements, provide the STDT information and metrics needed to make scientific tradeoff decisions, and support the STDT in assessing technology readiness and preparing technology development plans and roadmaps.

**IWG Chair**

- Mark Bautz, Massachusetts Institute of Technology

**IWG Co-Chairs**

- Simon Bandler, NASA/Goddard Space Flight Center, Microcalorimeters
- Abe Falcone, Penn State University, High-definition Imagers
- Enectali Figueroa-Feliciano, Northwestern University, Microcalorimeters
- Ralf Heilmann, Massachusetts Institute of Technology, Grating Spectrometers
- Ralph Kraft, Smithsonian Astrophysical Observatory, High-definition Imagers
- Randy McEntaffer, Penn State University, Grating Spectrometers

## H.2 Further Technical Details

This section provides further details of the engineering approach, analyses and structure for the *Lynx* concept as presented in this report.

### H.2.1 Ground Rules and Assumptions

The design analysis cycle ground rules and assumptions (GR&A) utilized by the MSFC Advanced Concepts Office (ACO) team for the *Lynx* X-ray Observatory (LXO) design presented in this report are summarized in **Table H-1**. The GR&A are organized by discipline. At the beginning of each design analysis cycle, the *Lynx* Study Office and STDT co-chairs work with ACO to identify the scope of the study, including mission-level trade studies and detailed design for critical aspects of the observatory. The team also identifies science, technology, and observatory configuration requirements. ACO utilizes NASA design standards and engineering judgment as further input to the design cycle GR&A. Once agreed upon by the *Lynx* team, the GR&A are set and the analysis cycle commences. As the design matures, science, technology, and observatory requirements are refined as necessary.

**Table H-1. Ground Rules and Assumptions.**

| 1.0 | Property | Value (as of 03/20/18) | Comments/Rationale |
|---|---|---|---|
| General | Mission | X-Ray Observatory | |
| | Approximate launch date | Early to mid 2030s | |
| | Destination | Sun-Earth L2 (SE-L2) | Orbit Trade Outcome (2017 Study) |
| | Mission duration | 5-year primary mission with 20 years of consumables and consider 20 years of degradation | |
| | Risk class | Modified Risk Class B with additional redundancy where necessary<br><br>Single fault tolerant on spacecraft, additional redundancy on instruments | |
| | Servicing interval | Assume minimal provisions for servicing | |
| | Configuration | Use heritage where appropriate with special consideration to *Chandra* | |
| | Launch environments | Launch Services Program (LSP) provided guidance for 2030s vehicle | To include SLS |
| | Operations | Use *Chandra* operational profile | See Guidance, Navigation, and Control (GN&C) section (5.0) for observing efficiency GR&A |
| | Dynamic environment | 1. No thruster firings occur during science<br>2. The only sources of vibration during science are reaction wheels and the LXM cryocooler<br>3. RWA's will be located and oriented very similarly to those on *Chandra*<br><br>Limit exported vibrations from GN&C components and the cryocooler, isolate vibration sensitive optical components as necessary, impose design requirements on structures to circumvent strong dynamic coupling, and potentially impose requirements on design (locations of components) to maximize optical performance. | |
| | | | |
| | Payload envelope | Per LSP guidelines for 2030s timeframe | To include SLS |
| | Mass allocation | Per LSP guidelines for 2030s timeframe | To include SLS |
| **2.0** | **Property** | **Value (as of 03/20/18)** | **Comments/Rationale** |
| Mission Analysis | Orbit | SE-L2 halo (same parameters as are currently planned for *James Webb Space Telescope* (*JWST*)) | |
| | Parking orbit vs direct ascent | Assume 185 km low-Earth orbit (LEO) parking orbit | |
| | Maximum eclipse period | Due to the orbital parameters for the SE-L2 halo, there will be no eclipses<br>During the outbound portion of the trajectory, once solar arrays are deployed, there should be no eclipses | |
| | Maximum time from launch to spacecraft separation | 121 minutes | See GN&C and Power GR&A for de-spin and SA deployment times, respectively |
| | Delta-V margin | Margin will be applied to each maneuver individually, with an appropriate value based on the data source (i.e., other missions such as *JWST*, previous studies, or ACO analysis)<br>A 5% Attitude Control System (ACS) tax will also be added to all maneuvers using the main propulsion system | |
| | Risk of human casualty/End of life (EOL) disposal | Risk of human casualty analysis is not required for this orbit<br>A small disposal maneuver will be included to ensure that the observatory does not re-enter the SE-L2 region post-disposal | |
| **3.0** | **Property** | **Value (as of 03/20/18)** | **Comments/Rationale** |
| Avionics—Command and Data Handling (C&DH) | Downlink frequency | 1–3 times/day; 1 hour each (Ka band) | *Chandra* downlink once every 8 hours, 60 minutes each |
| | Uplink frequency | 1–3 times/day; 1 hour each (X-band) | |
| | Uplink data rate | <1 Mbps | |
| | Total science data collection rate | 240 Gbits/day (2.78 Mbps) | Memo from Ralph Kraft, 2 March 2015; presentation "kraftHDXI_IndustryDay_May17_v1.1.pdf" of 22May 17; presentation "eckart*Lynx*_LXM_QuestionsAnswered_20171003_v1.pptx" by M. Eckart 3 Oct 2017. |
| | Total science memory storage | 48 hours of data (~500 Gbits) | |
| | Spacecraft bus baseline | Flight heritage design | Designed for long life in deep space environment, handles similar comm requirements |
| | Spacecraft bus storage | 1 Tbit, at 1.4 Gbps | Ref EADS Astrium Coreci mass memory unit; Computed as 2x data storage capacity |
| **4.0** | **Property** | **Value (as of 03/20/18)** | **Comments/Rationale** |
| Comm | SE-L2 baseline | Flight heritage design | Deep Space Network (DSN) link |
| | SE-L2 antenna configuration | Phased Array Antennas, no pointing mechanism | No pointing vibrations using PA antennas |
| | SE-L2 frequency bands | Ka-band for science, X-band for telemetry and navigation | TWTA high power required for SE-L2 distance |

| 5.0 | Property | Value (as of 03/20/18) | Comments/Rationale |
|---|---|---|---|
| GN&C | Observing efficiency | 85% (assume *Chandra* operations profile), goal of 90% | |
| | Observations | 1–20 targets per day; 1,000s–100,000s per observation | |
| | Momentum unloading calculation | Assume unloading when reaction wheels reach 85% of momentum capacity<br>Assume 70% of worst-case continuous solar pressure torque on observatory | |
| | Tip off rate | +/- 0.5 deg/sec (roll); +/- 1.5 deg/sec (pitch and yaw, each) | From Delta IV Heavy |
| | Time for de-tumble | 5 minutes | Conservative estimate based on de-spin burn time calculation for a Delta IV-Heavy upper stage |
| | **Pointing** | | |
| | Accuracy | 10 arcsecond (3 sigma) | Could be relaxed |
| | On board knowledge | 4 arcseconds | *Chandra* heritage |
| | Ground aspect knowledge | Post-facto ground aspect must be 1 arcsecond RMS absolute to sky<br>Image reconstruction to 0.2 arcseconds HPD diameter, within central 10 arcminutes radius | *Chandra* heritage |
| | Aspect reconstruction | Gyro requirements per *Chandra* heritage as initial input to design | *Chandra* heritage |
| | Star camera | Telemeters 8 images (e.g., 5 stars and 3 fid lights), every 2 seconds, with 0.5 arcseconds precision at S:N = 5<br>Readout and acquisition no worse than 4 minutes (average) | *Chandra* heritage |
| | Stability | +/- 1/6 arcseconds per sec, per axis (3 sigma) | Short-term stability is key<br>Allows exposure times as long as 1 sec with no impact on image reconstruction |
| | Dithering | Lissajous figure, up to +/- 30 arcseconds amplitude with 8 bits resolution<br>Periods 100 to 10,000 seconds subject to derived rate constraint<br>Arbitrary phase (8 bits: amplitude, rate, and phase are to be independently commanded in yaw and pitch) | Alternate dithering pattern may be a 2D linear raster scan or triangular one direction |
| | Roll angle restriction (during observations) | +/-15° | Minimizes solar impingement on thermal control radiators; minimizes off-nominal pointing for Solar Arrays. Solar arrays maintain sun-pointing during slew. |
| | Sun avoidance angle | 45° | From sunshade design |
| **6.0** | **Property** | **Value (as of 03/20/18)** | **Comments/Rationale** |
| Power | Power provision | Power System will store, generate, manage/condition, and distribute power to all subsystems and payloads on the vehicle | |
| | Checkout power | 126 Minutes prior to solar array deploy +30 min for Deployment = 156 min | Key driver for battery sizing. Survival mode is currently defined as an extreme safe mode that is entered in the event of a failure(s) that result |
| | | | in loss of SA pointing (including loss of control/tumbling) |
| | Survival power duration | 60 minutes | |
| | Bus voltage | 28 V Nominal | |
| | Power during initial checkout / solar array deployment | Power will be provided to all attached architecture elements during initial checkout (2.6hr) and solar array deployment per power schedule<br>Full power will remain available during final orbit insertion | |
| | Overload protection | Will be provided for all circuits | |
| | Solar irradiance (AM0) | 1,367 W per $m^2$ (at L2) | |
| | Ground reference | A common ground reference will be provided across all subsystems | |
| | Solar cell degradation (%/yr) | 3% | *Chandra* actuals are 2.2% per year |
| | Initial solar array de-rate | 0.9 | |
| | Secondary battery charge/discharge efficiency | 95% | |
| | Secondary battery max depth of discharge | 60% | Assumes 10 battery cycles |
| **7.0** | **Property** | **Value (as of 03/20/18)** | **Comments/Rationale** |
| Thermal | Spacecraft thermal control philosophy | Thermal control of the spacecraft shall utilize proven techniques<br>System to be designed with consideration of all phases of life cycle (ground testing and model validation, etc.)<br>Thermostats to be placed in proximity to the heaters they control | |
| | Optical bench temperature | 283K +/- 2 (mechanical interface temperature for optics) | All zones within Optical Bench Assembly (OBA) |
| | Spacecraft bus temperature | 283K max average | Best assembly temperature for the optics minus 10K |
| | Optics steady state temperature | 293K +/- 1 | Dependent on optics technology |
| | Optics survival temperature | TBD | Pending input from optics teams |
| | Optics maximum temperature | TBD | Pending input from optics teams |
| | Instrument survival temperature | TBD | Pending input from optics teams |
| | Instrument maximum temperature | TBD | Pending input from optics teams |
| | Optics heater power | 1,200W max | Pending review/memo from Mark Freeman |
| | Optics thermal gradient | TBD per mirror design | For optics developers—does not impact ACO analysis |
| | Vehicle orientation | Longitudinal axis not less than 45° from Sun with 5° exclusion tail to sun | |
| | Environmental heat loads | Solar flux at SE-L2: 1,367 W/m$^2$. | |
| | Science payload heat loads | Science payload (optics and instruments) thermally isolated to maximum extent possible from the spacecraft bus and OBA | All zones within OBA |
| **8.0** | **Property** | **Value (as of 03/20/18)** | **Comments/Rationale** |
| Structures | General | Primary structure will be designed to meet minimum strength requirements as stated in NASA-STD-5001B | |
| | Load cases | Telescope will be designed to withstand Delta IV Heavy launch loads (6g axial, 2g lateral) | Worst-case; Delta IV Heavy assumption is consistent with assumption made by Mission Analysis and GN&C. |
| | Components analyzed | Structural sizing to be performed for the following components: optical bench, spacecraft BUS, launch adapter, translation table | |
| | Factor of safety for composite materials | Ultimate factor of safety<br>FSu = 1.4 (Uniform Areas)<br>FSu = 2.0 (Areas with discontinuities) | Per NASA-STD-5001B. All environments affecting structure (temperature, moisture, time) should be considered<br>Lower factors of safety should be considered if dictated by requirements |
| | Factors of safety for optics technologies | Optics teams should meet internal design constraints and considerations<br>Note: may be more stringent than for spacecraft | For optics developers- does not impact current ACO analysis |
| | Factor of safety for metallic materials | FSu = 1.4<br>FSy = 1.25 | Per NASA-STD-5001B. |
| | Optical bench stiffness requirement | First Normal Mode of 30Hz or higher | Per Delta IV payload users guide |
| | *Lynx* stiffness requirements | First Constrained Mode >8 Hz Lateral, 15 Hz Axial | Per Delta IV payload users guide |
| | Secondary structures | Assume *Lynx* BUS secondary structures have a mass equal to 20% of the subsystem mass which attach to the BUS | |
| **9.0** | **Property** | **Value (as of 03/20/18)** | **Comments/Rationale** |
| Mechanisms – Translation Table | Instruments' focal plane location | Focal plane instruments translation plane perpendicular to optical axis | |
| | Grating arrays (GA) location | GA will be structurally mounted to the LMA (optics assembly) | |
| | GA mechanism fail safe | GA mechanism will fail safe in the "TBD" position | |
| | Grating contamination | GA needs will have contamination protection | |
| | TTA mechanism horizontal translation accuracy | 0.0100 inches | |
| | TTA mechanism translation stability | maintains position within 3 microns | |
| | TTA mechanism vertical translation distance | +/- 0.4 inches | |
| | X-Ray calorimeter instrumentation locations | All instruments (coolers, power, etc.) requiring to be less than 1 meter from Dewar assembly will reside on the translation table | |
| | Enclosure | Translation table and supporting instruments will be fully enclosed within an unpressurized enclosure to minimize external source interference | |
| | Launch locks | Used until telescope has decoupled from orbital insertion stage | |
| Inner Optics Door | On orbit service life | Single use on orbit | |
| | On ground service life | 20 cycles | |
| | Pressure | Optics enclosure does not need to hold pressure | |
| | Open/closed position | Opened door must reside within optical bench and outside of optical path | |
| | Door position monitoring | Secondary monitoring device will be used (*Chandra* heritage) | |
| Outer Optics Door | On orbit service life | Single use on orbit | |
| | On ground service life | 20 cycles | |
| | Open/closed position | Opened door must open beyond optical path and may serve as sunshade | |
| | Door position monitoring | Secondary monitoring device will be used (*Chandra* heritage) | |
| | Purge | Must support purge operations | |
| XGS Mechanisms | Operation range | Grating must be designed to regularly swing into and out of optical path depending on instrumentation used | |
| | Operational cycles | 10,000 cycles (5k on ground, 5k on orbit) | |
| | GA axis of operation | GA operating position will be perpendicular to optical axis | |
| | Accuracy and precision | TBD | |
| | Neighboring structure and mechanisms | Inner door will not obscure optical path during operation | |
| | Door position monitoring | Secondary monitoring device will be used (*Chandra* heritage) | |
| | Repeatability | TBD | |
| | Grating array size | 32,000 $cm^2$ (actual mirror area coverage) | Soft requirement; actual geometry TBD and related to resolving power science requirement |
| | Detector focus range | +/- 0.4 inches | |
| | Detector focus stability | Maintains position within 3 microns | |
| | Detector focus accuracy | 0.0100 inches | |
| **10.0** | **Property** | **Value (as of 03/20/18)** | **Comments/Rationale** |
| Propulsion | Main propulsion system | Responsible for orbit maneuvering, attitude control during the orbit maneuvering, on-orbit station keeping, and momentum unloading<br>May also be required to provide EOL disposal (TBD) | |
| | Residual propellant | 5% unusable propellant assumed | |
| | Initial mass assumed for sizing calculations | Predicted mass, with no propellant loaded | |
| | Propellant mass calculations | 15% tax assumed when calculating propellant mass from maneuver Delta-Vs | Note: Margin applied to calculated delta-vs is for the purpose of covering all potential maneuvers; Margin on propellant mass is separate and required to account for uncertainty in observatory mass and inefficiencies in the propulsion system. |
| **11.0** | **Property** | **Value (as of 03/20/18** | **Comments/Rationale** |
| Environment | Trapped radiation models (dose and single event effects) | AP8MIN and AE8MAX | AP8MIN (trapped protons, solar minimum) and AE8MAX (trapped electrons, solar maximum) are worst-case environments only needed for outbound trajectory radiation belt transits. Newer AE9/AP9 models may be used for environment definition document when needed |
| | Solar Particle Event model for total dose | ESP / PSYCHIC (same model as *JWST* specification paragraph 3.3.3.1) | 95% confidence for full mission length includes large events. |
| | Galactic Cosmic Rays and SPE single event effects model | CREME 97 (same model as *JWST* specification paragraph 3.3.3.3) | Single event effects environment |
| | Solar plasma for surface dose | *JWST* specification paragraph 3.3.3.4 | Low energy particle dose for surface coatings |
| | Meteoroid model | Meteoroid engineering model (MEM) | Hypervelocity impact risk (damage, attitude disturbances) |
| | SE-L2 plasma environment model | Plasma properties (temperatures, densities) | Probably not a design driver but available for completeness |
| | Solar ultraviolet radiation | Solar spectrum (*JWST* specification paragraph 3.4.6) | For external materials selection |
| | *JWST* specification is "Environmental Requirements *JWST* Observatory (SE-17) EV100074F", dated 4/16/09 | | |
| **12.0** | **Property** | **Value (as of 03/20/18)** | **Comments/Rationale** |
| Margin and MGA Philosophy | Mass contingency (spacecraft) | Use AIAA schedule at component level where applicable to take credit for heritage and calculate average for system | |
| | Mass contingency (instruments) | AIAA standard | |
| | Power contingency (spacecraft) | Use AIAA schedule at component level where applicable to take credit for heritage and calculate average for system | |
| | Power contingency (instruments) | AIAA standard | |

### H.2.2 SE Approach

This section will describe in more detail the SE approach and rationale for the *Lynx* concept study. It will provide figures of the SysML models developed for the ConOps content intended to drive cost, schedule, and requirements. This section is TBD for the Interim Report.

### H.2.3 Work Breakdown Structure

The *Lynx* X-ray Observatory (LXO) Work Breakdown Structure (WBS) was developed in accordance with guidance provided in the NASA WBS Handbook (NASA/SP-2016-3404). The WBS has been defined to level 3 for all elements, and level 5 for the X-ray telescope (XRT) and spacecraft element (SCE). The WBS, with summary key elements and definitions, follows. The WBS will support the mission cost and schedule allocations down to the element level, unambiguous cost reporting, and serve as the project organizational backbone. It also serves as the genesis for the Product Breakdown Structure (PBS), defines the end items to be developed, and serves as the overall organizational scheme for the cost model. The PBS is shown in **FO-3**. Furthermore, the WBS is included in the SysML model file structure and serves as the genesis for Block Definition Diagrams, Internal Block Diagrams, and the Observatory Requirements Tree. For more information on the SysML modeling activity, refer to Appendix **H.2.2.**

| Lynx X-ray Observatory Work Breakdown Structure | | |
|---|---|---|
| WBS Code | Level | WBS Elements |
| | **1** | **Lynx X-ray Observatory Project** |
| **.01** | **2** | **Project Management** |
| .01.01 | 3 | Project Management |
| .01.02 | 3 | Project Planning and Control |
| .01.03 | 3 | Configuration Management |
| **.02** | **2** | **Systems Engineering** |
| .02.01 | 3 | Systems Engineering Management |
| .02.02 | 3 | Requirements Development & Verification |
| .02.03 | 3 | System and Mission Analysis |
| **.03** | **2** | **Safety and Mission Assurance** |
| .03.01 | 3 | Safety and Mission Assurance Management |
| .03.02 | 3 | Mission Assurance |
| .03.03 | 3 | Systems Safety |
| .03.04 | 3 | Reliability |
| **.04** | **2** | **Science and Technology** |
| .04.01 | 3 | Science Operations |
| .04.02 | 3 | Science Data Analysis |
| .04.03 | 3 | Science Team and Support |
| .04.04 | 3 | Optics System Technology Development |
| .04.05 | 3 | X-Ray Gratings Spectrometer (XGS) Technology Development |
| .04.06 | 3 | X-Ray Microcalorimeter Spectrometer (XMIS) Technology Development |
| .04.07 | 3 | High Definition X-ray Imager (HDXI) Technology Development |
| **.05** | **2** | **X-ray Telescope** |
| .05.01 | 3 | X-ray Telescope Management |
| .05.02 | 3 | X-ray Telescope Systems Engineering |
| .05.03 | 3 | X-ray Telescope Integration and Test |
| .05.03.01 | 4 | X-ray Mirror Assembly (XMA) Assembly, Integration & Test |
| .05.03.02 | 4 | Lynx Mirror Assembly [XMA+ADA+TCS](LMA) Assembly, Integration & Test |
| .05.03.03 | 4 | Optical Bench Assembly (OBA) Assembly, Integration & Test |
| .05.03.04 | 4 | Integrated Science Instrument Module (ISIM) Assembly, Integration & Test |
| .05.03.05 | 4 | X-ray Telescope [LMA+OBA+ISIM] (XRT) Assembly, Integration & Test |
| .05.04 | 3 | Lynx Mirror Assembly (LMA) |
| .05.04.01 | 4 | LMA Integration & Test |
| .05.04.02 | 4 | LMA Systems Engineering |
| .05.04.03 | 4 | LMA Design and Fabrication |
| .05.04.04 | 4 | LMA Calibration |
| .05.04.05 | 4 | X-ray Mirror Assembly (XMA) |
| .05.04.05.01 | 5 | X-ray Mirror Assembly Optics |
| .05.04.05.02 | 5 | X-ray Mirror Assembly Structure |
| .05.04.06 | 4 | LMA Aft Door Assembly (ADA) |
| .05.04.06.01 | 5 | ADA Structural System |
| .05.04.06.02 | 5 | ADA Thermal Control System |
| .05.04.06.03 | 5 | ADA Power System |
| .05.04.06.04 | 5 | ADA Avionics System |
| .05.04.06.05 | 5 | ADA Motor Assembly |
| .05.04.07 | 4 | LMA Thermal Control System (TCS) |
| .05.04.07.01 | 5 | Pre-Collimator |
| .05.04.07.02 | 5 | Post-Collimator |
| .05.04.08 | 4 | LMA System Computer |

| Lynx X-ray Observatory Work Breakdown Structure | | |
|---|---|---|
| WBS Code | Level | WBS Elements |
| | **1** | **Lynx X-ray Observatory Project** |
| .05.05 | 3 | Optical Bench Assembly [MB+GA+IDA](OBA) |
| .05.05.01 | 4 | Magnetic Broom |
| .05.05.02 | 4 | Optical Bench Structural System |
| .05.05.02.01 | 5 | Optical Bench to TTA Mount |
| .05.05.03 | 4 | Optical Bench Thermal Control System |
| .05.05.03.01 | 5 | Optical Bench Heaters |
| .05.05.03.02 | 5 | Optical Bench MLI |
| .05.05.04 | 4 | Optical Bench Power System |
| .05.05.05 | 4 | Optical Bench Avionics System |
| .05.06 | 3 | Integrated Science Instrument Module (ISIM) |
| .05.06.01 | 4 | ISIM Integration & Test |
| .05.06.02 | 4 | ISIM Systems Engineering |
| .05.06.03 | 4 | ISIM Structural System |
| .05.06.04 | 4 | ISIM Thermal Control System |
| .05.06.04.01 | 5 | ISIM MLI |
| .05.06.05 | 4 | ISIM Electronics System |
| .05.06.06 | 4 | Translation Table Assembly (TTA) |
| .05.06.06.01 | 5 | TTA Mounting Plate |
| .05.06.06.02 | 5 | TTA Thermal Control System |
| .05.06.07 | 4 | TTA Mechanisms |
| .05.06.07.01 | 5 | Horizontal TTA mechanisms |
| .05.06.07.02 | 5 | Vertical TTA Mechanisms |
| .05.06.07.03 | 5 | Filter Wheel Assembly Motors |
| .05.07 | 3 | Science Instruments |
| .05.07.01 | 4 | Lynx X-ray Microcalorimeter (LXM) Instrument |
| .05.07.01.01 | 5 | LXM Filter Wheel Assembly |
| .05.07.01.02 | 5 | LXM Calibration Source Assembly |
| .05.07.01.03 | 5 | LXM Dewar Assembly |
| .05.07.01.04 | 5 | LXM Electronics System |
| .05.07.01.05 | 5 | LXM Harnesses |
| .05.07.01.06 | 5 | LXM Thermal Control System (heat pipes) |
| .05.07.01.07 | 5 | LXM Miscellaneous Hardware |
| .05.07.02 | 4 | High Definition X-ray Imager (HDXI) Instrument |
| .05.07.02.01 | 5 | HDXI Detector Assembly (DA) |
| .05.07.02.02 | 5 | HDXI Detector Electronics Unit (DEU) |
| .05.07.03 | 4 | X-ray Gratings Spectrometer (XGS) Instrument |
| .05.07.03.01 | 5 | XGS Detector Assembly (DA) |
| .05.07.03.02 | 5 | XGS Detector Electronics Unit (DEU) |
| .05.07.03.03 | 5 | XGS Grating Array (GA) |
| .05.08 | 3 | Reserved |
| .05.09 | 3 | Reserved |
| .05.10 | 3 | Calibration Facility (placeholder for upgrades as necessary) |
| **.06** | **2** | **Spacecraft Element (SCE)** |
| .06.01 | 3 | Spacecraft Management |
| .06.02 | 3 | Spacecraft Systems Engineering |
| .06.03 | 3 | Spacecraft Element (SCE) Assembly, Integration & Test |

| Lynx X-ray Observatory Work Breakdown Structure | | |
|---|---|---|
| WBS Code | Level | WBS Elements |
| | **1** | **Lynx X-ray Observatory Project** |
| .06.04 | 3 | Structural System |
| .06.04.01 | 4 | Spacecraft Bus |
| .06.04.02 | 4 | Secondary Structures |
| .06.04.03 | 4 | Spacecraft Adapter |
| .06.04.04 | 4 | LMA Forward Door Assembly (FDA) |
| .06.04.04.01 | 5 | FDA Structural System |
| .06.04.04.02 | 5 | FDA Thermal Control System |
| .06.04.04.03 | 5 | FDA Power System |
| .06.04.04.04 | 5 | FDA Avionics System |
| .06.04.04.05 | 5 | FDA Motors |
| .06.05 | 3 | Thermal Control System |
| .06.05.01 | 4 | Heaters, SCE Bus |
| .06.05.02 | 4 | Multilayer Insulation (MLI), SCE Bus |
| .06.05.03 | 4 | Radiator SCE Bus |
| .06.05.04 | 4 | MLI, Propulsion Tanks |
| .06.05.05 | 4 | Radiator, LXM Cryocooler Compressor |
| .06.05.06 | 4 | Radiator, LXM Electronics 1 |
| .06.05.07 | 4 | Radiator, LXM Electronics 2 |
| .06.05.08 | 4 | Radiator, HDXI Detector Assembly |
| .06.05.09 | 4 | Radiator, HDXI DEU |
| .06.05.10 | 4 | Radiator, XGS Detector Assembly |
| .06.05.11 | 4 | Radiator, XGS DEU |
| .06.06 | 3 | Electrical Power System (EPS) |
| .06.06.01 | 4 | Solar Array Wing (with Boom) |
| .06.06.02 | 4 | Solar Array Drive Actuator |
| .06.06.03 | 4 | Integrated Power Electronics |
| .06.06.04 | 4 | Secondary Batteries |
| .06.06.05 | 4 | Cabling |
| .06.07 | 3 | Avionics System |
| .06.07.01 | 4 | Command and Data Handling (C&DH) System |
| .06.07.01.01 | 5 | Flight Computer |
| .06.07.01.02 | 5 | Solid State Recorder |
| .06.07.01.03 | 5 | Data Acquisition Unit |
| .06.07.01.04 | 5 | RCS Controller |
| .06.07.01.05 | 5 | RW Controller |
| .06.07.01.06 | 5 | Instrumentation |
| .06.07.01.07 | 5 | Avionics Cabling |
| .06.07.02 | 4 | Communications |
| .06.07.02.01 | 5 | Ka Phased Array Antenna |
| .06.07.02.02 | 5 | X Transponder |
| .06.07.02.03 | 5 | Ka Transceiver |
| .06.07.02.04 | 5 | Ka Diplexer |
| .06.07.02.05 | 5 | X-Band TWTA |
| .06.07.02.06 | 5 | X-TWT Amp |
| .06.07.02.07 | 5 | Ka-Band TWT |
| .06.07.02.08 | 5 | Ka-Band TWT Amp |
| .06.07.02.09 | 5 | Waveguides |
| .06.07.02.10 | 5 | RF Combiner |

| Lynx X-ray Observatory Work Breakdown Structure | | |
|---|---|---|
| WBS Code | Level | WBS Elements |
| | **1** | **Lynx X-ray Observatory Project** |
| .06.07.02.11 | 5 | RF Switch 1-2 |
| .06.07.02.12 | 5 | RF Switch 1-3 |
| .06.07.02.13 | 5 | X-Band Conical Patch Antenna |
| .06.07.02.14 | 5 | X-Band MGA Array |
| .06.07.02.15 | 5 | Coax Cabling, Misc |
| .06.08 | 3 | Guidance, Navigation and Control (GN&C) System |
| .06.08.01 | 4 | Coarse Sun Sensor |
| .06.08.02 | 4 | Aspect Camera Assembly |
| .06.08.03 | 4 | Ultra Fine Sun Sensor |
| .06.08.04 | 4 | Intertial Measurement Unit |
| .06.08.05 | 4 | Reaction Wheel |
| .06.08.05.01 | 5 | Reaction Wheel Drive Elecctronics |
| .06.08.06 | 4 | Reaction Wheel Isolation System |
| .06.08.07 | 4 | Fiducial Light Assembly |
| .06.08.07.01 | 5 | Fiducial Light Controller Assembly |
| .06.08.08 | 4 | Periscope |
| .06.08.09 | 4 | Control Electronics Assembly |
| .06.09 | 3 | Propulsion System |
| .06.09.01 | 4 | Main Propulsion System (MPS) Engine |
| .06.09.02 | 4 | Reaction Control System (RCS)/Attitude Control System (ACS) Engine |
| .06.09.03 | 4 | Propellant Tank |
| .06.09.03.01 | 5 | Monoprop Propellant (N2H4) |
| .06.09.03.02 | 5 | Monoprop Pressurant (GN2) |
| .06.09.03.03 | 5 | Redidual Propellant (N2H4) |
| .06.09.04 | 4 | Feed System Components |
| .06.09.04.01 | 5 | Service Valve |
| .06.09.04.02 | 5 | Latch Valve |
| .06.09.04.03 | 5 | Flow Control Orifice |
| .06.09.04.04 | 5 | Filter |
| .06.09.04.05 | 5 | Pressure Transducer |
| .06.09.05 | 4 | Miscellaneous Mounts |
| .06.10 | 3 | Flight Software (FSW) |
| **.07** | **2** | **Mission Operations** |
| .07.01 | 3 | Mission Ops Management |
| .07.02 | 3 | Mission Operations |
| **.08** | **2** | **Launch Vehicle Services** |
| .08.01 | 3 | Launch Vehicle Management |
| .08.02 | 3 | Launch Vehicle Systems Engineering |
| .08.03 | 3 | Launch Vehicle Integration and Test |
| **.09** | **2** | **Ground Systems** |
| .09.01 | 3 | Mission Operations Center (MOC) |
| .09.02 | 3 | Communications/Network Infrastructure |
| **.10** | **2** | **Systems Integration and Test** |
| .10.01 | 3 | I&T Management |
| .10.02 | 3 | Lynx Observatory [XRT+SCE] Assembly, Integration & Test |
| .10.03 | 3 | Observatory AI&T Support Equipment |
| .10.04 | 3 | Observatory Ground Support Equipment (GSE) |
| **.11** | **2** | **Education and Public Outreach** |

### H.2.4 Assembly, Integration and Test

The system-level Assembly, Integration and Test (AI&T) and calibration activities are performed to verify system and project-level requirements during the build-up of the *Lynx* X-ray Observatory (LXO). The tests, along with other standard requirement verification methods, will be defined in detail in the *Lynx* Verification and Validation Plan and specific test planning documents. The LXO AI&T flow is modeled after that developed for the *Chandra* flight systems. The overall flow is provided in **FO-3**. This sequence of activities is indicated in the project schedule (Figure G-2) and is the critical path through launch.

The AI&T flow assumes that all sub-assemblies are delivered for integration into the next higher assembly fully tested and qualified to the extent required, and availability of necessary ground support equipment (GSE) to support testing. To reduce system risk and protect fight hardware life, vibration and acoustic environmental testing is limited on the X-ray Mirror Assembly (XMA), *Lynx* Mirror Assembly (LMA), and Integrated Science Instrument Module (ISIM) assemblies. The XMA consists of the X-ray optics and support structure. The LMA consists of the XMA plus the pre-collimator and aft door assembly. The ISIM includes the three scientific instruments and translation table assembly. These assemblies undergo environmental testing prior to calibration, and at the observatory assembly level.

The first major activity in the AI&T flow is ground calibration of the XMA and instruments to verify functional performance of the integrated assemblies. The XMA (**Figure FO3-1**)is evaluated with known sources and detectors first to calibrate mirror performance and then calibrated with the individual instruments already integrated into the ISIM (**Figure FO3-2**) to calibrate the system. For this calibration phase, the GA is mounted using test GSE to the XMA, and the XGS readout is integrated with the ISIM to calibrate the entire XGS instrument. Following this testing, the XMA is integrated with the GA, thermal control system (TCS), and Aft Door Assembly (ADA) to become the LMA

The next major phase of AI&T includes the integration of the OBA with the LMA. As the LMA was required for instrument calibration, completion of this integration activity is contingent upon LMA availability. Testing of the OBA/LMA configuration includes alignment, and pre- and post-functional testing. No environmental testing is planned for this configuration. **Figure FO3-3** shows this integration configuration.

The assembled OBA/LMA configuration is then integrated with the ISIM to create the XRT assembly (**Figure FO3-4**). This assembly is aligned, undergoes pre- and post-functional testing, and is integrated with the spacecraft **(Figure FO3-5)** by the prime contractor. Activities up to shipping the XRT are conducted during Phase C of the project life cycle. Integration of the XRT with the spacecraft is the subject of the mission-level System Integration Review (SIR).

The assembled XRT and spacecraft configurations create the LXO (**Figure FO3-6**). AI&T of the LXO is the responsibility of the prime contractor. The LXO undergoes vibration, acoustic, and cryovacuum environmental testing, electromagnetic interference/electromagnetic compatibility (EMI/EMC) testing, and pre- and post-functional testing to conclude the requirement verification and certification of the LXO prior to shipment to the launch vehicle provider for payload integration, checkout, and launch (**Figure FO3-7**).

# Systems Engineering and AI&T

XMA — 1 — Calibration
ISIM — 2 — GA
→ Calibration Configuration — Calibration

XMA, GA, TCS, ADA → LMA — Alignment, Pre/post functional, Vibe/acoustic, EMI/EMC, Cryo-vac

GA, Mag Broom, OBA Structure → OBA — Alignment, Pre/post functional

OBA, LMA → OBA/LMA Config — 3 — Alignment, Pre/post functional

OBA/LMA Config, ISIM → XRT — 4 — Alignment, Pre/post functional

XRT, SCE — 5 → LXO — 6 — Vibe/acoustic, EMI/EMC, Cryo-vac, Pre/post functional

LXO, LV → Integrated Vehicle — 7 — Electrical checks

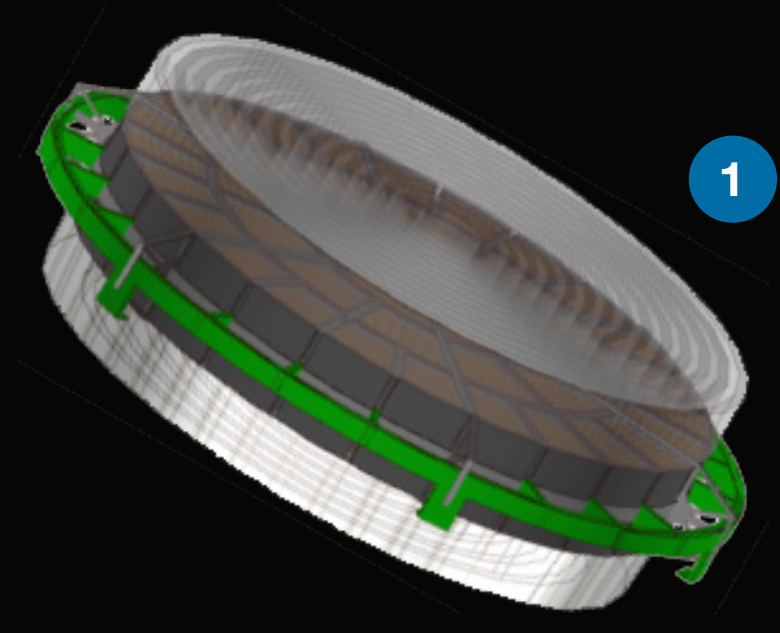

**Figure F03-1.** X-ray Mirror Assembly

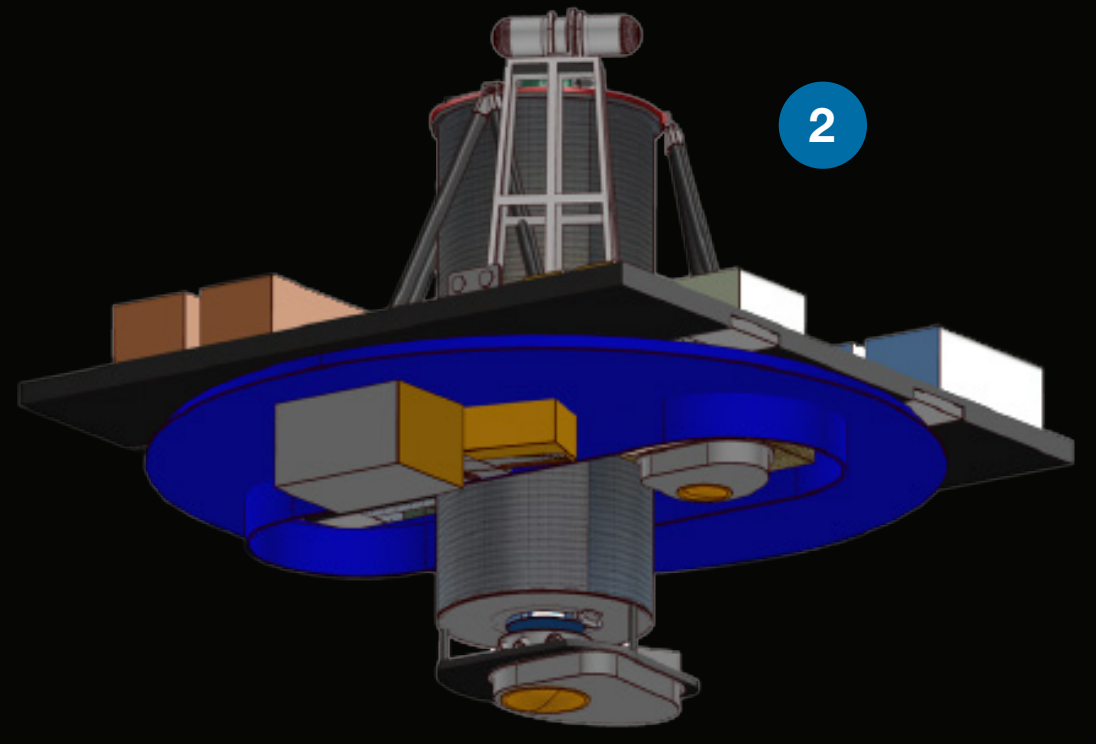

**Figure F03-2.** Integrated Science Instrument Module

**Table F03-1**

| AI&T Assumptions |
|---|
| Testing shown approximates Chandra process |
| All subassemblies delivered to next integration step fully qualified and accepted |
| XMA calibrated followed by calibration with each instrument on the ISIM |
| XRCF modified for LXO calibration |
| LMA and ISIM likely driving start of AI&T process |

**Table F03-2**

| Acronyms | |
|---|---|
| GA | Grating Array |
| HDXI | High-Definition X-ray Imager |
| ISIM | Integrated Science Inst. Module |
| LMA | *Lynx* Mirror Assembly |
| LV | Launch Vehicle |
| LXM | *Lynx* X-ray Microcalorimeter |
| LXO | *Lynx* X-ray Observatory |
| OBA | Optical Bench Assy |
| SCE | Spacecraft Element |
| TCS | Thermal Control System |
| TTA | Translation Table Assy |
| XGS | X-ray Grating Spectrometer |
| XMA | X-ray Mirror Assembly |
| XRT | X-ray Telescope |

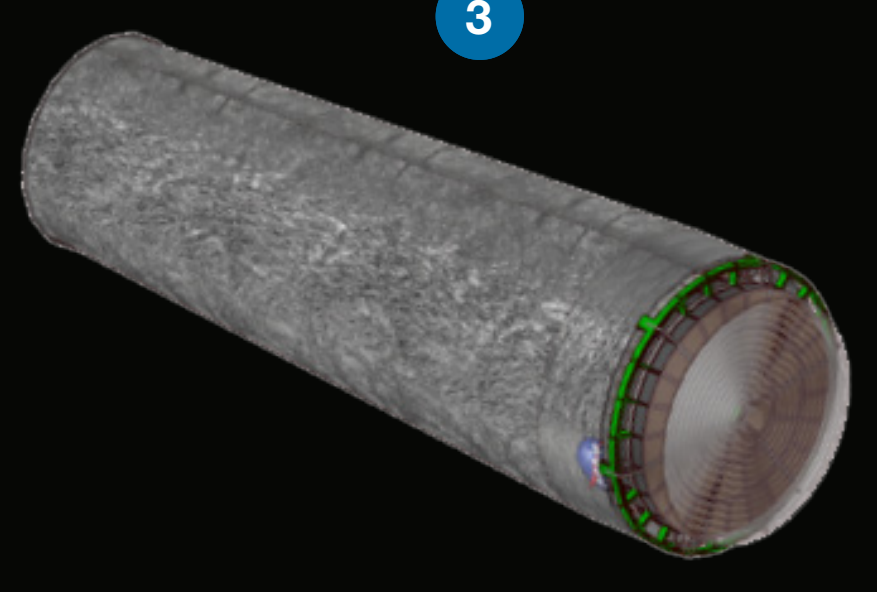

**Figure F03-3.** Optical Bench Assembly/*Lynx* Mirror Assembly

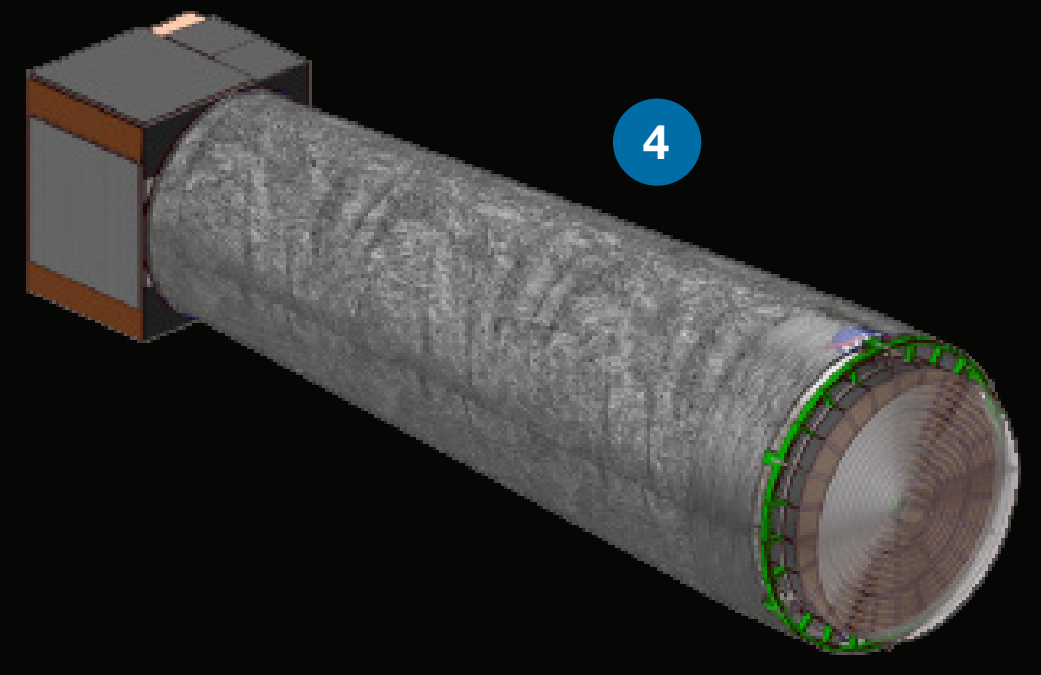

**Figure F03-4.** X-ray Telescope

***Lynx* X-ray Observatory**

- 05 XRT
  - 05.04 LMA
  - 05.05 OBA
  - 05.06 ISIM
  - 05.07 Science Instruments
    - 05.07.01 LXM Instrument
    - 05.07.02 HDXI Instrument
    - 05.07.03 XGS Instrument
  - 05.10 Calibration Facility
- 06 SCE
  - 06.04 Structural System
  - 06.05 Thermal Control System
  - 06.06 Electrical Power System
  - 06.07 Avionics System
  - 06.08 GN&C System
  - 06.09 Propulsion System
  - 06.10 Fight Software
- 09 Ground Systems
  - 09.01 Mission Ops Center
  - 09.02 Communications Network Infrastructure
- 10 Systems I&T
  - 10.02 LXO AI&T
  - 10.03 Observatory GSE

**Figure F03-8.** *Lynx* X-ray Observatory Product Breakdown Structure

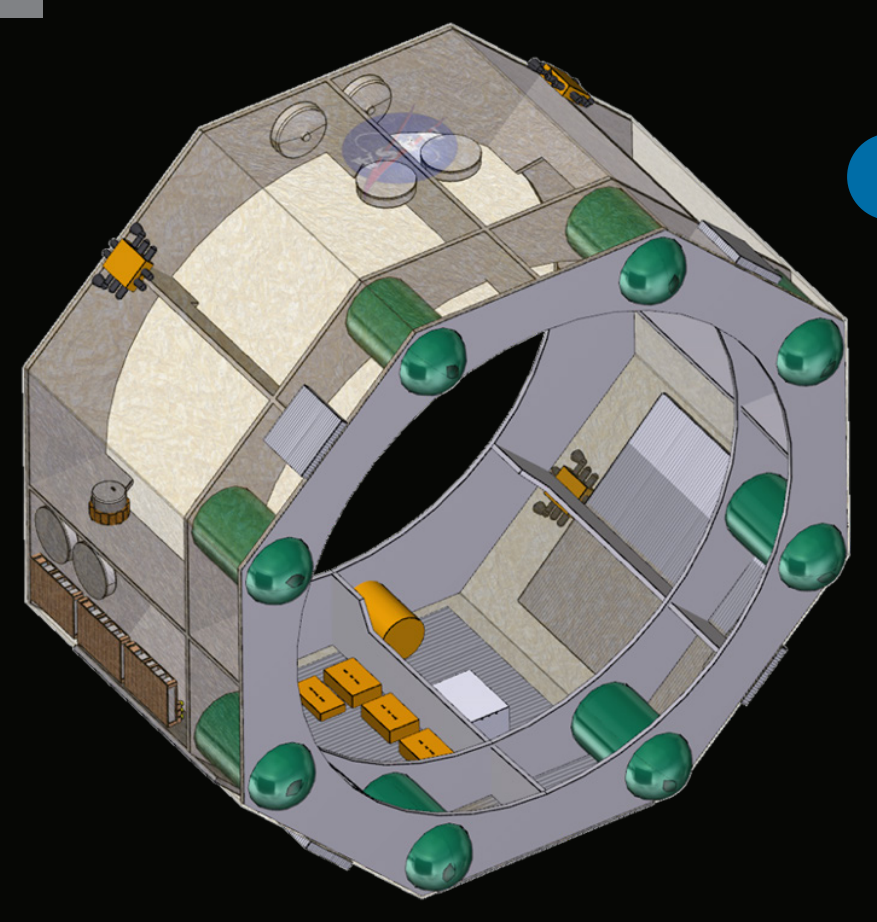

**Figure F03-5.** Spacecraft Element

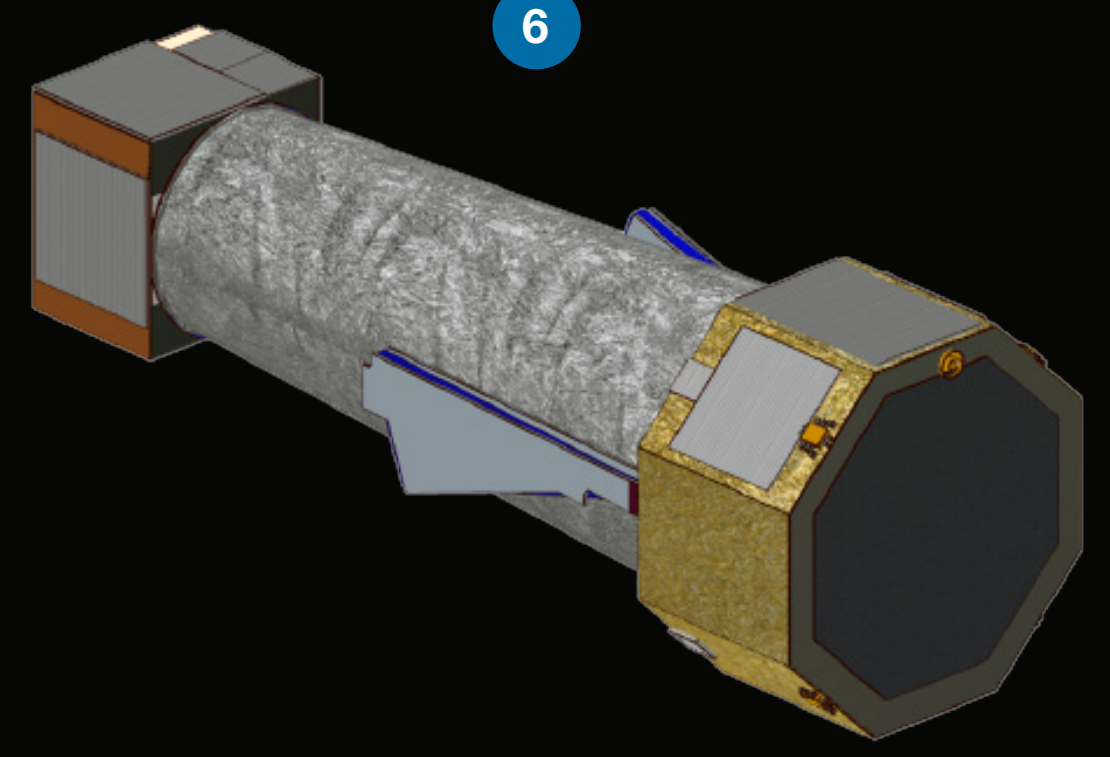

**Figure F03-6.** *Lynx* X-ray Observatory

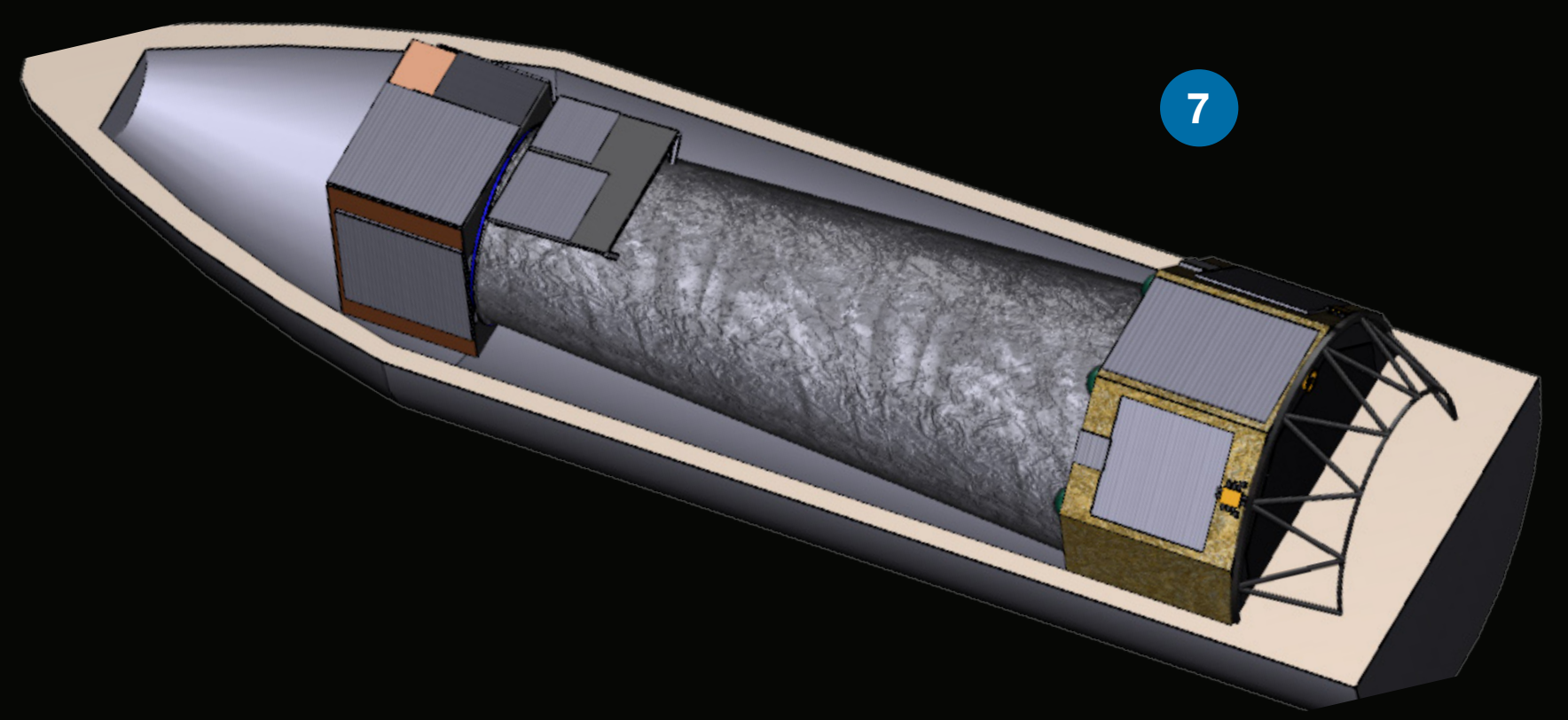

**Figure F03-7.** *Lynx* X-ray Observatory Integrated into Vehicle

### H.2.5 *Lynx* Trade Studies

The *Lynx* mission concept development depends upon objective comparisons with respect to performance, cost, schedule, and risk of identified realistic alternatives to achieving the *Lynx* scientific goals. These trade studies are an integral part of the decision-making activities carried out by the *Lynx* team.

This section outlines the trade studies undertaken by the *Lynx* team to achieve the most acceptable technical solutions among viable alternatives. The purpose is to document what trades were undertaken and their outcomes to enable traceability. The trade studies are organized into mission design, payload, and spacecraft trades that may range from formal trades (e.g., using the Kepner-Tregoe decision-analysis method to identify a mirror architecture for the *Lynx* DRM) to informal decisions made by the STDT through open discussion and to internal engineering team and study office evaluations or assessments of best practices, heritage, and similar considerations. Listed in the following are the primary/secondary trade category (science, payload, technology, cost, or schedule), the disposition (or status) of the trade study, references to specific location in the Interim Report where the trade is mentioned, and a brief summary of the trade framework.

#### H.2.5.1 Mission-Level Trades

**Configuration**

Category: payload

Disposition: *Chandra*-like with spacecraft bus forward encircling LMA

M4 Reference: none

Summary: Grazing-incidence X-ray telescope payloads with 2 reflections will be somewhat longer than their focal lengths with mass distributed primarily at both ends of the optical bench. Thermal management favors location of the spacecraft bus at the optics end of the optical bench. Therefore, like *Chandra, Lynx* design places the spacecraft bus surrounding the LMA. Trade was performed by the *Lynx* engineering team with disposition based on structural launch and operational thermal considerations.

**Orbit**

Category: science/payload

Disposition: Sun-Earth L2

M4 Reference: E.3.2 Launch to Orbit

Summary: The *Lynx* orbit trade study considered observing efficiency, radiation environment, launch vehicle and Delta-V requirements, thermal, and communication factors. The trade was performed by the *Lynx* engineering team and considered multiple orbits.

**Launch Vehicle**

Category: cost/payload

Disposition: ongoing

M4 Reference: E.2.1.6 OBA; E.3.1 Launch Vehicle

Summary: While details of the launch vehicle fleet available at the time of *Lynx* anticipated launch is highly uncertain, total observatory mass and geometry vs. anticipated launch vehicle mass and fairing size capabilities are central to mission cost and payload architecture. A trade study assessing intermediate, heavy, and SLS co-manifested options is being performed by the engineering team in accordance with NASA Launch Services Program recommendations.

### H.2.5.2 Optics

**Mirror designs: 3-m OD x 10-m f.l. vs 6x20 vs 3x5, etc.**
Category: science/payload
Disposition: 3 m OD x 10 m focal length only design being considered
M4 Reference: none
Summary: This trade was carried out by the STDT over a series of discussions along with numerous mock observations simulated to determine strengths of several combinations of mirror outer diameter and focal length. Additional performance properties were presented by and on behalf of the *Lynx* Optics Working Group.

**LMA 3x10 architecture (LMAT)**
Category: technology/schedule
Disposition: completed
M4 Reference: E.2.1.1 XMA
Summary: A detailed formal study using the Kepner-Tregoe decision-analysis method has been chartered and is underway to assess the 3 technologies currently under development in order to recommend one as the *Lynx* mirror architecture to the STDT to serve as the Reference Design for the *Lynx* mission concept for Milestone M6. Study is being performed by a team of experts recruited from government, academia, and industry including international participation.

### H.2.5.3 Other Optics Trades

**XMA high-E effective area**
Category: science
Disposition: ongoing
M4 Reference: E.2.1.3 LXM
Summary: The possibility of enhancing the science capabilities of *Lynx* by extending the performance range to higher X-ray energies is under discussion by the STDT.

**XMA fabrication schedule**
Category: schedule/cost
Disposition: ongoing
M4 Reference: G.2 Risks and Risk Mitigation (Risk 3)
Summary: The manufacturability and production of the mirror components has been recognized as a risk and, depending on the outcome of the LMAT study (**§H.2.5.2**), a study to identify areas to reduce the overall development schedule and cost for this portion of the project will be performed by the *Lynx* study office.

### H.2.5.4 Science Instruments—LXM

**Focal plane arrays**
Category: science/payload
Disposition: 3-pixel array types span the scientific needs of *Lynx*
M4 Reference: none
Summary: Trade of various focal plane arrays and sub-arrays needed to carry out *Lynx* science objectives was discussed by STDT including at face-to-face meeting in January 2018 where it was decided that three of the original five array types will cover the most important *Lynx* science while simplifying the demands on the instrument. Trade conducted by Instrument Working Group in conjunction with STDT.

**Count rate vs readout**
Category: science
Disposition: ongoing
M4 Reference: E.2.1.3
Summary: Trade under discussion by STDT concerning observations of bright sources (including extended sources) with LXM and consequent readout speeds required especially when using focal plane arrays with hydras.

**Cryocooler**
Category: payload/cost
Disposition: ongoing
M4 Reference: E.2.1.3
Summary: Numerous high-TRL cryocooler technologies are available for consideration. Study is being conducted by the *Lynx* study office in conjunction with industry developers and Instrument Working Group experts.

**Optical/infrared blocking filter**
Category: science
Disposition: ongoing
M4 Reference: E.2.1.3
Summary: Filters that block unwanted optical and IR light yet are thin enough to minimize soft X-ray attenuation are under study by the *Lynx* engineering team. Numerous high-TRL heritage is available for consideration although *Lynx* LXM has a large FOV and stringent low-E throughput (science) requirements that go beyond most previous mission's needs.

### H.2.5.5 Science Instruments—HDXI

**Focal plane FOV**
Category: science/cost
Disposition: 22 arcminute x 23 arcminute FOV
M4 Reference: E.2.1.3
Summary: The *Lynx* mirror assembly PSF degrades slowly with off-axis distance but certain scientific enhancements may be possible if the HDXI detector field-of-view extends beyond that needed to meet the *Lynx* grasp requirement for sub-arcsecond resolution. This trade was made by the STDT in consultation with the *Lynx* Instrument Working Group and their recommendations.

**Architecture**
Category: technology
Disposition: ongoing
M4 Reference: F.3.1
Summary: There are three technologies identified by the *Lynx* Instrument Working Group members for the HDXI that require similar resources from the spacecraft and that have similar development paths. The *Lynx* Instrument Working Group will document the strengths and weaknesses of each technology for consideration by the STDT.

**XGS Gratings Architectures**
Category: science/technology

Disposition: ongoing
M4 Reference: F.3.1
Summary: There are two technologies identified by the *Lynx* Instrument Working Group members for the X-ray grating spectrometer architecture. The IWG will document the strengths and weaknesses of each technology for consideration by the STDT.

#### H.2.5.6 Spacecraft

**General thermal model/OBA thermal control**
Category: payload
Disposition: ongoing
M4 Reference: E.2.2.4 Thermal/ E.2.1.6 OBA
Summary: Development of a detailed integrated thermal model of the observatory including the thermal control of the LMA and OBA is being performed by a *Lynx* Study Office-industry partnership and will be completed by the Spring/Summer 2018. A separate OBA trade comparing passive and active thermal control options concluded the use of heat pipes and multi-layer insulation alone (a purely passive control system) could not achieve the thermal environment necessary for the *Lynx* payload.

**Communications Trade**
Category: payload
Disposition: Ka-band for science data download and X-band for command and engineering up- and downlink
M4 Reference: none
Summary: A trade study performed by the *Lynx* engineering team, with guidance from NASA/HQ experts on future DSN communications capabilities, was concluded and reported in April 2017.

**Fixed vs. Extendable Bench, Bench Construction**
Category: payload
Disposition: ongoing
M4 Reference: E.2.1.6 OBA
Summary: Depending upon launch vehicle fairing constraints, the 10-meter focal length *Lynx* payload may best be accommodated using a one-time extendable, rather than fixed-length, OBA. Consideration of an extendable bench may facilitate launch vehicle flexibility with modest impacts on mass and cost. This study is under preliminary assessment by the *Lynx* Study Office in conjunction with the engineering team and industry partners.

**Focal Plane (ISIM) Translation Table vs. Movable Mirror Assembly**
Category: payload
Disposition: only ISIM translation table being considered
M4 Reference: none
Summary: The *Athena* concept study deemed a movable mirror assembly, as opposed to a movable instrument platform, was needed from a structural viewpoint. An early appraisal of the distribution of mass within the *Lynx* observatory indicated, similarly to *Chandra*, that this conclusion did not apply for *Lynx* with its sub-arcsecond angular resolution as well as the presence of the XGS instrument. A translating table affixed to the ISIM is baselined for further study by the *Lynx* engineering team and associated instrument studies.

## H.3 Acronyms

| | |
|---|---|
| AA | Associate Administrator |
| ACO | Advanced Concepts Office |
| ACS | Attitude Control System |
| ACT | Atacama Cosmology Telescope |
| ADA | Aft Door Assembly |
| ADR | Adiabatic Demagnetization Refrigerator |
| AGN | Active Galactic Nucleus |
| AI&T | Assembly Integration and Test |
| AIAA | American Institute of Aeronautics and Astronautics |
| Al | Aluminum |
| APD | Astrophysics Projects Division |
| APS | Active Pixel Sensor |
| ASCA | Advanced Satellite for Cosmology and Astrophysics |
| ASIC | Application-Specific Integrated Circuit |
| BBXRT | Broadband X-ray Telescope |
| BHMF | Black Hole Mass Function |
| C | Carbon |
| C&DH | Command and Data Handling |
| CADRe | Cost Analysis and Data Requirements |
| CAN | Cooperative Agreement Notification |
| CAP | Command Action Procedure |
| CAT | Critical Angle Transmission |
| CAT-XGS | Critical Angle Transmission Gratings |
| CC | Core Collapse |
| CCD | Charge-Coupled Device |
| CCO | Central Compact Object |
| CDR | Critical Design Review |
| CE | Chief Engineer |
| CER | Cost Estimating Relationship |
| CGM | Circumgalactic Medium |
| CME | Coronal Mass Ejection |
| CMOS | Complementary Metal-Oxide Semiconductor |
| CTO | Chandra-type Orbit |
| DD | Double-Degenerate |
| DDT&E | Design, Development, Test, and Evaluation |
| DEU | Detector Electronics Unit |
| DRM | Design Reference Mission |
| DSN | Deep Space Network |
| EM | Electromagnetic |
| EMC | Electromagnetic Compatibility |
| EMI | Electromagnetic Interference |
| EOL | End-of-Life |
| EPB | Event Processing Board |
| ERP | Event Recognition Processor |
| ESA | European Space Agency |
| EW | Equivalent Width |
| Fe | Iron |
| FOM | Figure of Merit |
| FOT | Flight Operations Team |
| FOV | Field of View |
| FPGA | Field Programmable Gate Array |
| FPA | Focal Plane Assembly |
| FSW | Flight Software |
| FWHM | Full Width at Half Maximum |
| FY | Fiscal Year |
| GA | Grating Array |
| GMC | Giant Molecular Cloud |
| GN&C | Guidance, Navigation, and Control |
| GO | General Observer |
| GOT | Ground Operations Team |
| GPR | Goddard Procedural Requirements |
| GR&A | Ground Rules and Assumptions |
| GSE | Ground Support Equipment |
| GSFC | Goddard Space Flight Center |
| GW | Gravitational Wave |
| HDXI | High Definition X-ray Imager |
| HEMT | High-Electron Mobility Transistor |
| HERA | Hydrogen Epoch of Reionization Array |
| HETG | High-Energy Transmission Grating |
| HOD | Halo Occupation Distribution |
| HPD | Half-Power Diameter |
| HQ | Headquarters |
| HR | Hertzsprung-Russell |
| I&T | Integration and Test |
| IGM | Intergalactic Medium |
| INAF | Instituto Nazionale Di Astrofisica |
| IR | Infrared |
| IRU | Inertial Reference Unit |
| Ir | Iridium |
| ISIM | Integrated Science Instrument Module |
| ISM | Interstellar Medium |
| IXO | International X-ray Observatory |
| JPL | Jet Propulsion Laboratory |
| JWST | James Webb Space Telescope |
| KDP | Key Decision Point |
| KSC | Kennedy Space Center |
| LDRO | Lunar Distant Retrograde Orbit |
| LEO | Low-Earth Orbit |
| LIGO | Laser Interferometer Gravitational-Wave Observatory |
| LISA | Laser Interferometer Space Antenna |
| LMA | *Lynx* Mirror Assembly |
| LMC | Large Magellanic Cloud |
| LSC | *Lynx* Science Center |
| LSP | Launch Services Program |
| LSST | Large Synoptic Survey Telescope |
| LV | Launch Vehicle |
| LXM | *Lynx* X-ray Microcalorimeter |
| LXO | *Lynx* X-ray Observatory |
| MAC | Molecular Absorber Coating |
| MCR | Mission Concept Review |
| MEM | Meteoroid Engineering Model |
| MGA | Mass Growth Allowance |
| MIRI | Mid-Infrared Instrument |
| MIT | Massachusetts Institute of Technology |
| MLI | Multilayer Insulation |
| MPE | Max Planck Institute for Extraterrestrial Physics |
| MPS | Main Propulsion System |
| MSFC | Marshall Space Flight Center |
| MW | Milky Way |
| MXS | Modulated X-ray Source |
| Ni | Nickel |
| NIR | Near-Infrared |
| NPR | NASA Procedural Requirement |
| NS | Neutron Star |
| OAB | Astronomical Observatory of Brera |
| OBA | Optical Bench Assembly |
| OBC | Onboard Computer |
| OBF | Optical Blocking Filter |
| OP | Off-Plane |
| OP-XGS | Off-Plane Reflection Gratings |
| ORR | Operational Readiness Review |
| OWG | Optics Working Group |
| PBS | Product Breakdown Structure |
| PCAD | Pointing Control and Aspect Determination |
| PCEC | Project Cost Estimating Capability |
| PCOS | Physics of the Cosmos |
| PDR | Preliminary Design Review |
| PS | Project Scientist |
| PSF | Point Spread Function |
| PSU | Pennsylvania State University |
| Pt | Platinum |
| PZT | Lead Zirconate Titanate |
| QE | Quantum Efficiency |
| RCS | Reaction Control System |
| REDSTAR | Resource Data Analysis and Retrieval |
| RMS | Root Mean Square |
| ROSAT | Roentgen Satellite |
| ROSES | Research Opportunities in Space and Earth Sciences |
| S&MA | Safety and Mission Assurance |
| SAO | Smithsonian Astrophysical Observatory |
| SAT | Strategic Astrophysics Technology |
| SCAN | Space Communications and Navigation |
| SCE | Spacecraft Element |
| SD | Single-Degenerate |
| SDSS | Sloan Digital Sky Surveys |
| SE | Systems Engineering |
| SE&I | Systems Engineering and Integration |
| SE-L2 | Sun-Earth L2 |
| SFR | Star Formation Rate |
| SI | Science Instrument |
| Si | Silicon |
| SIR | System Integration Review |
| SKA | Square Kilometer Array |
| SLS | Space Launch System |
| SMBH | Supermassive Black Hole |
| SMC | Small Magellanic Cloud |
| SMD | Science Mission Directorate |
| SN | Supernova |
| SNe | Supernovae |
| SNR | Supernova Remnants |
| SOT | Science Operations Team |
| SOTA | State of the Art |
| SQUID | Superconducting Quantum Interference Device |
| SRB | Standing Review Board |
| SRI | Sarnoff Research Institute |
| SRR | Systems Requirement Review |
| SSS | Shell Supporting Structure |
| STDT | Science and Technology Division Team |
| STM | Science Traceability Matrix |
| SWG | Science Working Group |
| SZ | Sunyaev-Zeldovich |
| TBD | To Be Determined |
| TBR | To Be Resolved |
| TCS | Thermal Control System |
| TES | Transition-Edge Sensor |
| TESS | Transiting Exoplanet Survey Satellite |
| TGCAT | Transmission Grating Data Archive and Catalog |
| ToO | Target of Opportunity |
| TRL | Technology Readiness Level |
| TTA | Translation Table Assembly |
| TT&C | Telemetry, Tracking, and Command |
| TTI | Transfer Trajectory Insertion |
| UAH | University of Alabama in Huntsville |
| UFO | Ultra-Fast Outflow |
| ULX | Ultraluminous X-ray |
| USAF | United States Air Force |
| UV | Ultraviolet |
| W-I | Wolter Type I |
| W-S | Wolter-Schwarzchild |
| WBS | Work Breakdown Structure |
| WFIRST | Wide-Field Infrared Survey Telescope |
| WSS | Wolter-Schwarzchild-Saha |
| XIFU | X-ray Integral Field Unit |
| XARM | X-ray Astronomy Recovery Mission |
| XGS | X-ray Grating Spectrometer |
| XIS | X-ray Imaging Spectrometer |
| XLF | X-ray Luminosity Function |
| XMA | X-ray Mirror Assembly |
| XMM | X-ray Multi-Mirror |
| XRB | X-ray Binaries |
| XRCF | X-ray Cryogenic Facility |
| XRT | X-ray Telescope |
| XUV | X-ray and Extreme Ultraviolet |
| ZnO | Zinc Oxide |

## H.4 References